\documentclass[PAPER, atlasdraft=false, UKenglish, numpmcorr=true, cernpreprint, texmf, orcidlogo]{atlasdoc}
\usepackage{atlaspackage}
\usepackage{atlasbiblatex}

\usepackage[misc=true]{atlasphysics}
\graphicspath{{logos/}{figures/}}

\usepackage{ANA-HIGP-2025-16-PAPER-defs}

\usepackage{bm}
\usepackage{rotating}
\usepackage{multirow}
\usepackage{dcolumn}
\usepackage{pifont}
\usepackage{ifthen}
\usepackage{tikz}
\usepackage{siunitx}

\AtlasTitle{Combination
of Higgs boson measurements
at $\sqrt{s} =$ 13 \TeV
and their interpretations
by the ATLAS experiment}

\AtlasAbstract{%
This paper presents combined measurements of Higgs boson production and decay using up to 140 fb$^{-1}$ of proton--proton collision data collected by the ATLAS experiment at the LHC at $\sqrt{s} = 13$ \TeV.
The inclusive Higgs boson event rate relative to its Standard Model prediction is determined to be $0.990 ^{+0.053}_{-0.051} = 0.990 \pm 0.027 \text{ (stat.) } \pm 0.024 \text{ (exp.) } ^{+0.037}_{-0.035} \text{ (sig. theo.) } ^{+0.016}_{-0.015} \text{ (bkg. theo.)}$, separating statistical, experimental, signal-theory and background-theory uncertainty components.
The signal theory uncertainty is the largest contribution, dominated by missing higher-order terms in QCD and the modelling of parton shower effects.
Inclusive production cross-sections and decay branching ratios are measured, with sensitivities improved by up to 35\% relative to the previous ATLAS publication.
Production cross-sections for individual decay channels are also reported.
The results are interpreted in terms of modifiers to Higgs boson couplings. The combination of single Higgs boson and Higgs boson pair production measurements yields $\kappa_\lambda = 1.3^{+3.1}_{-1.6}$ for the modifier to the Higgs boson self-coupling, with an interval of $[-1.5, 6.5]$ at 95\% confidence level.
The partial width in the $H\to b\bar{b}$ decay channel is used to obtain the first ATLAS determination of the running mass of the $b$-quark at the Higgs boson mass scale in the modified minimal subtraction scheme, found to be $m_b(m_H) = 2.73^{+0.27}_{-0.25}~\GeV$.
The total width of the Higgs boson is measured to be $\Gamma_H = 3.6^{+2.1}_{-1.6}~\MeV$ based on its on- and off-shell production rates. The kinematics of Higgs boson production are probed within the framework of Simplified Template Cross-sections in 44 kinematic regions, and the results are interpreted in the context of Standard Model Effective Field Theory models.
}

\AtlasRefCode{HIGP-2025-19}

\PreprintIdNumber{CERN-EP-2026-228}

\AtlasJournal{Rep. Prog. Phys.}
\AtlasCoverEgroupAnalysisTeam{atlas-higp-2025-19-analysis-team@cern.ch}

\hypersetup{pdftitle={ATLAS document},pdfauthor={The ATLAS Collaboration}}

\begin{document}

\maketitle

\tableofcontents
\clearpage

\section{Introduction}
\label{sec:introduction}


One of the main physics objectives of Run 2 of the LHC (2015-2018) is to measure with increasing precision the properties of the Higgs boson discovered by the ATLAS and CMS Collaborations in 2012~\cite{HIGG-2012-27,CMS-HIG-12-028}.  Measurements of these properties, which have so far been found to be compatible with Standard Model (SM) predictions, provide sensitive tests of the consistency of the SM and a window into possible new phenomena beyond it.

\begin{figure}[tbp]
\centering
\includegraphics[width=1.0\textwidth]{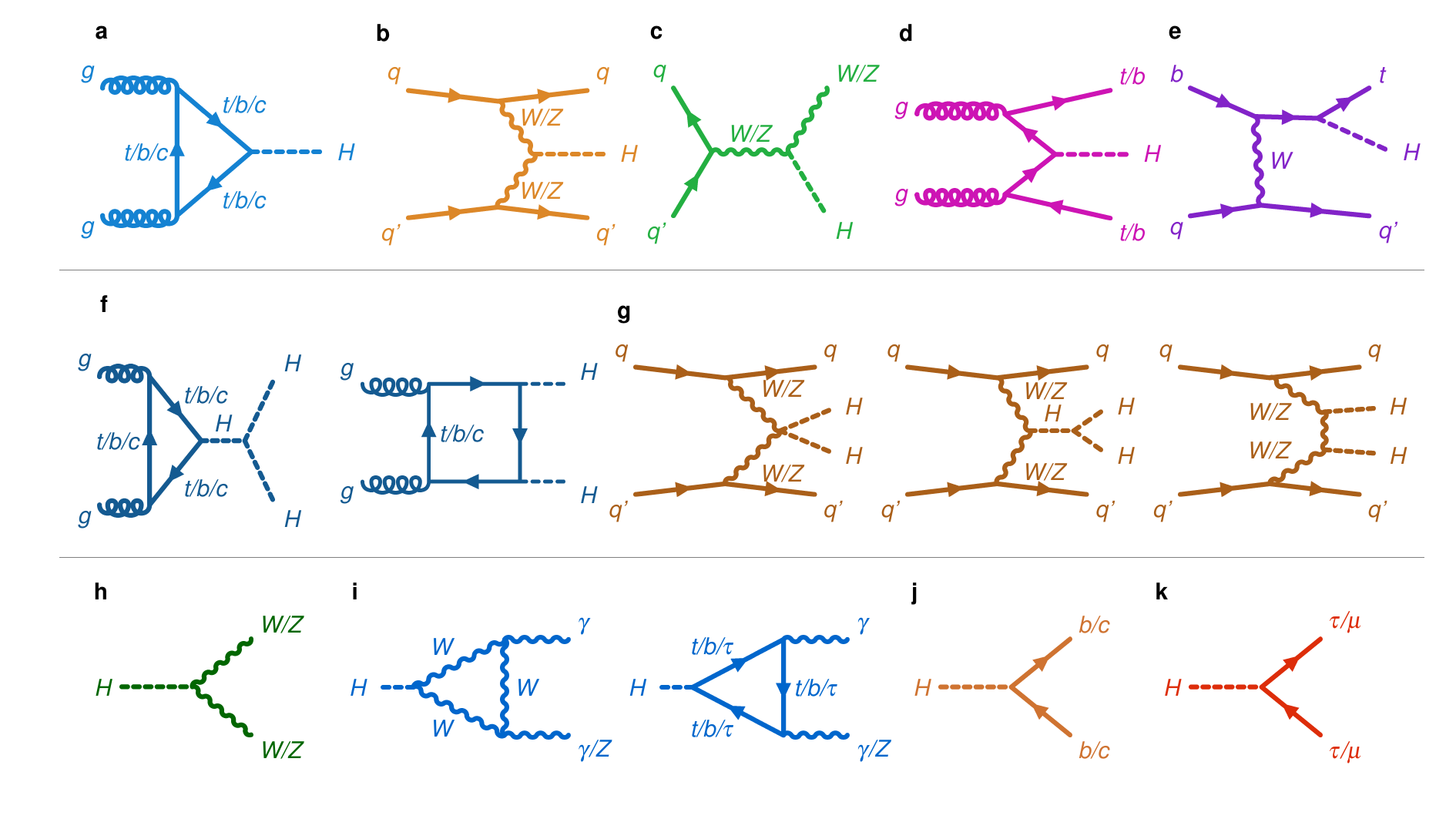}
\caption{
Examples of Feynman diagrams for single (top row) and double (middle row) Higgs boson production, and Higgs boson decay (bottom row):
Higgs boson production via gluon--gluon fusion (\ggF; a),
$W$ or $Z$ vector-boson fusion (\VBF; b), and associated production with $W$ or $Z$ bosons (\VH; c),
top- or $b$-quark pairs (\ttH, \bbH; d), or a single top quark (\tH; e);
Higgs boson pair production via gluon--gluon fusion (f), and
Higgs boson pair production via vector-boson fusion (g).
Higgs boson decays into a pair of $W$ or $Z$ bosons (h), a pair of photons, or a $Z$ boson and a photon (i), a pair of quarks (j), and a pair of charged leptons (k).
}
\label{fig:feynman_graphs}
\end{figure}

Within the SM, the couplings of the Higgs boson to fundamental SM particles and the rates of its production and decay processes are fully determined by the knowledge of SM parameters. ATLAS~\cite{HIGG-2021-23} and CMS~\cite{CMS-HIG-21-018} have performed measurements of these properties using analyses targeting a variety of Higgs boson production and decay channels. The production processes considered consist of gluon--gluon fusion (\ggF), $W$ or $Z$ vector-boson fusion (\VBF), associated production with a $W$ or $Z$ boson (\VH, with $V = W,Z$), associated production with a pair of top quarks (\ttH), a pair of $b$-quarks (\bbH) or a single top quark (\tH), and Higgs boson pair production ($HH$). The latter provides direct sensitivity to the Higgs boson trilinear self-coupling $\lambda_{HHH}$. The decay channels include \Hbb, \Hww, \Htt, \Hcc, \Hzz, \Hyy, \Hzy, and \Hmm.\footnote{No distinction is made between particles and anti-particles, and the same notation is used to refer to both. The symbol $\ell$ refers to electrons and muons.} These processes are illustrated in Figure~\ref{fig:feynman_graphs}.

This paper studies the production and decay rates of the Higgs boson, using a combination of analyses targeting individual production and decay processes. Compared to the previous publication of Ref.~\cite{HIGG-2021-23}, nine measurements are updated and several new measurements are included. The changes lead to significant improvements, in particular to the measurements of the \Htt, \Hww, \VHbbcc\ and \ttH\ processes.

All analyses use the full dataset of proton--proton ($pp$) collisions collected during Run 2, except for those of Refs.~\cite{HIGG-2019-04} and~\cite{HDBS-2019-29} that use triggers that were not available during part of the running period. The results assume the value $m_H = \qty{125.09 \pm 0.24}{\GeV}$ for the Higgs boson mass, obtained from the latest combination of ATLAS and CMS Higgs-boson mass measurements~\cite{HIGG-2014-14} and compatible with more recent measurements performed individually by ATLAS~\cite{HIGG-2022-20,HIGG-2020-07} and CMS~\cite{CMS-HIG-19-004,CMS-HIG-21-019}. More details on the analyses included in the combination are provided in Section~\ref{sec:combination}.

Measurements of Higgs boson processes generally target Higgs boson production event rates in specific decay channels. These measurements are either inclusive or differential with respect to kinematics associated with the Higgs boson production as established by the Simplified Template Cross-section (STXS) framework~\cite{YR4,Andersen:2016qtm,Berger:2019wnu,Amoroso:2020lgh}. Combining the production measurements across multiple decay channels as performed here provides a more comprehensive determination of Higgs boson properties from which further model-independent interpretations can be performed.

The measurements are compared with state-of-the-art theory predictions of the rates and kinematics of these processes within the SM. The precision of these predictions was significantly improved in recent years, in particular due to computations involving higher-order corrections in QCD and electroweak effects, which have pushed the boundaries of achievable precision at hadron colliders~\cite{YR4,
Nason:2004rx,Frixione:2007vw,Alioli:2010xd,Alioli:2008tz,
Hamilton:2013fea,Hamilton:2015nsa,Catani:2007vq,Hamilton:2012rf,
Anastasiou:2015ema,Anastasiou:2016cez,
Actis:2008ug,Anastasiou:2008tj,Grazzini:2013mca,Nason:2009ai,
Hamilton:2012np,
Ciccolini:2007jr,Ciccolini:2007ec,Bolzoni:2010xr,
Brein:2003wg,Denner:2011id,Altenkamp:2012sx,
Alwall:2014hca,Beenakker:2002nc,Dawson:2003zu,Yu:2014cka,Frixione:2014qaa,Frixione:2015zaa,Demartin:2015uha,
Dawson_1998,Borowka:2016ehy,Baglio_2019,de_Florian_2013,Shao_2013,deflorian2015higgs,Grazzini_2018,Baglio_2021,
Frederix_2014,Liu-Sheng:2014gxa,Dreyer:2018qbw,Dreyer:2018rfu}. The resulting theory uncertainties are generally comparable to or smaller than the corresponding experimental uncertainties, but form the dominant uncertainty in several important cases, in particular in measurements of inclusive event rates.

Results are reported in terms of measurements of the inclusive rates of the production and decay processes mentioned above, presented in Sections~\ref{sec:global_mu} and~\ref{sec:inclusive}, and interpreted as modifications to the Higgs boson couplings to other SM particles within the $\kappa$ framework~\cite{YR3} in Section~\ref{sec:kappa}. The latter also includes the first ATLAS determination of $m_b(m_H)$, the running mass of the $b$-quark at the Higgs boson mass scale, a measurement of the total Higgs boson width $\Gamma_H$, and a determination of the Higgs boson self-coupling. Measurements of Higgs boson production kinematics in the STXS framework~\cite{YR4,Andersen:2016qtm,Berger:2019wnu,Amoroso:2020lgh} are reported in Section~\ref{sec:stxs} and an interpretation in the context of Standard Model Effective Field Theory (SMEFT)~\cite{Brivio:2017vri,ATL-PHYS-PUB-2019-042} models is presented in Section~\ref{sec:eft}.

\FloatBarrier


\section{ATLAS detector}
\label{sec:atlas}

The ATLAS experiment~\cite{PERF-2007-01} at the LHC is a multipurpose particle detector with a forward--backward symmetric, cylindrical geometry and a near \(4\pi\) coverage in solid angle. The detector records digitized signals produced by the products of proton bunch collisions at the LHC, hereafter termed collision `events'. It is designed to identify a wide variety of particles and measure their momenta and energies. These particles include electrons, muons and $\tau$-leptons, photons, as well as gluons and quarks that produce collimated jets of particles in the detector. Since the jets from $b$-quarks and $c$-quarks contain hadrons with relatively long lifetimes, they can be identified by observing decay vertices that typically occur at a measurable distance from the collision point. The presence of particles that do not interact with the detector, such as neutrinos, can be inferred by summing the vector momenta of the visible particles in the plane transverse to the beam and imposing conservation of transverse momenta.

The detector components  closest to the collision point measure  charged-particle
trajectories and momenta. This inner spectrometer is surrounded by a thin superconducting solenoid
providing a \qty{2}{\tesla} axial magnetic field and by calorimeters that are
used in the identification of particles and in the measurement of their energies. The
calorimeters are in turn surrounded by an outer spectrometer
based on three large superconducting air-core toroidal magnets with eight coils each.
The field integral of the toroids ranges between \num{2.0} and \qty{6.0}{\tesla\metre}
across most of the detector.
The muon spectrometer includes a system of precision tracking chambers
dedicated to measuring the trajectories and
momenta of muons, the only charged particles to travel through the calorimeters. A two-level
trigger system was optimised for Run~2 data-taking~\cite{TRIG-2019-04} to select events of
interest at a rate of about \SI{1.25}{\kHz} from the proton bunch collisions occurring at a
rate of 40 MHz.
An extensive software suite~\cite{ATL-SOFT-PUB-2021-001} is used in the simulation, reconstruction and analysis of real and simulated data, in detector operations, and in the trigger and data acquisition systems of the experiment. The results presented in this paper use $pp$ collision data at $\sqrt{s} = \SI{13}{\TeV}$ collected during Run~2 (2015--2018), corresponding to an integrated luminosity of up to \SI{140}{\per\femto\barn}~\cite{DAPR-2021-01}.


\section{Input measurements and combination procedure}
\label{sec:combination}
\vspace{-5pt}

\subsection{Analyses included in the combination}
\label{sec:input_channels}

The measurements entering the combination are listed in Table \ref{tab:inputs}. A total of 26 measurements are included in the combination, each focusing on a specific set of Higgs boson production and decay processes. Some analyses are only used for a subset of the results shown in this paper. Measurements of off-shell Higgs boson production~\cite{HIGP-2024-14,HIGP-2024-05} and of Higgs boson decay into invisible final states~\cite{EXOT-2020-11,HIGG-2018-26,SUSY-2019-12,EXOT-2021-17} are only used in the Higgs boson coupling modifier measurements presented in Section~\ref{sec:kappa}, while the measurement of Higgs boson production at high transverse momentum in the \Hbb\ decay channel~\cite{HIGG-2021-08} is only used in the STXS measurements (Section~\ref{sec:stxs}) and their effective-field theory interpretations (Section~\ref{sec:eft}). The measurements targeting the \HZy~\cite{HIGG-2018-42} and \Hmm~\cite{HIGG-2019-14} decays and \VH\ production in the \Htt\ decay~\cite{HIGG-2018-20} are only used in the inclusive measurements of production and decay rates (Section~\ref{sec:inclusive}) and coupling modifier measurements  (Section~\ref{sec:kappa}).

All analyses use the full set of $pp$ collision data collected by ATLAS during Run 2. The integrated luminosity of this dataset is \SI{140}{fb^{-1}}, except for three analyses for which the trigger algorithms used to collect the data were only available for part of Run 2: the measurements of \VBF\ production in the \Hbb\ channel~\cite{HIGG-2019-04} and of Higgs boson pair production in the \HHbbbb\ channel~\cite{HDBS-2019-29}, which both use a dataset with an integrated luminosity of \SI{126}{fb^{-1}}, and the measurement of \Hbb\ production at high \ptH~\cite{HIGG-2021-08}, which uses a dataset of \SI{136}{fb^{-1}}. Compared to the previous ATLAS publication on combined Higgs boson measurements~\cite{HIGG-2021-23}, several channels are added and some of the existing channels are augmented with a larger dataset or an improved re-analysis of the dataset.

The analyses included in the STXS measurements are those which provide a modelling of the Higgs boson signal processes with separate event rates for each kinematic region of the STXS binning described in Section~\ref{sec:stxs}, which is based on the Stage 1.2 scheme~\cite{YR4,Andersen:2016qtm,Berger:2019wnu,Amoroso:2020lgh}. In the remaining analyses, the Higgs boson signal rate is modelled with a granularity corresponding to the production and decay modes described in Section~\ref{sec:introduction}.

In some cases these input analyses are modified compared to their individual publications. Analysis results using the previous determination of \qty{139}{fb^{-1}} for the integrated luminosity of the Run 2 dataset are modified to reflect the updated value of \qty{140}{fb^{-1}} and its associated relative uncertainty of 0.83\%~\cite{DAPR-2021-01}.  In analyses included in the STXS combination, electroweak corrections computed using HAWK~\cite{Denner:2011id} are applied to the \VBF\ and quark-initiated \VH\ production as a function of the variables defining the Stage 1.2 regions for these processes: the transverse momentum of the Higgs boson, the invariant mass of the two leading jets, and the transverse momentum of the $W$ or $Z$ boson in the case of \VH.

The signal-sensitive regions of all input analyses are non-overlapping by construction, except for the \HsZZ\ analysis of Ref.~\cite{HIGP-2024-14} and the
\ZH, \Hinv\ analysis of Ref.~\cite{HIGG-2018-26} where about 40\% of the events in the latter are also selected by the former. For this reason, only one of these analyses is used in each of the Higgs boson coupling modifier measurements presented in Section~\ref{sec:kappa}.
In some cases, the selections of the control regions used to constrain background processes overlap with signal or control regions in other analyses. The effect of these overlaps is however found to be small compared to uncertainties of the results and is not considered in the statistical treatment.

\begin{table}[tbp]
\centering
\caption{Input analyses to the combination, with in each case the targeted Higgs boson decay and production processes, the integrated luminosity of the dataset used ($\mathcal{L}$) and the reference to the original publication. Analyses targeting Higgs boson production at large transverse momentum are denoted by \emph{high \ptH}, and $V \to \text{leptons}$ refers to $W$ or $Z$ boson decays into leptons. Analyses initially reporting results corresponding to a Run 2 integrated luminosity of \qty{139}{fb^{-1}} are rescaled to the updated \qty{140}{fb^{-1}} value for the combination. In the sixth column, \emph{New} denotes analyses not present in the combination of Ref.~\cite{HIGG-2021-23}; \emph{Full Run 2} refers to analyses that used a partial Run 2 dataset and are updated to use the full dataset; and \emph{Updated} to cases where an improved analysis of the full Run 2 dataset is used. The last three columns indicate the classes of measurement that include each analysis: the column labelled Incl. indicates the analysis used inclusive measurements of Higgs boson production and decay of Sections~\ref{sec:global_mu} and~\ref{sec:proddecay}, the column labelled $\kappa$ those used in the coupling modifier measurements of Section~\ref{sec:kappa} and the column labelled STXS those used in the measurements of Sections~\ref{sec:stxs} and~\ref{sec:eft} making use of the STXS framework.}
\vspace{3pt}
\resizebox{\textwidth}{!}{%
\begin{tabular}{llclclccc}%
\toprule
Analysis & Prod.          & $\mathcal{L}$  & Ref.      & Comparison                    & \multicolumn{3}{c}{Measurements} \\
& modes          & [\si{fb^{-1}}] &           & with Ref.~\cite{HIGG-2021-23} & Incl. & $\kappa$ & STXS \\
\midrule
\Hllll               & All                                   & 140 & \cite{HIGG-2018-28} & Unchanged  & \checkmark & \checkmark & \checkmark \\
\Hlvlv               & \ggF, \VBF                            & 140 & \cite{HIGP-2024-07} & Updated    & \checkmark & \checkmark & \checkmark \\
\Hlvlv, $\ell\nu jj$ & \VH                                   & 140 & \cite{HIGG-2023-09} & Full Run 2 & \checkmark & \checkmark & \checkmark \\
\Hyy                 & All                                   & 140 & \cite{HIGG-2020-16} & Unchanged  & \checkmark & \checkmark & \checkmark \\
\HZy                 & All                                   & 140 & \cite{HIGG-2018-42} & Unchanged  & \checkmark & \checkmark &            \\
\Htt                 & All, except $(V \to \text{leptons})H$ & 140 & \cite{HIGG-2022-07} & Updated    & \checkmark & \checkmark & \checkmark \\
\Htt                 & $(V \to \text{leptons})H$             & 140 & \cite{HIGG-2018-20} & New        & \checkmark & \checkmark &            \\
\Hmm                 & All                                   & 140 & \cite{HIGG-2019-14} & Unchanged  & \checkmark & \checkmark &            \\
\Hbb                 & \VBF                                  & 126 & \cite{HIGG-2019-04} & Unchanged  & \checkmark & \checkmark & \checkmark \\
\Hbbcc               & $(V \to \text{leptons})H$             & 140 & \cite{HIGG-2020-20} & Updated    & \checkmark & \checkmark & \checkmark \\
\Hbb\ (high \ptH)    & All                                   & 136 & \cite{HIGG-2021-08} & Unchanged  &            &            & \checkmark \\
\HML                 & \ttH                                  & 140 & \cite{HIGP-2024-08} & Full Run 2 & \checkmark & \checkmark & \checkmark \\
\Hbb                 & \ttH                                  & 140 & \cite{HIGG-2020-24} & Updated    & \checkmark & \checkmark & \checkmark \\
\Hbb                 & \tH                                   & 140 & \cite{HIGP-2024-03} & Updated    & \checkmark & \checkmark & \checkmark \\
\Hinv                & \VBF                                  & 140 & \cite{EXOT-2020-11} & Unchanged  &            & \checkmark &            \\
\Hinv                & $\VBF+\gamma$                         & 140 & \cite{EXOT-2021-17} & New        &            & \checkmark &            \\
\Hinv                & $(V \to \text{leptons})H$             & 140 & \cite{HIGG-2018-26} & Unchanged  &            & \checkmark &            \\
\Hinv                & \ttH                                  & 140 & \cite{SUSY-2019-12} & New        &            & \checkmark &            \\
\HsZZ                & \ggF, \VBF                            & 140 & \cite{HIGP-2024-14} & Updated    &            & \checkmark &            \\
\HsWW                & \ggF, \VBF                            & 140 & \cite{HIGP-2024-05} & New        &            & \checkmark &            \\
\midrule
\HHbbbb              & \ggF, \VBF                            & 126 & \cite{HDBS-2019-29} & New        &            & \checkmark &            \\
\HHbbbb\ (high \ptH) & \VBF                                  & 140 & \cite{HDBS-2022-02} & New        &            & \checkmark &            \\
\HHbbtt              & \ggF, \VBF                            & 140 & \cite{HDBS-2019-27} & New        &            & \checkmark &            \\
\HHbbyy              & \ggF, \VBF                            & 140 & \cite{HDBS-2021-10} & New        &            & \checkmark &            \\
\HHbbll              & \ggF, \VBF                            & 140 & \cite{HDBS-2019-02} & New        &            & \checkmark &            \\
\HHML                & \ggF, \VBF                            & 140 & \cite{HDBS-2019-04} & New        &            & \checkmark &            \\
\bottomrule
\end{tabular}}
\label{tab:inputs}
\end{table}

%

\subsection{Treatment of systematic uncertainties}
\label{sec:uncertainties}

Uncertainties on the Higgs boson production cross-sections are applied to the SM prediction in each region of the STXS Stage 1.2 scheme, as described in Ref.~\cite{HIGG-2021-23}, and uncertainties in Higgs boson branching ratios are applied as prescribed in Ref.~\cite{YR4}. Theory uncertainties from identical sources are correlated across analyses, in particular uncertainties from missing higher-order terms and parton distribution function (PDF) uncertainties for analyses using the same PDF set. The uncertainty in the $\alpha_s$ parameter is correlated across the PDF and branching ratio schemes.

Experimental uncertainties associated with luminosity and pile-up modelling are correlated across the input analyses.
Trigger efficiency uncertainties are included in all analyses and are correlated across analyses where the same trigger algorithms and configurations are used.
Uncertainties related to the reconstruction, identification, and calibration of photons, electrons, muons, $\tau$-leptons, jets, and missing transverse momentum are also correlated when compatible reconstruction and calibration schemes are used; otherwise, they are left uncorrelated.
Hadronic jet definitions and energy calibration schemes vary across analyses, but the uncertainties in the jet energy scale and resolution are generally described using a common set of uncertainty sources across analyses.
These uncertainties are therefore correlated in most cases, and the same applies also to uncertainties in the missing transverse momentum. Uncertainties on the mass scale and resolution of large-radius jets are included in analyses of Higgs boson production at high transverse momentum making use of boosted topologies, and are treated as uncorrelated with the small-radius jet uncertainties.
Uncertainties on the flavour tagging of hadronic jets are generally treated as uncorrelated since the input analyses often use different tagging algorithms or working points and are correlated only when these choices are compatible. The only analyses satisfying this requirement are the on-shell and off-shell \Hzz\ and \Hww\ measurements.

In a small number of exceptions to the general rules above, uncertainties are kept uncorrelated across analyses if the uncertainty sources are not in exact correspondence in each case. However, in each of these cases the impact on the results of correlating or not correlating the uncertainty was found to be negligible. Some uncertainties initially implemented as correlated are also decorrelated in the final model as described in Section~\ref{sec:stats}.

%

\subsection{Combination procedure}
\label{sec:stats}

The measurement parameters, also referred to as parameters of interest (POIs), are used to express signal rates in the various analysis regions. These take the form of Higgs boson production cross-sections or decay branching ratios, or signal-strength parameters representing ratios of observed event rates to their SM expectations. In the latter case, uncertainties in the SM expectation are included in addition to the uncertainties in the event rate. These quantities are also expressed in terms of the parameters of specific models, such as Higgs boson coupling modifiers and SMEFT parameters for the interpretations shown in Sections~\ref{sec:kappa} and~\ref{sec:eft}, respectively.

The data are analysed using a combined likelihood model built from the likelihood functions describing each of the input measurements~\cite{ATL-PHYS-PUB-2011-011}. The effects of theoretical and experimental systematic uncertainties in the predicted signal and background yields are taken into account by including nuisance parameters (NPs) in the likelihood function, which are free to vary in the fit.
These NPs are also subject to a constraint in the likelihood function, representing an {\em auxiliary} measurement. The constraints used are represented by a normal distribution with a mean corresponding to the NP, or in some cases by a Poisson distribution. In most cases, the impact of Gaussian-constrained NPs on other parameters in the likelihood is implemented using an exponential form, making the procedure equivalent to the use of a log-normal constraint. For the small uncertainties typical of the systematic effects considered here, the log-normal and Gaussian constraints are numerically indistinguishable and this choice has negligible impact on the results. The {\em auxiliary observable} entering the constraints represents the observed value of the auxiliary measurement~\cite{ATL-PHYS-PUB-2011-011}. Correlated uncertainties across multiple analysis channels are implemented by using the same NP in each case; this allows individual analyses to constrain shared nuisance parameters, which in turn benefits other analyses in the combination.

The combined likelihood is obtained by including all the categories $c$ from each input analysis $a$, and the constraints for all constrained nuisance parameters. It is expressed as
\begin{equation}
\mathcal{L}(\bm{\mu}, \bm{\theta}; \text{data}) =
\prod\limits_{a=1}^{N_{\text{inputs}}} \prod\limits_{c=1}^{N_{\text{cats}}^{(a)}} \mathcal{L}^{(a)}_c(\bm{\mu}, \bm{\theta}; \text{data})
\prod\limits_{k=1}^{N_{\text{\text{cons}}}} \mathcal{G}\left(\tilde{\theta}_k; \theta_k\right),
\end{equation}

where $\bm{\mu}$ and $\bm{\theta}$ are the vectors of POIs and NPs, respectively, $N_{\text{inputs}}$ is the number of input analyses, $N_{\text{cats}}^{(a)}$ is the number of categories in analysis $a$, $\mathcal{L}^{(a)}_c$ is the likelihood function for category
$c$ of input analysis $a$, where $a$ labels one specific analysis in the combination. $N_{\text{cons}}$ is the overall number of constrained nuisance parameters, $\mathcal{G}$ denotes the constraint probability density function (not to be confused with parton distribution functions) and $\tilde{\theta}_{k}$ is the auxiliary observable corresponding to the $\theta_{k}$ nuisance parameter.
Combined likelihoods are built to perform the measurements reported in the following sections, using the subsets of input analyses described in Table~\ref{tab:inputs}.

A procedure to identify and remove NPs with negligible impact on the final results from the combined model is applied. In addition, uncertainty sources that were initially correlated across input analyses are decorrelated in cases where the NP describing the uncertainty is found to have significantly different behaviour in the combined statistical model in comparison with individual input analyses. The decorrelation procedure is applied if the best-fit value of the NP in the combined model shifts by more than $0.7\sigma$ when comparing fits to data using the combined model and fits with individual analyses, or if its uncertainty in the combined model is smaller than $0.5\sigma$. In both cases, $\sigma$ denotes the width of the constraint applied on the NP. The decorrelation is performed by implementing the uncertainty source using separate NPs in the various input analyses.  The affected parameters are predominantly jet energy scale and jet energy resolution uncertainties. In each such case the impact on the results of correlating or decorrelating the uncertainty was found to be negligible.

The resulting statistical models describe up to 663 analysis regions and contain about 6300 NPs. The RooFit framework~\cite{Verkerke:2003ir} is used to implement these models and perform statistical computations. The statistical test of a given signal hypothesis, used for the measurement of the parameters of interest, is performed with a test statistic based on the profile likelihood ratio~\cite{Cowan:2010st}, denoted by $-2\ln L$ in the following. The confidence intervals of the measured parameters are obtained using asymptotic formulae~\cite{Cowan:2010st}. The compatibility of the results and the SM predictions is estimated using a $p$-value $p\SM$, computed as described in Ref.~\cite{HIGG-2021-23}.

Uncertainties are decomposed into components corresponding to different classes of uncertainty using two methods. All results except the Higgs boson total width measurement of Section~\ref{sec:width} make use of the \emph{conditional uncertainty} method already employed in Ref.~\cite{HIGG-2021-23}. The uncertainty corresponding to a given set of NPs is obtained by fixing these NPs to their best-fit value, and subtracting in quadrature the uncertainty in the POI obtained in this configuration from the baseline fit uncertainty (i.e.\ the uncertainty from the nominal fit in which all NPs are free to vary). The statistical uncertainty component is obtained from a fit in which all the NPs representing systematic uncertainties are fixed to their best-fit values. The uncertainties in the results presented in Section~\ref{sec:width} use the \emph{shifted auxiliary observable} method described in Ref.~\cite{Pinto:2023yob}. In this case, the contributions from each systematic uncertainty source are obtained by shifting the auxiliary observable of the corresponding constraint by $\pm 1$ standard deviation and performing two fits to data per nuisance parameter to evaluate the resulting absolute change in the POI. The uncertainty associated with a group of systematic sources is then computed as the sum in quadrature of their individual contributions. The shifted auxiliary observable method is used for the total width measurement since it yields an uncorrelated decomposition of the uncertainty, enabling the detailed breakdown provided in Appendix~\ref{app:width}; the conditional uncertainty method is sufficient for all other results.

Expected results under the SM hypothesis are obtained from Asimov datasets~\cite{Cowan:2010st} generated with POIs set to their values in the SM and NPs set to their best-fit value in data; they represent the results that would be obtained in the absence of statistical fluctuations if the Higgs boson couplings exactly matched their SM predictions. When fitting the Asimov dataset, auxiliary observables are set to the best-fit value in data of the corresponding NPs.

%

\section{Measurement of the inclusive signal strength}
\label{sec:global_mu}


This section presents a test of the overall Higgs boson production and decay rate compared to its SM expectation.
The rate of Higgs boson production and decay processes is expressed using the quantity $\sigma_i \times
\BR_f$, where $\sigma_i$ denotes the cross-section of the production process $i$ and $\BR_f$ the branching ratio into the final state $f$.
The signal strength of the process is defined by the ratio $\mu_{if} = (\sigma_i \times \BR_f)/ (\sigma^\text{SM}_i \times  \BR_f^\text{SM})$ of the measured event rate to its SM prediction. The SM corresponds to values of unity for each of the $\mu_{if}$ parameters.

In this section a single inclusive signal strength $\mu$ scaling all event rates considered in the combination is measured. Following the procedure described in Section~\ref{sec:stats}, a combined likelihood model is constructed in which all the $\mu_{if}$ parameters are identified with the inclusive $\mu$. The best-fit value observed in data is
\begin{equation*}
\mu =  0.990^{+0.053}_{-0.051} = 0.990 \pm 0.027 \text{ (stat.) }  \pm 0.024 \text{ (exp.) } ^{+0.037}_{-0.035} \text{ (sig. theo.) } ^{+0.016}_{-0.015} \text{ (bkg. theo.)}
\end{equation*}
where (stat.) refers to the statistical component of the uncertainty, (exp.) to the contribution from experimental systematic uncertainties, (sig. theo.) to  theory uncertainties in the Higgs boson signal, and (bkg. theo.) to  theory uncertainties in the background processes. The signal theory uncertainty is significantly larger than both the experimental systematic uncertainty and the statistical uncertainty. This reflects the improved experimental precision of the present combination: signal theory uncertainties are comparable to those of the previous combination of Ref.~\cite{HIGG-2021-23}, in which the three components were of similar size.

\begin{figure}[tbp]
\centering
\includegraphics[width=.7\textwidth]{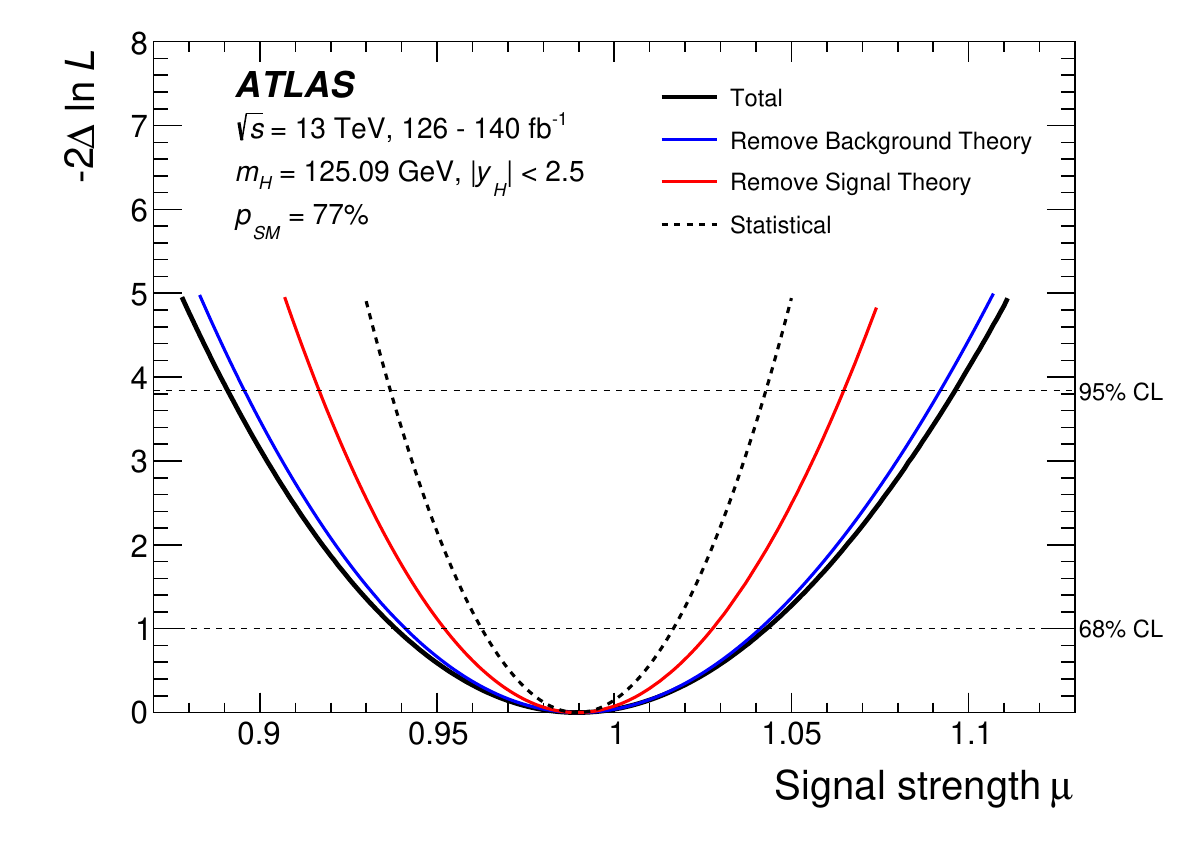}
\caption{Observed values of $-2\,\Delta \ln L$ as a function of the signal strength $\mu$, in the model with a single signal-strength scaling the event rate of all Higgs boson processes. The intersections with the horizontal dotted lines define the 68\% and 95\% confidence level intervals on $\mu$ under the asymptotic approximation. The solid black line (Total) corresponds to the full statistical model with all uncertainties included, and other lines to cases where a class of systematic uncertainties is removed from consideration: the background theory uncertainties (blue line), the signal theory uncertainties (red line), and all systematic uncertainties (dashed black line).}
\label{fig:mu_obs}
\end{figure}

The leading contributions to the signal theory uncertainty originate from missing higher-order QCD corrections ($^{+0.026}_{-0.024}$), with about 90\% of this uncertainty originating from the \ggF\ process. Other contributions include the uncertainties in the modelling of parton shower and underlying event effects ($^{+0.019}_{-0.018}$) and on the knowledge of parton distribution functions and $\alpha_s$ ($^{+0.017}_{-0.015}$). The observed profile log-likelihood scan is shown in Figure~\ref{fig:mu_obs}. The expected result under SM assumptions is $1.000 ^{+0.054}_{-0.052}$ and the expected scan is shown in Figure~\ref{fig:mu_exp} in Appendix~\ref{app:mu_global}.
The result is in good agreement with the SM, with a compatibility corresponding to a $p$-value $p\SM=77\%$.
The total uncertainty is reduced by about 15\%
compared to the results published in Ref.~\cite{HIGG-2021-23}.
The expected statistical uncertainty is reduced from $\pm 0.030$ in Ref.~\cite{HIGG-2021-23} to $\pm 0.027$ in the present combination. The improvement is driven by improved analysis techniques, in particular in the updated measurement of \ggF~and \VBF~in the \Hlvlv~final state~\cite{HIGP-2024-07} and the updated analysis of the \Htt\ decay~\cite{HIGG-2022-07}, as well as larger datasets. Improved analysis techniques also lead to a reduction in experimental uncertainties from $^{+0.032}_{-0.031}$ in Ref.~\cite{HIGG-2021-23} to $\pm 0.024$ in the present combination.

\FloatBarrier


\section{Measurements of production and decay rates}
\label{sec:inclusive}

\subsection{Production cross-section measurements}
\label{sec:prodXS}

A measurement of Higgs boson production cross-sections is performed in the fiducial volume $|y_H|<2.5$, where $y_H$ is the rapidity of the Higgs boson, which matches, to a good approximation, the detector region where the Higgs boson decay products can be reconstructed across the analyses in the combination. The measurement assumes that the branching ratios of Higgs boson decays are equal to their SM expectations, within SM theory uncertainties. The \bbH\ mode is considered together with \ggF\ since no dedicated analysis targeting this process is included in the combination and the acceptances for the \bbH\ and \ggF\ processes are similar in all the channels considered. The other production modes considered are \VBF, \WH, \ZH, \ttH\ and \tH. The \VH\ processes include both leptonic and hadronic decays of the $W$ and $Z$, and the gluon-initiated \ggZH\ process is considered as part of \ZH. The \tH\ process encompasses \tH\ production in association with a hadronic jet and in association with a $W$ boson, as well as the $s$-channel \tH\ production process.

\begin{figure}[tbp]
\begin{center}
\includegraphics[width=0.7\columnwidth]{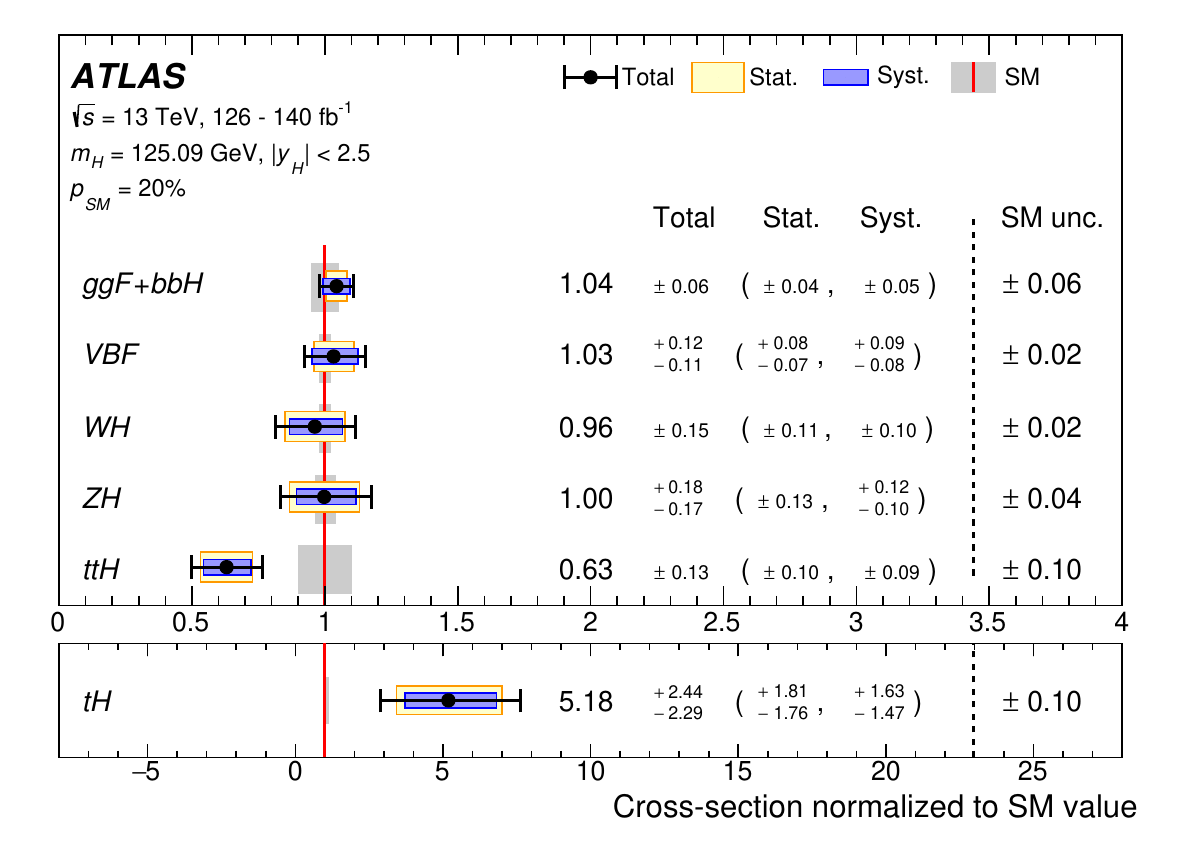}
\end{center}
\caption{Observed cross-section values for the main Higgs boson production modes, relative to their SM predictions. The \bbH\ mode is considered together with \ggF. Higgs boson decay branching ratios are assumed to be equal to their SM expectations within theory uncertainties. Total uncertainties in the measurements and their statistical and systematic components are shown. The vertical line and the shaded regions represent the SM predictions and their uncertainties.}
\label{fig:XS_obs}
\end{figure}

The observed results are shown in Figure~\ref{fig:XS_obs} and Table~\ref{tab:XS}. The observed \ttH\ cross-section shows a mild deficit relative to the SM prediction but results are in overall agreement with SM predictions, with a $p$-value $p\SM=20\%$.
Expected results under SM assumptions and the correlation matrices of the observed and expected measurements are shown respectively in Figures~\ref{fig:xs:exp} and~\ref{fig:xs:corr} in Appendix~\ref{app:xs_br}. Correlation coefficients are generally at 10\% or below in absolute value, except for the \ttH\ and \tH\ modes for which the observed (expected)  correlation is $-47\%$ ($-37\%$) due to the similar experimental signature between the two processes. An observed (expected) 95\% confidence level (CL) upper bound on the \tH\ cross-section is set at 10.1 (5.2) times its SM value.
Compared to the previous combination of Ref.~\cite{HIGG-2021-23}, uncertainties in $\sigma_{\WH}$ and $\sigma_{\ZH}$ are reduced by about 30\% and 20\%, respectively, mainly as a result of the updated measurement of \VH, \Hbbcc~\cite{HIGG-2020-20}. Similarly, the uncertainty in $\sigma_{\ttH}$ is reduced by about 40\%, mainly due to the updated measurements of \ttH\ in the \Hbb~\cite{HIGG-2020-24} and \HML~\cite{HIGP-2024-08} final states.
\begin{table}[tbp]
\caption{Measured values of Higgs boson production cross-sections, assuming branching ratios match their expectations in the SM within theory uncertainties. Observed and expected uncertainties are shown along with their contributions from data statistics (stat.) and systematic uncertainties (syst.). The last column indicates the corresponding SM predictions together with their theory uncertainties. }
\label{tab:XS}
\centering
\renewcommand{\arraystretch}{1.4}
\resizebox{0.8\textwidth}{!}{
\begin{tabular}{l S[table-format=2.2,round-mode=none] llllll S[table-format=2.3,round-mode=none] l}
\toprule
\multirow{2}{*}{Production mode} & \multicolumn{1}{c}{Observed}    & \multicolumn{3}{c}{ Observed uncertainty [pb]} & \multicolumn{3}{c}{ Expected uncertainty [pb]} & \multicolumn{2}{c}{SM prediction} \\
&  \multicolumn{1}{c}{value [pb]}  & Total &  Stat. & Syst. & Total & Stat. & Syst. & \multicolumn{2}{c}{[pb]}   \\
\midrule
$\ggF + \bbH$ & 46.7 & \errRP{1}{2.9}{-2.8} & ${\scriptstyle \pm \numRP[1]{1.8}}$ & ${\scriptstyle \pm \numRP[1]{2.2}}$ & ${\scriptstyle \pm \numRP[1]{2.8}}$ & ${\scriptstyle \pm \numRP[1]{1.8}}$ & \errRP{1}{2.2}{-2.1} & 44.8 & ${\scriptstyle \pm \numRP[1]{2.6}}$ \\
$\VBF$ & 3.6 & ${\scriptstyle \pm \numRP[1]{0.4}}$ & ${\scriptstyle \pm \numRP[1]{0.3}}$ & ${\scriptstyle \pm \numRP[1]{0.3}}$ & ${\scriptstyle \pm \numRP[1]{0.4}}$ & ${\scriptstyle \pm \numRP[1]{0.3}}$ & ${\scriptstyle \pm \numRP[1]{0.3}}$ & 3.50 & ${\scriptstyle \pm \numRP[2]{0.07}}$ \\
$\WH$ & 1.17 & \errRP{2}{0.19}{-0.18} & ${\scriptstyle \pm \numRP[2]{0.14}}$ & \errRP{2}{0.13}{-0.12} & \errRP{2}{0.19}{-0.18} & ${\scriptstyle \pm \numRP[2]{0.14}}$ & \errRP{2}{0.13}{-0.12} & 1.216 & ${\scriptstyle \pm \numRP[3]{0.024}}$ \\
$\ZH$ & 0.79 & \errRP{2}{0.14}{-0.13} & \errRP{2}{0.11}{-0.1} & \errRP{2}{0.1}{-0.08} & \errRP{2}{0.14}{-0.13} & \errRP{2}{0.11}{-0.1} & \errRP{2}{0.09}{-0.08} & 0.796 & ${\scriptstyle \pm \numRP[3]{0.029}}$ \\
$\ttH$ & 0.32 & ${\scriptstyle \pm \numRP[2]{0.07}}$ & ${\scriptstyle \pm \numRP[2]{0.05}}$ & \errRP{2}{0.05}{-0.04} & ${\scriptstyle \pm \numRP[2]{0.07}}$ & ${\scriptstyle \pm \numRP[2]{0.05}}$ & ${\scriptstyle \pm \numRP[2]{0.05}}$ & 0.50 & ${\scriptstyle \pm \numRP[2]{0.05}}$ \\
$\tH$ & 0.44 & \errRP{2}{0.21}{-0.19} & ${\scriptstyle \pm \numRP[2]{0.15}}$ & \errRP{2}{0.14}{-0.12} & \errRP{2}{0.17}{-0.15} & \errRP{2}{0.14}{-0.13} & \errRP{2}{0.11}{-0.09} & 0.085 & \errRP{3}{0.005}{-0.011} \\
\bottomrule
\end{tabular}}
\end{table}

\FloatBarrier


\subsection{Decay branching ratio measurements}
\label{sec:decaymodes}

\begin{figure}[tbp]
\begin{center}
\includegraphics[width=0.7\columnwidth]{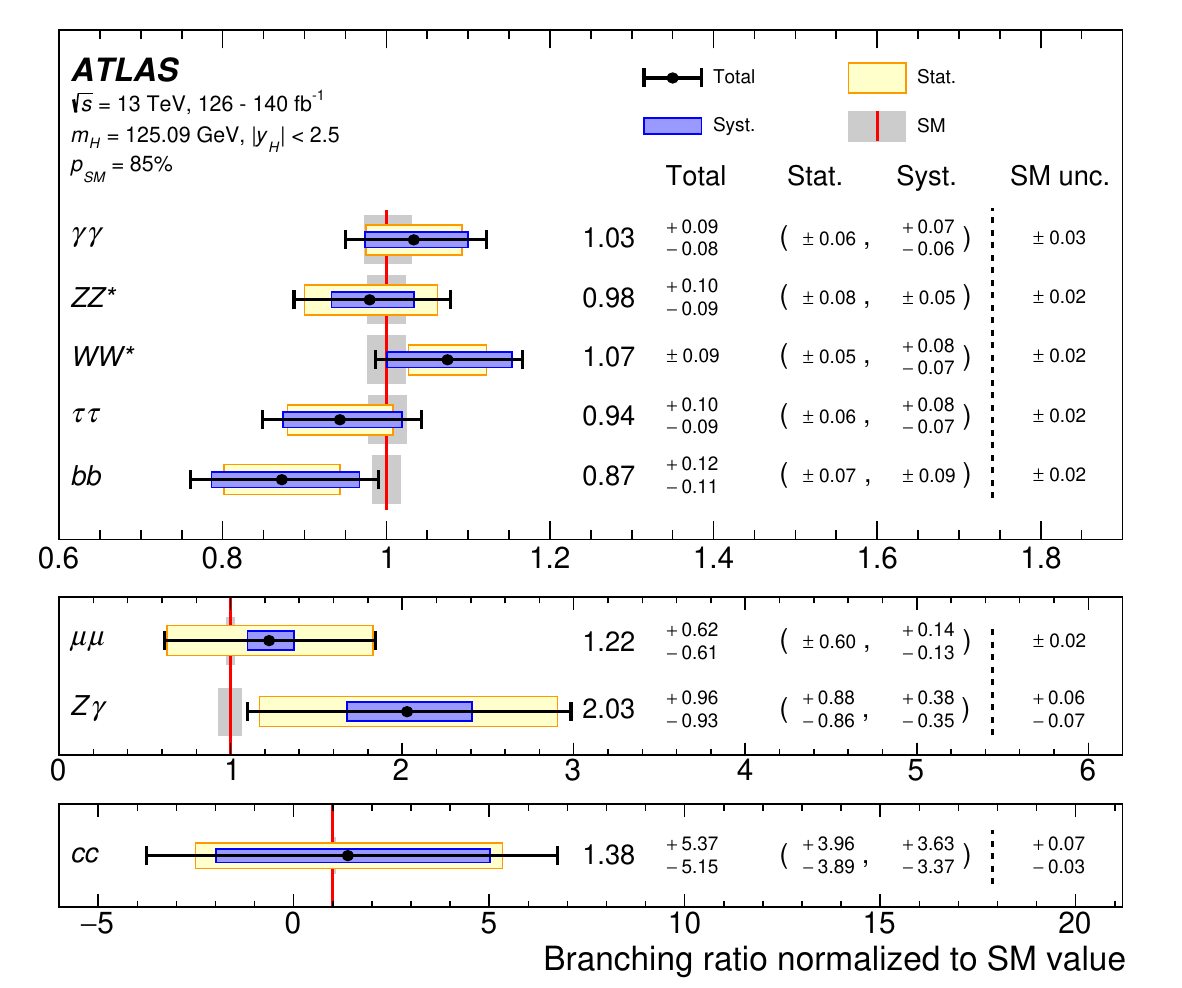}
\end{center}
\caption{Observed branching ratio values in the \Hyy, \Hzz, \Hww, \Htt, \Hbb, \Hmm, \HZy\ and \Hcc\ decay modes, relative to their SM predictions. Higgs boson production cross-sections are assumed to be equal to their SM expectations within theory uncertainties. Total uncertainties in the measurements and their statistical and systematic components are shown. The vertical line and the shaded regions represent the SM predictions and their uncertainties.}
\label{fig:BR_obs}
\end{figure}

A measurement of Higgs boson branching ratios is performed under the assumption that the production cross-sections are equal to their SM expectations, within SM theory uncertainties. The Higgs boson decay processes considered are \Hbb, \Hww, \Htt, \Hcc, \Hzz, \Hyy, \HZy\ and \Hmm. The rates of other decay processes are assumed to match their SM expectations. Higgs boson processes are considered in the fiducial volume $|y_H|<2.5$.
Observed results are shown in Figure~\ref{fig:BR_obs} and Table~\ref{tab:BR}. The results are compatible with SM predictions, with a $p$-value $p\SM=85\%$.
Expected results under SM assumptions are provided in Appendix~\ref{app:xs_br}.
\begin{table}[tbp]
\caption{Measured values of Higgs boson decay branching ratios, assuming production cross-sections to match their expectations in the SM within theory uncertainties. Observed and expected uncertainties are shown along with their contributions from data statistics (stat.) and systematic uncertainties (syst.). The last column indicates the corresponding SM predictions together with their theory uncertainties.}
\label{tab:BR}
\centering
\renewcommand{\arraystretch}{1.4}
\resizebox{0.8\textwidth}{!}{
\begin{tabular}{l S[table-format=2.3,round-mode=none] llllll S[table-format=2.4,round-mode=none] l}
\toprule
\multirow{2}{*}{Decay mode} & \multicolumn{1}{c}{Observed}    & \multicolumn{3}{c}{ Observed uncertainty [\%]} & \multicolumn{3}{c}{ Expected uncertainty [\%]} & \multicolumn{2}{c}{SM prediction} \\
&  \multicolumn{1}{c}{value [\%]}  & Total &  Stat. & Syst. & Total & Stat. & Syst. & \multicolumn{2}{c}{[\%]}   \\
\midrule
$\Hbb$ & 51 & \errRP{0}{7.0}{-6.0} & ${\scriptstyle \pm \numRP[0]{4.0}}$ & \errRP{0}{6.0}{-5.0} & ${\scriptstyle \pm \numRP[0]{7.0}}$ & ${\scriptstyle \pm \numRP[0]{4.0}}$ & \errRP{0}{6.0}{-5.0} & 58.1 & ${\scriptstyle \pm \numRP[1]{0.7}}$ \\
$\Hww$ & 23 & ${\scriptstyle \pm \numRP[0]{2.0}}$ & ${\scriptstyle \pm \numRP[0]{1.0}}$ & ${\scriptstyle \pm \numRP[0]{2.0}}$ & ${\scriptstyle \pm \numRP[0]{2.0}}$ & ${\scriptstyle \pm \numRP[0]{1.0}}$ & ${\scriptstyle \pm \numRP[0]{2.0}}$ & 21.5 & ${\scriptstyle \pm \numRP[1]{0.3}}$ \\
$\Htt$ & 5.9 & ${\scriptstyle \pm \numRP[1]{0.6}}$ & ${\scriptstyle \pm \numRP[1]{0.4}}$ & \errRP{1}{0.5}{-0.4} & ${\scriptstyle \pm \numRP[1]{0.6}}$ & ${\scriptstyle \pm \numRP[1]{0.4}}$ & \errRP{1}{0.5}{-0.4} & 6.26 & ${\scriptstyle \pm \numRP[2]{0.08}}$ \\
$\Hcc$ & 4 & ${\scriptstyle \pm \numRP[0]{15.0}}$ & ${\scriptstyle \pm \numRP[0]{11.0}}$ & ${\scriptstyle \pm \numRP[0]{10.0}}$ & \errRP{0}{15.0}{-14.0} & ${\scriptstyle \pm \numRP[0]{11.0}}$ & \errRP{0}{10.0}{-9.0} & 2.9 & ${\scriptstyle \pm \numRP[1]{0.2}}$ \\
$\Hzz$ & 2.6 & \errRP{1}{0.3}{-0.2} & ${\scriptstyle \pm \numRP[1]{0.2}}$ & ${\scriptstyle \pm \numRP[1]{0.1}}$ & \errRP{1}{0.3}{-0.2} & ${\scriptstyle \pm \numRP[1]{0.2}}$ & ${\scriptstyle \pm \numRP[1]{0.1}}$ & 2.64 & ${\scriptstyle \pm \numRP[2]{0.03}}$ \\
$\Hyy$ & 0.23 & ${\scriptstyle \pm \numRP[2]{0.02}}$ & ${\scriptstyle \pm \numRP[2]{0.01}}$ & \errRP{2}{0.02}{-0.01} & ${\scriptstyle \pm \numRP[2]{0.02}}$ & ${\scriptstyle \pm \numRP[2]{0.01}}$ & ${\scriptstyle \pm \numRP[2]{0.01}}$ & 0.227 & ${\scriptstyle \pm \numRP[3]{0.004}}$ \\
$\Hzy$ & 0.31 & \errRP{2}{0.15}{-0.14} & \errRP{2}{0.14}{-0.13} & \errRP{2}{0.06}{-0.05} & \errRP{2}{0.14}{-0.13} & ${\scriptstyle \pm \numRP[2]{0.13}}$ & \errRP{2}{0.05}{-0.04} & 0.154 & ${\scriptstyle \pm \numRP[3]{0.008}}$ \\
$\Hmm$ & 0.027 & ${\scriptstyle \pm \numRP[3]{0.013}}$ & ${\scriptstyle \pm \numRP[3]{0.013}}$ & ${\scriptstyle \pm \numRP[3]{0.003}}$ & ${\scriptstyle \pm \numRP[3]{0.013}}$ & ${\scriptstyle \pm \numRP[3]{0.013}}$ & \errRP{3}{0.003}{-0.002} & 0.0217 & ${\scriptstyle \pm \numRP[4]{0.0003}}$ \\
\bottomrule
\end{tabular}}
\end{table}
Compared to Ref.~\cite{HIGG-2021-23}, the uncertainty in the \Hbb\ branching ratio is reduced by approximately 20\%, mainly due to the higher sensitivity of the updated analyses of the \VH, \Hbbcc\ and \ttH, \Hbb\ processes~\cite{HIGG-2020-20, HIGG-2020-24}. The uncertainties in the \Hww\ and \Htt\ branching ratios are reduced by respectively about 20\% and 10\% due to the updated analyses of Refs.~\cite{HIGP-2024-07} and~\cite{HIGG-2022-07}, as well as to a somewhat lesser degree the analyses of Refs.~\cite{HIGG-2023-09} and~\cite{HIGG-2018-20}.

\FloatBarrier


\subsection{Measurements of Higgs boson production per decay channel}
\label{sec:proddecay}

This section presents measurements of Higgs boson production rates for individual decay channels. The \Hbb, \Hww, \Htt, \Hcc, \Hzz, \Hyy, \Hzy, and \Hmm\ processes are considered within the fiducial volume $|y_H|<2.5$. The measured production modes are defined in the same way as in Section~\ref{sec:prodXS} with some adjustments driven by the limited statistical sensitivity in some combinations of production and decay: in the \Hbb\ mode, the $\ggF+\bbH$ and $\VBF$ modes are considered as a single process; in the $\Hzz$ decay, the \ttH\ and \tH\ modes and the \WH\ and \ZH\ modes are similarly merged; in the \Hmm\ decay, the $\ggF+\bbH$ and $\ttH+\tH$ modes are considered together, as are the $\VBF$, $\WH$ and $\ZH$ modes; in the \HZy\ mode all production processes are merged; and finally in the \Hcc\ mode, only the \WH\ and \ZH\ processes are considered, while the others are fixed to their SM expectation. In total, 32 combinations $(\sigma_i \times \BR_f)$ of production cross-sections $\sigma_i$ and branching ratios $\BR_f$ are reported.

\begin{sidewaysfigure}
\centering
\includegraphics[width=\textheight]{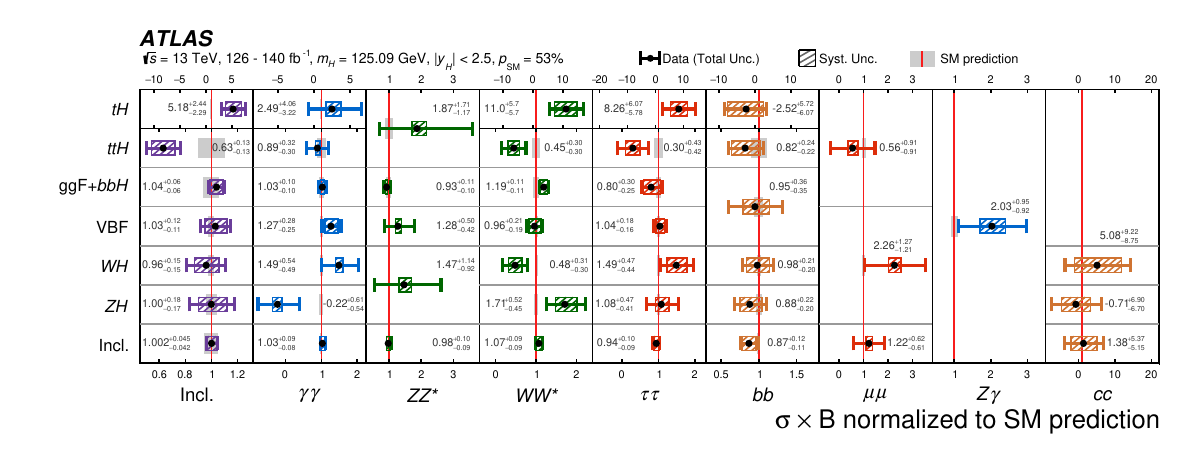}
\caption{Measured values of $(\sigma_i \times \BR_f)$ for the measured combinations of production cross-sections and branching ratios, relative to their SM predictions. The production modes considered are indicated on the vertical axis, and the decay processes on the horizontal axis. In each case \emph{Incl.} refers to an inclusive result over all production or decay processes; the \emph{Incl./Incl.} entry corresponds to the overall ratio of the total measured event rate to its SM prediction. Error bars and hashed regions represent respectively the total uncertainty and its systematic uncertainty component. Vertical lines indicate the SM expectation, and the shaded regions the corresponding SM theory uncertainty. Merged cells indicate that the measurement is performed inclusively over the corresponding combination of production modes.}
\label{fig:PxD_illustration}
\end{sidewaysfigure}

Results are shown in Figure~\ref{fig:PxD_illustration} and Table~\ref{tab:PxD}. The results are all in agreement with SM predictions, with a $p$-value $p\SM=53\%$. The inclusive $(\sigma_i \times \BR_f)$ result, obtained by summing over all production and decay processes, is $1.002^{+0.045}_{-0.042} = 1.002 \pm 0.028\text{ (stat.) }^{+0.026}_{-0.021}\text{ (exp.) }^{+0.021}_{-0.020}\text{ (sig. theo.) }\pm 0.013\text{ (bkg. theo.)}$ relative to the SM prediction. This value differs from the inclusive signal strength of Section~\ref{sec:global_mu} ($\mu = 0.990\,^{+0.053}_{-0.052}$) primarily because signal theory uncertainties in Higgs boson event rates enter in the denominator of the signal strength, which leads to a small shift between the central values of the two measurements and a larger signal theory uncertainty component for the signal strength.
\begin{table}[tbp]
\caption{Measured values of $(\sigma_i \times \BR_f)$ for different combinations of production processes and decay modes. The last two columns list the SM prediction for the measurement and its theory uncertainty. Results are expressed in \unit{pb} in the top part of the table, and in \unit{fb} for the rarer processes in the bottom part. The values for the $\tH(\Hbb)$, $ZH(\rightarrow \yy)$ and $ZH(\rightarrow \cc)$ processes are measured to be negative due to an observed event count that is below the predicted background level in the corresponding signal-sensitive regions.}
\label{tab:PxD}
\centering
\renewcommand{\arraystretch}{1.3}
\resizebox{0.9\textwidth}{!}{
\begin{tabular}{ll S[table-format=-3.3,round-mode=none] llllll S[table-format=3.4,round-mode=none] l}
\toprule
Decay mode & Prod. mode & \multicolumn{1}{c}{Observed} & \multicolumn{3}{c}{Observed uncertainty [pb]} & \multicolumn{3}{c}{Expected uncertainty [pb]} & \multicolumn{2}{c}{SM prediction} \\
&&                      \multicolumn{1}{c}{value [pb]}  & Total &  Stat. & Syst. & Total & Stat. & Syst. & \multicolumn{2}{c}{[pb]}   \\
\midrule
\multirow{5}{*}{\Hbb} & $\ggF + \bbH + \VBF$ & 26 & ${\scriptstyle \pm \numRP[0]{10.0}}$ & ${\scriptstyle \pm \numRP[0]{9.0}}$ & ${\scriptstyle \pm \numRP[0]{5.0}}$ & ${\scriptstyle \pm \numRP[0]{10.0}}$ & ${\scriptstyle \pm \numRP[0]{9.0}}$ & ${\scriptstyle \pm \numRP[0]{5.0}}$ & 28.0 & ${\scriptstyle \pm \numRP[1]{1.5}}$ \\
& $\WH$ & 0.69 & \errRP{2}{0.15}{-0.14} & ${\scriptstyle \pm \numRP[2]{0.1}}$ & \errRP{2}{0.11}{-0.1} & \errRP{2}{0.16}{-0.14} & ${\scriptstyle \pm \numRP[2]{0.1}}$ & \errRP{2}{0.12}{-0.1} & 0.706 & ${\scriptstyle \pm \numRP[3]{0.016}}$ \\
& $\ZH$ & 0.41 & \errRP{2}{0.1}{-0.09} & ${\scriptstyle \pm \numRP[2]{0.07}}$ & \errRP{2}{0.07}{-0.06} & \errRP{2}{0.11}{-0.1} & ${\scriptstyle \pm \numRP[2]{0.07}}$ & \errRP{2}{0.08}{-0.07} & 0.462 & ${\scriptstyle \pm \numRP[3]{0.018}}$ \\
& $\ttH$ & 0.24 & \errRP{2}{0.07}{-0.06} & ${\scriptstyle \pm \numRP[2]{0.04}}$ & \errRP{2}{0.06}{-0.05} & ${\scriptstyle \pm \numRP[2]{0.07}}$ & ${\scriptstyle \pm \numRP[2]{0.04}}$ & ${\scriptstyle \pm \numRP[2]{0.06}}$ & 0.290 & ${\scriptstyle \pm \numRP[3]{0.029}}$ \\
& $\tH$ & -0.12 &\errRP{2}{0.28}{-0.3} & \errRP{2}{0.15}{-0.15} & \errRP{2}{0.24}{-0.26} & \errRP{2}{0.29}{-0.29} & \errRP{2}{0.16}{-0.16} & \errRP{2}{0.24}{-0.24} & 0.049 & \errRP{3}{0.003}{-0.006} \\
\midrule
\multirow{6}{*}{\Hww} & $\ggF + \bbH$ & 11.4 & ${\scriptstyle \pm \numRP[1]{1.1}}$ & ${\scriptstyle \pm \numRP[1]{0.6}}$ & ${\scriptstyle \pm \numRP[1]{0.9}}$ & \errRP{1}{1.1}{-1.0} & ${\scriptstyle \pm \numRP[1]{0.6}}$ & ${\scriptstyle \pm \numRP[1]{0.9}}$ & 9.6 & ${\scriptstyle \pm \numRP[1]{0.6}}$ \\
& $\VBF$ & 0.72 & \errRP{2}{0.16}{-0.14} & ${\scriptstyle \pm \numRP[2]{0.1}}$ & \errRP{2}{0.12}{-0.1} & \errRP{2}{0.16}{-0.14} & ${\scriptstyle \pm \numRP[2]{0.1}}$ & \errRP{2}{0.12}{-0.1} & 0.753 & ${\scriptstyle \pm \numRP[3]{0.018}}$ \\
& $\WH$ & 0.13 & ${\scriptstyle \pm \numRP[2]{0.08}}$ & ${\scriptstyle \pm \numRP[2]{0.07}}$ & ${\scriptstyle \pm \numRP[2]{0.04}}$ & ${\scriptstyle \pm \numRP[2]{0.09}}$ & \errRP{2}{0.08}{-0.07} & ${\scriptstyle \pm \numRP[2]{0.05}}$ & 0.262 & ${\scriptstyle \pm \numRP[3]{0.006}}$ \\
& $\ZH$ & 0.29 & \errRP{2}{0.09}{-0.08} & ${\scriptstyle \pm \numRP[2]{0.07}}$ & \errRP{2}{0.05}{-0.04} & \errRP{2}{0.08}{-0.07} & \errRP{2}{0.07}{-0.06} & \errRP{2}{0.04}{-0.03} & 0.171 & ${\scriptstyle \pm \numRP[3]{0.007}}$ \\
& $\ttH$ & 0.05 & ${\scriptstyle \pm \numRP[2]{0.03}}$ & ${\scriptstyle \pm \numRP[2]{0.03}}$ & ${\scriptstyle \pm \numRP[2]{0.02}}$ & ${\scriptstyle \pm \numRP[2]{0.03}}$ & ${\scriptstyle \pm \numRP[2]{0.03}}$ & ${\scriptstyle \pm \numRP[2]{0.02}}$ & 0.108 & ${\scriptstyle \pm \numRP[3]{0.011}}$ \\
& $\tH$ & 0.20  & \errRP{2}{0.1}{-0.1} & \errRP{2}{0.08}{-0.08} & \errRP{2}{0.07}{-0.07} & \errRP{2}{0.09}{-0.09} & \errRP{2}{0.07}{-0.07} & \errRP{2}{0.06}{-0.06} & 0.018 & \errRP{3}{0.001}{-0.002} \\
\midrule
\multirow{6}{*}{\Htt} & $\ggF + \bbH$ & 2.2 & \errRP{1}{0.8}{-0.7} & ${\scriptstyle \pm \numRP[1]{0.4}}$ & \errRP{1}{0.7}{-0.6} & \errRP{1}{0.9}{-0.8} & ${\scriptstyle \pm \numRP[1]{0.4}}$ & \errRP{1}{0.8}{-0.7} & 2.80 & ${\scriptstyle \pm \numRP[2]{0.17}}$ \\
& $\VBF$ & 0.228 & \errRP{3}{0.04}{-0.036} & \errRP{3}{0.028}{-0.027} & \errRP{3}{0.028}{-0.023} & \errRP{3}{0.038}{-0.035} & \errRP{3}{0.028}{-0.027} & \errRP{3}{0.026}{-0.021} & 0.219 & ${\scriptstyle \pm \numRP[3]{0.005}}$ \\
& $\WH$ & 0.113 & \errRP{3}{0.036}{-0.034} & \errRP{3}{0.028}{-0.027} & \errRP{3}{0.022}{-0.02} & \errRP{3}{0.034}{-0.032} & \errRP{3}{0.027}{-0.026} & \errRP{3}{0.021}{-0.018} & 0.0761 & ${\scriptstyle \pm \numRP[4]{0.0018}}$ \\
& $\ZH$ & 0.054 & \errRP{3}{0.024}{-0.021} & \errRP{3}{0.021}{-0.019} & \errRP{3}{0.011}{-0.008} & \errRP{3}{0.022}{-0.019} & \errRP{3}{0.02}{-0.018} & \errRP{3}{0.01}{-0.007} & 0.0498 & ${\scriptstyle \pm \numRP[4]{0.002}}$ \\
& $\ttH$ & 0.009 & \errRP{3}{0.014}{-0.013} & ${\scriptstyle \pm \numRP[3]{0.012}}$ & ${\scriptstyle \pm \numRP[3]{0.006}}$ & \errRP{3}{0.014}{-0.013} & ${\scriptstyle \pm \numRP[3]{0.012}}$ & ${\scriptstyle \pm \numRP[3]{0.006}}$ & 0.0313 & \errRP{4}{0.0032}{-0.0031} \\
& $\tH$ & 0.044 & \errRP{3}{0.032}{-0.031} & \errRP{3}{0.028}{-0.026} & ${\scriptstyle \pm \numRP[3]{0.016}}$ & \errRP{3}{0.03}{-0.029} & \errRP{3}{0.027}{-0.025} & \errRP{3}{0.015}{-0.014} & 0.0053 & \errRP{4}{0.0003}{-0.0007} \\
\midrule
\multirow{4}{*}{\Hzz} & $\ggF + \bbH$ & 1.1 & \errRP{2}{0.13}{-0.12} & ${\scriptstyle \pm \numRP[2]{0.12}}$ & ${\scriptstyle \pm \numRP[2]{0.04}}$ & \errRP{2}{0.13}{-0.12} & ${\scriptstyle \pm \numRP[2]{0.12}}$ & ${\scriptstyle \pm \numRP[2]{0.04}}$ & 1.18 & ${\scriptstyle \pm \numRP[2]{0.07}}$ \\
& $\VBF$ & 0.12 & \errRP{2}{0.05}{-0.04} & \errRP{2}{0.05}{-0.04} & ${\scriptstyle \pm \numRP[2]{0.01}}$ & \errRP{2}{0.05}{-0.04} & \errRP{2}{0.05}{-0.04} & ${\scriptstyle \pm \numRP[2]{0.01}}$ & 0.0924 & ${\scriptstyle \pm \numRP[4]{0.0022}}$ \\
& $\VH$ & 0.08 & \errRP{2}{0.06}{-0.05} & \errRP{2}{0.06}{-0.05} & ${\scriptstyle \pm \numRP[2]{0.01}}$ & \errRP{2}{0.05}{-0.04} & \errRP{2}{0.05}{-0.04} & \errRP{2}{0.01}{-0.0} & 0.0531 & ${\scriptstyle \pm \numRP[4]{0.0014}}$ \\
& $\ttH + \tH$ & 0.029 & \errRP{3}{0.026}{-0.018} & \errRP{3}{0.026}{-0.018} & \errRP{3}{0.003}{-0.0} & \errRP{3}{0.02}{-0.012} & \errRP{3}{0.02}{-0.012} & \errRP{3}{0.002}{-0.001} & 0.0154 & \errRP{4}{0.0013}{-0.0014} \\
\midrule
Decay mode & Prod. mode & \multicolumn{1}{c}{Observed} & \multicolumn{3}{c}{Observed uncertainty [pb]} & \multicolumn{3}{c}{Expected uncertainty [pb]} & \multicolumn{2}{c}{SM prediction} \\
&&                      \multicolumn{1}{c}{value [fb]}  & Total &  Stat. & Syst. & Total & Stat. & Syst. & \multicolumn{2}{c}{[fb]}   \\
\midrule
\multirow{2}{*}{\Hcc} & $\WH$ & 180 & \errRP{0}{320}{-310} & \errRP{0}{220.0}{-220.0} & \errRP{0}{230.0}{-220.0} & \errRP{0}{300.0}{-290.0} & \errRP{0}{210.0}{-210.0} & \errRP{0}{210.0}{-200.0} & 35.1 & ${\scriptstyle \pm \numRP[1]{2.0}}$ \\
& $\ZH$ & -20 & \errRP{0}{160.0}{-150.0} & \errRP{0}{130.0}{-120.0} & \errRP{0}{90.0}{-90.0} & \errRP{0}{160.0}{-150.0} & \errRP{0}{130.0}{-130.0} & \errRP{0}{90.0}{-90.0} & 23.0 & ${\scriptstyle \pm \numRP[1]{1.5}}$ \\
\midrule
\multirow{6}{*}{\Hyy} & $\ggF + \bbH$ & 105 & ${\scriptstyle \pm \numRP[0]{10.0}}$ & ${\scriptstyle \pm \numRP[0]{8.0}}$ & \errRP{0}{6.0}{-5.0} & ${\scriptstyle \pm \numRP[0]{10.0}}$ & ${\scriptstyle \pm \numRP[0]{8.0}}$ & \errRP{0}{6.0}{-5.0} & 102 & ${\scriptstyle \pm \numRP[0]{6.0}}$ \\
& $\VBF$ & 10.1 & \errRP{1}{2.2}{-2.0} & ${\scriptstyle \pm \numRP[1]{1.5}}$ & \errRP{1}{1.6}{-1.3} & \errRP{1}{2.0}{-1.7} & \errRP{1}{1.5}{-1.4} & \errRP{1}{1.3}{-1.0} & 7.94 & ${\scriptstyle \pm \numRP[2]{0.2}}$ \\
& $\WH$ & 4.1 & \errRP{1}{1.5}{-1.4} & \errRP{1}{1.5}{-1.3} & ${\scriptstyle \pm \numRP[1]{0.3}}$ & \errRP{1}{1.3}{-1.2} & \errRP{1}{1.3}{-1.2} & \errRP{1}{0.2}{-0.1} & 2.76 & ${\scriptstyle \pm \numRP[2]{0.07}}$ \\
& $\ZH$ & -0.4 & \errRP{1}{1.1}{-1.0} & \errRP{1}{1.1}{-0.9} & \errRP{1}{0.2}{-0.3} & \errRP{1}{1.2}{-1.1} & \errRP{1}{1.2}{-1.1} & \errRP{1}{0.3}{-0.2} & 1.81 & ${\scriptstyle \pm \numRP[2]{0.07}}$ \\
& $\ttH$ & 1.01 & \errRP{2}{0.36}{-0.34} & \errRP{2}{0.35}{-0.33} & \errRP{2}{0.08}{-0.06} & \errRP{2}{0.35}{-0.33} & \errRP{2}{0.34}{-0.32} & \errRP{2}{0.09}{-0.06} & 1.13 & ${\scriptstyle \pm \numRP[2]{0.11}}$ \\
& $\tH$ & 0.5 & \errRP{1}{0.8}{-0.6} & \errRP{1}{0.7}{-0.6} & ${\scriptstyle \pm \numRP[1]{0.2}}$ & \errRP{1}{0.6}{-0.5} & \errRP{1}{0.6}{-0.4} & ${\scriptstyle \pm \numRP[1]{0.1}}$ & 0.192 & \errRP{3}{0.013}{-0.025} \\
\midrule
\multirow{1}{*}{\HZy} & All & 159 & \errRP{0}{75.0}{-72.0} & \errRP{0}{69.0}{-68.0} & \errRP{0}{29.0}{-25.0} & \errRP{0}{70.0}{-68.0} & \errRP{0}{66.0}{-64.0} & \errRP{0}{24.0}{-22.0} & 78 & ${\scriptstyle \pm \numRP[0]{6.0}}$ \\
\midrule
\multirow{2}{*}{\Hmm} & $\ggF + \bbH + \ttH + \tH$ & 6 & ${\scriptstyle \pm \numRP[0]{9.0}}$ & ${\scriptstyle \pm \numRP[0]{9.0}}$ & ${\scriptstyle \pm \numRP[0]{2.0}}$ & ${\scriptstyle \pm \numRP[0]{9.0}}$ & ${\scriptstyle \pm \numRP[0]{9.0}}$ & ${\scriptstyle \pm \numRP[0]{2.0}}$ & 9.8 & ${\scriptstyle \pm \numRP[1]{0.6}}$ \\
& $\VBF + \VH$ & 2.7 & ${\scriptstyle \pm \numRP[1]{1.5}}$ & \errRP{1}{1.5}{-1.4} & ${\scriptstyle \pm \numRP[1]{0.3}}$ & \errRP{1}{1.4}{-1.3} & \errRP{1}{1.4}{-1.3} & ${\scriptstyle \pm \numRP[1]{0.3}}$ & 1.197 & \errRP{3}{0.026}{-0.027} \\
\bottomrule
\end{tabular}}
\end{table}

\FloatBarrier


\section{Measurements of coupling parameters in the $\kappa$ framework}
\label{sec:kappa}

\newenvironment{DIFnomarkup}{}{}
The measurements presented in Section~\ref{sec:proddecay} are interpreted in the context of Higgs boson coupling parameters, defined within the $\kappa$-framework~\cite{YR3}.
Multiplicative real modifiers $\kappa_p$ are introduced for the couplings of the Higgs boson to the elementary SM particles $p = W,Z,t,b,c,\tau,\mu$, with the SM hypothesis corresponding to a unit value for all $\kappa_p$. The modifiers $\kappa_g$, $\kappa_\gamma$ and  $\kappa_{Z\gamma}$ are also introduced to describe effective Higgs boson interactions with gluons ($Hgg$), photons ($H\gamma\gamma$) and the combined interaction with a photon and a $Z$ boson ($HZ\gamma$). The event rates of Higgs boson processes not probed by the analyses included in the combination are assumed to correspond to their SM expectations, and only SM decays of the Higgs boson are allowed, except in Section~\ref{sec:kappa:bsm_decays} where invisible and undetected decay modes are also included. Higgs boson production cross-sections and decay rates are expressed in terms of the coupling modifiers using several parameterisations, each based on a set of physics assumptions. In each case, the modifications are applied to the SM predictions; the uncertainties in these predictions therefore enter the uncertainties in the measured values of the modifiers.

\subsection{Measurement of Higgs boson couplings to weak vector bosons and fermions}
\label{sec:kappa:kvkf}

A first model considers only two coupling modifiers: the bosonic coupling parameters $\kappa_W$ and $\kappa_Z$ are assumed to be equal to a single modifier $\kappa_V$, while a modifier $\kappa_F$ is associated to all fermion couplings. Both parameters are assumed to be positive: for $\kappa_V$ this is by convention, and a negative value of the relative sign between $\kappa_F$ and $\kappa_V$ is experimentally excluded~\cite{HIGG-2015-07}.
The effective couplings $\kappa_g$, $\kappa_\gamma$ and  $\kappa_{Z\gamma}$ are expressed as a function of $\kappa_F$ and $\kappa_V$, using expressions derived from the corresponding leading-order loop processes in the SM. The parameterisations used for this model and the ones shown below are described in detail in Table~\ref{tab:kappa_param} in Appendix~\ref{sec:kappa_defs}.

\begin{figure}[tbp]
\centering
\includegraphics[width=0.8\columnwidth]{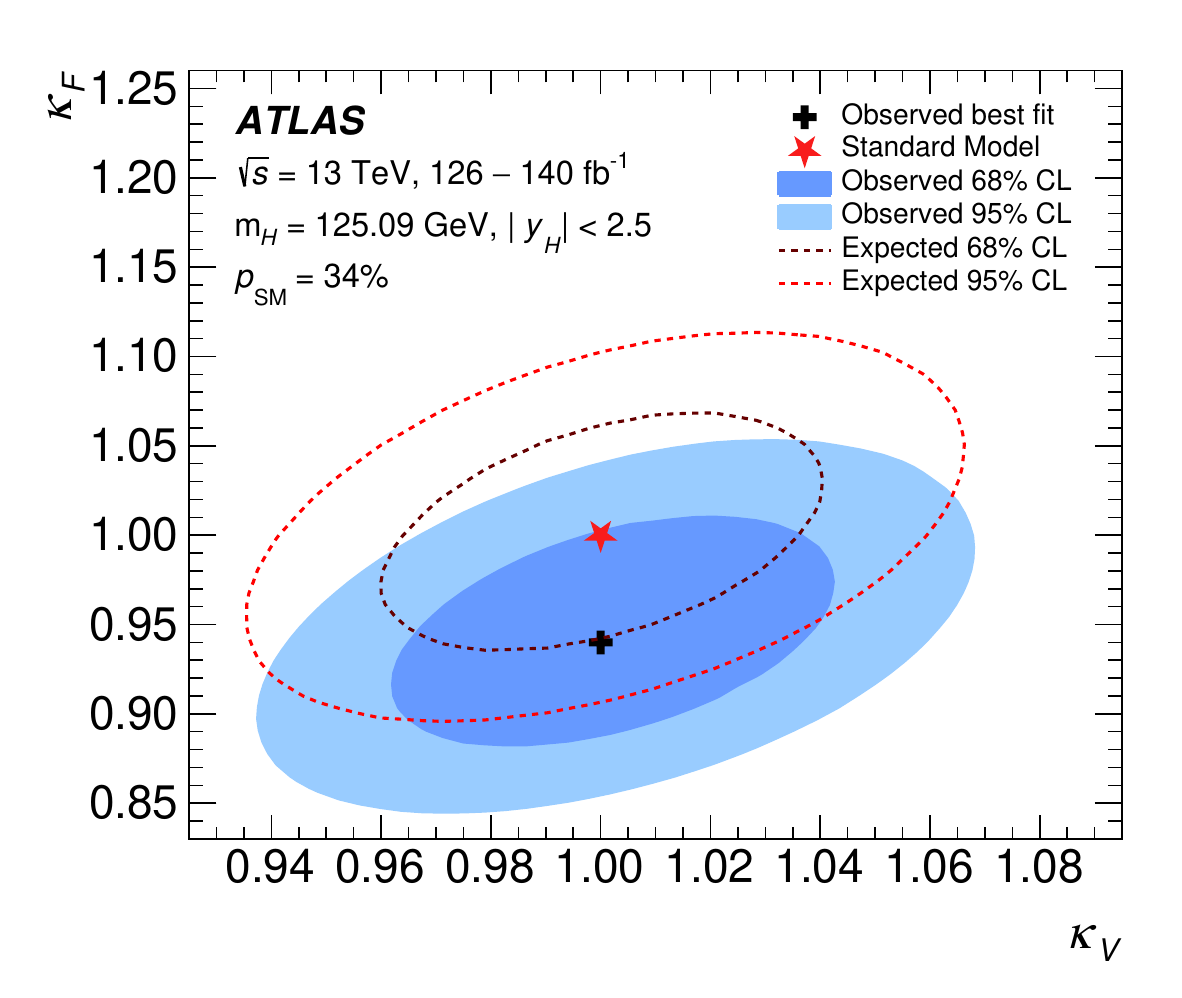}
\caption{Contours of $-2\,\Delta\ln L$, the profile log-likelihood ratio, in the plane $(\kappa_V, \kappa_F)$ of modifiers to Higgs boson fermionic and bosonic coupling modifiers. The observed 68\% and 95\% CL regions are shown respectively as dark and light blue shaded areas, while the corresponding expected regions under the SM hypothesis are shown as black and red dashed lines. The cross indicates the best-fit point, while the SM expectation is represented by the red star. The data are obtained from a combined fit assuming no deviations from the SM in Higgs boson decays except for those encoded in the coupling modifiers.}
\label{fig:kV_kF_scan}
\end{figure}

The observed (expected) values of the coupling modifiers are $\kappa_V = 1.002 ^{+0.027}_{-0.026}$ ($1.000\,{\scriptstyle \pm 0.027}$) and $\kappa_F = 0.945 ^{+0.043}_{-0.041}$ ($1.000$\numpmerr{+0.045}{-0.043}). The observed (expected) correlation coefficient between the two parameters is 40\% (38\%). Profiled log-likelihood contours at 68\% and 95\% CL in the $(\kappa_V, \kappa_F)$ plane are shown in Figure~\ref{fig:kV_kF_scan}. Observed results are consistent with the SM, with a $p$-value $p_{\text{SM}} = 34\%$.

\FloatBarrier
\subsection{Measurement of Higgs boson couplings assuming only SM decays}
\label{sec:kappa:no_Biu}

A more general model introduces independent $\kappa$ modifiers for Higgs boson couplings to each of $W$ and $Z$ bosons, top, bottom and charm quarks, $\tau$-leptons and muons. A positive sign is assumed by convention for $\kappa_W$. For $\kappa_Z$ and $\kappa_t$ only positive values are considered, since the negative regions are experimentally excluded~\cite{HIGG-2021-21,CMS-HIG-23-007,HIGG-2020-16}. For $\kappa_b$, $\kappa_\tau$ and $\kappa_\mu$ the combination is sensitive to the absolute values of the modifiers but has almost no sensitivity to their signs. The fit is allowed to explore both positive and negative values; the absolute value of the envelope of the negative and positive regions of the confidence interval is reported. The $\kappa_c$ modifier is however allowed to range over both positive and negative values since it is only weakly constrained in the combination and the $\kappa_c=0$ scenario cannot be excluded at the considered confidence levels.

Two scenarios are explored for the effective modifiers $\kappa_g$, $\kappa_\gamma$ and $\kappa_{Z\gamma}$. Firstly, a \emph{resolved} parameterisation similar to the first model above, in which the effective modifiers are expressed as functions of the other modifiers using expressions derived from the corresponding leading-order loop contributions within the SM. Secondly, an \emph{effective} parameterisation in which they are instead considered as independent parameters in the fit, providing sensitivity to potential contributions from new phenomena beyond the SM (BSM) in the loop processes that would not be captured by the resolved expressions. In the resolved parameterisation, $\kappa_c$ is either included as a measurement parameter, or set equal to $\kappa_t$. The assumption $\kappa_c = \kappa_t$ corresponds to treating the charm and top quarks as having equal coupling modifiers, motivated by the fact that both form the up-type components of $SU(2)_L$ doublets, respectively of the second and third quark generations.

\begin{figure}[tbp]
\centering
\includegraphics[width=0.8\columnwidth]{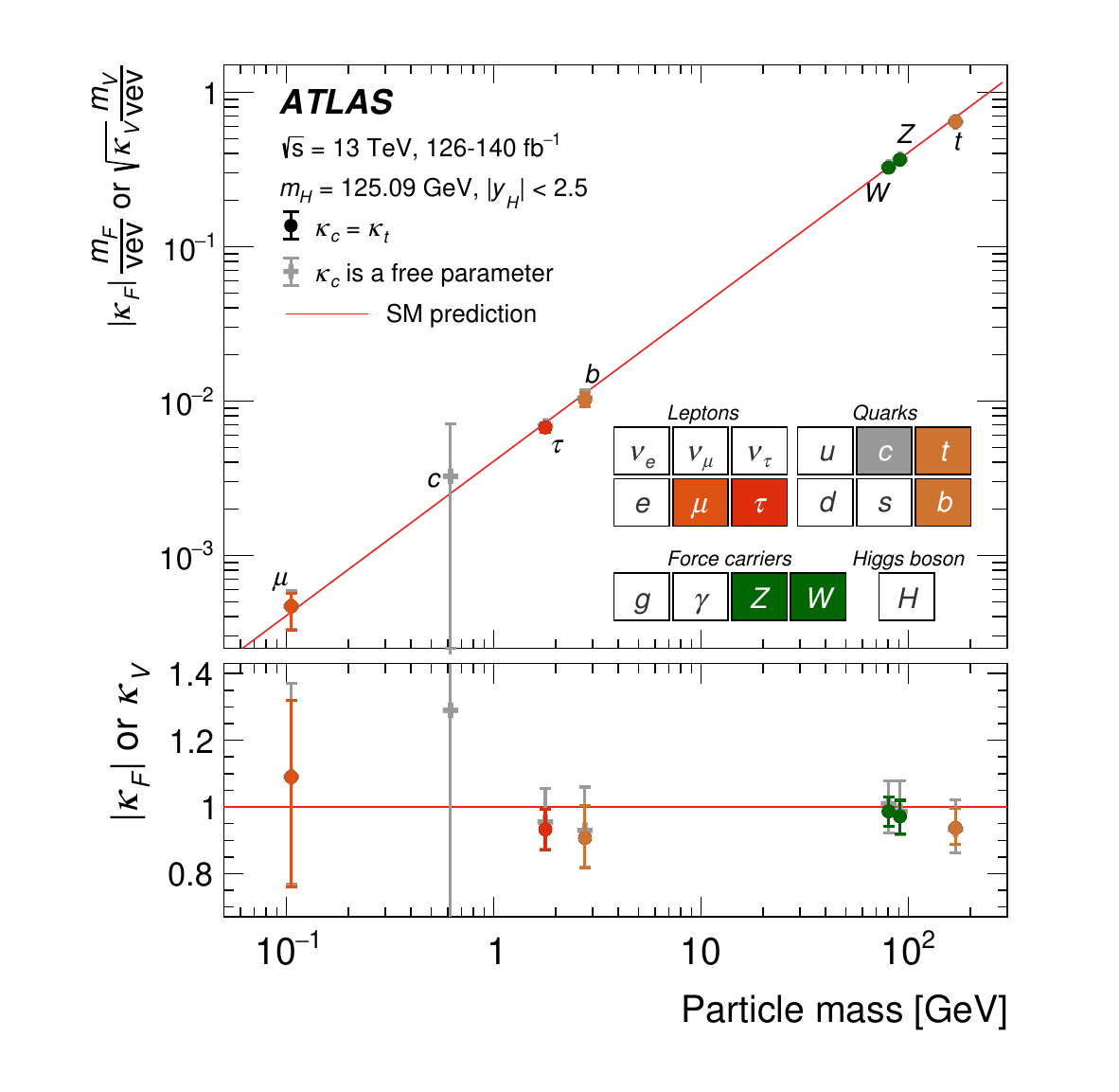}
\caption{Observed values of Higgs boson coupling modifiers as a function of the corresponding particle mass. In the top panel, the values $|\kappa_F| m_F / v$ and $\sqrt{\kappa_V} m_V/v$ are shown respectively for fermion and boson couplings, where $\kappa_F$ and $\kappa_V$ are the coupling modifiers, $m_F$ and $m_V$ are the particle masses, and $v = \SI{246}{\GeV}$ is the vacuum expectation value of the Higgs field. Quark masses are evaluated in the $\overline{\text{MS}}$ scheme at the scale $m_H$, while physical masses are used in other cases. The bottom panel shows the raw coupling modifiers. Points in light grey colour show the results for the case in which the $\kappa_c$ modifier is a free parameter in the model, while points in dark colours correspond to the case in which $\kappa_c = \kappa_t$. The resolved parameterisation described in the text is used in both cases and no BSM contributions to Higgs boson decays are considered except for those encoded in the coupling modifiers.}
\label{fig:kappas_mass}
\end{figure}

\begin{figure}[tbp]
\centering
\includegraphics[width=0.8\columnwidth]{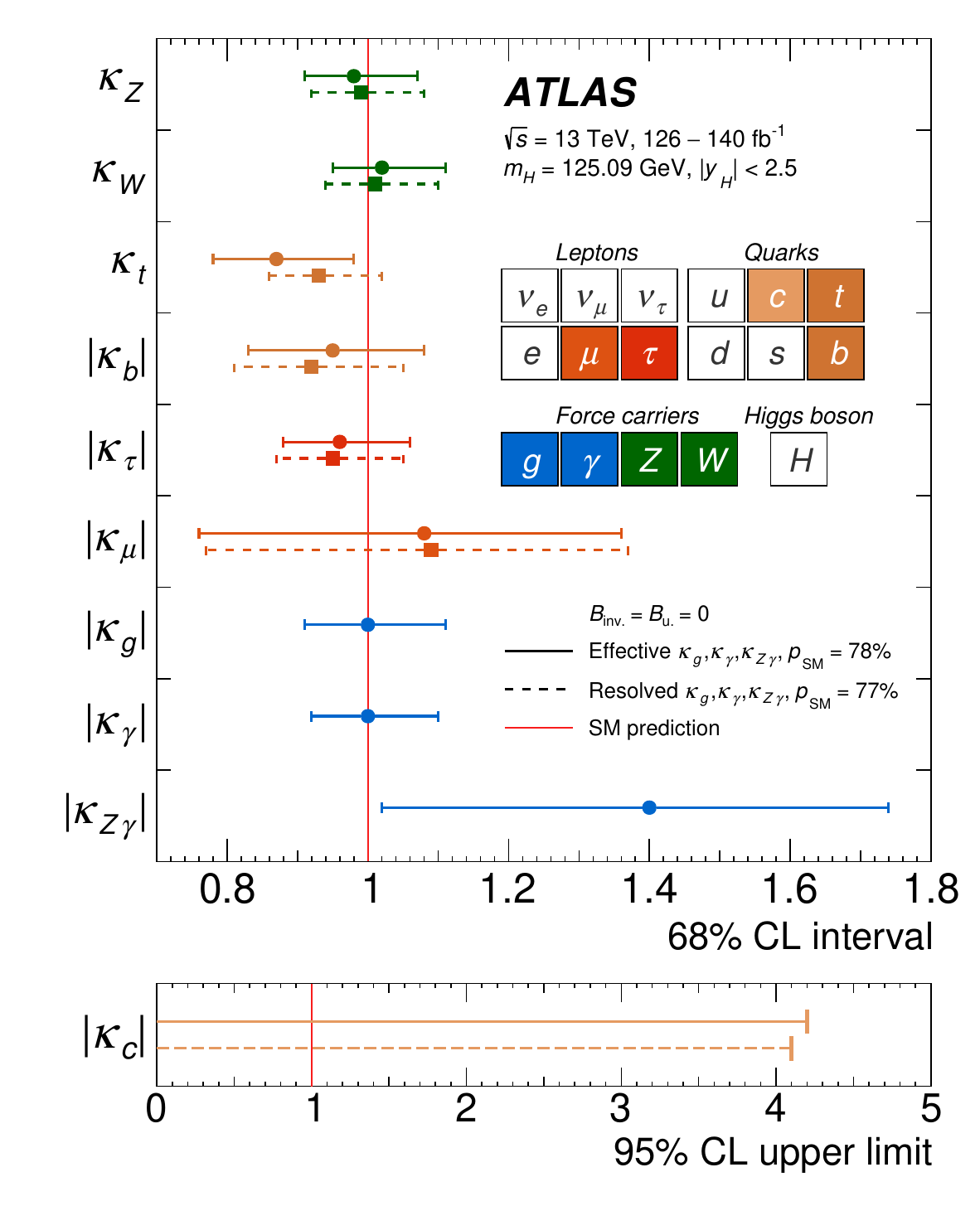}
\caption{Observed values of the Higgs boson coupling modifiers for the case in which $\kappa_c$ is included as a free parameter and no BSM contributions to Higgs boson decays are considered except for those encoded in the coupling modifiers.
Results are shown for the effective parameterisation of the loop-induced couplings $\kappa_g$,
$\kappa_\gamma$ and $\kappa_{Z\gamma}$ (round markers and solid error bars), and for the resolved parameterisation in which these
effective couplings are expressed in terms of the coupling modifiers of the particles contributing
to the corresponding loop processes in the SM (square markers and dashed error bars). The intervals correspond to 68\% CL uncertainties.
The SM prediction is indicated by the vertical red line.}
\label{fig:kappa_effective}
\end{figure}

The observed values of the coupling modifiers in the resolved parameterisation are shown as a function of the corresponding particle masses in Figure~\ref{fig:kappas_mass}.
The measured values for both parameterisations are summarised in Table~\ref{tab:kappas_obs} and illustrated in Figure~\ref{fig:kappa_effective} for the case in which $\kappa_c$ is included as a measurement parameter. The case of $\kappa_c = \kappa_t$ is shown in Figure~\ref{fig:kappa:generic_effective_resolved} in Appendix~\ref{app:kappa}. The results are consistent with SM expectations, with $p_{\text{SM}} = 77\%$ ($69\%$) for the resolved parameterisation with (without) $\kappa_c$ as a free parameter, and $p_{\text{SM}} = 78\%$ ($70\%$) for the effective parameterisation with (without) $\kappa_c$ as a free parameter. The uncertainty in the charm coupling modifier $\kappa_c$ is reduced by a factor of about two compared to the results of Ref.~\cite{HIGG-2021-23}, mainly due to the improved constraints on the \VHcc\ process from the analysis of Ref.~\cite{HIGG-2020-20}. The uncertainty in $\kappa_t$ is improved by about 15\% relative to Ref.~\cite{HIGG-2021-23}, mainly due to the reanalysed \ttH, \Hbb\ ~\cite{HIGG-2020-24} and the full Run~2 \ttH, \HML~\cite{HIGP-2024-08} analyses.
\begin{DIFnomarkup}
\begin{table}[tbp]
\caption{Observed values of Higgs boson coupling modifiers. The second and third columns correspond to the resolved parameterisation, in which the modifiers $\kappa_g, \kappa_\gamma$ and $\kappa_{Z\gamma}$ are expressed as functions of the other modifiers. The fourth and fifth columns correspond to the effective parameterisation in which they are treated as independent parameters. In the second and fourth columns, $\kappa_c$ is treated as a free parameter, while in the third and fifth columns it is set equal to $\kappa_t$.}
\label{tab:kappas_obs}
\renewcommand{\arraystretch}{1.5}
\begin{center}
\begin{tabular}{lcccc}
\toprule
\multirow{2}{*}{Parameter} & \multicolumn{2}{c}{Resolved $\kappa_g, \kappa_\gamma, \kappa_{Z\gamma}$} &
\multicolumn{2}{c}{Effective $\kappa_g, \kappa_\gamma, \kappa_{Z\gamma}$} \\ \cmidrule{2-5}
& Free $\kappa_c$ & $\kappa_c = \kappa_t$ & Free $\kappa_c$ & $\kappa_c = \kappa_t$  \\ \midrule
$\kappa_Z$            & $0.99$\numpmerr{+0.09}{-0.07} & $0.98\,{\scriptstyle \pm 0.05}$        & $0.98$\numpmerr{+0.09}{-0.07}   & $ 0.97\,{\scriptstyle \pm 0.05}$        \\
$\kappa_W$            & $1.01$\numpmerr{+0.09}{-0.07} & $1.00\,{\scriptstyle \pm 0.04}$        & $1.02$\numpmerr{+0.09}{-0.07}   & $1.01$\numpmerr{+0.05}{-0.04}        \\ \midrule
$\kappa_t$            & $0.93$\numpmerr{+0.09}{-0.07} & $0.93$\numpmerr{+0.06}{-0.05} & $0.87$\numpmerr{+0.11}{-0.09}   & $ 0.86\,{\scriptstyle \pm 0.08}$        \\
$|\kappa_b|$          & $0.92$\numpmerr{+0.13}{-0.11} & $0.91$\numpmerr{+0.10}{-0.09} & $0.95$\numpmerr{+0.13}{-0.12}   & $ 0.93\,{\scriptstyle \pm 0.10}$ \\
$|\kappa_\tau|$       & $0.95$\numpmerr{+0.10}{-0.08} & $0.94\,{\scriptstyle \pm 0.06}$        & $0.96$\numpmerr{+0.10}{-0.08}   & $0.95$\numpmerr{+0.07}{-0.06}        \\
$|\kappa_\mu|$        & $1.09$\numpmerr{+0.28}{-0.32} & $1.08$\numpmerr{+0.25}{-0.31} & $1.08$\numpmerr{+0.28}{-0.32}   & $1.07$\numpmerr{+0.25}{-0.31} \\ \midrule
$|\kappa_g|$          &                    --- &                    --- & $1.00$\numpmerr{+0.11}{-0.09}   & $ 0.99\,{\scriptstyle \pm 0.07}$ \\
$|\kappa_\gamma|$     &                    --- &                    --- & $1.00$\numpmerr{+0.10}{-0.08}   & $ 0.99\,{\scriptstyle \pm 0.06}$ \\
$|\kappa_{Z\gamma}|$  &                    --- &                    --- & $1.40$\numpmerr{+0.34}{-0.38}   & $1.38$\numpmerr{+0.31}{-0.36} \\ \midrule
$|\kappa_c|$ (95\% CL)& $< 4.1$                &                     ---& $<4.2$         &                 ---     \\
\bottomrule
\end{tabular}
\end{center}
\end{table}
\end{DIFnomarkup}

Tables~\ref{tab:kappas_obs_breakdown} and~\ref{tab:kappas_5B5F_breakdown} show the decomposition of the uncertainties in the observed coupling modifiers, respectively for the resolved and effective parameterisations and considering both the free-$\kappa_c$ and $\kappa_c = \kappa_t$ models. Uncertainty contributions from data statistics (stat.), experimental systematic uncertainties (exp.), signal theory uncertainties (sig. theo.) and background theory uncertainties (bkg. theo.) are considered. The observed values of $\kappa_W$ and $\kappa_Z$ shown in this section are all somewhat lower than could be expected from the value of the common modifier $\kappa_V$ obtained in Section~\ref{sec:kappa:kvkf}. This is related to the best-fit value of $\kappa_b$ being below unity in all cases, which is associated with a computed value of the total Higgs boson width that is below its SM expectation. Since the total width enters the denominator of the expressions for the Higgs boson event rates, this also leads to smaller values for the coupling modifiers entering in the numerator.
\begin{DIFnomarkup}
\begin{table}[tbp]
\caption{Decomposition of the observed uncertainties in the coupling modifiers into contributions from data statistics (stat.), experimental systematic uncertainties (exp.), signal theory uncertainties (sig.\ theo.) and background theory uncertainties (bkg.\ theo.), for the resolved parameterisation with $\kappa_c$ as a free parameter and $\kappa_c = \kappa_t$, respectively. The decomposition for $|\kappa_c|$ shows one-sided uncertainties at 95\% CL; for all other rows, the uncertainties are shown at 68\% CL. Central values are given in Table~\ref{tab:kappas_obs}.}
\label{tab:kappas_obs_breakdown}
\centering
\renewcommand{\arraystretch}{1.5}
\resizebox{\textwidth}{!}{%
\begin{tabular}{lcccc|cccc}
\toprule
& \multicolumn{4}{c}{Resolved $\kappa_g, \kappa_\gamma, \kappa_{Z\gamma}$ — Free $\kappa_c$} & \multicolumn{4}{c}{Resolved $\kappa_g, \kappa_\gamma, \kappa_{Z\gamma}$ — $\kappa_c = \kappa_t$} \\
\midrule
Parameter & Stat. & Exp. & Sig.\ theo. & Bkg.\ theo. & Stat. & Exp. & Sig.\ theo. & Bkg.\ theo. \\
\midrule
$\kappa_Z$      & \numpmerr{+0.07}{-0.06} & \numpmerr{+0.04}{-0.024} & \numpmerr{+0.028}{-0.023} & \numpmerr{+0.04}{-0.026} & ${\scriptstyle \pm 0.04}$ & ${\scriptstyle \pm 0.018}$ & \numpmerr{+0.022}{-0.020} & \numpmerr{+0.019}{-0.019}\\
$\kappa_W$      & \numpmerr{+0.06}{-0.05} & \numpmerr{+0.04}{-0.025} & ${\scriptstyle \pm 0.022}$ & \numpmerr{+0.04}{-0.025} & ${\scriptstyle \pm 0.029}$ & ${\scriptstyle \pm 0.018}$ & \numpmerr{+0.020}{-0.019} & \numpmerr{+0.018}{-0.018}\\
\midrule
$\kappa_t$      & \numpmerr{+0.06}{-0.05} & \numpmerr{+0.04}{-0.028} & \numpmerr{+0.04}{-0.03} & \numpmerr{+0.04}{-0.026} & ${\scriptstyle \pm 0.03}$ & \numpmerr{+0.022}{-0.023} & \numpmerr{+0.03}{-0.029} & \numpmerr{+0.021}{-0.021}\\
$|\kappa_b|$    & ${\scriptstyle \pm 0.08}$ & ${\scriptstyle \pm 0.06}$ & \numpmerr{+0.05}{-0.011} & \numpmerr{+0.06}{-0.05} & ${\scriptstyle \pm 0.06}$ & ${\scriptstyle \pm 0.04}$ & \numpmerr{+0.05}{-0.04} & \numpmerr{+0.05}{-0.04}\\
$|\kappa_\tau|$ & \numpmerr{+0.07}{-0.06} & \numpmerr{+0.05}{-0.04} & ${\scriptstyle \pm 0.03}$ & \numpmerr{+0.04}{-0.025} & ${\scriptstyle \pm 0.04}$ & \numpmerr{+0.030}{-0.029} & \numpmerr{+0.030}{-0.026} & \numpmerr{+0.020}{-0.019}\\
$|\kappa_\mu|$  & \numpmerr{+0.26}{-0.3} & \numpmerr{+0.07}{-0.06} & \numpmerr{+0.04}{-0.03} & \numpmerr{+0.06}{-0.03} & \numpmerr{+0.24}{-0.3} & \numpmerr{+0.05}{-0.06} & \numpmerr{+0.04}{-0.028} & \numpmerr{+0.026}{-0.013}\\
\midrule
$|\kappa_c|$ (95\% CL) & $+2.3$        & $ +1.4$              & $+0.6$              & $ +1.3$ & \multicolumn{4}{c}{---} \\
\bottomrule
\end{tabular}}
\end{table}
\end{DIFnomarkup}
\begin{DIFnomarkup}
\begin{table}[tbp]
\caption{Decomposition of the observed uncertainties in the coupling modifiers into contributions from data statistics (stat.), experimental systematic uncertainties (exp.), signal theory uncertainties (sig.\ theo.) and background theory uncertainties (bkg.\ theo.), for the effective parameterisation with $\kappa_c$ as a free parameter and $\kappa_c = \kappa_t$, respectively. The decomposition for $|\kappa_c|$ shows one-sided uncertainties at 95\% CL; for all other rows, the uncertainties are shown at 68\% CL. Central values are given in Table~\ref{tab:kappas_obs}.}
\label{tab:kappas_5B5F_breakdown}
\centering
\renewcommand{\arraystretch}{1.5}
\resizebox{\textwidth}{!}{%
\begin{tabular}{lcccc|cccc}
\toprule
& \multicolumn{4}{c}{Effective $\kappa_g, \kappa_\gamma, \kappa_{Z\gamma}$ — Free $\kappa_c$} & \multicolumn{4}{c}{Effective $\kappa_g, \kappa_\gamma, \kappa_{Z\gamma}$ — $\kappa_c = \kappa_t$} \\
\midrule
Parameter & Stat. & Exp. & Sig.\ theo. & Bkg.\ theo. & Stat. & Exp. & Sig.\ theo. & Bkg.\ theo. \\
\midrule
$\kappa_Z$             & \numpmerr{+0.07}{-0.06} & \numpmerr{+0.04}{-0.024} & \numpmerr{+0.025}{-0.022} & \numpmerr{+0.04}{-0.024} & ${\scriptstyle \pm 0.04}$ & \numpmerr{+0.019}{-0.018} & \numpmerr{+0.022}{-0.019} & \numpmerr{+0.019}{-0.019}\\
$\kappa_W$             & \numpmerr{+0.07}{-0.06} & \numpmerr{+0.05}{-0.028} & ${\scriptstyle \pm 0.024}$ & \numpmerr{+0.04}{-0.028} & ${\scriptstyle \pm 0.03}$ & ${\scriptstyle \pm 0.021}$ & \numpmerr{+0.021}{-0.020} & \numpmerr{+0.021}{-0.021}\\
\midrule
$\kappa_t$             & ${\scriptstyle \pm 0.07}$ & ${\scriptstyle \pm 0.04}$ & \numpmerr{+0.05}{-0.04} & \numpmerr{+0.05}{-0.04} & \numpmerr{+0.05}{-0.05} & \numpmerr{+0.026}{-0.03} & \numpmerr{+0.05}{-0.04} & \numpmerr{+0.03}{-0.03}\\
$|\kappa_b|$           & \numpmerr{+0.09}{-0.08} & \numpmerr{+0.06}{-0.05} & \numpmerr{+0.05}{-0.04} & \numpmerr{+0.06}{-0.05} & ${\scriptstyle \pm 0.07}$ & ${\scriptstyle \pm 0.04}$ & \numpmerr{+0.04}{-0.03} & ${\scriptstyle \pm 0.05} $\\
$|\kappa_\tau|$        & \numpmerr{+0.07}{-0.06} & \numpmerr{+0.05}{-0.04} & ${\scriptstyle \pm 0.03}$ & \numpmerr{+0.04}{-0.026} & ${\scriptstyle \pm 0.04}$ & ${\scriptstyle \pm 0.03}$ & \numpmerr{+0.03}{-0.027} & \numpmerr{+0.021}{-0.019}\\
$|\kappa_\mu|$         & \numpmerr{+0.26}{-0.3} & \numpmerr{+0.07}{-0.05} & ${\scriptstyle \pm 0.04}$ & \numpmerr{+0.06}{-0.04} & \numpmerr{+0.24}{-0.3} & \numpmerr{+0.05}{-0.06} & \numpmerr{+0.04}{-0.026} & \numpmerr{+0.026}{-0.012}\\
\midrule
$|\kappa_g|$             & \numpmerr{+0.07}{-0.06} & \numpmerr{+0.05}{-0.04} & ${\scriptstyle \pm 0.04}$ & \numpmerr{+0.05}{-0.03} & ${\scriptstyle \pm 0.04}$ & ${\scriptstyle \pm 0.029}$ & ${\scriptstyle \pm 0.04}$ & \numpmerr{+0.024}{-0.023}\\
$|\kappa_\gamma|$        & \numpmerr{+0.07}{-0.06} & \numpmerr{+0.05}{-0.03} & \numpmerr{+0.03}{-0.027} & \numpmerr{+0.04}{-0.025} & ${\scriptstyle \pm 0.04}$ & \numpmerr{+0.028}{-0.026} & \numpmerr{+0.026}{-0.024} & \numpmerr{+0.020}{-0.019}\\
$|\kappa_{Z\gamma}|$  & \numpmerr{+0.30}{-0.3} & \numpmerr{+0.12}{-0.13} & \numpmerr{+0.08}{-0.06} & \numpmerr{+0.07}{-0.05} & \numpmerr{+0.28}{-0.3} & \numpmerr{+0.10}{-0.13} & \numpmerr{+0.08}{-0.05} & \numpmerr{+0.03}{-0.012}\\
\midrule
$|\kappa_c|$ (95\% CL)  & $+2.2$             & $ +1.3$              & $+0.6$              & $ +1.3$ & \multicolumn{4}{c}{---} \\
\bottomrule
\end{tabular}}
\end{table}
\end{DIFnomarkup}

\FloatBarrier
\subsection{Constraints on new phenomena in Higgs boson decays}
\label{sec:kappa:bsm_decays}

\begin{figure}[tbp]
\centering
\includegraphics[width=0.8\columnwidth]{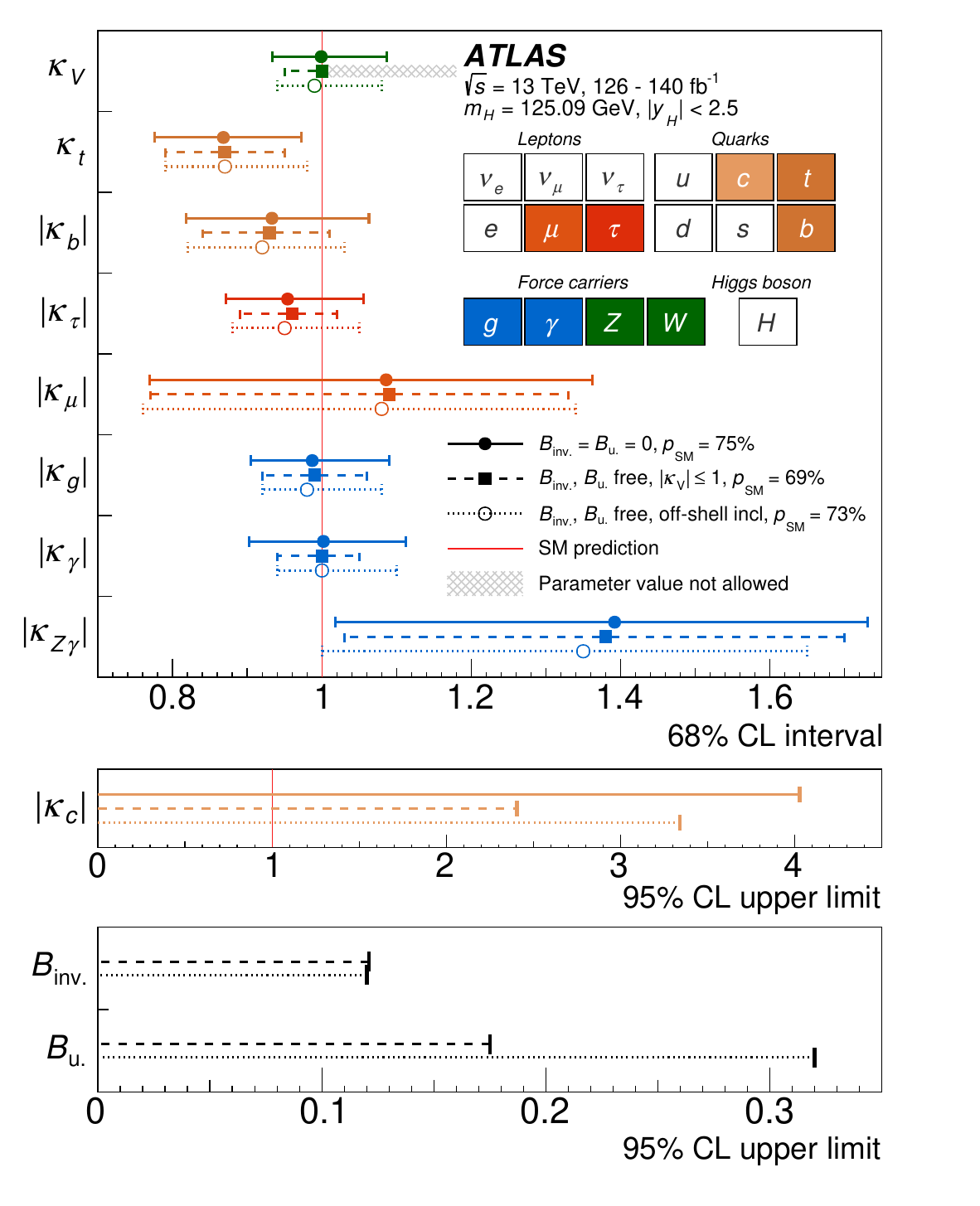}
\caption{Observed values of Higgs boson coupling modifiers in the effective parameterisation with unified vector boson couplings ($\kappa_V = \kappa_W = \kappa_Z$) and $\kappa_c$ free. Three scenarios are shown: no BSM contributions to Higgs boson decays except those encoded in the coupling modifiers ($B_{\text{inv}} = B_{\text{u}} = 0$, solid circles); invisible ($B_{\text{inv}}$) and undetected ($B_{\text{u}}$) branching ratios free and $\kappa_V \leq 1$ (solid squares); $B_{\text{inv}}$ and $B_{\text{u}}$ free with \Hinv, off-shell Higgs boson production included, and no $\kappa_V \leq 1$ constraint (open circles). The SM expectation is indicated by the vertical red line.}
\label{fig:kappaV_effective}
\end{figure}

This section considers the presence of BSM contributions to Higgs boson decays into final states that are either invisible (decays producing particles that are invisible to the detector components) or undetected (decays producing detectable particles that are not targeted by any analysis included in the combination). The corresponding branching ratios are denoted respectively as $B_{\text{inv}}$ and $B_{\text{u}}$. While in the previous section separate modifiers $\kappa_W$ and $\kappa_Z$ were considered, the models described here make use of a single parameter $\kappa_V = \kappa_W = \kappa_Z$. This is due to the fact that some of the analyses included in this model that target off-shell Higgs boson production are designed with a unified vector-boson coupling modifier; modifying these analyses to enable the use of separate $\kappa_W$ and $\kappa_Z$ parameters is not feasible within the scope of this combination. Three scenarios are considered in the effective parameterisation, which is adopted here to account for potential BSM contributions to the loop-induced couplings ($\kappa_g$, $\kappa_\gamma$ and $\kappa_{Z\gamma}$) that would not be captured by the resolved loop expressions:
\begin{enumerate}
\item \textit{Scenario~1} ($B_{\text{inv}} = B_{\text{u}} = 0$): no BSM contributions to Higgs boson decays are considered except for those encoded in the coupling modifiers. This scenario differs from the results of Section~\ref{sec:kappa:no_Biu} solely through the unification of the vector boson coupling modifiers into a common $\kappa_V = \kappa_W = \kappa_Z$, and serves as a reference case connecting this section to the previous one.
\item \textit{Scenario~2} ($B_{\text{inv}}$ and $B_{\text{u}}$ free, $\kappa_V \leq 1$): the invisible and undetected branching ratios are left free, with the constraint $\kappa_V \leq 1$ imposed to lift the intrinsic degeneracy between the changes to the coupling modifiers and to the total Higgs boson width when only on-shell measurements are considered. The analyses targeting \Hinv~\cite{EXOT-2020-11,EXOT-2021-17,HIGG-2018-26,SUSY-2019-12} are included in the combination in order to constrain $B_{\text{inv}}$.
\item \textit{Scenario~3} ($B_{\text{inv}}$ and $B_{\text{u}}$ free, no $\kappa_V \leq 1$ constraint): $B_{\text{inv}}$ and $B_{\text{u}}$ are both left free without the $\kappa_V \leq 1$ assumption. The analyses targeting \Hinv~\cite{EXOT-2020-11,EXOT-2021-17,HIGG-2018-26,SUSY-2019-12} and off-shell Higgs boson production~\cite{HIGP-2024-14,HIGP-2024-05} are both included, with the latter providing a constraint on the total width that plays a similar role to the $\kappa_V \leq 1$ assumption.
\end{enumerate}
The measured coupling modifiers for these three scenarios are summarised in Figure~\ref{fig:kappaV_effective}.
The observed coupling modifier values and branching ratio limits for all three scenarios are summarised in Table~\ref{tab:kappas_bsm}.
In Scenario~2, the observed (expected) 95\% CL upper limit on the branching ratio to invisible final states is $B_{\text{inv}} < 12\%$ (7.6\%) and to undetected final states $B_{\text{u}} < 18\%$ (20\%). In Scenario~3, the corresponding observed (expected) upper limits are $B_{\text{inv}} < 12\%$ (7.5\%) and $B_{\text{u}} < 32\%$ (45\%). No correlation between $B_{\text{inv}}$ and $B_{\text{u}}$ is quoted since the best-fit value of $B_{\text{u}}$ lies at the boundary of its physical range, making the covariance matrix ill-defined.
\begin{table}[tbp]
\caption{Observed values of Higgs boson coupling modifiers in the effective parameterisation with unified vector boson couplings ($\kappa_V = \kappa_W = \kappa_Z$) and $\kappa_c$ free, for the three scenarios shown in Figure~\ref{fig:kappaV_effective}. \textit{Scenario~1}: no BSM contributions to Higgs boson decays except those encoded in the coupling modifiers. \textit{Scenario~2}: invisible ($B_{\text{inv}}$) and undetected ($B_{\text{u}}$) branching ratios free, $\kappa_V \leq 1$ assumed, and \Hinv\ (Hinv) search results included. \textit{Scenario~3}: $B_{\text{inv}}$ and $B_{\text{u}}$ branching ratios free, Hinv and off-shell Higgs boson production measurements included, no $\kappa_V \leq 1$ constraint. Uncertainties are given at 68\% CL and upper limits at 95\% CL.}
\label{tab:kappas_bsm}
\centering
\renewcommand{\arraystretch}{1.5}
\begin{tabular}{lccc}
\toprule
Parameter & Scenario 1 ($B_{\text{inv}}=B_{\text{u}}=0$) & Scenario 2 ($\kappa_V\leq 1$, Hinv) & Scenario 3 (Hinv+off-shell) \\
\midrule
$\kappa_V$        & $1.00$\numpmerr{+0.09}{-0.07} & $1.00$\numpmerr{+0.00}{-0.05} & $0.99$\numpmerr{+0.09}{-0.05} \\
$\kappa_t$        & $0.87$\numpmerr{+0.10}{-0.09} & $0.87\,{\scriptstyle \pm 0.08}$ & $0.87$\numpmerr{+0.11}{-0.08} \\
$|\kappa_b|$      & $0.93$\numpmerr{+0.13}{-0.12} & $0.93$\numpmerr{+0.08}{-0.09} & $0.92$\numpmerr{+0.11}{-0.10} \\
$|\kappa_\tau|$   & $0.95$\numpmerr{+0.10}{-0.08} & $0.96$\numpmerr{+0.06}{-0.07} & $0.95$\numpmerr{+0.10}{-0.07} \\
$|\kappa_\mu|$    & $1.09$\numpmerr{+0.28}{-0.32} & $1.09$\numpmerr{+0.24}{-0.32} & $1.08$\numpmerr{+0.26}{-0.32} \\
$|\kappa_g|$        & $0.99$\numpmerr{+0.10}{-0.08} & $0.99\,{\scriptstyle \pm 0.07}$ & $0.98$\numpmerr{+0.10}{-0.06} \\
$|\kappa_\gamma|$   & $1.00$\numpmerr{+0.10}{-0.08} & $1.00$\numpmerr{+0.05}{-0.06} & $1.00$\numpmerr{+0.10}{-0.06} \\
$|\kappa_{Z\gamma}|$& $1.39$\numpmerr{+0.34}{-0.37} & $1.38$\numpmerr{+0.32}{-0.35} & $1.35$\numpmerr{+0.30}{-0.35} \\
$|\kappa_c|$      & $< 4.0$               & $< 2.4$                & $< 3.3$               \\
\midrule
$B_{\text{inv}}$  & $0$ (fixed)  & $< 12\%$   & $< 12\%$ \\
$B_{\text{u}}$    & $0$ (fixed)  & $< 18\%$   & $< 32\%$ \\
\bottomrule
\end{tabular}
\end{table}

In Scenario~3, the inclusion of the off-shell Higgs boson production measurement provides a constraint on the total Higgs boson width through the assumption that the coupling modifier $\kappa_V$ is the same in the on-shell and off-shell kinematic regimes ($\kappa_V^{\text{on-shell}} = \kappa_V^{\text{off-shell}}$).
In addition, the kinematics of on- and off-shell Higgs boson production are assumed to match their SM predictions. This implies in particular that the particles responsible for any non-zero $B_{\text{u}}$ should couple only feebly to quarks and gluons, motivated by the expectation that a BSM particle with sizeable couplings to quarks or gluons would already have been observed in direct searches. This constraint makes it possible to simultaneously determine the Higgs boson coupling modifiers and allow for non-zero $B_{\text{inv}}$ and $B_{\text{u}}$ without requiring the $\kappa_V \leq 1$ assumption, and without a significant deterioration in the precision of the coupling measurements. These results demonstrate that precise determinations of the Higgs boson coupling modifiers are achievable with only mild model assumptions, even in the absence of a direct measurement of the total width.

\FloatBarrier
\subsection{Model-independent coupling-strength ratios}
\label{sec:kappa:ratio}

The coupling modifiers of Section~\ref{sec:kappa:no_Biu} can be re-expressed as ratios that are independent of any assumption on the total Higgs boson width, providing the most model-independent measurements achievable within the $\kappa$-framework~\cite{YR3}. The measurement parameters are the overall coupling scale $\kappa_{gZ} \equiv \kappa_g\kappa_Z/\kappa_H$, where $\kappa_H \equiv \sqrt{\Gamma_H/\Gamma_H^{\text{SM}}}$ parametrizes the total Higgs boson decay width, and ratios of individual couplings, as defined in Table~\ref{tab:kappa_ratio_obs}. All parameters are assumed to take positive values. Two models are presented: one fixing $\kappa_c = \kappa_t$ (the minimal flavour assumption described in Section~\ref{sec:kappa:no_Biu}) and one in which the ratio $\lambda_{cb} \equiv \kappa_c/\kappa_b$ is a free parameter, providing a direct test of the universality of the second- and third-generation down-type Yukawa couplings. The measurement parameters form the complete set of width-independent combinations of individual coupling modifiers to which the analyses included in the combination are sensitive.

\begin{figure}[tbp]
\centering
\includegraphics[width=0.8\columnwidth]{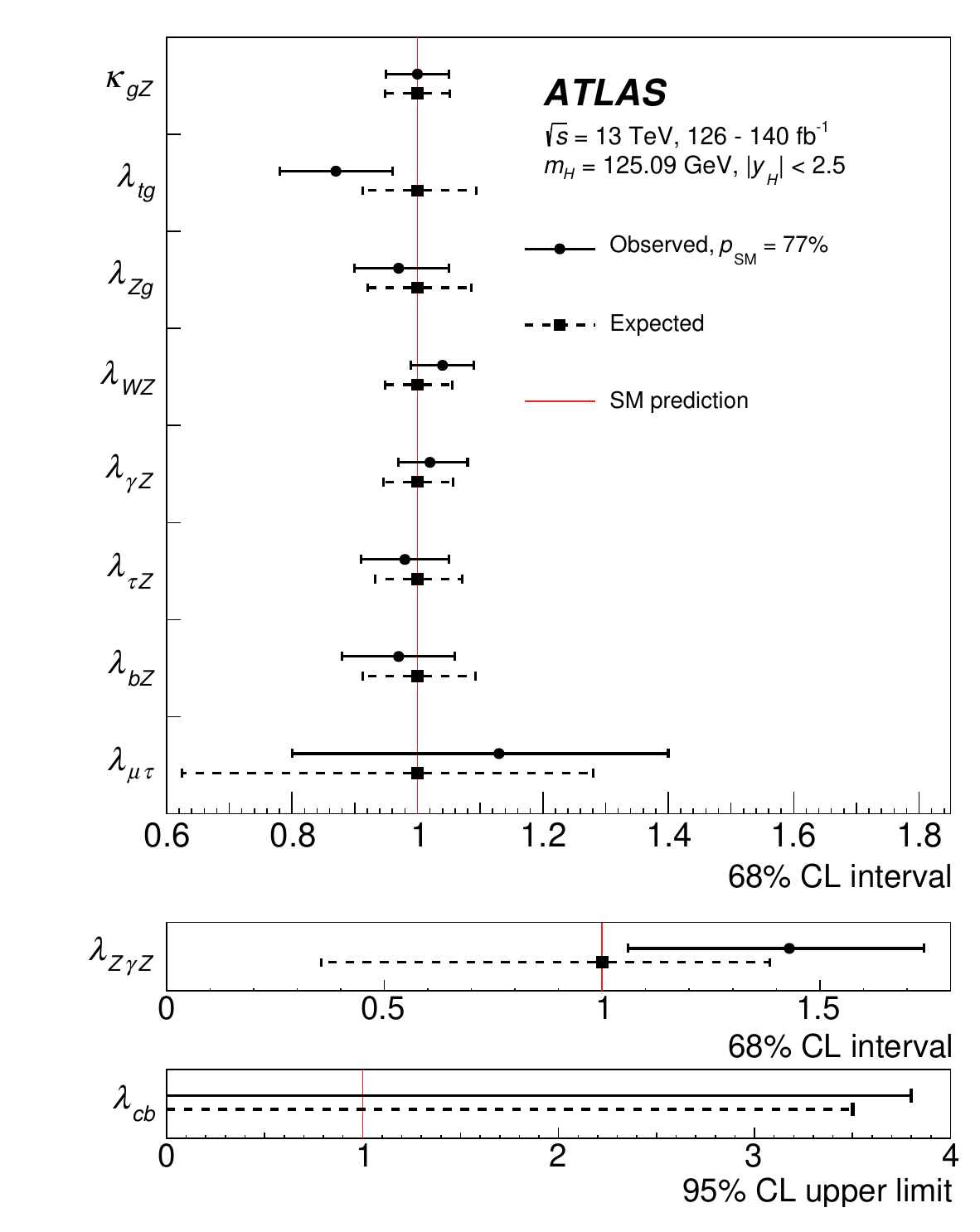}
\caption{Observed (solid circles) and expected (dashed squares) values of the model-independent coupling-ratio parameters for the model with $\lambda_{cb} \equiv \kappa_c/\kappa_b$ free. The main panel shows 68\% CL intervals for eight parameters on a common axis; $\lambda_{Z\gamma Z}$ is shown in a separate panel due to its wider uncertainty range. The bottom panel shows the 95\% CL upper limit on $|\lambda_{cb}|$. The SM expectation is indicated by the vertical red line. All results are consistent with SM expectations, with $p_{\text{SM}} = 77\%$.}
\label{fig:kappa_ratio_summary}
\end{figure}

The observed and expected results for the model with free $\lambda_{cb}$ are shown in Figure~\ref{fig:kappa_ratio_summary} and Table~\ref{tab:kappa_ratio_obs}, together with the full decomposition of observed uncertainties into statistical (stat.), experimental (exp.), signal theory (sig.\ theo.) and background theory (bkg.\ theo.) components. All parameters are consistent with SM expectations, with $p_{\text{SM}} = 77\%$. The ratio $\lambda_{cb} \equiv \kappa_c/\kappa_b$ is constrained to $\lambda_{cb} < 3.8$ at 95\% CL, providing a direct test of the universality of the charm and bottom Yukawa couplings. Results for the model with $\kappa_c = \kappa_t$ ($p_{\text{SM}} = 70\%$) are given in Table~\ref{tab:kappa_ratio_obs_noRcb} in Appendix~\ref{app:kappa}; the remaining coupling-ratio parameters in this model are consistent with those obtained from the free-$\lambda_{cb}$ fit.
\begin{table}[tbp]
\caption{Definitions and observed values of the model-independent coupling-ratio parameters for the model with free $\lambda_{cb} \equiv \kappa_c/\kappa_b$, together with the expected total uncertainty (the SM central value is 1 for all parameters by construction). For $\lambda_{cb}$, the third and last columns show respectively the observed and expected 95\% CL upper limit in this parameter.
Observed uncertainties are decomposed into contributions from data statistics (stat.), experimental systematic uncertainties (exp.), signal theory uncertainties (sig.\ theo.) and background theory uncertainties (bkg.\ theo.). One-sided 95\% CL uncertainty decompositions are shown for $\lambda_{cb}$, while 68\% CL decompositions are shown for the remaining coupling-ratio parameters. All results are consistent with the SM.}
\label{tab:kappa_ratio_obs}
\centering
\renewcommand{\arraystretch}{1.5}
\resizebox{\textwidth}{!}{%
\begin{tabular}{llcccccc}
\toprule
Parameter & Definition & Observed & Stat. & Exp. & Sig.\ theo. & Bkg.\ theo. & Expected unc. \\
\midrule
$\kappa_{gZ}$         & $\kappa_g\kappa_Z/\kappa_H$ & $1.001$\numpmerr{+0.054}{-0.049} & ${\scriptstyle \pm 0.038}$ & ${\scriptstyle \pm 0.014}$ & \numpmerr{+0.035}{-0.025} & \numpmerr{+0.008}{-0.009} & ${\scriptstyle \pm 0.05}$ \\
$\lambda_{tg}$        & $\kappa_t/\kappa_g$         & $0.866$\numpmerr{+0.090}{-0.086} & ${\scriptstyle \pm 0.060}$ & \numpmerr{+0.030}{-0.031} & \numpmerr{+0.052}{-0.044} & \numpmerr{+0.031}{-0.032} & ${\scriptstyle \pm 0.09}$ \\
$\lambda_{Zg}$        & $\kappa_Z/\kappa_g$         & $0.973$\numpmerr{+0.084}{-0.075} & \numpmerr{+0.054}{-0.052} & ${\scriptstyle \pm 0.032}$ & \numpmerr{+0.046}{-0.040} & \numpmerr{+0.031}{-0.015} & \numpmerr{+0.09}{-0.08} \\
$\lambda_{WZ}$        & $\kappa_W/\kappa_Z$         & $1.042$\numpmerr{+0.055}{-0.054} & \numpmerr{+0.045}{-0.042} & \numpmerr{+0.024}{-0.019} & \numpmerr{+0.014}{-0.019} & \numpmerr{+0.016}{-0.020} & \numpmerr{+0.06}{-0.05} \\
$\lambda_{\gamma Z}$  & $\kappa_\gamma/\kappa_Z$    & $1.024$\numpmerr{+0.058}{-0.054} & \numpmerr{+0.050}{-0.048} & \numpmerr{+0.024}{-0.022} & ${\scriptstyle \pm 0.013}$ & \numpmerr{+0.011}{-0.006} & \numpmerr{+0.06}{-0.05} \\
$\lambda_{\tau Z}$    & $\kappa_\tau/\kappa_Z$      & $0.979$\numpmerr{+0.072}{-0.066} & \numpmerr{+0.054}{-0.051} & \numpmerr{+0.031}{-0.032} & \numpmerr{+0.031}{-0.025} & \numpmerr{+0.018}{-0.012} & ${\scriptstyle \pm 0.07}$ \\
$\lambda_{bZ}$        & $\kappa_b/\kappa_Z$         & $0.966$\numpmerr{+0.094}{-0.087} & \numpmerr{+0.069}{-0.065} & \numpmerr{+0.034}{-0.031} & \numpmerr{+0.038}{-0.034} & \numpmerr{+0.037}{-0.034} & ${\scriptstyle \pm 0.09}$ \\
$\lambda_{\mu\tau}$   & $\kappa_\mu/\kappa_\tau$    & $1.131$\numpmerr{+0.269}{-0.334} & \numpmerr{+0.257}{-0.323} & \numpmerr{+0.000}{-0.062} & \numpmerr{+0.086}{-0.044} & \numpmerr{+0.020}{-0.032} & \numpmerr{+0.28}{-0.38} \\
$\lambda_{Z\gamma Z}$ & $\kappa_{Z\gamma}/\kappa_Z$ & $1.427$\numpmerr{+0.314}{-0.374} & \numpmerr{+0.290}{-0.347} & \numpmerr{+0.098}{-0.131} & \numpmerr{+0.073}{-0.051} & \numpmerr{+0.001}{-0.000} & \numpmerr{+0.4}{-0.6} \\
\midrule
$\lambda_{cb}$ (95\% CL) & $\kappa_c/\kappa_b$ & $< 3.8$ & $+2.0$ & $+1.0$ & $+0.4$ & $+1.0$ & $< 3.5$ \\
\bottomrule
\end{tabular}}
\end{table}

\FloatBarrier
\subsection{Determination of the mass of the $b$-quark from Higgs boson couplings}
\label{sec:mb}

In Section~\ref{sec:kappa:no_Biu}, measurements of fermion masses were used to determine the SM expectations for the Higgs boson Yukawa couplings that are used to define the corresponding coupling modifiers. The agreement between these reference values and the measured values of the coupling modifiers is shown in Figure~\ref{fig:kappas_mass}. In this section, the logic is reversed and a measurement of $m_b(m_H)$, the running mass of the $b$-quark at the Higgs boson mass scale in the modified minimal subtraction (\MSbar) scheme, is inferred from Higgs boson coupling properties to the $b$-quark assuming that other Higgs boson properties match their SM predictions.

The measurement uses the same combination of analyses as described in Section~\ref{sec:proddecay}. Production cross-sections and decay branching ratios are parameterised using the production cross-sections $\sigma_i^{ZZ}$ in the \Hzz\ final state and the ratios of partial widths ${\Gamma_f}/{\Gamma_{\zz}}$. The production modes considered are \ggF, \VBF, \WH, \ZH, \ttH\ and \tH, defined as in Section~\ref{sec:prodXS} and the decay processes are \Hbb, \Hww, \Htt, \Hcc, \Hyy, \HZy\ and \Hmm. The rates of other production and decay processes are assumed to match their SM expectations. Under the narrow-width approximation, the event rate for production process $i$ and decay process $f$ can then be expressed as
\begin{equation}
\sigma_i^f = \sigma_i^{\zz} \left(\frac{\Gamma_f}{\Gamma_{\zz}}\right).
\end{equation}
A measurement of $m_b(m_H)$ is obtained by expressing ${\Gamma_{bb}}/{\Gamma_{\zz}}$ as a function of this parameter. The other ratios of partial widths and cross-sections depend only weakly on $m_b(m_H)$ and are left free to vary in the fit in order to reduce the sensitivity of the result to potential deviations from SM predictions.

The expression for ${\Gamma_{bb}}/{\Gamma_{\zz}}$ is obtained using HDECAY 6.61~\cite{Djouadi:1997yw,Spira:1997dg,Djouadi:2006bz}, which provides a computation of $\Gamma_{bb}$ to fifth order (N4LO) in QCD and next-to-leading-order (NLO) in electroweak corrections and allows $m_b(m_H)$ to be specified as an input parameter. Values of $\Gamma_{bb}$ are obtained by scanning the range $2.2 \le m_b(m_H) \le \qty{4.0}{\GeV}$ in steps of \qty{0.5}{\MeV}. The value $\alpha_s(m_Z) = 0.118$ is assumed for the strong coupling constant. A multiplicative correction is applied to $\Gamma_{bb}$ so that its value matches the recommendation from Ref.~\cite{YR4} at $m_b^{\text{ref}}(m_H) = \qty{2.787}{\GeV}$, which corresponds to the value $m_b(m_b) = \qty{4.18}{\GeV}$ assumed in the other results reported in this paper. The dependence is parameterised as
\begin{equation}
\frac{\Gamma_{bb}}{\Gamma_{\zz}} = \left(\frac{\Gamma_{bb}}{\Gamma_{\zz}}\right)_{\text{SM}} \left(1 + \sum\limits_{n=1}^5 a_n \left[\left(\frac{m_b(m_H)}{m_b^{\text{ref}}(m_H)}\right)^n - 1\right]\right)
\label{eq:GammabbGammaZZ}
\end{equation}
with the polynomial coefficients given in Table~\ref{tab:mb:param} in Appendix~\ref{app:mb}.
The fifth-order polynomial is an empirical choice providing a sufficiently flexible interpolation; the leading dependence is quadratic, originating from the dependence of $\Gamma_{bb}$ on the Yukawa coupling of the $b$-quark, but higher-order terms are needed to account for subleading effects. The result is stable under variations of the polynomial order from 3 to 6, with changes in $m_b(m_H)$ that are negligible compared with the experimental uncertainties.
The experimental and theory uncertainties are implemented as described in Section~\ref{sec:uncertainties}.
The observed value is
\begin{equation}
m_b(m_H) = \qty[parse-numbers=false]{2.73^{+0.27}_{-0.25}}{\GeV}
= \qty[parse-numbers=false]{2.73^{+0.22}_{-0.20}\text{ (stat.)}^{+0.15}_{-0.14}\text{ (syst.)}^{+0.07}_{-0.06}\text{ (theo.)}}{\GeV},
\end{equation}
where the uncertainty components correspond respectively to data statistics, experimental systematic uncertainty, and theory uncertainties. The results of the fit where ${\Gamma_{bb}}/{\Gamma_{\zz}}$ is considered as a free parameter rather than as a function of $m_b(m_H)$ are shown in Figure~\ref{fig:mb:Gamma:obs} in Appendix~\ref{app:kappa}.

Figure~\ref{fig:mb:obs} shows the measurement alongside determinations of the \MSbar\ mass $m_b$ at other energy scales.
The running of $m_b$ is probed using the parameterisation $m_b(\mu; x, m_b(m_b)) = x \left[ m_b^{\text{SM}}(\mu; m_b(m_b)) - m_b(m_b) \right] + m_b(m_b)$, following Refs.~\cite{Aparisi:2021tym,CMS-TOP-19-007}. The quantity $m_b^{\text{SM}}(\mu; m_b(m_b))$ is the prediction obtained from REvolver~\cite{Hoang:2021fhn} and $x$ is a parameter that allows to smoothly interpolate between the SM ($x=1$) and the scenario where $m_b$ does not run ($x=0$). The parameters $m_b(m_b)$ and $x$ are obtained from a fit to a dataset consisting of the value of $m_b(m_H)$ obtained in this section, the values of $m_b(m_Z)$ measured at LEP and SLC~\cite{Barate:2000ab,Abdallah:2008ac,Abbiendi:2001tw,Brandenburg:1999nb} and the PDG world average for $m_b(m_b)$~\cite{Zyla:2020zbs}. The result is shown in Figure~\ref{fig:mb:obs}. The best-fit values are $m_b(m_b) = \qty[parse-numbers=false]{4.18 \pm 0.02}{\GeV}$ and $x = 1.04 \pm 0.12\text{ (exp.)} \pm 0.05\text{ }(\alpha_s)$, where the first uncertainty component corresponds to the experimental uncertainties in the inputs and the second to the effect of varying $\alpha_s(m_Z)$ by $\pm 0.004$~\cite{Aparisi:2021tym}. The value of $m_b(m_b)$ is in agreement with the PDG average of $m_b(m_b) = \qty[parse-numbers=false]{4.18 ^{+0.03}_{-0.02}}{\GeV}$. The value of $x$ agrees with the SM expectation ($x=1$) and confirms the exclusion of the no-running scenario with a significance of more than 8 standard deviations.
\begin{figure}[tbp]
\centering
\includegraphics[width=.7\textwidth]{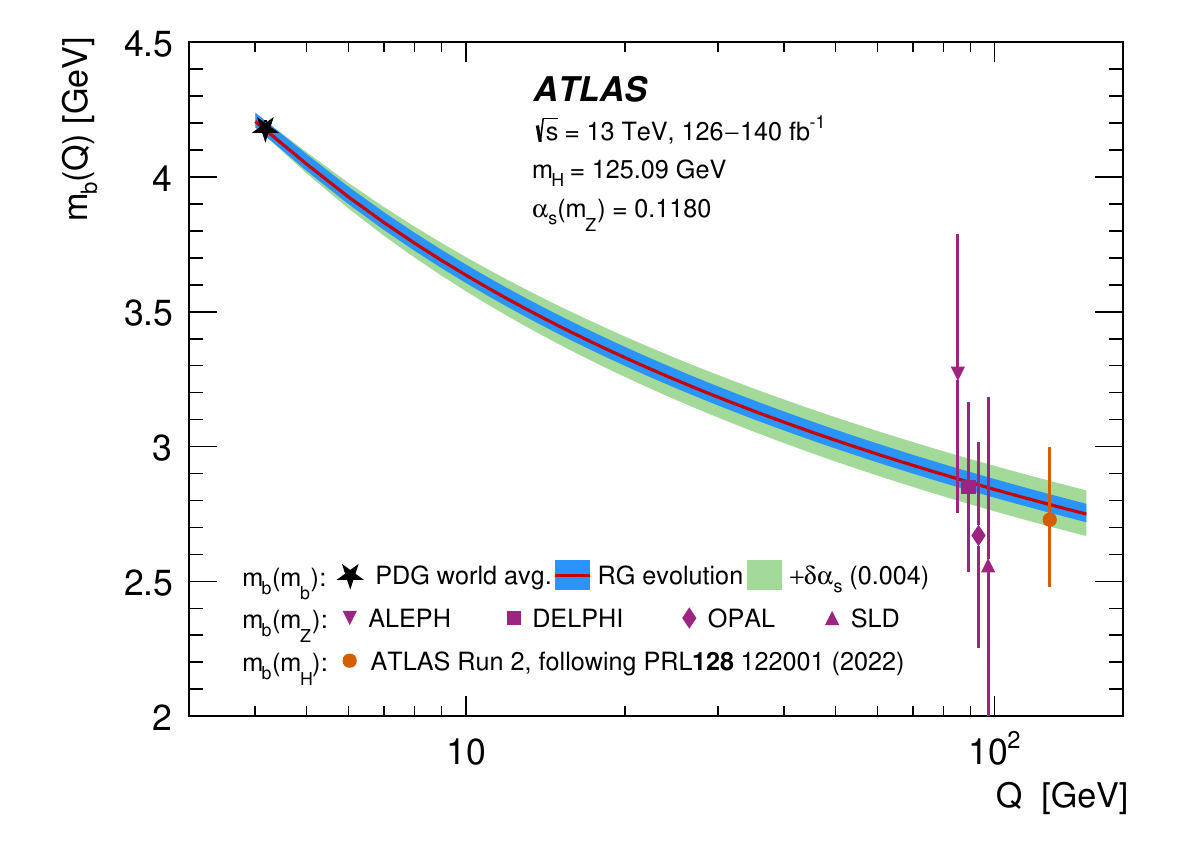}
\caption{Value of the \MSbar \, running mass  $m_b(Q)$ at various energy scales $Q$. The value of $m_b(m_H)$ obtained in the present work is shown alongside values of $m_b(m_Z)$ obtained at the Z pole~\cite{Brandenburg:1999nb,Barate:2000ab,Abbiendi:2001tw,Abdallah:2008ac} and the PDG world average of $m_b(m_b)$~\cite{Zyla:2020zbs}. The line shows the renormalisation-group evolution of $m_b(Q)$ obtained using REvolver~\cite{Hoang:2021fhn} with the PDG world average of $m_b(m_b)$ taken as reference. Uncertainties in the evolution due to missing higher orders and the value of the strong coupling constant are also shown.}
\label{fig:mb:obs}
\end{figure}

\FloatBarrier
\subsection{Constraint on the total width of the Higgs boson}
\label{sec:width}

\begin{figure}[tbp]
\centering
\subfloat[]{\includegraphics[width=0.49\columnwidth]{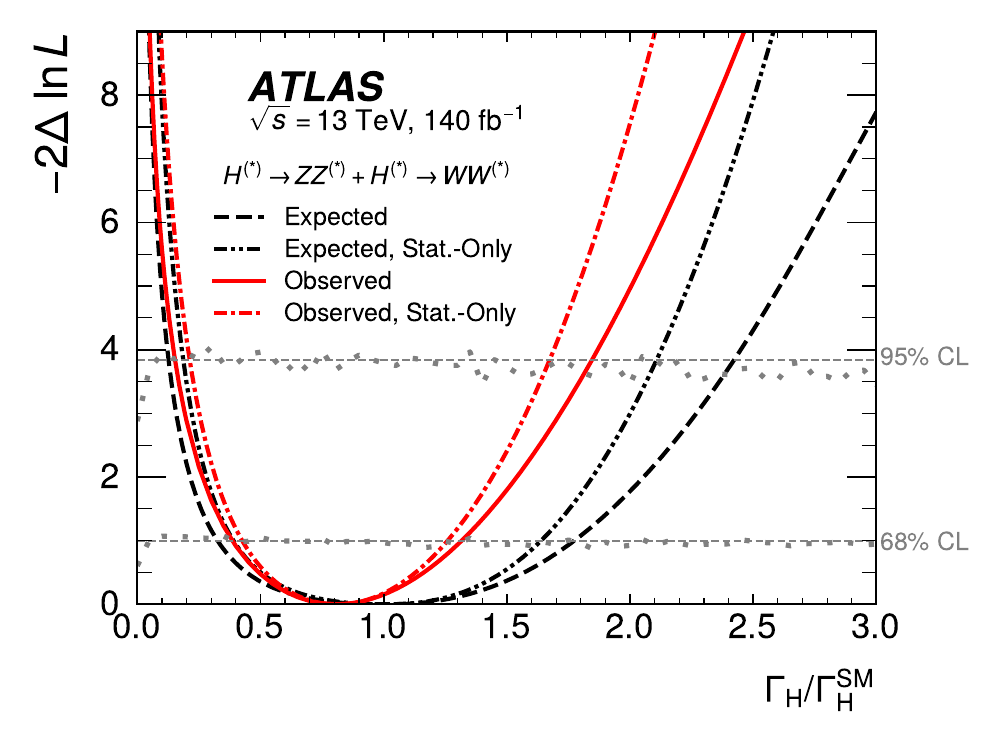}
\label{fig:width-1}}
\subfloat[]{\includegraphics[width=0.49\columnwidth]{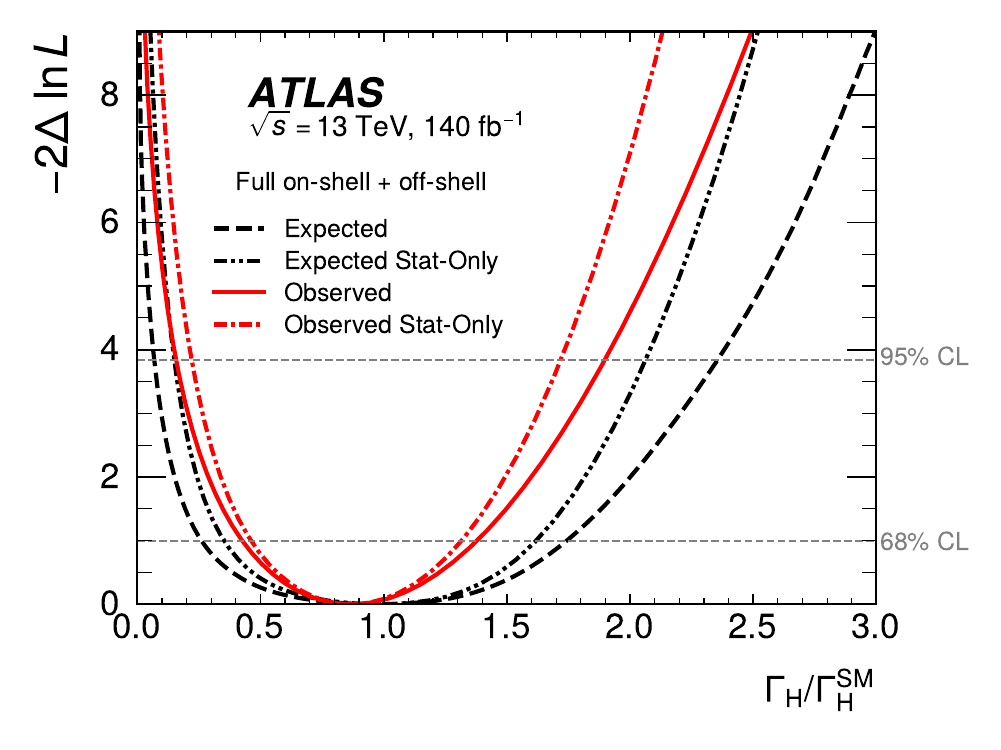}
\label{fig:width-2}}
\caption{Observed and expected negative log-likelihood scans as a function of the ratio of the total width of the Higgs boson to its SM prediction, $\Gamma_H/\Gamma_H^{\text{SM}}$, from the combination of \HsZZ, \HsWW\ and on-shell Higgs boson production and decay measurements. Panel~(a) shows the results using only the measurements of on-shell production in the \Hzz\ and \Hww\ final states; the dashed horizontal lines show the levels $-2\,\Delta\ln L = 1$ and $3.84$ used to estimate the 68\% and 95\% CL intervals in the asymptotic limit, respectively. The dots show the 68\% and 95\% CL intervals obtained from a Neyman construction. Panel~(b) shows the results obtained using all measurements of on-shell production, except for those in the \Hinv\ channel; due to the small difference (at most 2\%) between the asymptotic limit and the Neyman construction observed in the first panel, only the asymptotic result is shown for this panel.}
\label{fig:width}
\end{figure}

By combining off-shell Higgs boson production analyses with the on-shell Higgs boson measurements, a constraint on the total width of the Higgs boson can be obtained within the $\kappa$-framework~\cite{Caola:2013yja,Campbell:2013una}.
This combination makes the same assumptions as for Scenario 3 of Section~\ref{sec:kappa:bsm_decays}: that modifications to Higgs boson couplings to $W$ and $Z$ bosons are the same in the on-shell and off-shell kinematic regimes, and that the kinematics of on-shell and off-shell Higgs boson production match their SM predictions. In particular, the relative fraction of Higgs boson production in the region $m_{VV} \gg m_H$ probed by off-shell production analyses is assumed to correspond to its SM value.
The full set of measurements of off-shell Higgs boson production in the \HsZZ~\cite{HIGP-2024-14} and \HsWW~\cite{HIGP-2024-05} channels is considered in this interpretation without removing any region. Concerning on-shell measurements, all production and decay channels are considered except for the \Hinv\ decay, due to the overlap between analyses of this process and those of \HsZZ\ that is described in Section~\ref{sec:input_channels}. Two scenarios are considered: one combines the measurements of off-shell and on-shell Higgs boson production in the \HsZZ\ and \HsWW\ final states, while the other combines all available measurements of off-shell and on-shell production. The second scenario provides the most comprehensive result, but the first scenario is also retained as its simpler model allows performing two additional studies: first the explicit validation of the asymptotic approximation via a Neyman construction (Figure~\ref{fig:width-1}), and second a detailed decomposition of systematic uncertainty contributions (Appendix~\ref{app:width}). Both of these would be computationally prohibitive in the more complex model of the second scenario.

Negative log-likelihood scans as a function of $\Gamma_H/\Gamma_H^{\text{SM}}$ for the first scenario are shown in Figure~\ref{fig:width-1}. In this scenario, only the modifiers for vector boson ($\kappa_V$) and gluon ($\kappa_g$) couplings are considered as free parameters in the fit. All other coupling modifiers are found to have a negligible impact on the results and are fixed to their SM values. The comparison of the 68\% and 95\% CL intervals obtained in the asymptotic approximation and from a Neyman construction shows differences of at most 2\%. The difference between the two methods is much reduced compared to Ref.~\cite{HIGP-2024-14}, as the larger dataset reduces the statistical uncertainty relative to the systematic component, improving the accuracy of the asymptotic approximation. The observed (expected) result is
\begin{equation}
\Gamma_H = 3.3^{+2.1}_{-1.7}\text{ (obs.)} \quad \left(4.1^{+3.2}_{-2.7}\text{ (exp.)}\right) \MeV
\end{equation}
at 68\% CL, showing good agreement between the measurement and the SM prediction.
The impact of different sources of systematic uncertainties in the measurement of the Higgs boson width is shown in Appendix~\ref{app:width}. The uncertainties are evaluated using two complementary methods: the conditional uncertainty method and the shifted auxiliary observable method, both described in Section~\ref{sec:stats}. Consistent values of the total uncertainties are found with the two methods; the values obtained using the shifted auxiliary observable method are reported as they allow contributions from independent sources to be summed in quadrature. The observed (expected) statistical significance of Higgs boson off-shell production is found to be 3.7 (2.9) standard deviations using the combination of on-shell and off-shell \HsZZ\ and \HsWW\ analyses. The observed (expected) 68\% CL interval on $\Gamma_H$ is reduced by 17\%~(14\%) compared to the result obtained using \HsZZ\ alone (the most sensitive input to the combination), while making use of the same measurement assumptions.
This improves upon the previous ATLAS result of $\Gamma_H = \qty[parse-numbers=false]{4.3^{+2.7}_{-1.9}}{\MeV}$ from Ref.~\cite{HIGP-2024-14}, obtained from a combination of analyses targeting on-shell and off-shell Higgs boson production in the \HsZZ\ decay, for which the expected significance was 2.4 standard deviations.

Negative log-likelihood scans as a function of $\Gamma_H/\Gamma_H^{\text{SM}}$ for the second scenario are shown in Figure~\ref{fig:width-2}. In this scenario, the on-shell input is extended to include all analyses of Higgs boson production and decay, rather than only the ones targeting \Hzz\ and \Hww\ as in the first scenario. The coupling modifiers that are left free are now the same set as in Section~\ref{sec:kappa:no_Biu}, in the effective parameterisation and with $\kappa_c$ considered as a free parameter. No additional assumptions beyond those of Section~\ref{sec:kappa:no_Biu} are required. A Neyman construction is not computationally feasible for the full combination model; instead, the validity of the asymptotic approximation is inferred from the first scenario. The larger dataset entering the full combination further improves the accuracy of the approximation, so that only the asymptotic result is shown. The observed result of $\Gamma_H$ at 68\% CL is:
\begin{equation}
\Gamma_H = \qty[parse-numbers=false]{3.6^{+2.1}_{-1.6}}{\MeV} = 3.6^{+1.9}_{-1.4} \text{ (stat.) } ^{+0.9}_{-0.8} \text{ (syst.)} \MeV.
\end{equation}
The expected result at 68\% CL is:
\begin{equation}
\Gamma_H = \qty[parse-numbers=false]{4.1 \pm 3.0}{\MeV} = 4.1 ^{+2.5}_{-2.3} \text{ (stat.) } ^{+1.7}_{-2.0} \text{ (syst.) } \MeV.
\end{equation}
This result is the most sensitive determination to date of $\Gamma_H$ in this scenario. The larger statistical uncertainty compared with the first scenario reflects the introduction of additional degrees of freedom from the coupling modifiers profiled in the fit; this effect outweighs the gain in statistical precision from the additional input analyses. The observed value is again in good agreement with the SM prediction.

\FloatBarrier
\subsection{Constraints on the Higgs boson self-coupling}
\label{sec:kappa:kl}

Single-Higgs boson production processes acquire sensitivity to the trilinear self-coupling modifier \kl\ through NLO electroweak corrections to production cross-sections and branching ratios~\cite{Degrassi:2016wml, Maltoni:2017ims}. The dependence on \kl\ is parameterised using process- and kinematic-dependent $C_1$ coefficients and $K_{\text{EW}}$ factors, provided by the LHC Higgs Working Group~\cite{LHCHWG-2022-002}. These corrections encode the effect of potential BSM phenomena coupling to the SM via the Higgs potential and affecting mainly the self-coupling of the Higgs boson.

The total width of the Higgs boson is fixed to its SM value. In the primary fit, \kl\ is the only free coupling modifier, with all other coupling modifiers ($\kappa_V$, $\kappa_t$, $\kappa_b$, $\kappa_\tau$) fixed to their SM values in the resolved parameterisation, so that $\kappa_g$, $\kappa_\gamma$ and $\kappa_{Z\gamma}$ are also fixed to their SM values through the corresponding loop expressions. The single-$H$ channels are combined with a combination of six analyses of $HH$ production (\HHbbyy, \HHbbtt, \HHbbbb, \HHbbbb (high \ptH), \HHbbll, and \HHML) following the procedure described in Ref.~\cite{ATLAS:2024ish}. Relative to Ref.~\cite{ATLAS:2024ish}, two modifications are applied to the $HH$ inputs: the \kl-dependent scale and $m_\text{top}$ uncertainty parameterisation of the \ggF\ $HH$ production are updated, and the dependence of the single-$H$ processes on \kl\ is incorporated into all the $HH$ analyses except \HHbbtt, which already included this effect.

The result labelled ``Di-Higgs'' in Table~\ref{tab:kl_results:obs} reports the $HH$-only constraint with these modifications applied, and therefore does not reproduce exactly the result of Ref.~\cite{ATLAS:2024ish}. Combining the single-$H$ and $HH$ inputs requires the treatment of systematic uncertainties common to both sets of analyses. A correlation scheme is adopted in which shared nuisance parameters are identified and correlated, with the exception of parameters found to be strongly pulled or constrained in either input, which are left decorrelated to avoid propagating tensions across the analyses as explained in Section~\ref{sec:stats}. Overlaps between the event selections in the analyses of single-$H$ and $HH$ production, studied in particular for the \HHbbtt\ and \Htt\ analyses, are found to have a negligible impact on the results.

\begin{figure}[tbp]
\centering
\subfloat[]{\includegraphics[width=0.49\columnwidth]{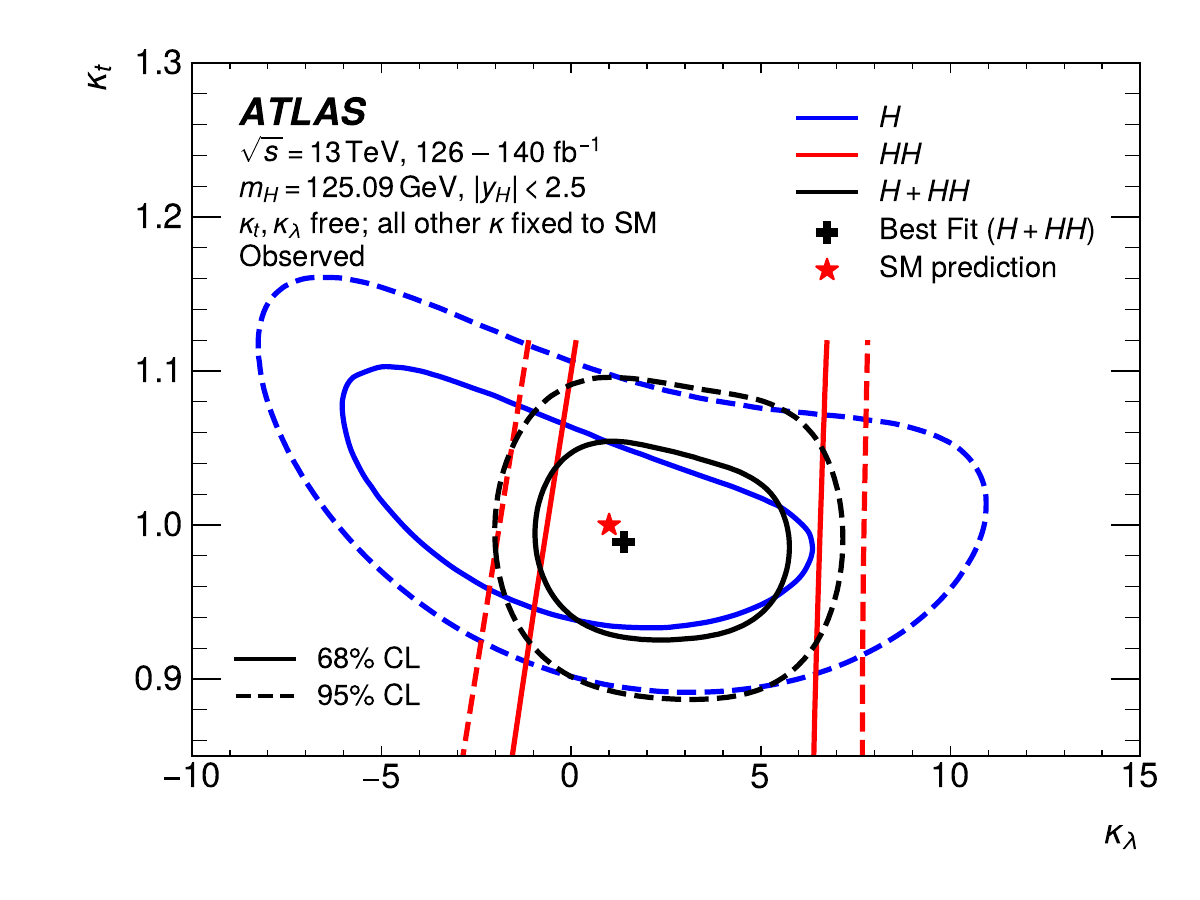}}
\subfloat[]{\includegraphics[width=0.49\columnwidth]{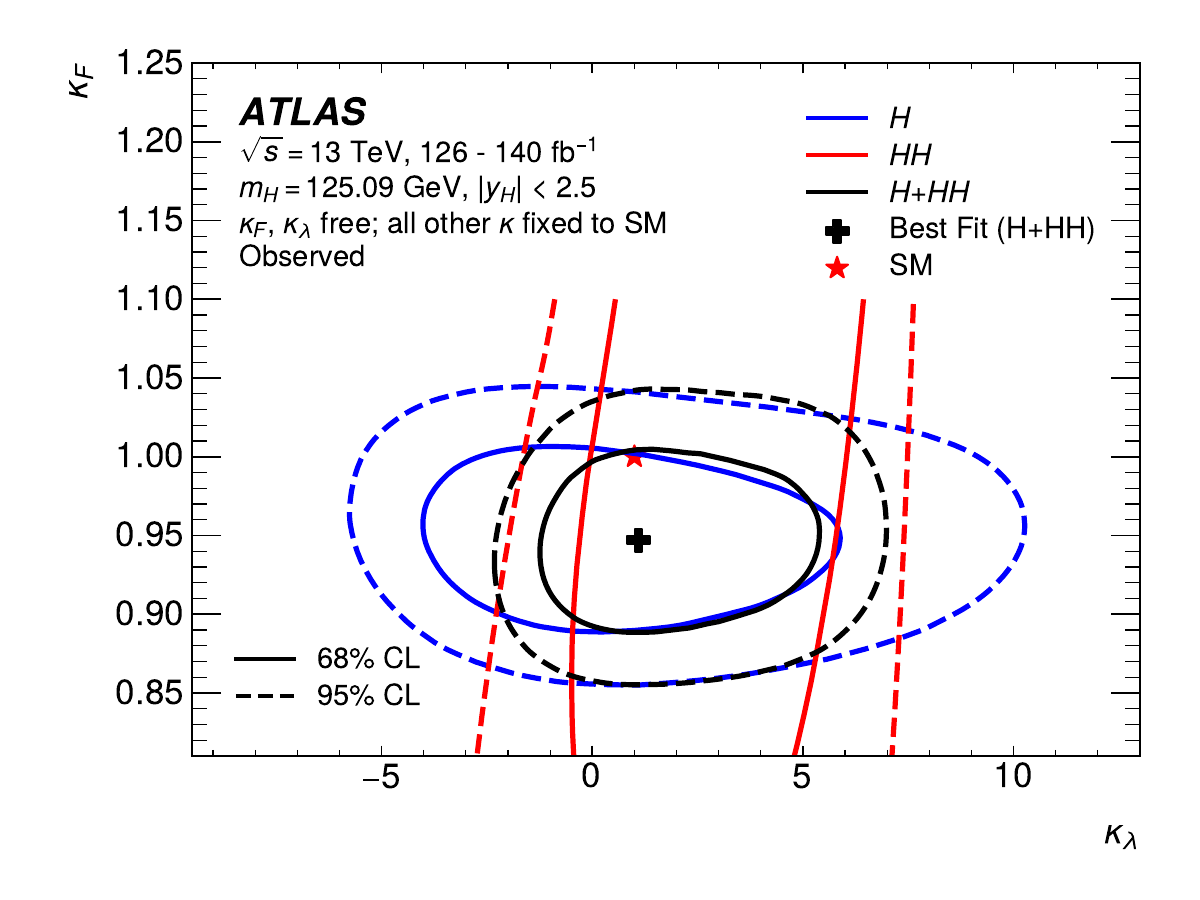}}
\caption{Observed constraints in the  (a) $(\kl,\,\kappa_t)$ and (b) $(\kl,\,\kappa_F)$ planes with all other coupling modifiers fixed to their SM values, for the single-$H$, $HH$ and combined fits.}
\label{fig:HHH_kl_2Dcont}
\end{figure}

The observed value of \kl\ in the combination of single-$H$ and $HH$ analyses is
\begin{equation*}
\kl = 1.3\,^{+3.1}_{-1.6},
\end{equation*}
with a 68\% CL interval of $[-0.3,\,4.4]$ and a 95\% CL interval of $[-1.5,\,6.5]$. The expected result under the SM hypothesis is $\kl = 1.0\,^{+3.7}_{-1.4}$, with a 68\% CL interval $[-0.5,\,4.6]$ and a 95\% CL interval $[-1.5,\,6.8]$. Observed results for the individual single-$H$ and $HH$ inputs, and for a generic model in which $\kappa_V$, $\kappa_t$, $\kappa_b$ and $\kappa_\tau$ are also profiled alongside \kl, are summarised in Table~\ref{tab:kl_results:obs}. The observed likelihood scans of \kl\ are shown in Figure~\ref{fig:HHH_kl_scan}. The corresponding expected results are shown in Table~\ref{tab:kl_results:exp} in Appendix~\ref{app:kappa_lambda}.
\begin{table}[tbp]
\caption{Observed best-fit values of \kl\ and 68\% and 95\% CL intervals in various measurement configurations. In the first three rows (Single-Higgs, Di-Higgs, H+HH others fixed), the remaining coupling modifiers ($\kappa_V$, $\kappa_t$, $\kappa_b$, $\kappa_\tau$) are fixed to their SM values. In the fourth row (H+HH generic) these modifiers are free in the fit alongside \kl; the shift in the best-fit value between these two scenarios is due to the correlations between the \kl\ and the other coupling modifiers through their impact on single-$H$ event rates.}
\label{tab:kl_results:obs}
\centering
\renewcommand{\arraystretch}{1.2}
\begin{tabular}{lccc}
\toprule
Fit & Best fit & 68\% CL interval & 95\% CL interval \\
\midrule
Single-Higgs         & $-0.9$\numpmerr{+3.9}{-2.5} & $[-3.4,\; 3.0]$ & $[-5.3,\;8.8]$ \\
Di-Higgs             &  $3.5$\numpmerr{+2.4}{-3.4} & $[ 0.1,\; 5.9]$ & $[-1.3,\;7.2]$ \\
H+HH (others fixed)  &  $1.3$\numpmerr{+3.1}{-1.6} & $[-0.3,\; 4.4]$ & $[-1.5,\;6.5]$ \\
H+HH (generic)       &  $0.6$\numpmerr{+2.5}{-1.5} & $[-0.9,\; 3.1]$ & $[-2.1,\;5.9]$ \\
\bottomrule
\end{tabular}
\end{table}
\begin{figure}[tbp]
\centering
\includegraphics[width=.7\textwidth]{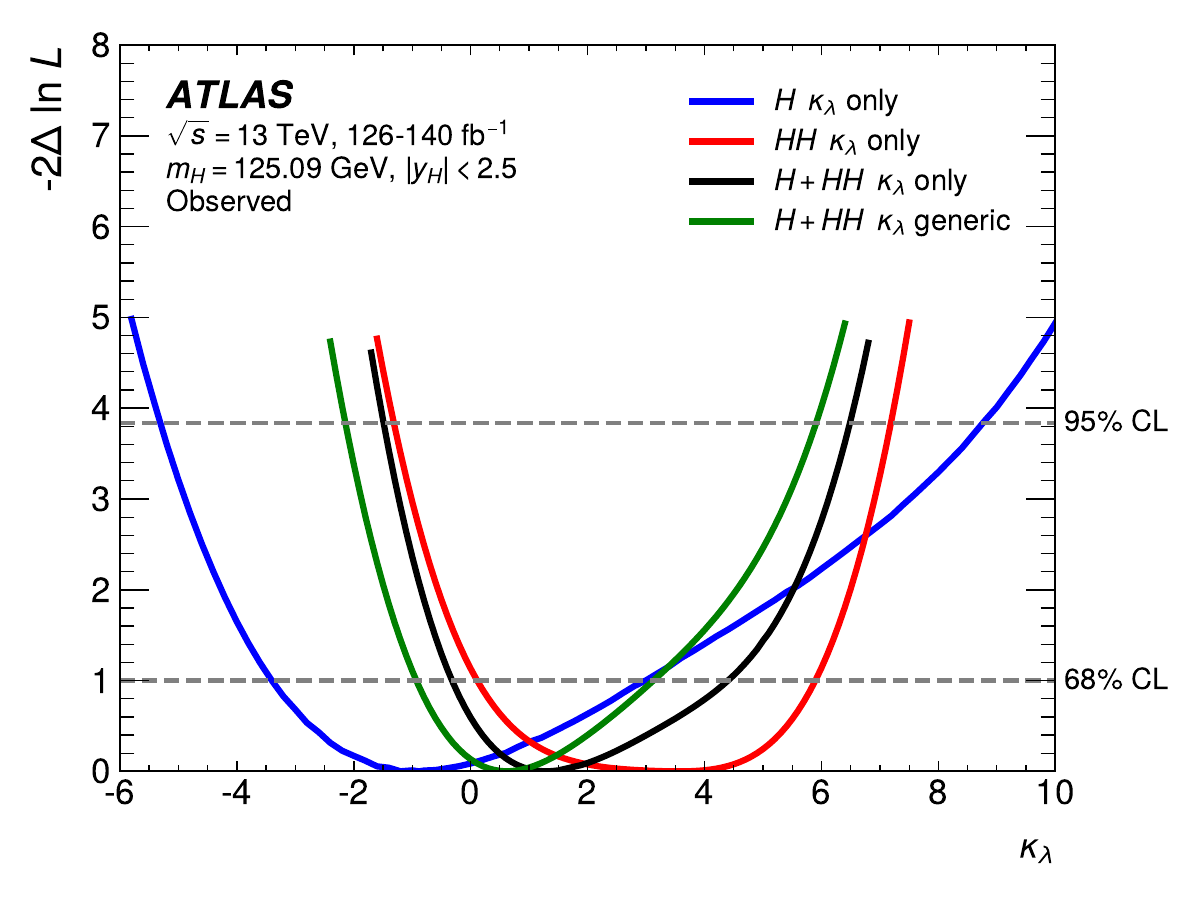}
\caption{Observed likelihood scans of $\kl$ with all other coupling modifiers fixed to their SM values, for single-$H$ channels, $HH$ channels and their combination. A fit in which $\kappa_b$, $\kappa_\tau$, $\kappa_V$ and $\kappa_t$ are profiled is also shown for the combined result.}
\label{fig:HHH_kl_scan}
\end{figure}
Observed constraints in the two-dimensional planes $(\kl,\,\kappa_t)$ and $(\kl,\,\kappa_F)$ with all other coupling modifiers fixed to their SM values are shown in Figure~\ref{fig:HHH_kl_2Dcont}.

\begin{figure}[tbp]
\centering
\subfloat[]{\includegraphics[width=0.49\columnwidth]{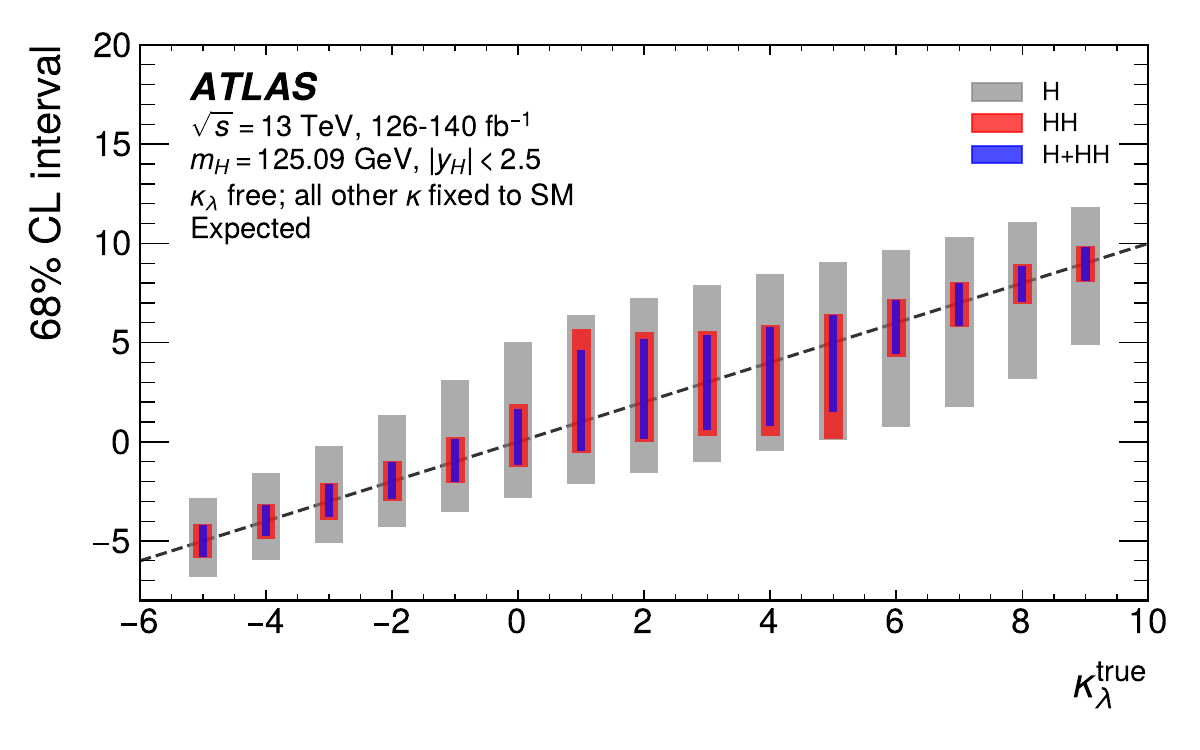}}
\subfloat[]{\includegraphics[width=0.49\columnwidth]{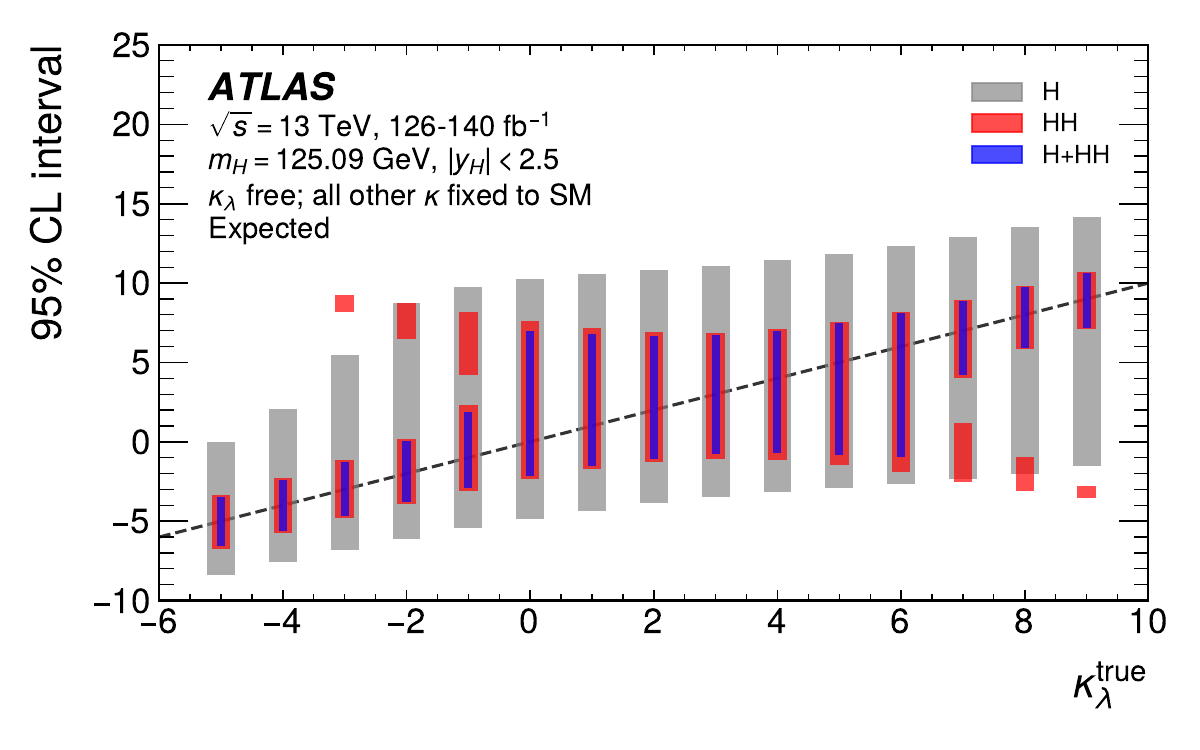}}
\caption{(a) Expected 68\% CL interval and (b) Expected 95\% CL interval on \kl\ as a function of the true value of \kl, obtained from single-$H$ channels, $HH$ channels and their combination. The dashed line shows the ideal case in which the measured value of \kl\ is equal to the true value.}
\label{fig:kllimits}
\end{figure}

The expected 68\% and 95\% CL intervals on $\kl$ as a function of the true value of $\kl$ are shown in Figure~\ref{fig:kllimits}, illustrating the relative contribution of the single-$H$ and $HH$ results. For some values of the true $\kl$, the single-$H$ results have the effect of lifting a degeneracy in \kl\ intervals at 95\% CL that are obtained from the $HH$ results alone.

The combined result $\kl = 1.3 \,^{+3.1}_{-1.6}$ is consistent with the SM prediction. The single-$H$ and $HH$ inputs provide complementary sensitivities: $HH$ production is sensitive to \kl\ at leading order through the trilinear Higgs boson vertex in the \ggF\ $HH$ amplitude and dominates the combined constraint, particularly at large values of $|\kl|$ where the $HH$ production rate differs significantly from the SM expectation. Single-$H$ processes acquire sensitivity to \kl\ only through NLO electroweak corrections parameterised by the $C_1$ coefficients; among single-$H$ production modes, $\ttH$ has the largest $C_1$ coefficient and the greatest individual sensitivity to \kl, but these corrections are at most a few percent for order-unity deviations on \kl\ from unity. The combination of the two inputs yields a significant improvement in sensitivity with respect to either input alone.

In the generic model where $\kappa_V$, $\kappa_t$, $\kappa_b$ and $\kappa_\tau$ are profiled alongside \kl, the expected upper and lower bounds on \kl\ at 68\% CL are looser when compared to the case where other modifiers are fixed to the SM, respectively by 1\% and 5\%. The complementarity of the $HH$ and single-$H$ inputs takes a qualitatively different form in this model: the \kl-induced shift in single-$H$ rates can largely be absorbed by adjustments to the profiled coupling modifiers, and the primary role of the single-$H$ measurements in this model is to constrain $\kappa_t$, $\kappa_V$, $\kappa_b$ and $\kappa_\tau$. This constraint resolves an intrinsic degeneracy between \kl\ and $\kappa_t$ in the \ggF\ $HH$ amplitude, and the combination therefore yields a more model-independent determination of \kl\ than $HH$ production alone.

\FloatBarrier


\section{Measurements of the kinematics of Higgs boson production in the STXS framework}
\label{sec:stxs}

Results presented in the previous sections focused on inclusive event rates within each production and decay channel, but further information on Higgs boson properties can be obtained through the study of the kinematics of these processes. In this section, measurements are reported within the STXS framework~\cite{YR4,Andersen:2016qtm,Berger:2019wnu,Amoroso:2020lgh}, defined jointly with the CMS experiment and the theory community, which defines a set of non-overlapping measurement regions in the phase space of the main Higgs boson production processes.
The STXS binning is designed to isolate regions providing high sensitivity to Higgs boson properties or potential BSM effects, with associated theory uncertainties that can be reliably estimated. The regions are defined using selections in simple kinematic variables as described below. They are designed to closely match reconstruction-level event selections, but are not in exact correspondence since these selections vary across the analyses in the combination and employ a variety of multivariate techniques.
The results presented in this section are based on the Stage 1.2 binning~\cite{Berger:2019wnu}. Its standard definition introduces the following production processes and kinematic selections:
\begin{itemize}
\item The \ggtoH\ process includes \ggF\ production and \ggZH\ events where the $Z$ decays into hadrons. Measurement regions are defined in the transverse momentum of the Higgs boson (\ptH), the number of jets (\Njets), the invariant mass of the two  highest-$p_T$ (leading) jets (\mjj), and the transverse momenta of the systems formed by the Higgs boson and the leading jet (\ptHj) and the Higgs boson and the two leading jets (\ptHjj)
\item The \ewqqH\ process includes \VBF\ production and \VH\ production in which the vector boson decays into hadrons. Measurement regions are defined in \ptH, \Njets, \mjj\ and \ptHjj.
\item The \VH\ process considers \VH\ production with a leptonic decay of the vector boson, encompassing \WH\ production (\qqtoHln), quark-initiated \ZH\ production (\qqtoHllnn) and gluon-initiated \ZH\ production (\ggtoHllnn). Measurement regions are defined in the transverse momentum of the vector boson (\ptV) and in \Njets.
\item The \ttH\ process is considered in bins of \ptH.
\item The \bbH, \tHq\ and \tHW\ processes each correspond to a single region.
\end{itemize}
The definitions of the STXS measurement regions used in this paper are modified with respect to Ref.~\cite{Berger:2019wnu} to more closely correspond to the sensitivity of the analyses included in the combination. In cases where the analyses are not sensitive to the separation between certain regions, these regions are merged together into a single bin. Uncertainties on the efficiency and acceptance factors for the resulting bin are obtained from those that are merged together, under the assumption that the fraction of events in each merged bin follows its SM prediction. Conversely, some Stage 1.2 bins are split into smaller regions in cases where this is permitted by the analysis sensitivity. These changes are as follows:

\begin{itemize}
\item In the \ggtoH\ process, regions split in \ptHj\ and \ptHjj\ are merged together. The regions within $\Njets \ge 2$, $\mjj < \qty{350}{\GeV}$ and $\ptH < \qty{120}{\GeV}$ are merged into a single bin, as well as the regions within $\Njets \ge 2$, $\ptH < \qty{200}{\GeV}$, $\mjj > \qty{350}{\GeV}$. In each case, this removes separations within the Stage 1.2 definition to which none of the analyses in the combination are sensitive.
\item In the \ewqqH\ process, all \ptHjj\ splits are removed, and the bins $\Njets=0$ and $\Njets=1$ are merged into a single $\Njets \le 1$ region. Within the region $\Njets \ge 2$, $\mjj < \qty{350}{\GeV}$, the $60 \le \mjj < \qty{120}{\GeV}$ bin is denoted by \emph{VH-enriched} since it is dominated by \VH\ production with hadronic vector boson decays. The remaining bins $\mjj < \qty{60}{\GeV}$ and $\mjj \ge \qty{120}{\GeV}$ are merged together into a region denoted by \emph{VBF-enriched}.
\item In the \VH\ process, the regions defined for \qqtoHllnn\ and \ggtoHllnn\ production are merged together pairwise into regions corresponding to a single \pptoHllnn\ process. The $\ptV \ge \qty{400}{\GeV}$ regions in the \qqtoHln\ and \pptoHllnn\ processes are both split into separate bins for $400 \le \ptV < \qty{600}{\GeV}$ and $\ptV \ge \qty{600}{\GeV}$ to take advantage of the sensitivity provided by the \VH, \Hbbcc\ analysis of Ref.~\cite{HIGG-2020-20} at high values of \ptV. The expected yields and theory uncertainties in these bins are derived following the same procedure as followed for the original Stage 1.2 scheme~\cite{ATL-PHYS-PUB-2018-035}. Separations in \Njets\ are retained for the $75 \le \ptV < \qty{150}{\GeV}$, $150 \le \ptV < \qty{250}{\GeV}$ and $250 \le \ptV < \qty{400}{\GeV}$ bins of the \pptoHllnn\ process and removed elsewhere.
\item The \bbH\ process is considered together with the \ggtoH\ process, since these processes cannot be distinguished by the analyses in the combination.
\item The \tHq\ and \tHW\ processes are merged together into a single \tH\ process, since the combination has insufficient sensitivity to distinguish between them.
\end{itemize}
Only the regions in the range $|y_H| < 2.5$ of the Higgs boson rapidity $y_H$ are considered, since no significant sensitivity outside this range is provided in the combination. Production rates in the region $|y_H| > 2.5$ are assumed to follow SM predictions.

\begin{figure}[tbp]
\centering
\includegraphics[width=0.8\textwidth]{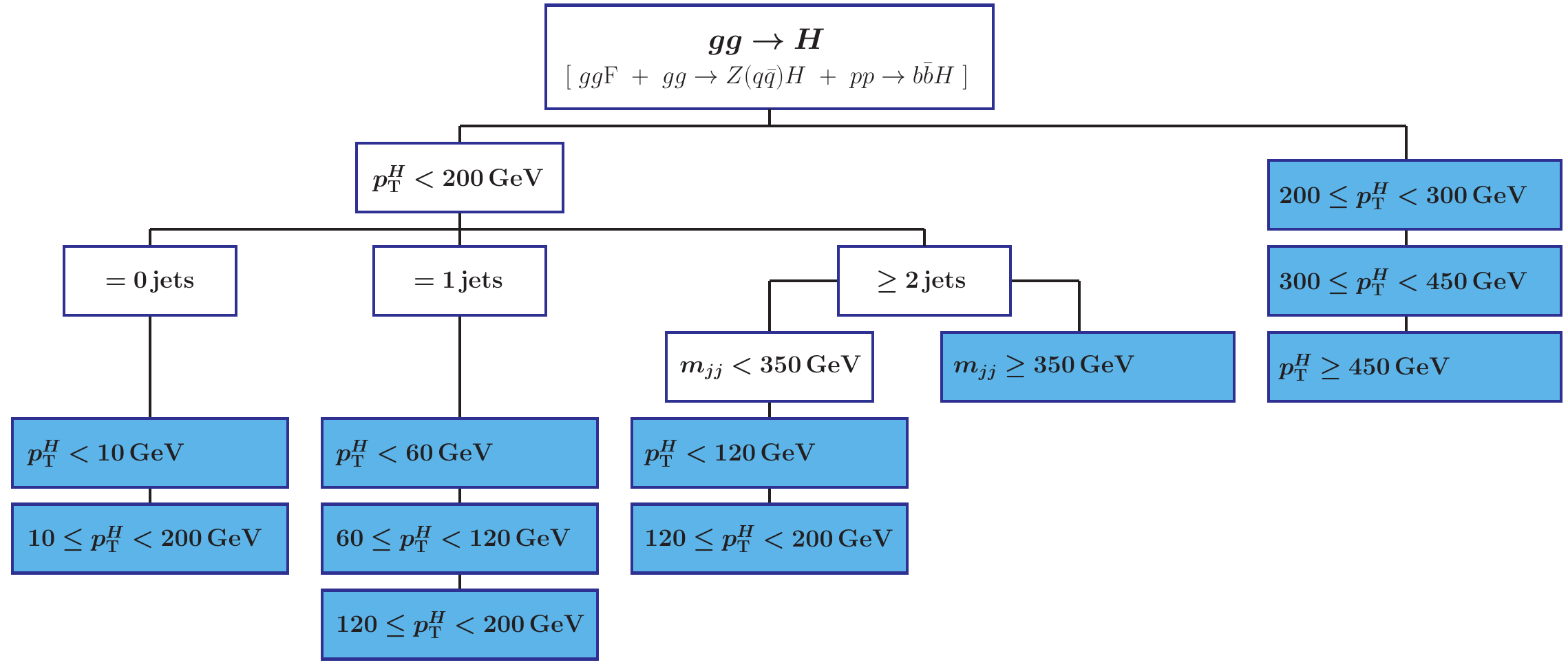}\\
\vspace{0.5cm}
\includegraphics[width=0.8\textwidth]{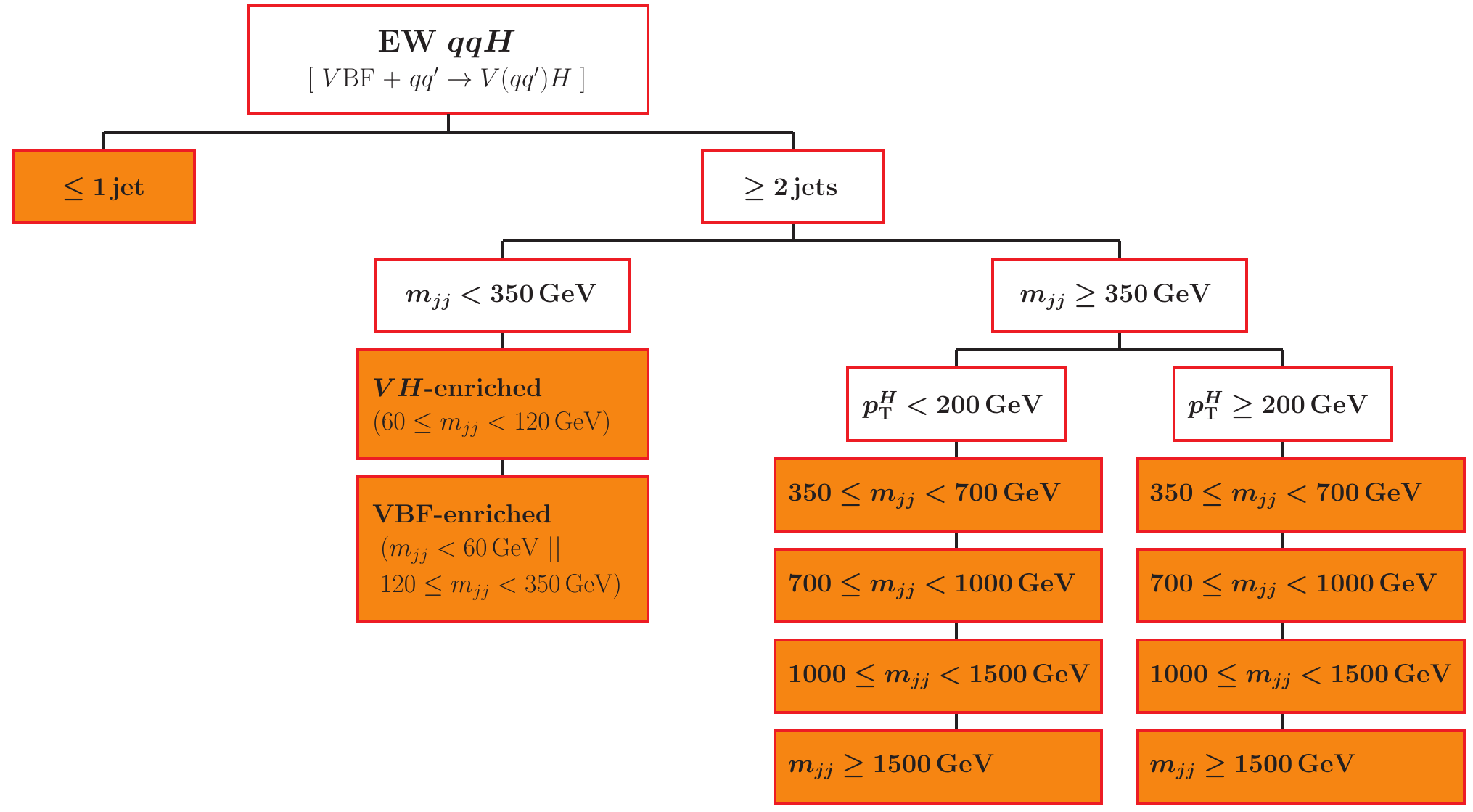}\\
\vspace{0.5cm}
\includegraphics[width=0.50\textwidth]{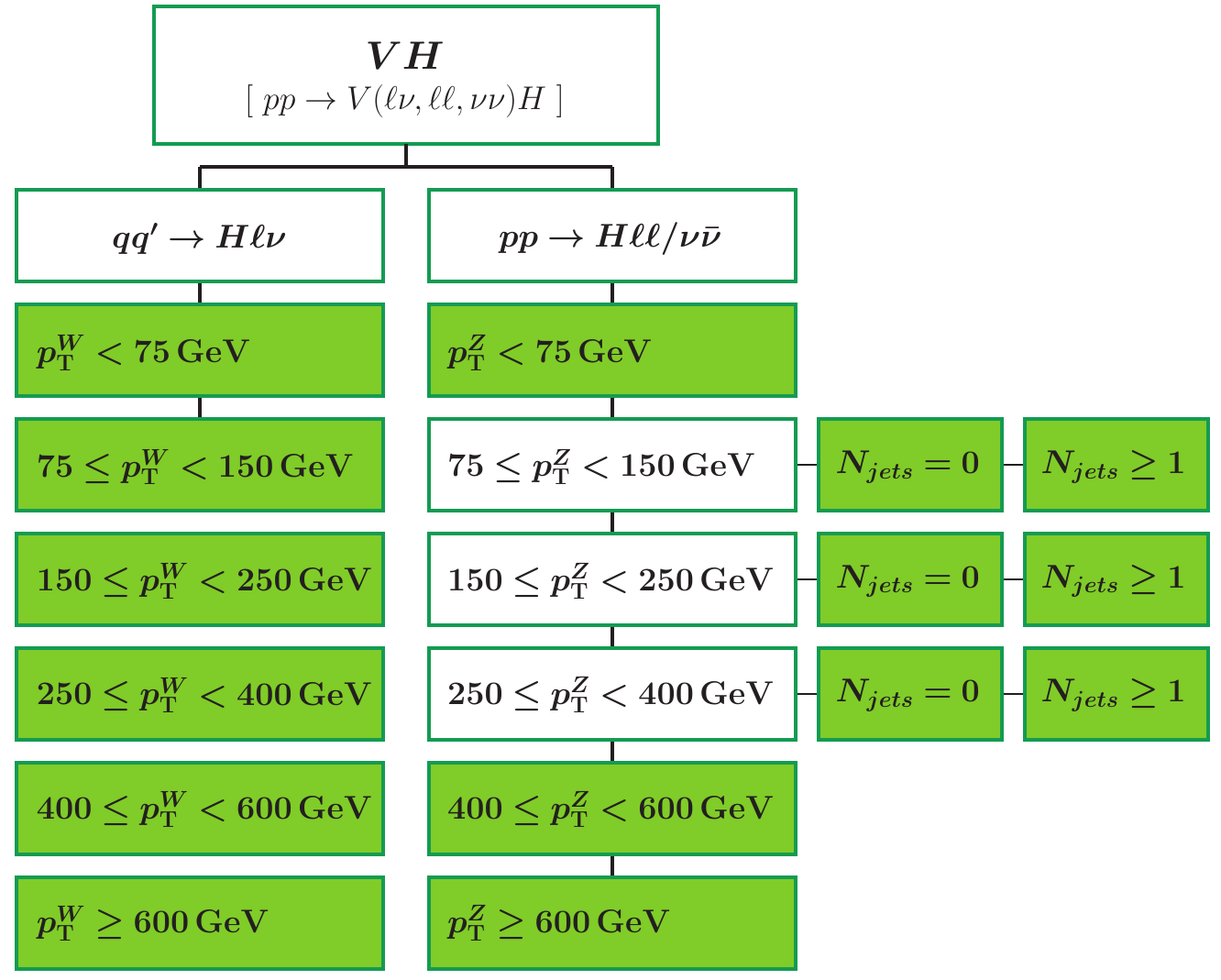}
\includegraphics[width=0.45\textwidth]{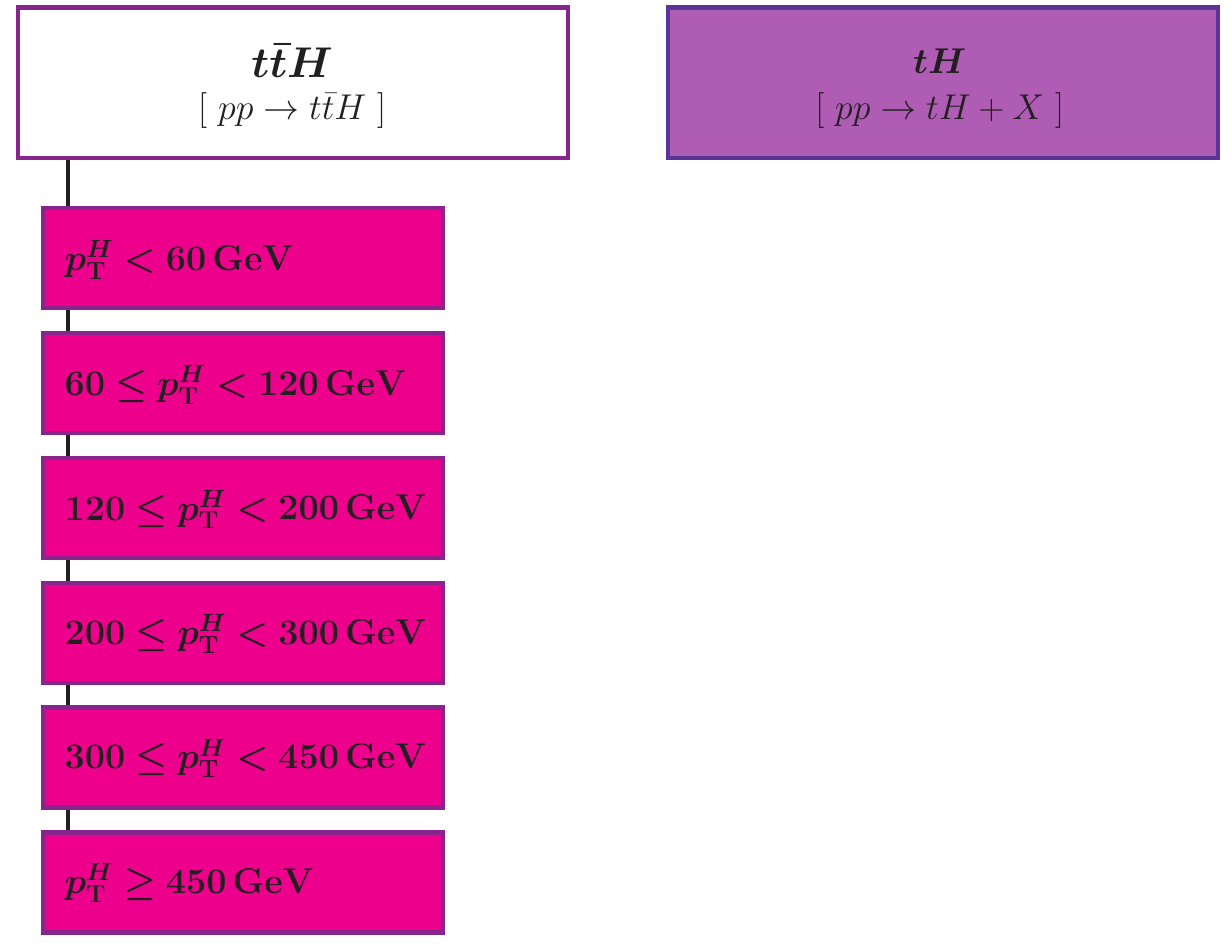}
\caption{STXS measurement regions in the modified Stage 1.2 scheme used in this work for the \ggtoH\ (top), \ewqqH\ (middle), \VH\ (bottom left), \ttH\ (bottom centre) and \tH\ (bottom right) processes. In each case the top-level box indicates the targeted production modes, clear boxes show intermediate selection criteria and solid boxes show the measurement regions.
}
\label{fig:stxs:diagrams}
\end{figure}

The resulting set of 44 measurement regions is presented in Figure~\ref{fig:stxs:diagrams}.

\begin{figure}[tbp]
\centering
\includegraphics[width=0.90\columnwidth]{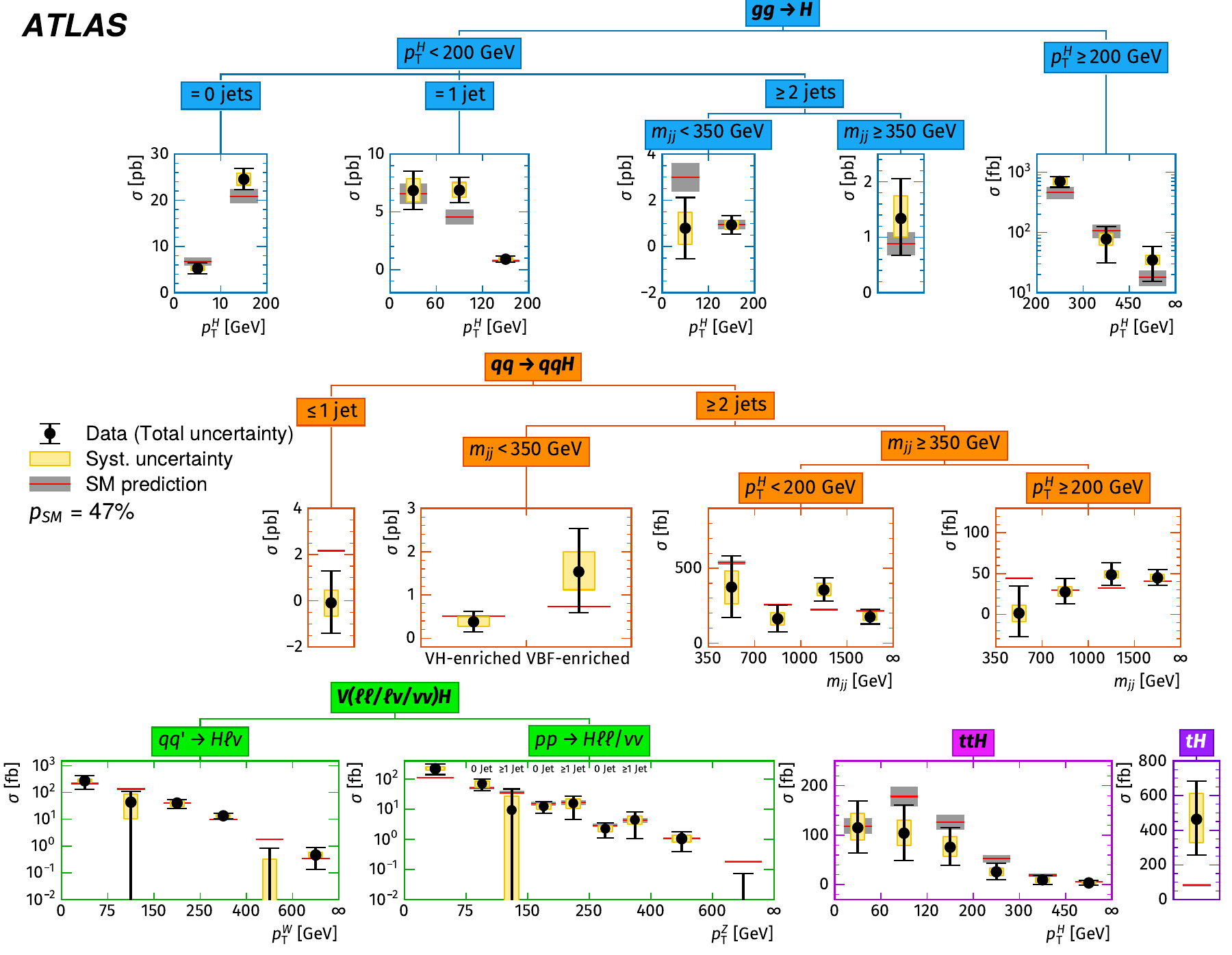}
\caption{Observed production cross-sections in 44 STXS regions for the \ggtoH, \ewqqH, \VH, \ttH\ and \tH\ processes. Points with error bars represent the best-fit values and total measurement uncertainties, and hashed bands show the systematic uncertainty component. The SM prediction in each region is indicated by vertical lines, and the associated theory uncertainty by the shaded area.}
\label{fig:stxs:fixedBR:results}
\end{figure}

Results are reported under two sets of assumptions. In the first model, the branching ratios of Higgs boson decays are assumed to match their SM predictions, within the uncertainties described in Section~\ref{sec:uncertainties}. The measured and expected production cross-sections in each region are presented in Table~\ref{tab:stxs:fixedBR:obs} and illustrated in Figure~\ref{fig:stxs:fixedBR:results}.

\begin{table}[ht]
\caption{
Best-fit values and uncertainties for the production cross-section in each measured STXS region, assuming SM values for the Higgs boson branching ratios. The values for the \ggtoH\ process also include the contributions from \bbH\ production.
The total uncertainties are decomposed into components for data statistics (Stat.) and systematic uncertainties (Syst.).
SM predictions are also shown for each quantity with their total uncertainties.
}
\centering
\renewcommand{\arraystretch}{1.2}
\resizebox{0.8\textwidth}{!}{
\begin{tabular}{l S[table-format=-5.1,round-mode=none] lll S[table-format=5.2,round-mode=none] l}
\toprule
\multirow{2}{*}{STXS region} & \multicolumn{1}{c}{Value} & \multicolumn{3}{c}{Uncertainty [fb]} & \multicolumn{2}{c}{SM prediction} \\
&  \multicolumn{1}{c}{[fb]}                & Total   & Stat.                & Syst.   & \multicolumn{2}{c}{[fb]}   \\
\midrule
\ggHjPt{0}{}{10}{} &  5200 & \errRP{-2}{1200.0}{-1100.0} & ${\scriptstyle \pm \numRP[-2]{1100.0}}$ & \errRP{-2}{500.0}{-400.0} &  6600 &  ${\scriptstyle \pm \numRP[-2]{900.0}}$ \\
\ggHjPt{0}{10}{}{} &  24300 & \errRP{-2}{2300.0}{-2200.0} & ${\scriptstyle \pm \numRP[-2]{1800.0}}$ & \errRP{-2}{1400.0}{-1300.0} &  20600 &  ${\scriptstyle \pm \numRP[-2]{1500.0}}$ \\
\ggHjPt{1}{}{60}{} &  6800 & \errRP{-2}{1700.0}{-1600.0} & ${\scriptstyle \pm \numRP[-2]{1300.0}}$ & ${\scriptstyle \pm \numRP[-2]{1000.0}}$ &  6500 &  ${\scriptstyle \pm \numRP[-2]{900.0}}$ \\
\ggHjPt{1}{60}{120}{} &  6800 & \errRP{-2}{1100.0}{-1000.0} & ${\scriptstyle \pm \numRP[-2]{900.0}}$ & \errRP{-2}{700.0}{-600.0} &  4500 &  ${\scriptstyle \pm \numRP[-2]{600.0}}$ \\
\ggHjPt{1}{120}{200}{} &  900 & \errRP{-2}{300.0}{-200.0} & ${\scriptstyle \pm \numRP[-2]{200.0}}$ & ${\scriptstyle \pm \numRP[-2]{100.0}}$ &  750 &  ${\scriptstyle \pm \numRP[-1]{130.0}}$ \\
\ggHmPt{}{350}{}{120} &  800 & ${\scriptstyle \pm \numRP[-2]{1300.0}}$ & ${\scriptstyle \pm \numRP[-2]{1100.0}}$ & ${\scriptstyle \pm \numRP[-2]{700.0}}$ &  3000 &  ${\scriptstyle \pm \numRP[-2]{600.0}}$ \\
\ggHmPt{}{350}{120}{200} &  900 & ${\scriptstyle \pm \numRP[-2]{400.0}}$ & ${\scriptstyle \pm \numRP[-2]{300.0}}$ & ${\scriptstyle \pm \numRP[-2]{200.0}}$ &  900 &  ${\scriptstyle \pm \numRP[-2]{200.0}}$ \\
\ggHmPt{350}{}{}{200} &  1300 & ${\scriptstyle \pm \numRP[-2]{700.0}}$ & ${\scriptstyle \pm \numRP[-2]{600.0}}$ & \errRP{-2}{400.0}{-300.0} &  900 &  ${\scriptstyle \pm \numRP[-2]{200.0}}$ \\
\ggHPt{200}{300}{} &  690 & \errRP{-1}{150.0}{-140.0} & ${\scriptstyle \pm \numRP[-1]{110.0}}$ & \errRP{-1}{90.0}{-80.0} &  500 &  ${\scriptstyle \pm \numRP[-2]{100.0}}$ \\
\ggHPt{300}{450}{} &  80 & ${\scriptstyle \pm \numRP[-1]{50.0}}$ & ${\scriptstyle \pm \numRP[-1]{40.0}}$ & ${\scriptstyle \pm \numRP[-1]{20.0}}$ &  110 &  ${\scriptstyle \pm \numRP[-1]{30.0}}$ \\
\ggHPt{450}{}{} &  30 & ${\scriptstyle \pm \numRP[-1]{20.0}}$ & ${\scriptstyle \pm \numRP[-1]{20.0}}$ & \errRP{-1}{10.0}{-0.0} &  18 &  ${\scriptstyle \pm \numRP[0]{5.0}}$ \\
\midrule
\ewqqH, $\le 1$-jet &  \num{-100} & \errRP{-2}{1400.0}{-1300.0} & \errRP{-2}{1300.0}{-1200.0} & \errRP{-2}{500.0}{-600.0} &  2160 &  ${\scriptstyle \pm \numRP[-1]{60.0}}$ \\
\ewqqH, \VBF-enriched &  1500 & \errRP{-2}{1000.0}{-900.0} & \errRP{-2}{900.0}{-800.0} & \errRP{-2}{500.0}{-400.0} &  740 &  ${\scriptstyle \pm \numRP[-1]{20.0}}$ \\
\ewqqH, \VH-enriched &  400 & ${\scriptstyle \pm \numRP[-2]{200.0}}$ & ${\scriptstyle \pm \numRP[-2]{200.0}}$ & ${\scriptstyle \pm \numRP[-2]{100.0}}$ &  510 &  ${\scriptstyle \pm \numRP[-1]{20.0}}$ \\
\HqqmPt{350}{700}{}{200} &  400 & ${\scriptstyle \pm \numRP[-2]{200.0}}$ & ${\scriptstyle \pm \numRP[-2]{200.0}}$ & ${\scriptstyle \pm \numRP[-2]{100.0}}$ &  540 &  ${\scriptstyle \pm \numRP[-1]{10.0}}$ \\
\HqqmPt{700}{1000}{}{200} &  160 & ${\scriptstyle \pm \numRP[-1]{90.0}}$ & ${\scriptstyle \pm \numRP[-1]{80.0}}$ & ${\scriptstyle \pm \numRP[-1]{40.0}}$ &  260 &  ${\scriptstyle \pm \numRP[-1]{10.0}}$ \\
\HqqmPt{1000}{1500}{}{200} &  360 & ${\scriptstyle \pm \numRP[-1]{80.0}}$ & ${\scriptstyle \pm \numRP[-1]{70.0}}$ & ${\scriptstyle \pm \numRP[-1]{40.0}}$ &  220 &  ${\scriptstyle \pm \numRP[-1]{10.0}}$ \\
\HqqmPt{1500}{}{}{200} &  180 & ${\scriptstyle \pm \numRP[-1]{50.0}}$ & ${\scriptstyle \pm \numRP[-1]{40.0}}$ & \errRP{-1}{30.0}{-20.0} &  220 &  ${\scriptstyle \pm \numRP[-1]{10.0}}$ \\
\HqqmPt{350}{700}{200}{} &  1 & \errRP{0}{33.0}{-29.0} & \errRP{0}{32.0}{-27.0} & ${\scriptstyle \pm \numRP[0]{10.0}}$ &  44 &  ${\scriptstyle \pm \numRP[0]{1.0}}$ \\
\HqqmPt{700}{1000}{200}{} &  27 & \errRP{0}{16.0}{-15.0} & \errRP{0}{15.0}{-14.0} & \errRP{0}{6.0}{-5.0} &  29 &  ${\scriptstyle \pm \numRP[0]{1.0}}$ \\
\HqqmPt{1000}{1500}{200}{} &  49 & \errRP{0}{14.0}{-13.0} & \errRP{0}{14.0}{-13.0} & \errRP{0}{5.0}{-4.0} &  33 &  ${\scriptstyle \pm \numRP[0]{1.0}}$ \\
\HqqmPt{1500}{}{200}{}{} &  45 & \errRP{0}{10.0}{-9.0} & ${\scriptstyle \pm \numRP[0]{9.0}}$ & \errRP{0}{4.0}{-3.0} &  41 &  ${\scriptstyle \pm \numRP[0]{1.0}}$ \\
\midrule
\HlnPt{}{75}{} &  271 & \errRP{0}{150.0}{-139.0} & \errRP{0}{142.0}{-132.0} & \errRP{0}{48.0}{-42.0} &  215 &  ${\scriptstyle \pm \numRP[0]{8.0}}$ \\
\HlnPt{75}{150}{} &  43 & \errRP{0}{64.0}{-56.0} & \errRP{0}{49.0}{-46.0} & \errRP{0}{41.0}{-33.0} &  134 &  ${\scriptstyle \pm \numRP[0]{5.0}}$ \\
\HlnPt{150}{250}{} &  40 & \errRP{0}{15.0}{-14.0} & \errRP{0}{11.0}{-10.0} & \errRP{0}{11.0}{-10.0} &  41 &  ${\scriptstyle \pm \numRP[0]{2.0}}$ \\
\HlnPt{250}{400}{} &  13 & ${\scriptstyle \pm \numRP[0]{3.0}}$ & ${\scriptstyle \pm \numRP[0]{3.0}}$ & ${\scriptstyle \pm \numRP[0]{2.0}}$ &  10.0 &  ${\scriptstyle \pm \numRP[1]{0.4}}$ \\
\HlnPt{400}{600}{} &  \num{-0.2} & \errRP{1}{1.0}{-0.9} & ${\scriptstyle \pm \numRP[1]{0.8}}$ & ${\scriptstyle \pm \numRP[1]{0.5}}$ &  1.8 &  ${\scriptstyle \pm \numRP[1]{0.1}}$ \\
\HlnPt{600}{}{} &  0.5 & \errRP{1}{0.4}{-0.3} & \errRP{1}{0.4}{-0.3} & ${\scriptstyle \pm \numRP[1]{0.1}}$ &  0.34 &  ${\scriptstyle \pm \numRP[2]{0.02}}$ \\
\midrule
\HllnnPt{}{75}{} &  221 & \errRP{0}{99.0}{-81.0} & \errRP{0}{94.0}{-78.0} & \errRP{0}{33.0}{-22.0} &  112 &  ${\scriptstyle \pm \numRP[0]{8.0}}$ \\
\HllnnPtj{75}{150}{0} &  70 & \errRP{0}{30.0}{-28.0} & ${\scriptstyle \pm \numRP[0]{24.0}}$ & \errRP{0}{17.0}{-16.0} &  51 &  ${\scriptstyle \pm \numRP[0]{4.0}}$ \\
\HllnnPtj{75}{150}{1+} &  9 & \errRP{0}{37.0}{-36.0} & \errRP{0}{33.0}{-32.0} & \errRP{0}{18.0}{-17.0} &  36 &  ${\scriptstyle \pm \numRP[0]{5.0}}$ \\
\HllnnPtj{150}{250}{0} &  12 & ${\scriptstyle \pm \numRP[0]{5.0}}$ & \errRP{0}{5.0}{-4.0} & \errRP{0}{3.0}{-2.0} &  15 &  ${\scriptstyle \pm \numRP[0]{2.0}}$ \\
\HllnnPtj{150}{250}{1+} &  16 & \errRP{0}{12.0}{-11.0} & ${\scriptstyle \pm \numRP[0]{10.0}}$ & \errRP{0}{6.0}{-5.0} &  17 &  ${\scriptstyle \pm \numRP[0]{3.0}}$ \\
\HllnnPtj{250}{400}{0} &  2.3 & \errRP{1}{1.3}{-1.2} & \errRP{1}{1.2}{-1.1} & ${\scriptstyle \pm \numRP[1]{0.5}}$ &  2.9 &  ${\scriptstyle \pm \numRP[1]{0.4}}$ \\
\HllnnPtj{250}{400}{1+} &  4.5 & \errRP{1}{3.6}{-3.4} & \errRP{1}{3.3}{-3.2} & \errRP{1}{1.5}{-1.3} &  4.2 &  ${\scriptstyle \pm \numRP[1]{0.8}}$ \\
\HllnnPt{400}{600}{} &  1.1 & ${\scriptstyle \pm \numRP[1]{0.7}}$ & \errRP{1}{0.7}{-0.6} & \errRP{1}{0.3}{-0.2} &  1.1 &  ${\scriptstyle \pm \numRP[1]{0.1}}$ \\
\HllnnPt{600}{}{} &  \num{-0.2} & ${\scriptstyle \pm \numRP[1]{0.2}}$ & ${\scriptstyle \pm \numRP[1]{0.2}}$ & ${\scriptstyle \pm \numRP[1]{0.1}}$ &  0.18 &  ${\scriptstyle \pm \numRP[2]{0.01}}$ \\
\midrule
\ttHPt{}{60}{} &  120 & ${\scriptstyle \pm \numRP[-1]{50.0}}$ & \errRP{-1}{50.0}{-40.0} & ${\scriptstyle \pm \numRP[-1]{30.0}}$ &  120 &  ${\scriptstyle \pm \numRP[-1]{20.0}}$ \\
\ttHPt{60}{120}{} &  100 & \errRP{-1}{60.0}{-50.0} & ${\scriptstyle \pm \numRP[-1]{50.0}}$ & \errRP{-1}{30.0}{-20.0} &  180 &  ${\scriptstyle \pm \numRP[-1]{20.0}}$ \\
\ttHPt{120}{200}{} &  80 & ${\scriptstyle \pm \numRP[-1]{40.0}}$ & ${\scriptstyle \pm \numRP[-1]{30.0}}$ & ${\scriptstyle \pm \numRP[-1]{20.0}}$ &  130 &  ${\scriptstyle \pm \numRP[-1]{20.0}}$ \\
\ttHPt{200}{300}{} &  26 & \errRP{0}{17.0}{-16.0} & \errRP{0}{15.0}{-14.0} & \errRP{0}{8.0}{-7.0} &  53 &  ${\scriptstyle \pm \numRP[0]{7.0}}$ \\
\ttHPt{300}{450}{} &  10 & ${\scriptstyle \pm \numRP[0]{9.0}}$ & \errRP{0}{8.0}{-7.0} & ${\scriptstyle \pm \numRP[0]{5.0}}$ &  19 &  ${\scriptstyle \pm \numRP[0]{3.0}}$ \\
\ttHPt{450}{}{} &  4 & ${\scriptstyle \pm \numRP[0]{5.0}}$ & ${\scriptstyle \pm \numRP[0]{4.0}}$ & ${\scriptstyle \pm \numRP[0]{2.0}}$ &  5 &  ${\scriptstyle \pm \numRP[0]{1.0}}$ \\
\tH &  460 & \errRP{-1}{220.0}{-210.0} & ${\scriptstyle \pm \numRP[-1]{160.0}}$ & \errRP{-1}{150.0}{-130.0} &  80 &  ${\scriptstyle \pm \numRP[-1]{10.0}}$ \\
\bottomrule
\end{tabular}}
\label{tab:stxs:fixedBR:obs}
\end{table}


The correlation matrix of the measurement is shown in Figure~\ref{fig:stxs:fixedBR:corr} in Appendix~\ref{app:stxs}.
The measurement provides the most detailed description to date of Higgs boson production in $pp$ collisions. Compared to Ref.~\cite{HIGG-2021-23}, the $\ptH > \qty{200}{\GeV}$ region of the \ewqqH\ process has finer binning in \mjj, mainly due to the sensitivity provided by the improved analyses in the \Hww~\cite{HIGP-2024-07} and \Htt~\cite{HIGG-2022-07} decay modes.
In the \qqtoHln\ and \pptoHllnn\ processes, an additional region is defined for $\ptV \ge \qty{600}{\GeV}$.
In the \pptoHllnn\ process, bins corresponding to $\Njets = 0$ and $\Njets \ge 1$ are now separately measured in all the regions in the $75 \le \ptV < \qty{400}{\GeV}$ range.
These new measurements are made possible mainly by the increased sensitivity provided by the improved analysis of the $\VH, \Hbb$ process~\cite{HIGG-2020-20}.
Finally, a measurement is now provided in the region $\ptV < \qty{75}{\GeV}$ of the \pptoHllnn\ process, mainly due to the introduction of the $\VH, \Hww$ measurement~\cite{HIGG-2023-09}. The results are in agreement with SM predictions, with a $p$-value $p\SM=47\%$.

In the second model, the cross-section $\sigma_r^f$ for Higgs boson production and decay into the final state $f$ in STXS region $r$ is expressed as
\begin{equation}
\sigma_r^f = \sigma_r^{\zz} \left(\frac{B_f}{B_{\zz}}\right),
\end{equation}
in terms of the cross-sections $\sigma_r^{\zz}$ in the \Hzz\ final state and the ratios of branching ratios $B_f/B_{\zz}$ for $f=\yy,\ww,\bb$ and \tautau. The
\Hzz\ decay process is chosen as the reference due to its high experimental sensitivity and in particular small systematic uncertainties. Other branching ratios are set to their SM expectations. The observed and expected measurement results are shown in Table~\ref{tab:stxs:ratioBR:obs}. The results are in agreement with SM predictions, with a $p$-value $p\SM=56\%$.

\begin{table}[ht]
\caption{
Best-fit values and uncertainties for the production cross-section in each measured STXS region in the \Hzz\ decay process and ratios of branching ratios to $B_{\zz}$. The values for the \ggtoH\ process also include the contributions from \bbH\ production.
The total uncertainties are decomposed into components for data statistics (Stat.) and systematic uncertainties (Syst.).
SM predictions are also shown for each quantity with their total uncertainties. The relative uncertainty in $B_{\ww} / B_{\zz}$ is smaller than $10^{-3}$ and is not shown in the table.
}
\centering
\renewcommand{\arraystretch}{1.2}
\resizebox{0.7\textwidth}{!}{
\begin{tabular}{l S[table-format=-3.4,round-mode=none] lll S[table-format=3.5,round-mode=none] l}
\toprule
\multirow{2}{*}{Parameter} & \multicolumn{1}{c}{Value} & \multicolumn{3}{c}{Uncertainty} & \multicolumn{2}{c}{SM prediction} \\
&                 & Total   & Stat.  & Syst.   &  &  \\
\midrule
$B_{\yy} / B_{\zz}$ &  0.092 & \errRP{3}{0.011}{-0.01} & \errRP{3}{0.01}{-0.009} & \errRP{3}{0.005}{-0.004} &  0.086 &  ${\scriptstyle \pm \numRP[3]{0.001}}$ \\
$B_{\ww} / B_{\zz}$ &  9.2 & \errRP{1}{1.1}{-1.0} & \errRP{1}{0.9}{-0.8} & ${\scriptstyle \pm \numRP[1]{0.6}}$ &  8.1 &  - \\
$B_{\bb} / B_{\zz}$ &  21.6 & \errRP{1}{5.1}{-4.3} & \errRP{1}{4.1}{-3.4} & \errRP{1}{3.1}{-2.6} &  22.0 &  ${\scriptstyle \pm \numRP[1]{0.7}}$ \\
$B_{\tautau} / B_{\zz}$ &  2.23 & \errRP{2}{0.38}{-0.33} & \errRP{2}{0.3}{-0.26} & \errRP{2}{0.23}{-0.2} &  2.37 &  ${\scriptstyle \pm \numRP[2]{0.02}}$ \\
\midrule
\multirow{2}{*}{Region} & \multicolumn{1}{c}{$\sigma \times B_{\zz}$} & \multicolumn{3}{c}{ Uncertainty [fb]} & \multicolumn{2}{c}{SM prediction} \\
&  \multicolumn{1}{c}{[fb]}                & Total   & Stat.                & Syst.   & \multicolumn{2}{c}{[fb]}   \\
\midrule
\ggHjPt{0}{}{10}{} &  128 & \errRP{0}{33.0}{-30.0} & \errRP{0}{30.0}{-28.0} & \errRP{0}{12.0}{-11.0} &  175 &  ${\scriptstyle \pm \numRP[0]{23.0}}$ \\
\ggHjPt{0}{10}{}{} &  607 & \errRP{0}{72.0}{-67.0} & \errRP{0}{64.0}{-60.0} & \errRP{0}{33.0}{-29.0} &  545 &  ${\scriptstyle \pm \numRP[0]{41.0}}$ \\
\ggHjPt{1}{}{60}{} &  175 & \errRP{0}{43.0}{-41.0} & \errRP{0}{36.0}{-34.0} & ${\scriptstyle \pm \numRP[0]{23.0}}$ &  172 &  \errRP{0}{23.0}{-24.0} \\
\ggHjPt{1}{60}{120}{} &  168 & \errRP{0}{31.0}{-28.0} & \errRP{0}{26.0}{-25.0} & \errRP{0}{15.0}{-14.0} &  119 &  ${\scriptstyle \pm \numRP[0]{16.0}}$ \\
\ggHjPt{1}{120}{200}{} &  22.4 & \errRP{1}{6.8}{-6.3} & \errRP{1}{5.9}{-5.5} & \errRP{1}{3.4}{-3.0} &  19.7 &  ${\scriptstyle \pm \numRP[1]{3.4}}$ \\
\ggHmPt{}{350}{}{120} &  21 & \errRP{0}{34.0}{-33.0} & ${\scriptstyle \pm \numRP[0]{28.0}}$ & ${\scriptstyle \pm \numRP[0]{18.0}}$ &  78 &  ${\scriptstyle \pm \numRP[0]{16.0}}$ \\
\ggHmPt{}{350}{120}{200} &  24 & \errRP{0}{11.0}{-10.0} & ${\scriptstyle \pm \numRP[0]{9.0}}$ & \errRP{0}{5.0}{-4.0} &  24.9 &  ${\scriptstyle \pm \numRP[1]{5.7}}$ \\
\ggHmPt{350}{}{}{200} &  35 & \errRP{0}{18.0}{-17.0} & ${\scriptstyle \pm \numRP[0]{15.0}}$ & \errRP{0}{10.0}{-8.0} &  23.2 &  ${\scriptstyle \pm \numRP[1]{5.5}}$ \\
\ggHPt{200}{300}{} &  17.4 & \errRP{1}{4.1}{-3.7} & \errRP{1}{3.5}{-3.2} & \errRP{1}{2.2}{-1.9} &  12.1 &  ${\scriptstyle \pm \numRP[1]{2.7}}$ \\
\ggHPt{300}{450}{} &  1.9 & ${\scriptstyle \pm \numRP[1]{1.2}}$ & \errRP{1}{1.2}{-1.1} & ${\scriptstyle \pm \numRP[1]{0.4}}$ &  2.81 &  ${\scriptstyle \pm \numRP[2]{0.71}}$ \\
\ggHPt{450}{}{} &  0.86 & \errRP{2}{0.6}{-0.49} & \errRP{2}{0.58}{-0.48} & \errRP{2}{0.17}{-0.12} &  0.47 &  ${\scriptstyle \pm \numRP[2]{0.14}}$ \\
\midrule
\ewqqH, $\le 1$-jet &  -2 & \errRP{0}{34.0}{-33.0} & \errRP{0}{31.0}{-30.0} & \errRP{0}{13.0}{-14.0} &  57.1 &  ${\scriptstyle \pm \numRP[1]{1.7}}$ \\
\ewqqH, \VBF-enriched &  41 & \errRP{0}{26.0}{-24.0} & \errRP{0}{23.0}{-22.0} & \errRP{0}{11.0}{-10.0} &  19.4 &  ${\scriptstyle \pm \numRP[1]{0.6}}$ \\
\ewqqH, \VH-enriched &  9.8 & \errRP{1}{6.4}{-6.0} & \errRP{1}{5.7}{-5.3} & \errRP{1}{3.0}{-2.8} &  13.5 &  ${\scriptstyle \pm \numRP[1]{0.5}}$ \\
\HqqmPt{350}{700}{}{200} &  9.1 & \errRP{1}{5.4}{-5.1} & \errRP{1}{4.6}{-4.3} & \errRP{1}{2.7}{-2.8} &  14.1 &  ${\scriptstyle \pm \numRP[1]{0.4}}$ \\
\HqqmPt{700}{1000}{}{200} &  4.0 & \errRP{1}{2.4}{-2.2} & \errRP{1}{2.1}{-1.9} & \errRP{1}{1.1}{-1.0} &  6.77 &  ${\scriptstyle \pm \numRP[2]{0.2}}$ \\
\HqqmPt{1000}{1500}{}{200} &  9.0 & \errRP{1}{2.4}{-2.1} & \errRP{1}{2.1}{-1.9} & \errRP{1}{1.2}{-1.0} &  5.92 &  ${\scriptstyle \pm \numRP[2]{0.17}}$ \\
\HqqmPt{1500}{}{}{200} &  4.6 & \errRP{1}{1.5}{-1.3} & \errRP{1}{1.2}{-1.1} & \errRP{1}{0.8}{-0.6} &  5.70 &  ${\scriptstyle \pm \numRP[2]{0.18}}$ \\
\HqqmPt{350}{700}{200}{} &  0.13 & \errRP{2}{0.9}{-0.77} & \errRP{2}{0.85}{-0.73} & \errRP{2}{0.28}{-0.26} &  1.169 &  ${\scriptstyle \pm \numRP[3]{0.032}}$ \\
\HqqmPt{700}{1000}{200}{} &  0.72 & \errRP{2}{0.45}{-0.39} & \errRP{2}{0.42}{-0.37} & \errRP{2}{0.17}{-0.13} &  0.778 &  ${\scriptstyle \pm \numRP[3]{0.022}}$ \\
\HqqmPt{1000}{1500}{200}{} &  1.29 & \errRP{2}{0.43}{-0.37} & \errRP{2}{0.4}{-0.35} & \errRP{2}{0.16}{-0.12} &  0.860 &  ${\scriptstyle \pm \numRP[3]{0.025}}$ \\
\HqqmPt{1500}{}{200}{}{} &  1.2 & ${\scriptstyle \pm \numRP[1]{0.3}}$ & ${\scriptstyle \pm \numRP[1]{0.3}}$ & ${\scriptstyle \pm \numRP[1]{0.1}}$ &  1.07 &  ${\scriptstyle \pm \numRP[2]{0.03}}$ \\
\midrule
\HlnPt{}{75}{} &  6.4 & \errRP{1}{3.8}{-3.4} & \errRP{1}{3.5}{-3.2} & \errRP{1}{1.3}{-1.0} &  5.68 &  \errRP{2}{0.21}{-0.22} \\
\HlnPt{75}{150}{} &  1.0 & \errRP{1}{1.6}{-1.4} & \errRP{1}{1.3}{-1.2} & \errRP{1}{1.0}{-0.8} &  3.55 &  ${\scriptstyle \pm \numRP[2]{0.15}}$ \\
\HlnPt{150}{250}{} &  1.0 & \errRP{1}{0.5}{-0.4} & ${\scriptstyle \pm \numRP[1]{0.3}}$ & ${\scriptstyle \pm \numRP[1]{0.3}}$ &  1.087 &  \errRP{3}{0.046}{-0.047} \\
\HlnPt{250}{400}{} &  0.35 & \errRP{2}{0.13}{-0.1} & \errRP{2}{0.11}{-0.09} & \errRP{2}{0.07}{-0.05} &  0.265 &  ${\scriptstyle \pm \numRP[3]{0.012}}$ \\
\HlnPt{400}{600}{} &  -0.004 & \errRP{3}{0.026}{-0.025} & \errRP{3}{0.023}{-0.021} & \errRP{3}{0.013}{-0.014} &  0.047 &  ${\scriptstyle \pm \numRP[3]{0.002}}$ \\
\HlnPt{600}{}{} &  0.012 & \errRP{3}{0.012}{-0.009} & \errRP{3}{0.011}{-0.008} & \errRP{3}{0.004}{-0.002} &  0.0090 &  ${\scriptstyle \pm \numRP[4]{0.0006}}$ \\
\midrule
\HllnnPt{}{75}{} &  5.3 & \errRP{1}{2.5}{-2.0} & \errRP{1}{2.3}{-1.9} & \errRP{1}{0.9}{-0.5} &  2.95 &  ${\scriptstyle \pm \numRP[2]{0.21}}$ \\
\HllnnPtj{75}{150}{0} &  1.82 & \errRP{2}{0.9}{-0.77} & \errRP{2}{0.74}{-0.65} & \errRP{2}{0.5}{-0.41} &  1.35 &  ${\scriptstyle \pm \numRP[2]{0.12}}$ \\
\HllnnPtj{75}{150}{1+} &  0.2 & ${\scriptstyle \pm \numRP[1]{1.0}}$ & ${\scriptstyle \pm \numRP[1]{0.9}}$ & ${\scriptstyle \pm \numRP[1]{0.5}}$ &  0.94 &  ${\scriptstyle \pm \numRP[2]{0.12}}$ \\
\HllnnPtj{150}{250}{0} &  0.3 & \errRP{1}{0.2}{-0.1} & ${\scriptstyle \pm \numRP[1]{0.1}}$ & ${\scriptstyle \pm \numRP[1]{0.1}}$ &  0.408 &  ${\scriptstyle \pm \numRP[3]{0.061}}$ \\
\HllnnPtj{150}{250}{1+} &  0.41 & \errRP{2}{0.33}{-0.3} & \errRP{2}{0.29}{-0.26} & \errRP{2}{0.16}{-0.14} &  0.443 &  ${\scriptstyle \pm \numRP[3]{0.084}}$ \\
\HllnnPtj{250}{400}{0} &  0.060 & \errRP{3}{0.039}{-0.032} & \errRP{3}{0.035}{-0.029} & \errRP{3}{0.018}{-0.012} &  0.077 &  ${\scriptstyle \pm \numRP[3]{0.01}}$ \\
\HllnnPtj{250}{400}{1+} &  0.12 & \errRP{2}{0.1}{-0.09} & \errRP{2}{0.09}{-0.08} & \errRP{2}{0.04}{-0.03} &  0.112 &  ${\scriptstyle \pm \numRP[3]{0.021}}$ \\
\HllnnPt{400}{600}{} &  0.028 & \errRP{3}{0.022}{-0.018} & \errRP{3}{0.02}{-0.016} & \errRP{3}{0.01}{-0.007} &  0.0284 &  ${\scriptstyle \pm \numRP[4]{0.0023}}$ \\
\HllnnPt{600}{}{} &  -0.0043 & \errRP{4}{0.0063}{-0.005} & \errRP{4}{0.006}{-0.0046} & \errRP{4}{0.0018}{-0.002} &  0.00481 &  ${\scriptstyle \pm \numRP[5]{0.00032}}$ \\
\midrule
\ttHPt{}{60}{} &  2.91 & \errRP{2}{1.5}{-1.34} & \errRP{2}{1.3}{-1.18} & \errRP{2}{0.76}{-0.64} &  3.12 &  ${\scriptstyle \pm \numRP[2]{0.42}}$ \\
\ttHPt{60}{120}{} &  2.62 & \errRP{2}{1.56}{-1.43} & \errRP{2}{1.41}{-1.28} & \errRP{2}{0.68}{-0.62} &  4.70 &  ${\scriptstyle \pm \numRP[2]{0.52}}$ \\
\ttHPt{120}{200}{} &  1.96 & \errRP{2}{1.08}{-0.97} & \errRP{2}{0.93}{-0.84} & \errRP{2}{0.55}{-0.48} &  3.34 &  ${\scriptstyle \pm \numRP[2]{0.41}}$ \\
\ttHPt{200}{300}{} &  0.698 & \errRP{3}{0.473}{-0.42} & \errRP{3}{0.417}{-0.373} & \errRP{3}{0.224}{-0.192} &  1.39 &  ${\scriptstyle \pm \numRP[2]{0.2}}$ \\
\ttHPt{300}{450}{} &  0.258 & \errRP{3}{0.262}{-0.234} & \errRP{3}{0.222}{-0.2} & \errRP{3}{0.14}{-0.122} &  0.503 &  \errRP{3}{0.082}{-0.081} \\
\ttHPt{450}{}{} &  0.098 & \errRP{3}{0.14}{-0.124} & \errRP{3}{0.121}{-0.108} & \errRP{3}{0.071}{-0.061} &  0.142 &  ${\scriptstyle \pm \numRP[3]{0.026}}$ \\
\tH &  12.17 & \errRP{2}{5.91}{-5.38} & \errRP{2}{4.46}{-4.13} & \errRP{2}{3.87}{-3.44} &  2.24 &  \errRP{2}{0.15}{-0.29} \\
\bottomrule
\end{tabular}}
\label{tab:stxs:ratioBR:obs}
\end{table}


\FloatBarrier


\section{Constraints on effective field theory models}
\label{sec:eft}

Effective-field theory (EFT) models provide a framework to describe the effects of BSM phenomena occurring at energy scales much larger than the scales of the physics process considered. These effects usually depend on the kinematics of the process, with typically larger effects at higher momentum transfers. This section presents a reinterpretation of measurements of Higgs boson kinematic properties in the STXS framework similar to that presented in Section~\ref{sec:stxs}, in the context of Standard Model Effective-field theory (SMEFT)~\cite{Brivio:2017vri,ATL-PHYS-PUB-2019-042}.

These models supplement the SM Lagrangian with additional interactions of canonical mass dimension larger than 4, built up from the usual SM fields and respecting SM symmetries. The SMEFT Lagrangian is expressed as
\begin{equation}
\mathcal{L} = \mathcal{L}_{\text{SM}} + \sum\limits_{d > 4} \sum\limits_{n} \frac{c_n^{(d)}}{\Lambda^{d-4}} O_n^{(d)},
\end{equation}
where the $O^{(d)}_n$ are operators of canonical dimension $d$ describing the EFT interactions, the $c^{(d)}_n$ are real coupling parameters specifying the interaction strengths, and $\Lambda$ is the EFT mass scale. The SM corresponds to the case where all $c^{(d)}_n$ vanish. In this paper, operators conserving baryon and lepton number are considered, which leads to the leading-order corrections to the SM occurring at $d = 6$. EFT contributions with $d > 6$ are omitted in the rest of this work, and the superscripts on the $O^{(6)}_n$ and  $c^{(6)}_n$ terms are dropped. Operators are defined in the Warsaw basis~\cite{Warsaw}, which defines a complete and independent basis for the $d=6$ sector. Only CP-even operators are considered. The \texttt{top} flavour scheme is used, in which the quark fields of the first and second generation, which cannot be distinguished by the measurements included in the combination, are related through $U(2)$ symmetries while the third generation quark fields remain independent. Effective interactions involving first and second generation quarks are therefore described using a common set of operators in terms of the fields $q$, $u$ and $d$ representing respectively left-handed quarks, right-handed up-type quarks and right-handed down-type quarks of the first and second generation. The corresponding third generation quarks are respectively denoted by $Q$, $t$ and $b$. No flavour assumption is applied in the lepton sector and independent fields are used to describe leptons in each generation. Left-handed leptons and right-handed electrons are respectively designated as $l_i$ and $e_i$, where $i=1,2,3$ denotes the lepton generation.

SMEFT effects in Higgs boson processes are described using amplitudes containing at most one SMEFT operator insertion. The event rate for a given process is then proportional to
\begin{equation}
\int d\Omega \left| \mathcal{M}_{\text{SM}} + \sum\limits_{n} \frac{c_n}{\Lambda^2}  \mathcal{M}_n  \right|^2 =
\int d\Omega \left| \mathcal{M}_{\text{SM}} \right|^2 \left(1 + \sum\limits_{n}  c_n  A_n + \sum\limits_{mn}  c_n c_m B_{mn}\right)
\label{eq:eft:xs}
\end{equation}
where $\mathcal{M}_{\text{SM}}$ is the SM amplitude for the process and $\mathcal{M}_n$ the amplitude for one insertion of the operator $O_n$. The sums run over all SMEFT operators contributing to the process, and the integral runs over the phase space considered. The expression is quadratic in the EFT parameters $c_n$ and the event rate is expressed in terms of the linear coefficients $A_n$ and the quadratic coefficients $B_{mn}$; their values are computed separately for each production and decay process as described below. The first term on the right-hand side is the SM contribution to the event rate, while the following terms correspond to the interference between SM and EFT amplitudes, which is linear in the EFT parameters, and to purely EFT contributions that scale as products of two EFT parameters. They are respectively referred to as \emph{linear} and \emph{quadratic} terms in the following.

Measurements of the parameters $c_n$ are obtained from the description of Higgs boson production rates and kinematics provided by the STXS framework, described in Section~\ref{sec:stxs}. Signal strengths relative to SM predictions are measured for 116 STXS regions defined separately in the \Hbb, \Hww, \Htt, \Hcc, \Hzz, \Hyy, \Hzy\ and \Hmm\ decay channels. The STXS binning in each decay mode is adapted to match the sensitivity provided by the corresponding analyses, and presented in Table~\ref{tab:stxs:br_scheme} of Appendix~\ref{app:stxs_br}. Contributions from other STXS regions are assumed to match their SM expectations. Reference SM predictions are taken to be the highest-order prediction available, and theory uncertainties are included as described in Section~\ref{sec:uncertainties}.
The measurement is described in detail in Appendix~\ref{app:stxs_br}.

The parameterisation of Eq.~\eqref{eq:eft:xs} is applied separately to the production cross-sections $\sigma_r$ in STXS region $r$, the partial decay widths $\Gamma_f$ in decay mode $f$ and the Higgs boson total width $\Gamma_H$. Under the narrow-width approximation, the signal strength $\mu_r^f$ for Higgs boson production in STXS region $r$ and decay in final state $f$ is then expressed in terms of the set $\mathbf{c}$ of all $c_n$ parameters as
\begin{equation}
\mu_r^f(\mathbf{c}) = \left(\sigma_r(\mathbf{c})/\sigma_r^{\text{SM}}\right) \frac{\Gamma_f(\mathbf{c})/\Gamma_f^{\text{SM}}}{\Gamma_H(\mathbf{c})/\Gamma_H^{\text{SM}}}  = \frac{
\left(1 + \sum\limits_{n} c_n A^{\sigma_r}_n   + \sum\limits_{mn} c_n c_m B^{\sigma_r}_{mn} \right)
\left(1 + \sum\limits_{n} c_n A^{\Gamma_f}_n   + \sum\limits_{mn} c_n c_m B^{\Gamma_f}_{mn} \right) }{
\left(1 + \sum\limits_{n} c_n A^{\Gamma}_n     + \sum\limits_{mn} c_n c_m B^{\Gamma}_{mn}   \right)
}.
\label{eq:eft:param_mu}
\end{equation}
where quantities with an SM superscript refer to SM predictions. Higher-order QCD corrections are assumed to be the same for EFT and SM processes.
The coefficients $A^{\sigma_r}_n$, $A^{\Gamma_f}_n$ and $A^{\Gamma}_n$ respectively define the linear contributions to $\sigma_r$, $\Gamma_f$ and $\Gamma_H$ for the operator $O_n$. The coefficients $B^{\sigma_r}_{mn}$, $B^{\Gamma_f}_{mn}$ and $B^{\Gamma}_{mn}$ similarly define the quadratic contributions corresponding to the operators $O_m$ and $O_n$.

Two versions of the parameterisation are considered. In the \emph{linear} parameterisation only the linear terms are considered while the $B$ terms are set to zero, which corresponds to a truncation of the EFT expansion at order $\Lambda^{-2}$ separately for each factor in Eq~\ref{eq:eft:param_mu}.
The denominator expression $\left(1 + \sum_n c_n A^{\Gamma}_n\right)^{-1}$ in Eq.~\eqref{eq:eft:param_mu} is not linearized, since this leads to a significant change in the expression for the values of the EFT parameters probed in this interpretation. As described
below, the Jacobian used to obtain the basis of the measurement parameters is however
computed using the fully linearized approximation. Full details of this
procedure are described in Ref.~\cite{HIGG-2022-17}.
In the \emph{linear+quadratic} parameterisation, both linear and quadratic terms are considered. The quadratic terms occur at order $\Lambda^{-4}$, but this parameterisation does not correspond to a full description of the SMEFT at this order since contributions originating from the interference between the SM and $d=8$ SMEFT operators that also enter at order $\Lambda^{-4}$ are not considered. The comparison of the results obtained in the two scenarios provides an estimate of the impact of terms beyond order $\Lambda^{-2}$ in the EFT expansion, and both sets are presented in this work.

The $A$ and $B$ coefficients of Eq.~\eqref{eq:eft:param_mu} are computed using a procedure similar to that of Ref.~\cite{HIGG-2022-17}. Samples of simulated Higgs boson production and decay processes are produced for various values of the EFT coefficients $c_n$, and the reported cross-sections in each case are used to obtain the values of the $A$ and $B$ parameters. The event generation is performed at particle level using SMEFT@NLO~\cite{SMEFTatNLO} and SMEFTsim~\cite{SMEFTsim} in the $(G_F, m_W, m_Z)$ scheme, using the as universal \feynrules\ output (UFO) functionality within \amc~\cite{mg5aMCatNLO}, interfaced to PYTHIA8~\cite{Sjostrand:2007gs} with the A14 tune~\cite{ATL-PHYS-PUB-2014-021} to handle parton showering and hadronisation. SMEFT@NLO is used to model the \ggtoH, \ggtoZH\ and \Hgg\ loop processes at NLO in QCD, while SMEFTsim is used for other processes. Charm quark contributions are not included in the loop processes modelled using SMEFT@NLO, so that the effect of operators involving charm quarks is not included in the parameterisation of these processes. The parameterisations of the \Hyy\ and \Hzy\ decays are taken from theory predictions obtained at NLO in electroweak effects, respectively Ref.~\cite{Dedes:2018seb} and~\cite{Dedes:2019bew}. The parameterisation of $\Gamma_H$ is obtained by considering all Higgs boson decays with up to four elementary particles in the final state. Compared to Ref.~\cite{HIGG-2022-17}, the parameterisation of the \VH\ process is modified by the addition of a split in the STXS binning at $\pTV = \qty{600}{\GeV}$; separate parameterisations for quark induced and gluon induced \ZH\ production are used; and the renormalisation and factorisation scales used in the generation of the \VH\ and \ggtoH\ processes are set to fixed values to conform with SMEFT@NLO recommendations~\cite{SMEFTatNLO}. In the \Hzz\ and \Hww\ decays, non-negligible changes to the experimental acceptance are found for variations in \wc{HW},  \wc{HB},  \wc{HWB}\ and \wc[3]{Hl}. These changes are accounted for by applying the same corrections to the $A_n^{\Hzz}$ and $A_n^{\Hww}$ coefficients as described in Ref.~\cite{HIGG-2022-17}.

\begin{table}[tbp]
\caption{Operators considered in the SMEFT analysis. The $l_i$ and $e_i$ are respectively the left-handed and right-handed lepton fields for generation $i$. Third-generation quark fields are denoted as $Q$, $t$ and $b$, while $q$, $u$ and $d$ refer to first and second generation quarks. Higgs field derivatives are defined as $\phi^\dag i \overleftrightarrow{D}_\mu \phi = \phi^\dag (i D_\mu \phi) -  (i D_\mu \phi)^\dag \phi$ and $\phi^\dag i \overleftrightarrow{D}^a_\mu \phi = \phi^\dag \sigma^a(i D_\mu \phi) -  (i D_\mu \phi)^\dag \sigma^a \phi$.}
\centering
\renewcommand{\arraystretch}{1.4}
\begin{tabular}[t]{lc}
\toprule
Name & Operator \\
\midrule
\wc{H\Box}   & \opHBox             \\
\wc{HDD}     & \opHDD              \\
\midrule
\wc{G}       & \opG                \\
\midrule
\wc{HG}      & \opHG               \\
\wc{HW}      & \opHW               \\
\wc{HB}      & \opHB               \\
\wc{HWB}     & \opHWB              \\
\midrule
\wc{eH,22}   & \opeH{2}{2}         \\
\wc{eH,33}   & \opeH{3}{3}         \\
\wc{uH}      & \opuH               \\
\wc{tH}      & \optH               \\
\wc{bH}      & \opbH               \\
\midrule
\wc{tG}      & \optG               \\
\wc{tW}      & \optW               \\
\wc{tB}      & \optB               \\
\midrule
\wc[1]{Hl,11} & \opHlOne{1}{1}     \\
\wc[3]{Hl,11} & \opHlThree{1}{1}   \\
\wc[1]{Hl,22} & \opHlOne{2}{2}     \\
\wc[3]{Hl,22} & \opHlThree{2}{2}   \\
\wc[1]{Hl,33} & \opHlOne{3}{3}     \\
\wc[3]{Hl,33} & \opHlThree{3}{3}   \\
\wc{He,11} & \opHe{1}{1}        \\
\wc{He,22} & \opHe{2}{2}        \\
\wc{He,33} & \opHe{3}{3}        \\
\bottomrule
\end{tabular}
\begin{tabular}[t]{lc}
\toprule
Name & Operator \\
\midrule
\wc[1]{Hq} & \opHqOne    \\
\wc[3]{Hq} & \opHqThree  \\
\wc{Hu} & \opHu       \\
\wc{Hd} & \opHd       \\
\wc[1]{HQ} & \opHQOne    \\
\wc[3]{HQ} & \opHQThree  \\
\wc{Ht} & \opHt       \\
\wc{Hb} & \opHb       \\
\midrule
\wc{ll,1221} & \opll{1}{2}{2}{1} \\
\midrule
\wc[1,1]{Qq} & \opqqSS{Q}{q} \\
\wc[3,1]{Qq} & \opqqST{Q}{q} \\
\wc[1,8]{Qq} & \opqqOS{Q}{q} \\
\wc[3,8]{Qq} & \opqqOT{Q}{q} \\
\wc[1]{Qu} & \opqqSS{Q}{u} \\
\wc[8]{Qu} & \opqqOS{Q}{u} \\
\wc[1]{Qd} & \opqqSS{Q}{d} \\
\wc[8]{Qd} & \opqqOS{Q}{d} \\
\midrule
\wc[1]{tq} & \opqqSS{t}{q} \\
\wc[8]{tq} & \opqqOS{t}{q} \\
\wc[1]{tu} & \opqqSS{t}{u} \\
\wc[8]{tu} & \opqqOS{t}{u} \\
\wc[8]{td} & \opqqOS{t}{d} \\
\bottomrule
\end{tabular}
\label{tab:eft:ops}
\end{table}


The operators considered are those for which the absolute value of at least one $A$ coefficient is larger than 0.1\%. The resulting 46 operators, and their associated EFT parameters $c_n$, are listed in Table~\ref{tab:eft:ops}.

\begin{figure}[tbp]
\centering
\includegraphics[width=\textwidth]{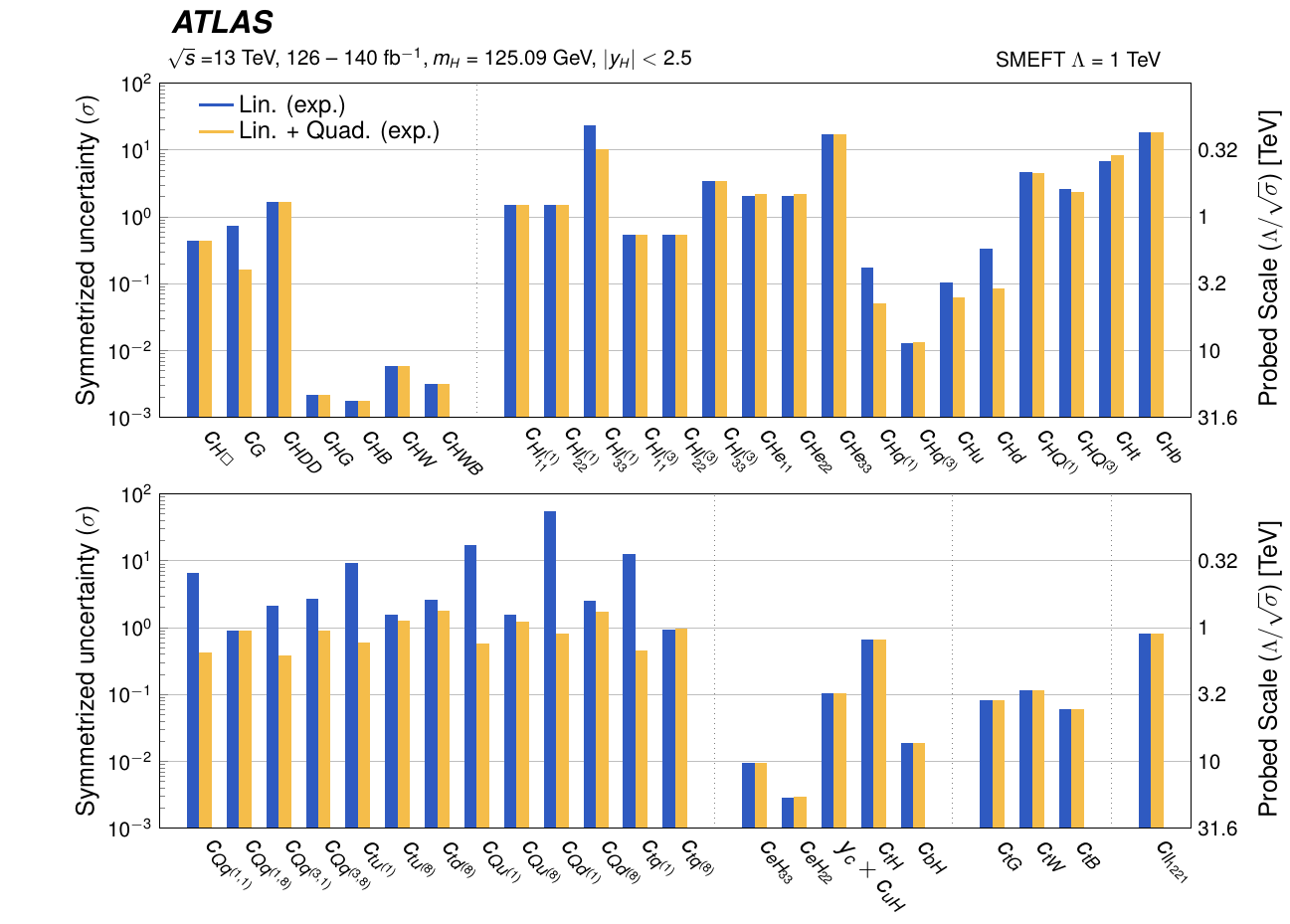}
\caption{Symmetrised expected 68\% CL uncertainty in EFT parameters in the Warsaw basis for the linear and linear+quadratic parameterisations, shown in two panels for readability. Each EFT parameter is measured while fixing all other parameters to their SM value of zero. The left axis shows the uncertainty $\sigma$ on the parameter assuming $\Lambda = \qty{1}{\TeV}$ for the EFT scale, while the right axis shows the probed EFT scale estimated as $\Lambda/\sqrt{\sigma}$.}
\label{fig:eft:indiv:exp}
\end{figure}

A qualitative estimate of the experimental sensitivity to individual EFT parameters is obtained in scenarios where one parameter is allowed to vary while all others are fixed to their SM expectation of 0. The symmetrised expected uncertainty to each parameter is shown in Figure~\ref{fig:eft:indiv:exp}. The sensitivity is reported as the uncertainty $\sigma$ on the parameter, assuming an EFT scale $\Lambda = \qty{1}{\TeV}$, or conversely as the sensitivity scale $\Lambda/\sqrt{\sigma}$ which corresponds to the EFT scale to which the combination would be sensitive in a scenario where $c_n=1$. Expected sensitivities are shown for both the linear and linear+quadratic models. The results show widely different sensitivities for the different EFT parameters, with expected uncertainties in the $c_n$ ranging from about $10^{-3}$ to over $10$ and sensitivity scales reaching about \qty{20}{\TeV} for the most precisely measured parameters.

Measurements of individual EFT parameters cannot usually be matched to specific models of new phenomena beyond the SM, since these typically lead to deviations from zero in  multiple parameters. All the EFT parameters presented in Table~\ref{tab:eft:ops} cannot however be left to vary freely, since some combinations of these parameters are not constrained by the analyses included in the measurement. The measurement is therefore performed using a reduced set of measurement parameters, obtained using the same procedure as described in Ref.~\cite{HIGG-2022-17}, which excludes directions with no experimental sensitivity. First a measurement of the $\mu_r^f$ parameters is performed, using the forms given in Eq.~\eqref{eq:eft:param_mu} to express the parameters of the STXS combination described in Appendix~\ref{app:stxs_br}. The Hessian matrix $(C^{-1})^{\mu}$ of this measurement is then used to define
\begin{equation}
(C^{-1})^{c}_{mn} = \sum\limits_{(r,f)} \sum\limits_{(r',f')}
\left(A_{c \to \mu}\right)^{(r,f)}_m (C^{-1})^{\mu}_{(r,f),(r',f')} \left(A_{c \to \mu}\right)^{(r',f')}_n
\label{eq:Fisher}
\end{equation}
where
\begin{equation}
\left(A_{c \to \mu}\right)^{(r,f)}_n = A^{\sigma_r}_n + A^{\Gamma_f}_n - A^{\Gamma}_n
\end{equation}
and the sums run over all the combinations of production and decay processes $(r,f)$ included in the STXS measurement.
The matrix $(C^{-1})^{c}_{mn}$ corresponds to the Hessian matrix of the measurement of the $c_n$, using a linear approximation of the dependence of the $\mu_r^f$ on the $c_n$. It cannot be reliably inverted, but its principal component analysis (PCA) provides information about the sensitivity of the combination of EFT parameters in the Warsaw basis: linear combinations of the $c_n$ corresponding to large eigenvalues are well measured, while small eigenvalues correspond to weakly constrained "flat" directions in the space of the $c_n$. As in Ref.~\cite{HIGG-2022-17}, the PCA is not performed on the full Hessian but on blocks corresponding to groups of $c_n$ with similar impact on the measurements included in the combination. These groups correspond to the set of parameters impacting mainly the \ggF\ and \ttH\ event rates, designated as $ggH$; parameters primarily impacting the \Hyy\ and \Hzy\ decays (\HyyHzy); parameters mainly affecting \WH\ and \ZH\ production ($ZH$); parameters impacting mainly the \Hllll\ decay process (\Hl); and finally a set of parameters with a global impact on all measured event rates (\glob). The remaining EFT parameters are considered individually. A separate PCA is performed in each of these sectors, and linear combinations with an eigenvalue of at least $0.01$ are retained in each case. The remaining linear combinations are assumed to have a negligible impact on the results and their values are fixed to zero in the measurement. The procedure is not intended to produce a fully diagonal covariance matrix in the space of the measurement parameters, since correlations between blocks are not considered. Its main objective is to remove flat directions from consideration so that the measurement is well-conditioned.

\begin{figure}[tbp]
\centering
\includegraphics[height=0.55\columnwidth,angle=90]{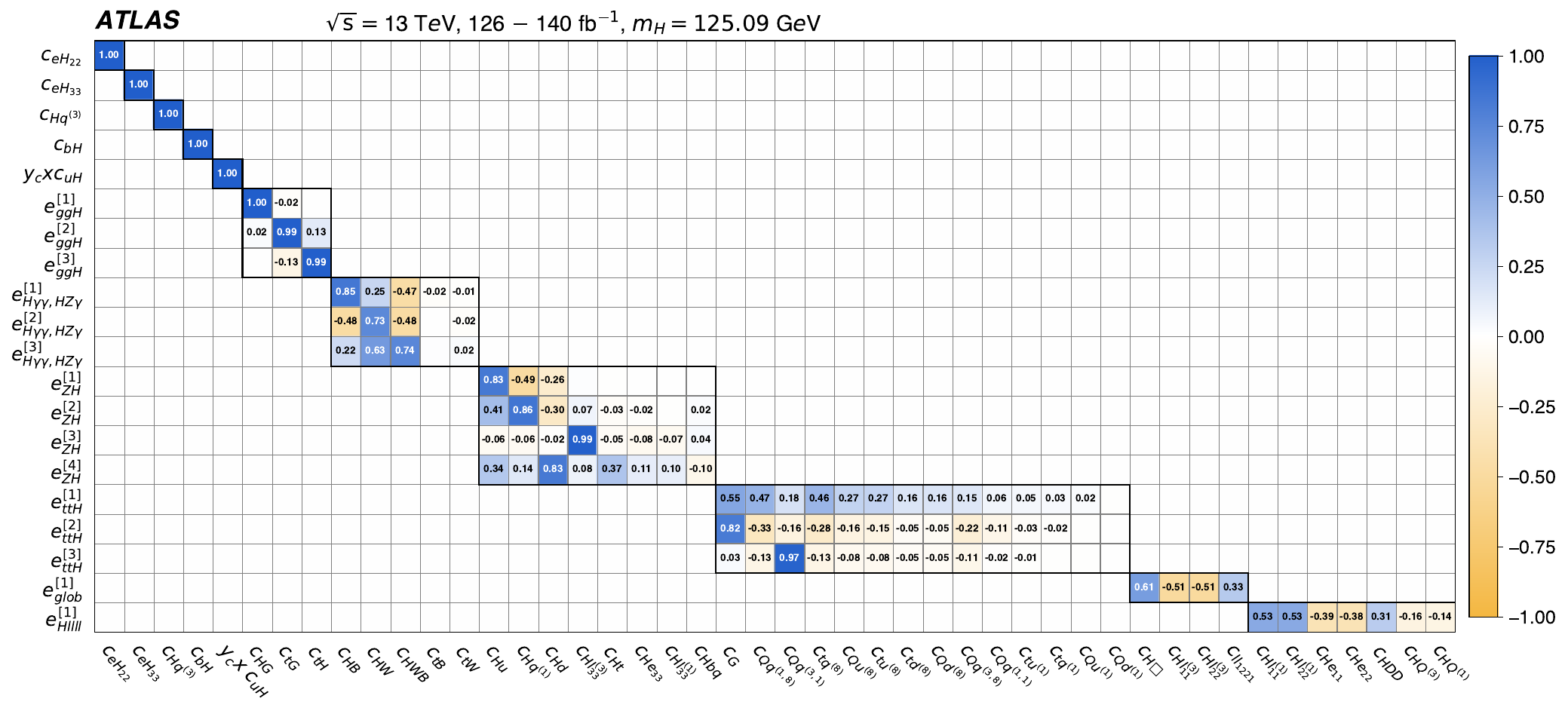}
\caption{Definition of the measurement parameters for the EFT measurement, each corresponding to a linear combination of EFT parameters defined in the Warsaw basis. Each cell corresponds to the coefficient of the Warsaw basis parameter given on the horizontal axis, for the measurement parameter given in the vertical axis. Coefficient values below $0.01$ are omitted for clarity.}
\label{fig:eft:evs}
\end{figure}

\begin{figure}[tbp]
\centering
\includegraphics[width=0.90\textwidth]{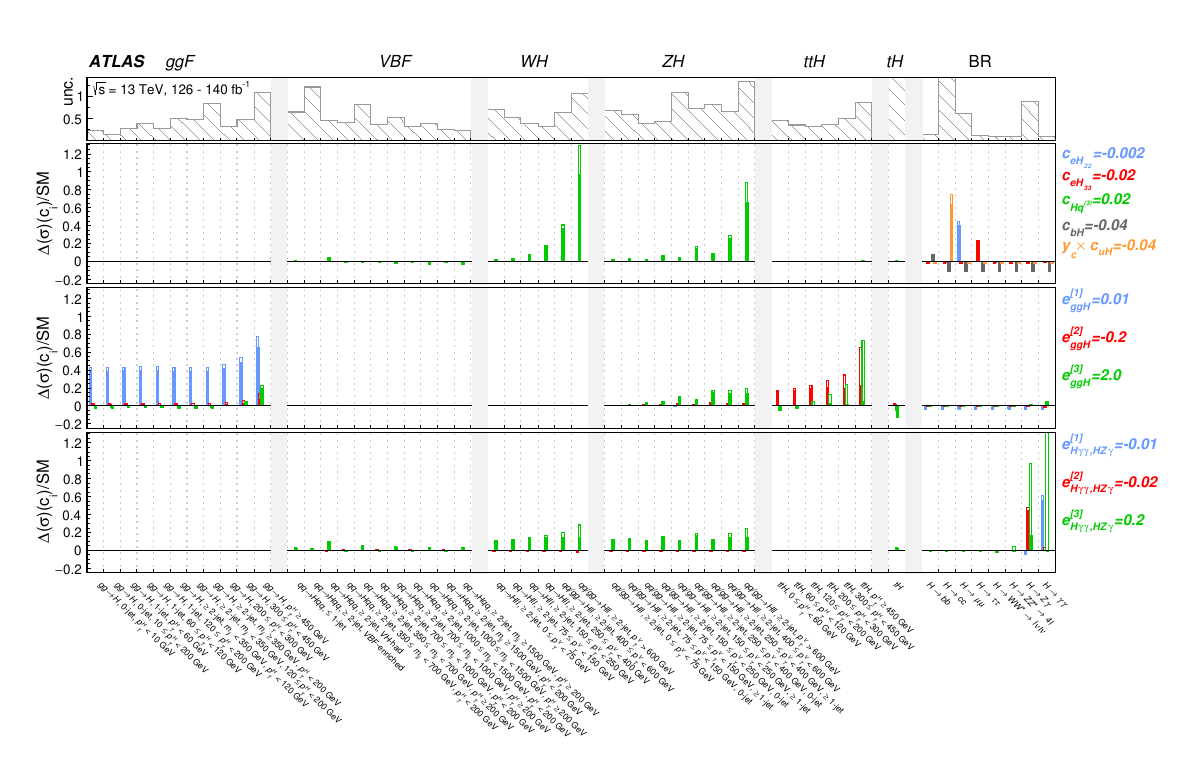} \\
\includegraphics[width=0.90\textwidth]{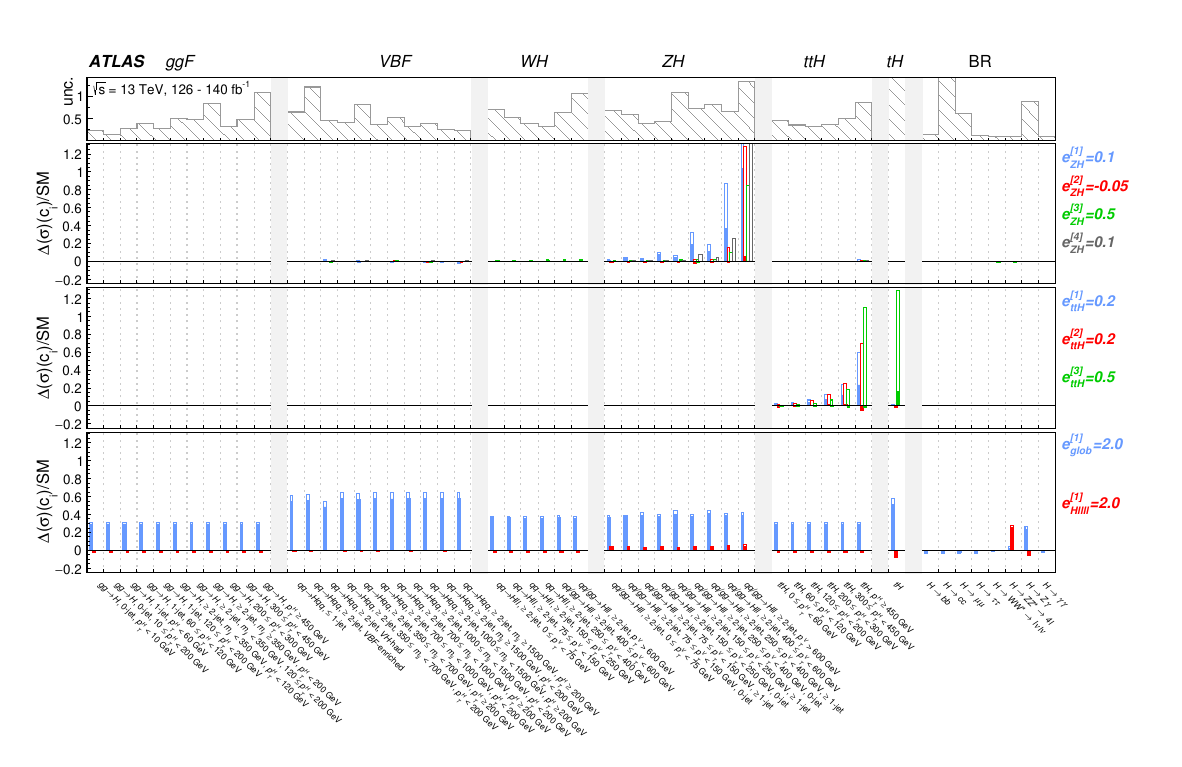}
\caption{Relative change in the rate of each production and decay process entering the combination compared to the SM reference, for parameter variations equal to the symmetrised 68\% CL expected uncertainty in each parameter. Filled bars indicate the changes due to linear terms, while the empty bars also include the contributions of quadratic terms.
The symmetrised statistical uncertainty for each observable is shown in the top panel.
The EFT scale is assumed to be  $\Lambda = \qty{1}{\TeV}$.}
\label{fig:eft:impacts:ev}
\end{figure}

The resulting set of 20 measurement parameters consists of the linear combinations of Warsaw-basis EFT parameters shown in Figure~\ref{fig:eft:evs}. Figure~\ref{fig:eft:impacts:ev} illustrates the impacts of each measurement parameter on the Higgs boson production cross-sections and branching ratios that enter the combination.
The parameter \wc{uH} is defined with an additional charm-Yukawa coupling factor $y_c$. In the \texttt{top} flavour scheme this parameter encompasses operators involving both the $u$ and $c$ quarks, and is constrained mainly through the rate of the \Hcc\ decay, which is impacted by its $c$-quark component. With the extra factor $y_c$, the parameter corresponds to the definition that one would have obtained for a $c$-quark operator that would be defined separately from the $u$-quark component.

\begin{figure}[tbp]
\centering
\includegraphics[width=.95\textwidth]{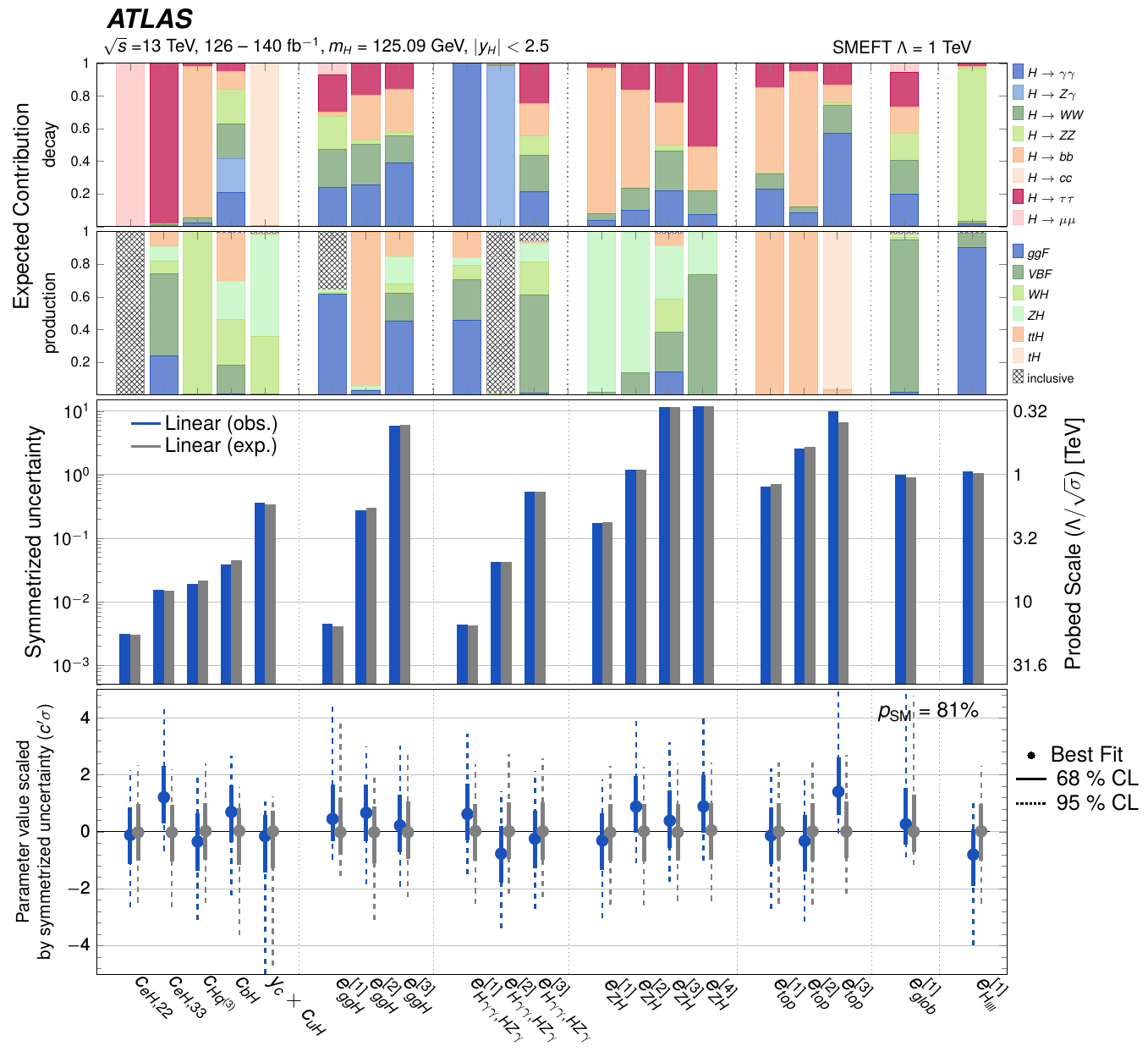}
\caption{Fitted values of the EFT measurement parameters. The top panel shows the contributions from each production and decay process, the middle panel the symmetrised uncertainty in each parameter, and the bottom panel the best-fit value and uncertainty in each parameter divided by its symmetrised uncertainty. In the middle panel, observed and expected results are shown using blue (left) and grey (right) bars. The left axis shows the uncertainty $\sigma$ on the parameter assuming $\Lambda = \qty{1}{\TeV}$ for the EFT scale, while the right axis shows the probed EFT scale evaluated as $\Lambda/\sqrt{\sigma}$. In the bottom panel, confidence intervals are shown at 68\% CL (solid lines) and 95\% CL (dashed lines).}
\label{fig:eft:global:obs}
\end{figure}

Results are obtained from a fit in which the measurement parameters are simultaneously left to vary. No bounds are applied on the parameters.
The observed values in the linear model are shown in Figure~\ref{fig:eft:global:obs}. The sensitivities for each parameter are expressed either
as the symmetrised uncertainty $\sigma$ on each parameter, assuming $\Lambda = \qty{1}{\TeV}$, or as the corresponding sensitivity scale $\Lambda/\sqrt{\sigma}$.
The contribution to each measurement from individual Higgs boson production and decay processes is evaluated by computing separate $(C^{-1})^{c, (i)}_{mn}$ matrices for each production or decay process $i$ using the same method as shown in Eq.~\eqref{eq:Fisher}.
The contribution of process $i$ to the measurement of $c_n$ is then computed as $(C^{-1})^{c, (i)}_{nn}/\sum_i (C^{-1})^{c, (i)}_{nn}$, with the sum running over all production or decay processes.
All results are in good agreement with the SM, with a compatibility corresponding to a $p$-value of 81\%. The best-constrained coupling parameters
are \evp{eH,22}, which is driven by a measurement of the \Hmm\ process that sets stringent limits on its event rate; \evp[1]{ggH}, driven by the
measurement of the \ggF\ production rate; and \evp[1]{\HyyHzy} that is mainly constrained by the rate of the \Hyy\ decay. Each of
these has a symmetrised uncertainty in the coupling between $10^{-2}$ and $10^{-3}$, corresponding to a sensitivity to EFT scales ranging from 10 to
\qty{30}{\TeV}. In general better constraints are seen when the SM processes are suppressed relative to the corresponding EFT processes: in the case of \Hmm, this suppression occurs due to the small value of the Yukawa coupling of the muon, while the \ggtoH\ and \Hyy\ processes are loop-suppressed in the SM.
The observed and expected correlation coefficients between the parameters are shown in Figure~\ref{fig:eft:corr} in Appendix~\ref{app:eft}.
Compared to Ref.~\cite{HIGG-2022-17}, the combination benefits from the improved analysis of the \VH, $H\rightarrow bb,cc$ process~\cite{HIGG-2020-20}. The improved sensitivity to \Hcc\ allows to include \wc{uH}\ among the measured parameters for the first time in EFT interpretations of LHC data.
The combination of large measurement uncertainties and a relatively large branching ratio for \Hcc\ leads to its rate having a sizeable impact on the total width of the Higgs boson which also impacts other branching ratios. This leads to weaker constraints on the \evp[1]{\glob} parameter, which has a similar impact on the observables, and a correlation of $-80\%$ between the two parameters.
This effect is however mitigated by the precision of the measurement of the \Hcc\ rate in Ref.~\cite{HIGG-2020-20} and future improvements in the measurement of the \Hcc\ process are expected to further reduce the impact of \wc{uH}\ on the other measurement parameters.
The analysis of Ref.~\cite{HIGG-2020-20} also provides improved sensitivity to the \Hbb\ process, in particular by providing the first measurement in the $\pTV > \qty{600}{\GeV}$ region.
This benefits the sensitivity to EFT operators that have a larger impact for high transverse momenta of the $W$ or $Z$ boson. This leads in particular to significantly stronger sensitivity to \wc[3]{HQ}, \wc{Hb}\ and the \evp[i]{ZH}\ parameters, in comparison to Ref.~\cite{HIGG-2022-17}. Improvements to the \ttH\ measurements in the \Hbb~\cite{HIGG-2020-24} and \HML~\cite{HIGP-2024-08} processes provide increased sensitivity to operators modifying top-quark couplings. Overall, the measurement basis contains 20 parameters compared to 19 in Ref.~\cite{HIGG-2022-17}, with $y_c \times \wc{uH}$ as the additional parameter as described above.

\begin{figure}[tbp]
\centering
\includegraphics[width=.95\textwidth]{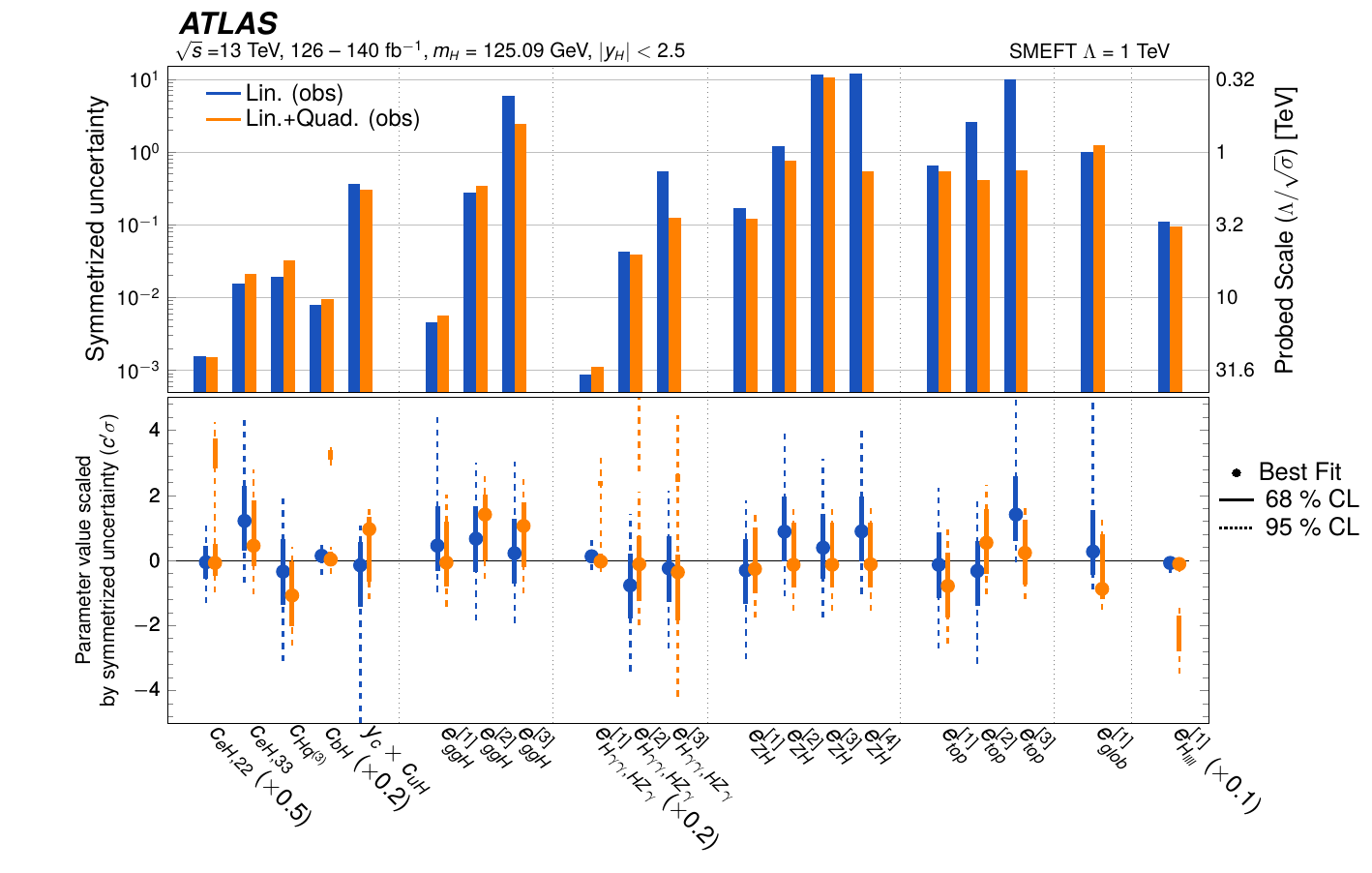}
\caption{Fitted values of the EFT measurement parameters for the linear and linear+quadratic models. The top panel shows the symmetrised uncertainty in each parameter, and the bottom panel the best-fit value and uncertainty of each parameter divided by its symmetrised uncertainty. Confidence intervals are shown at 68\% CL (solid lines) and 95\% CL (dashed lines). In the top panel, the left axis shows the uncertainty $\sigma$ on the parameter assuming $\Lambda = \qty{1}{\TeV}$ for the EFT scale, while the right axis shows the probed EFT scale estimated as $\Lambda/\sqrt{\sigma}$.}
\label{fig:eft:quad:obs}
\end{figure}

A comparison of observed results obtained for the linear and linear+quadratic parameterisations is shown in Figure~\ref{fig:eft:quad:obs}. In most cases similar results are observed for
the two models, indicating that contributions from higher-order terms in the EFT expansion are generally small. However some differences are observed in particular for parameters with weak constraints in the linear case, for which the linear+quadratic parameterisation can in some cases lift degeneracies present in the linear case. Measurements of Higgs boson production at high transverse momentum play an important role in the quadratic parameterisation, significantly reducing the uncertainties in \pT-dependent operators. A comparison of observed and expected results in the linear+quadratic parameterisation is shown in Figure~\ref{fig:eft:quad:exp}.
A simplified likelihood for the measurement, described in Appendix~\ref{app:mvg}, is found to closely reproduce the results obtained with the full likelihood.

\FloatBarrier


%
%
\FloatBarrier

\section{Conclusion}

Higgs boson production and decay rates are measured using up to \qty{140}{fb^{-1}} of $pp$ collision data collected at $\sqrt{s} = \qty{13}{\TeV}$ by the ATLAS detector during Run 2 of the LHC, combining the final analyses of each process using Run 2 data.
The inclusive Higgs boson signal strength relative to its SM expectation is determined to be $0.990^{+0.053}_{-0.051} = 0.990 \pm 0.027 \text{ (stat.) } \pm 0.024 \text{ (exp.) } ^{+0.037}_{-0.035} \text{ (sig. theo.) } ^{+0.016}_{-0.015} \text{ (bkg. theo.)}$, in agreement with SM predictions. The theory uncertainty in the SM prediction of the Higgs boson signal rates is dominant compared to the components originating from statistical uncertainties, experimental systematic uncertainties, and theory uncertainties in background processes. Its leading contributions originate from missing higher-order QCD terms in the SM prediction and the modelling of parton shower effects.

Cross-sections of Higgs boson production processes are measured to be compatible with the SM, with relative uncertainties ranging from 6\% to 21\% for the observed processes, while the \tH\ process still eludes observation. Observed uncertainties in the rates for the \WH, \ZH, and \ttH\ processes are reduced by about 30\%, 20\% and 40\%, respectively, compared to Ref.~\cite{HIGG-2021-23}, while expected uncertainties are reduced by up to 35\%. Branching ratios into the main Higgs boson decay modes (\Hyy, \Hzz, \Hww, \Hbb, \Htt) are measured with relative uncertainties in the 8\% -- 13\% range, and in agreement with SM expectations. The uncertainties in the \Hbb\ and \Hww\ branching ratios are reduced by about 20\% compared to Ref.~\cite{HIGG-2021-23}, and the uncertainty in the \Htt\ branching ratio by 10\%. Products of cross-sections and branching ratios for 32 combinations of production and decay channels are also reported.

Interpretations of these results in terms of Higgs boson coupling modifiers are presented. Higgs boson couplings to $W$ and $Z$ bosons, $t$ and $b$-quarks and $\tau$-leptons are measured with uncertainties between  4\% and 10\%, with improvements ranging from 15\% to 25\% compared to Ref.~\cite{HIGG-2021-23}, assuming only SM contributions to loop processes and to Higgs boson processes not probed by the analyses considered here, and that modifications to the $c$-quark coupling match those of the $t$-quark. The absolute value of the coupling to the muon is measured with an uncertainty of about 25\%. When an independent modifier for the $c$-quark coupling is also included, it is constrained to $|\kappa_c| < 4.1$ at 95\% CL, improving by about a factor of two compared to Ref.~\cite{HIGG-2021-23}. The total width of the Higgs boson is constrained to be $\Gamma_H = \qty[parse-numbers=false]{3.6^{+2.1}_{-1.6}}{\MeV}$ based on the comparison of its on-shell and off-shell production rates. The self-coupling of the Higgs boson is measured in the combination of analyses targeting both single and pair production of Higgs bosons, leading to $\kl = 1.3\,^{+3.1}_{-1.6}$, with a 95\% CL interval of $[-1.5,\,6.5]$. The measurement of the partial width of the Higgs boson in the \Hbb\ decay channel is used to obtain the first ATLAS direct determination of $m_b(m_H)$, the running mass of the $b$-quark at the scale $m_H$ in the \MSbar\ scheme, which is found to be $m_b(m_H) = \qty[parse-numbers=false]{2.73^{+0.27}_{-0.25}}{\GeV}$.

The kinematics of Higgs boson production at the LHC are probed within the STXS framework, and production cross-sections in 44 kinematic regions of Higgs boson production phase space are measured. An interpretation of STXS measurements in the framework of SM effective field theory models is presented. Constraints are set on 20 combinations of EFT parameters, including for the first time the parameter $y_c \times \wc{uH}$ relative to the coupling of the Higgs boson and charm quark.
All results are found to be in agreement with SM predictions.

%

%
%
%
%
%
\section*{Acknowledgments}
%

%

%
%

%
%

We thank CERN for the very successful operation of the LHC and its injectors, as well as the support staff at
CERN and at our institutions worldwide without whom ATLAS could not be operated efficiently.

The crucial computing support from all WLCG partners is acknowledged gratefully, in particular from CERN, the ATLAS Tier-1 facilities at TRIUMF/SFU (Canada), NDGF (Denmark, Norway, Sweden), CC-IN2P3 (France), KIT/GridKA (Germany), INFN-CNAF (Italy), NL-T1 (Netherlands), PIC (Spain), RAL (UK) and BNL (USA), the Tier-2 facilities worldwide and large non-WLCG resource providers. Major contributors of computing resources are listed in Ref.~\cite{ATL-SOFT-PUB-2026-001}.

We gratefully acknowledge the support of ANPCyT, Argentina; YerPhI, Armenia; ARC, Australia; BMWFW and FWF, Austria; ANAS, Azerbaijan; CNPq and FAPESP, Brazil; NSERC, NRC and CFI, Canada; CERN; ANID, Chile; CAS, MOST, NSFC and FRFCU, China; Minciencias, Colombia; MEYS CR, Czech Republic; DNRF and DNSRC, Denmark; IN2P3-CNRS and CEA-DRF/IRFU, France; SRNSFG, Georgia; BMFTR, HGF and MPG, Germany; GSRI, Greece; RGC and Hong Kong SAR, China; ICHEP and Academy of Sciences and Humanities, Israel; INFN, Italy; MEXT and JSPS, Japan; CNRST, Morocco; NWO, Netherlands; RCN, Norway; MNiSW, Poland; FCT, Portugal; MNE/IFA, Romania; MSTDI, Serbia; MSSR, Slovakia; ARIS and MVZI, Slovenia; DSI/NRF, South Africa; MICIU/AEI, Spain; SRC and Wallenberg Foundation, Sweden; SERI, SNSF and Cantons of Bern and Geneva, Switzerland; NSTC, Taipei; TENMAK, T\"urkiye; STFC/UKRI, United Kingdom; DOE and NSF, United States of America.

Individual groups and members have received support from BCKDF, CANARIE, CRC and DRAC, Canada; CERN-CZ and PRIMUS, Czech Republic; COST, ERC, ERDF, Horizon 2020 and Marie Sk{\l}odowska-Curie Actions, European Union; Investissements d'Avenir Labex, Investissements d'Avenir Idex and ANR, France; DFG and AvH Foundation, Germany; Herakleitos, Thales and Aristeia programmes co-financed by EU-ESF and the Greek NSRF, Greece; BSF-NSF and MINERVA, Israel; NCN and NAWA, Poland; La Caixa Banking Foundation, CERCA and AGAUR programs from Generalitat de Catalunya and PROMETEO and GenT Programmes Generalitat Valenciana, Spain; G\"{o}ran Gustafssons Stiftelse, Sweden; The Royal Society and Leverhulme Trust, United Kingdom; Eric and Wendy Schmidt Fund for Strategic Innovation, United States of America.

In addition, individual members wish to acknowledge support from Chile: Agencia Nacional de Investigaci\'on y Desarrollo (ANID FONDECYT reg. 1230987, FONDECYT 1230812, FONDECYT 1240864, Fondecyt 3240661, Fondecyt Regular 1240721); China: Fundamental Research Funds for the Central Universities (010-63263105), Chinese Ministry of Science and Technology (MOST-2023YFA1605700, MOST-2023YFA1609300), National Natural Science Foundation of China (NSFC 12275265, NSFC-W2543005); Czech Republic: Czech Science Foundation (GACR - 24-11373S), Ministry of Education Youth and Sports (ERC-CZ-LL2327, FORTE CZ.02.01.01/00/22\_008/0004632); EU: H2020 European Research Council (ERC - 101002463); European Union: European Research Council (BARD No. 101116429, ERC - 101219398, ERC 101089007), European Regional Development Fund (HE COFUND GA No.101081355, ERDF), Marie Sklodowska-Curie Actions (GAP-101168829); France: Agence Nationale de la Recherche (ANR-22-EDIR-0002, ANR-24-CE31-0504-01); Germany: Deutsche Forschungsgemeinschaft (DFG - 469666862); China: Research Grants Council (GRF); Italy: Istituto Nazionale di Fisica Nucleare (LHC-MIUR - 28003/2025), Ministero dell'Università e della Ricerca (NextGenEU I53D23000820006 M4C2.1.1, SOE2024\_0000023); Japan: Japan Society for the Promotion of Science (JSPS KAKENHI  JP25H0063, JSPS KAKENHI JP22H04944, JSPS KAKENHI JP24K23939, JSPS KAKENHI JP24KK0251, JSPS KAKENHI JP25H00650, JSPS KAKENHI JP25H01291, JSPS KAKENHI JP25K01011, JSPS KAKENHI JP25K01023, JSPS KAKENHI JP25KK0047); Poland: Polish National Science Centre (NCN 2021/42/E/ST2/00350, NCN OPUS 2023/51/B/ST2/02507, NCN OPUS nr 2022/47/B/ST2/03059, NCN UMO-2019/34/E/ST2/00393, UMO-2022/47/O/ST2/00148, UMO-2023/49/B/ST2/04085, UMO-2023/51/B/ST2/00920, UMO-2024/53/N/ST2/00869); Spain: Agència de Gestió d'Ajuts Universitaris i de Recerca. (AGAUR - 2023 BP 00141), Ministry of Science and Innovation (RYC2019-028510-I, RYC2020-030254-I, RYC2021-031273-I, RYC2022-038164-I), Ministerio de Ciencia, Innovación y Universidades/Agencia Estatal de Investigaci\'on (EU NextGenerationEU (PRTR-C17.I1), PID2022-142604OB-C22); Sweden: Carl Trygger Foundation (Carl Trygger Foundation CTS 22:2312), Swedish Research Council (Swedish Research Council 2023-04654, VR 2021-03651, VR 2022-03845, VR 2022-04683, VR 2023-03403, VR 2024-05451, VR 2025-05940), Knut and Alice Wallenberg Foundation (KAW 2023.0366); United Kingdom: The Binks Trust, Royal Society (NIF-R1-231091); United States of America: U.S. Department of Energy (ECA DE-AC02-76SF00515), John Templeton Foundation (John Templeton Foundation 63206), Neubauer Family Foundation.

%
%


%
\clearpage
\appendix
\part*{Appendix}
\addcontentsline{toc}{part}{Appendix}

\FloatBarrier
\section{Expected signal strength results}
\label{app:mu_global}

\begin{figure}[tbp]
\centering
\includegraphics[width=.7\textwidth]{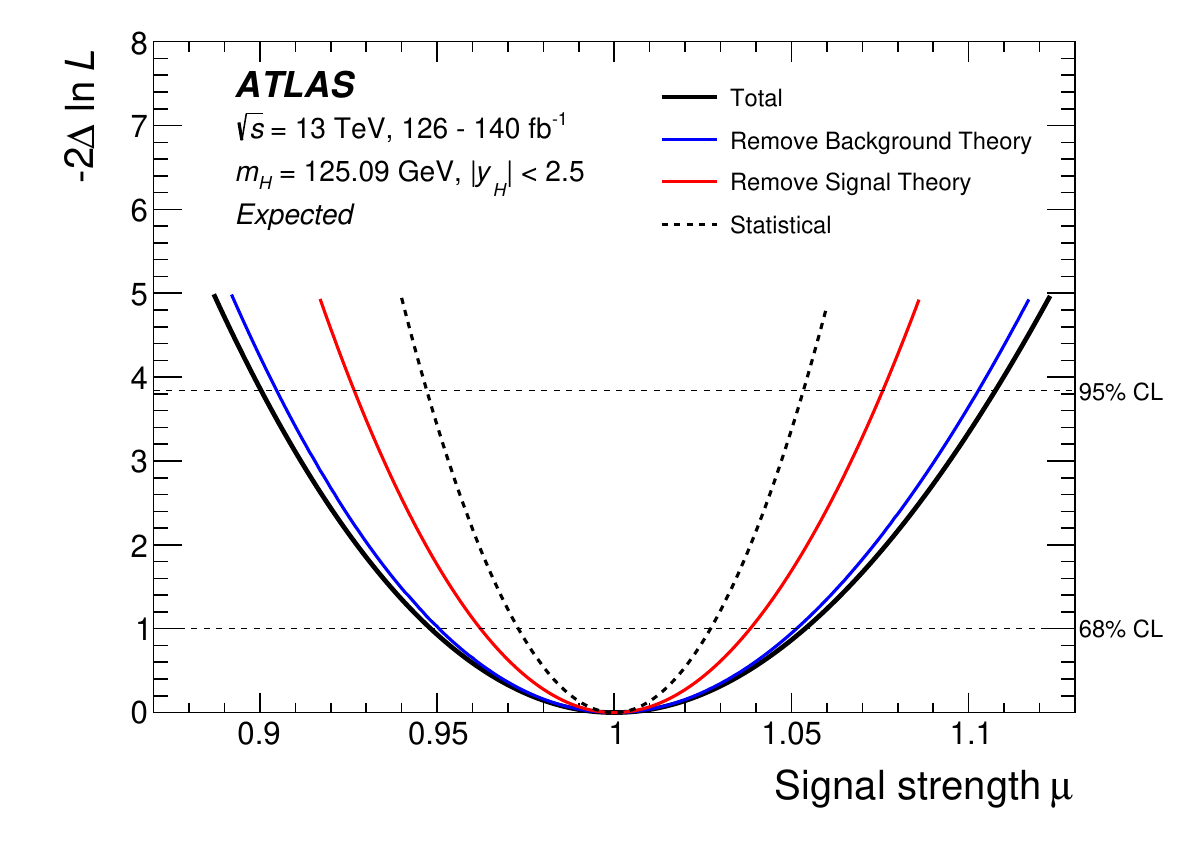}
\caption{Expected values of $-2\,\Delta \ln L$ as a function of the signal strength $\mu$, in the model with a single signal-strength scaling all processes. The intersections with the horizontal dotted lines define the 68\% and 95\% CL intervals on the measurement. The different curves correspond to the full statistical model and to fits in which the background theory uncertainties, the signal theory uncertainties, or all systematic uncertainties are removed.}
\label{fig:mu_exp}
\end{figure}

The expected value of the global signal strength under SM assumptions is
\begin{equation*}
\mu =  1.000 ^{+0.054}_{-0.052} = 1.000 \pm 0.027 \text{ (stat.) } \pm 0.024 \text{ (exp.) } ^{+0.038}_{-0.035} \text{ (sig. theo.) } ^{+0.012}_{-0.011} \text{ (bkg. theo.) }.
\end{equation*}
The corresponding profile log-likelihood scan is shown in Figure~\ref{fig:mu_exp}.

\FloatBarrier
\clearpage

\section{Additional production cross-section and branching ratio results}
\label{app:xs_br}

\begin{figure}[tbp]
\begin{center}
\includegraphics[width=0.7\columnwidth]{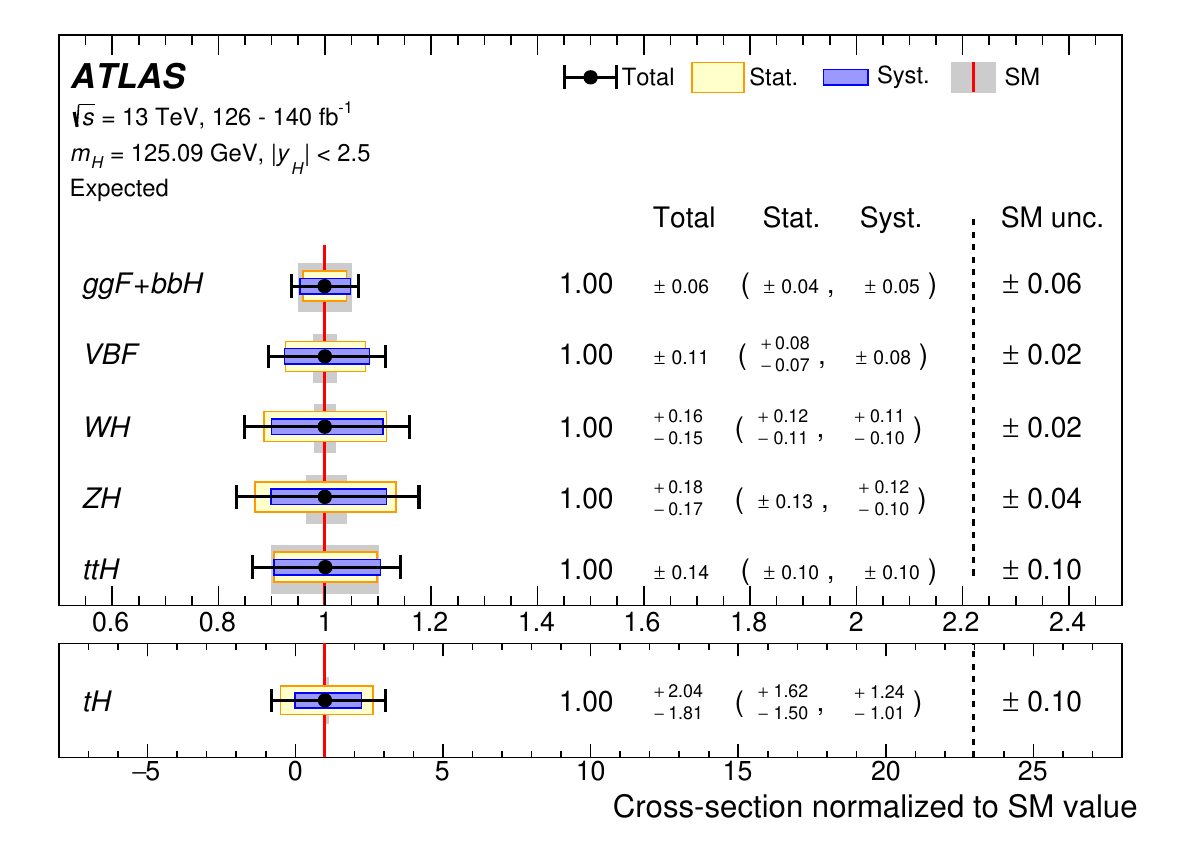}
\end{center}
\caption{Expected values of the cross-sections for the main Higgs boson production modes, relative to their SM predictions. The \bbH\ mode is considered together with \ggF. Higgs boson decay branching ratios are assumed to be equal to their SM predictions.}
\label{fig:xs:exp}
\end{figure}

\begin{figure}[tbp]
\begin{center}
\includegraphics[width=0.7\columnwidth]{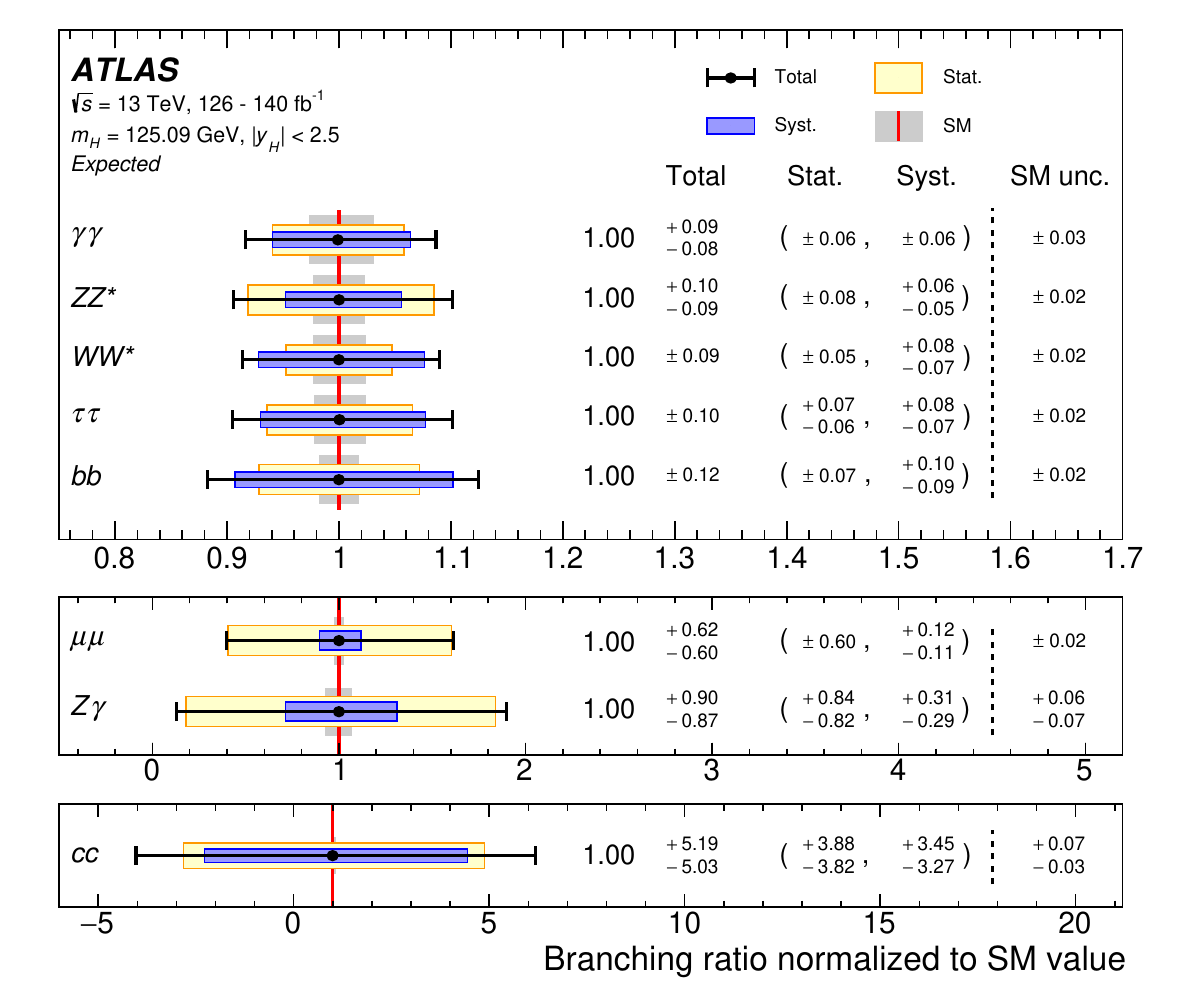}
\end{center}
\caption{Expected values of the branching ratios in the \Hyy, \Hzz, \Hww, \Htt, \Hbb, \Hmm, \HZy\ and \Hcc\ decay modes, relative to their SM predictions. Higgs boson production cross-sections are assumed to be equal to their SM predictions.}
\label{fig:br:exp}
\end{figure}

\begin{figure}[tbp]
\begin{center}
\subfloat[]{\includegraphics[width=0.49\columnwidth]{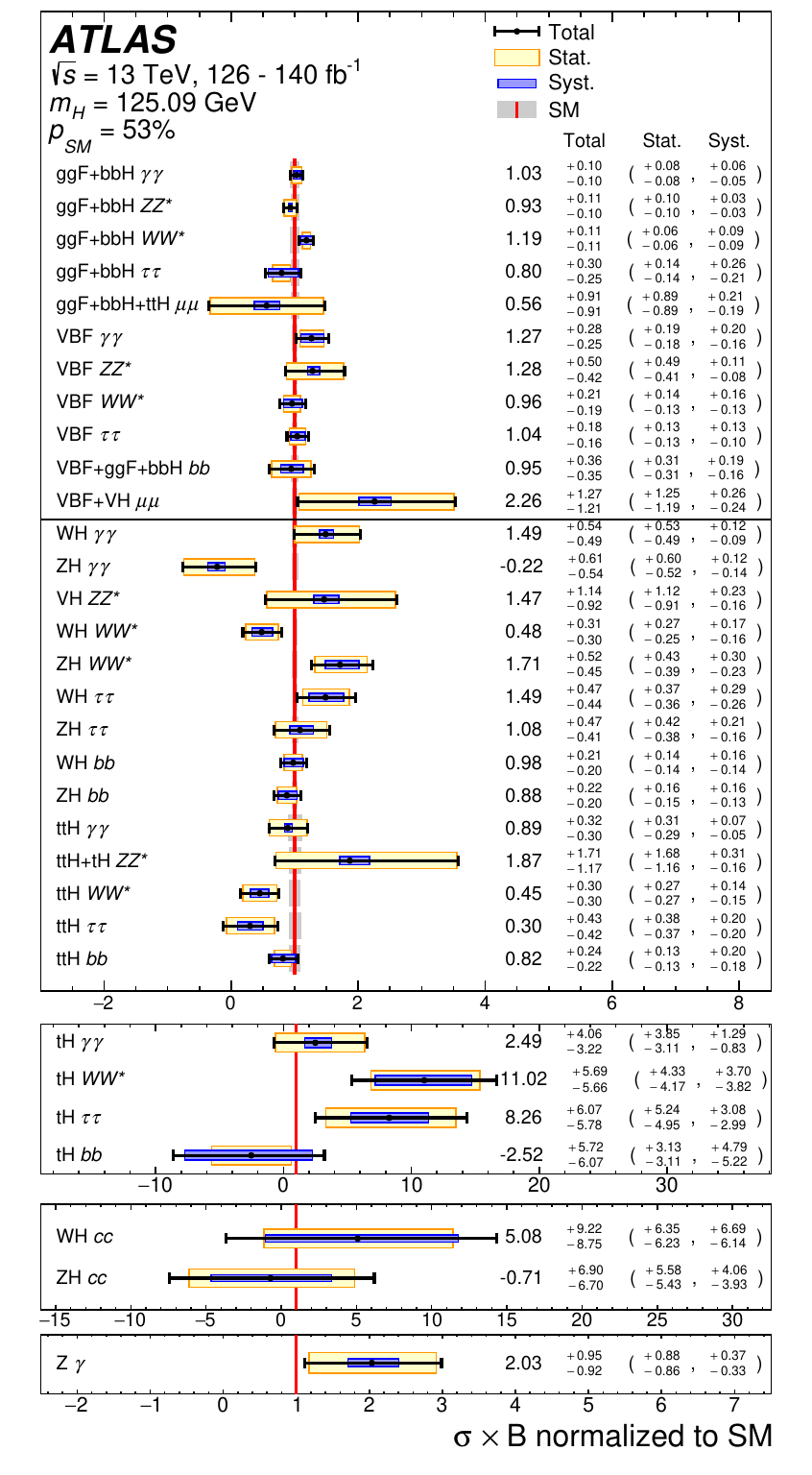}}
\subfloat[]{\includegraphics[width=0.49\columnwidth]{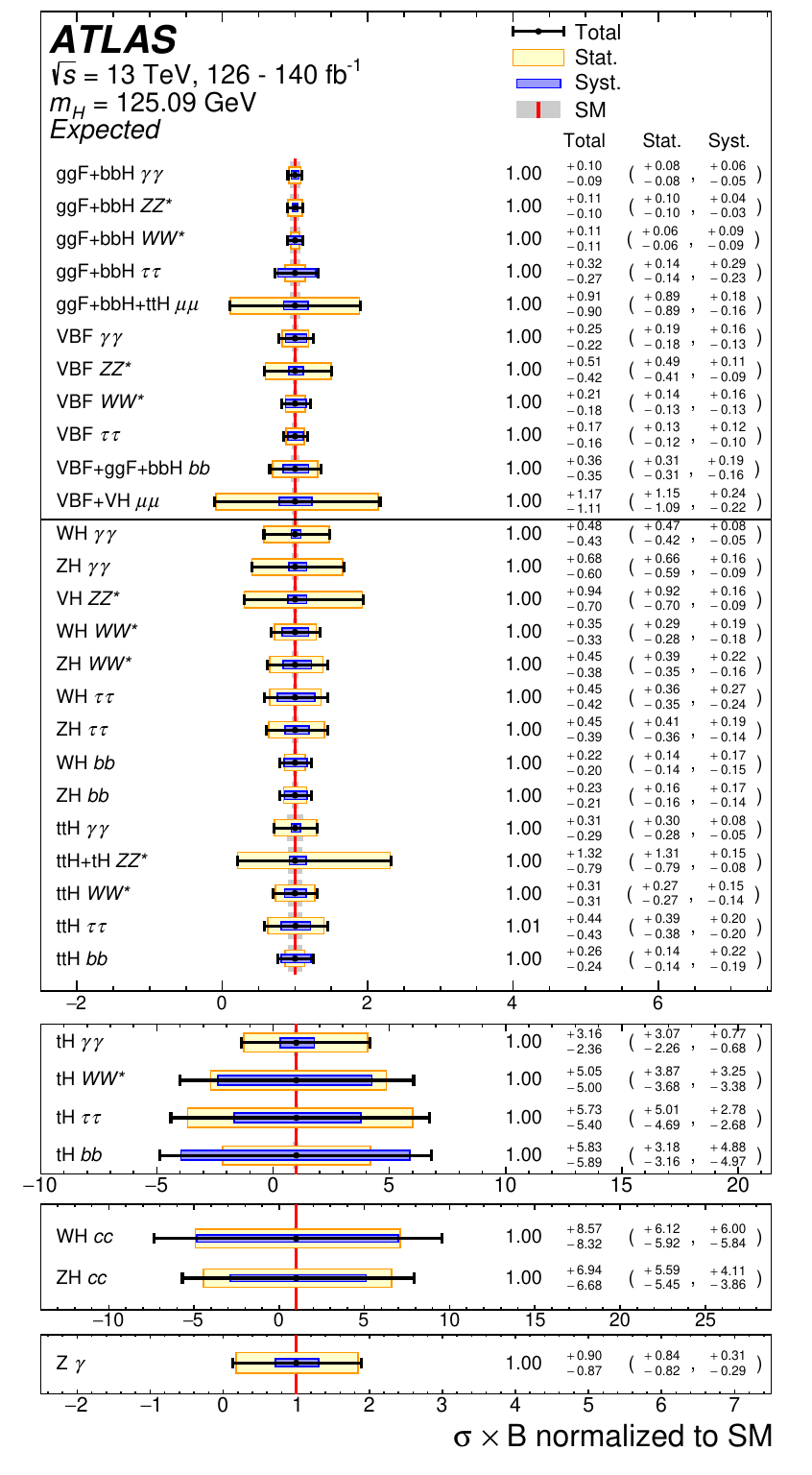}}
\end{center}
\caption{(a) Observed and (b) expected values of the measurements of products of production cross-sections and branching ratios, relative to their SM predictions. The total uncertainties (solid bars) are shown along with their statistical (light shaded regions) and systematic (dark shaded regions) components. Numerical values are shown rounded to two digits after the decimal point.}
\label{fig:PxD:details}
\end{figure}

\begin{figure}[tbp]
\begin{center}
\subfloat[]{\includegraphics[width=0.49\columnwidth]{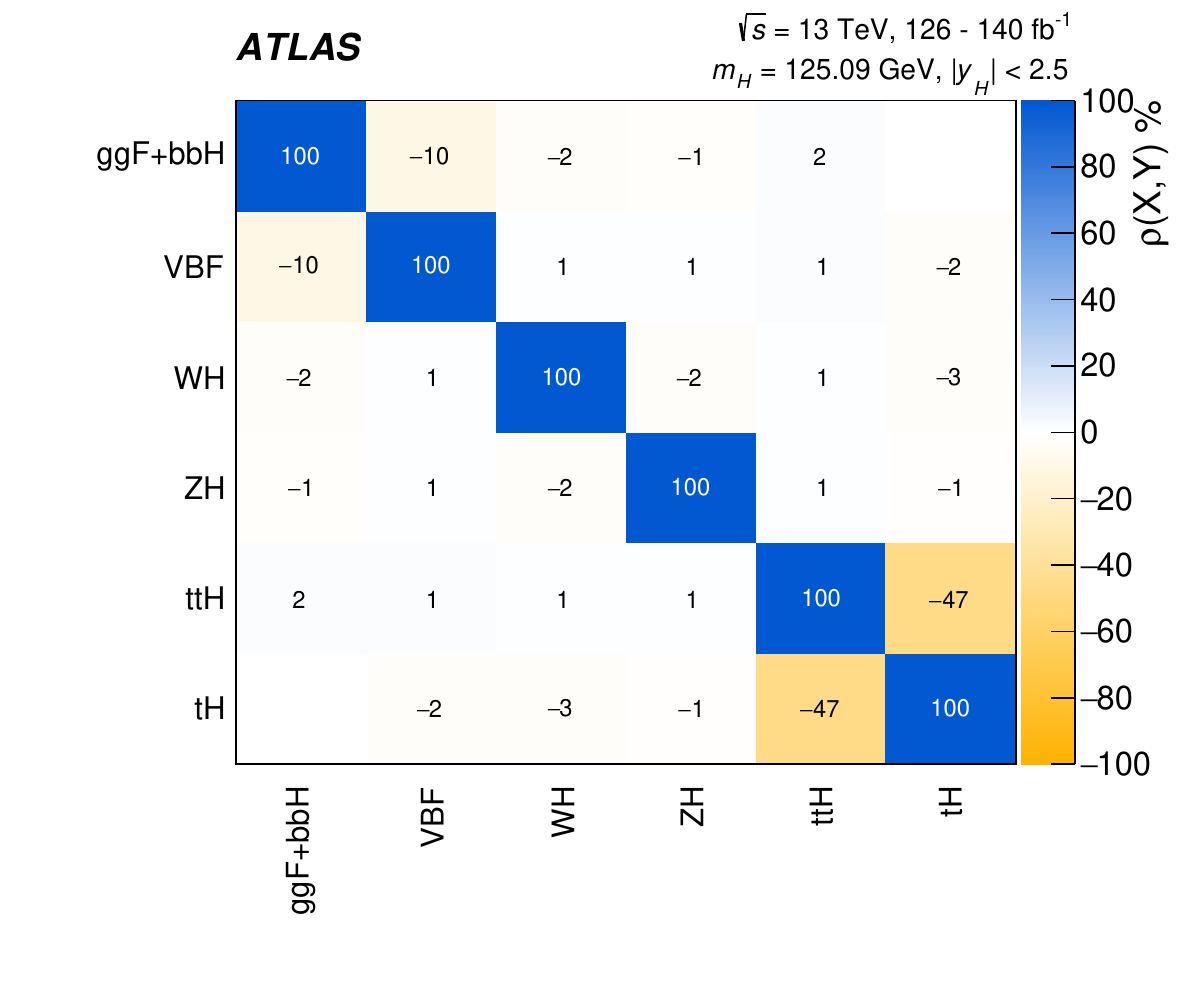}}
\subfloat[]{\includegraphics[width=0.49\columnwidth]{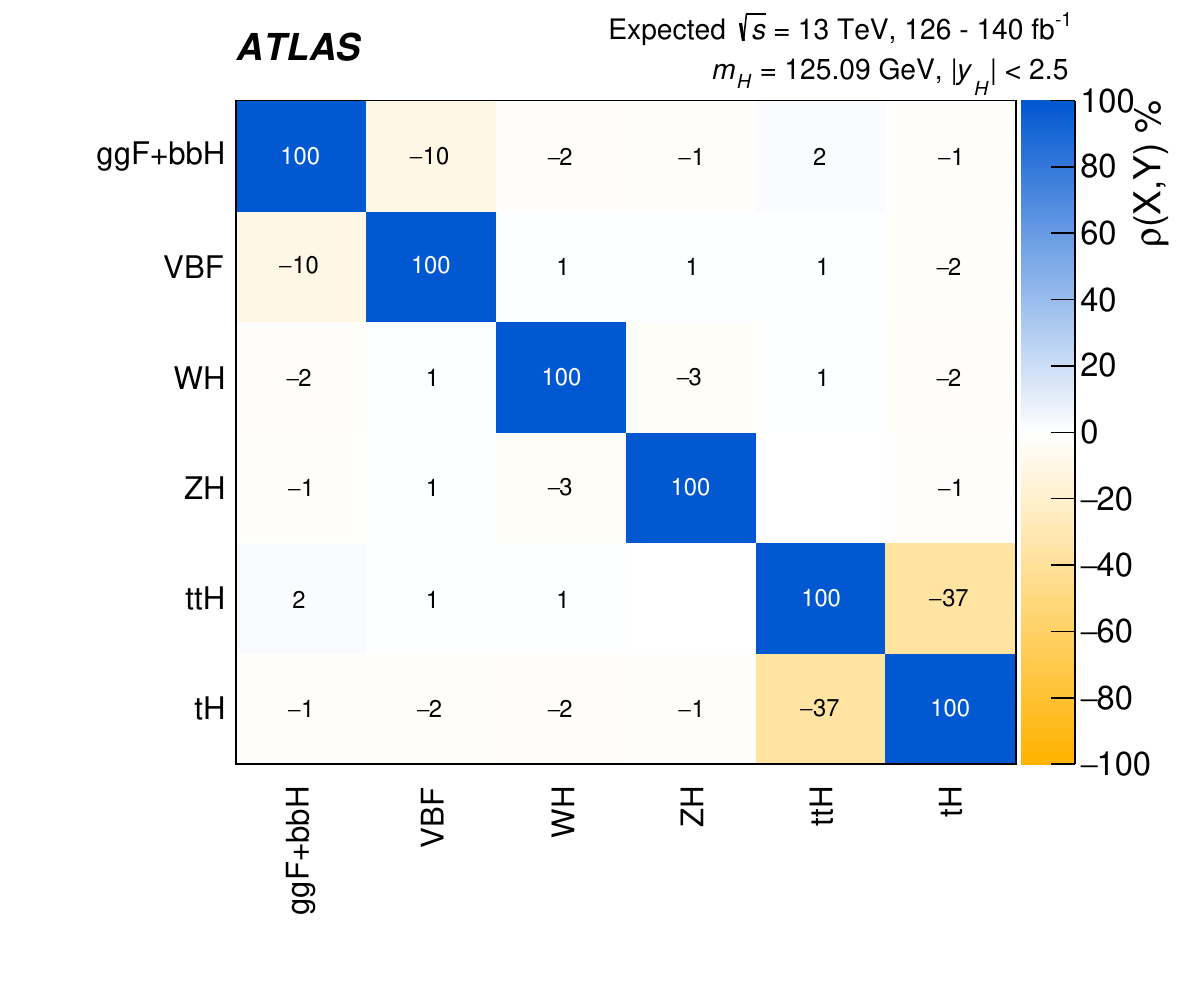}}
\end{center}
\caption{(a) Observed and (b) expected correlation matrices for the measurement of production cross-sections.}
\label{fig:xs:corr}
\end{figure}

\begin{figure}[tbp]
\begin{center}
\subfloat[]{\includegraphics[width=0.49\columnwidth]{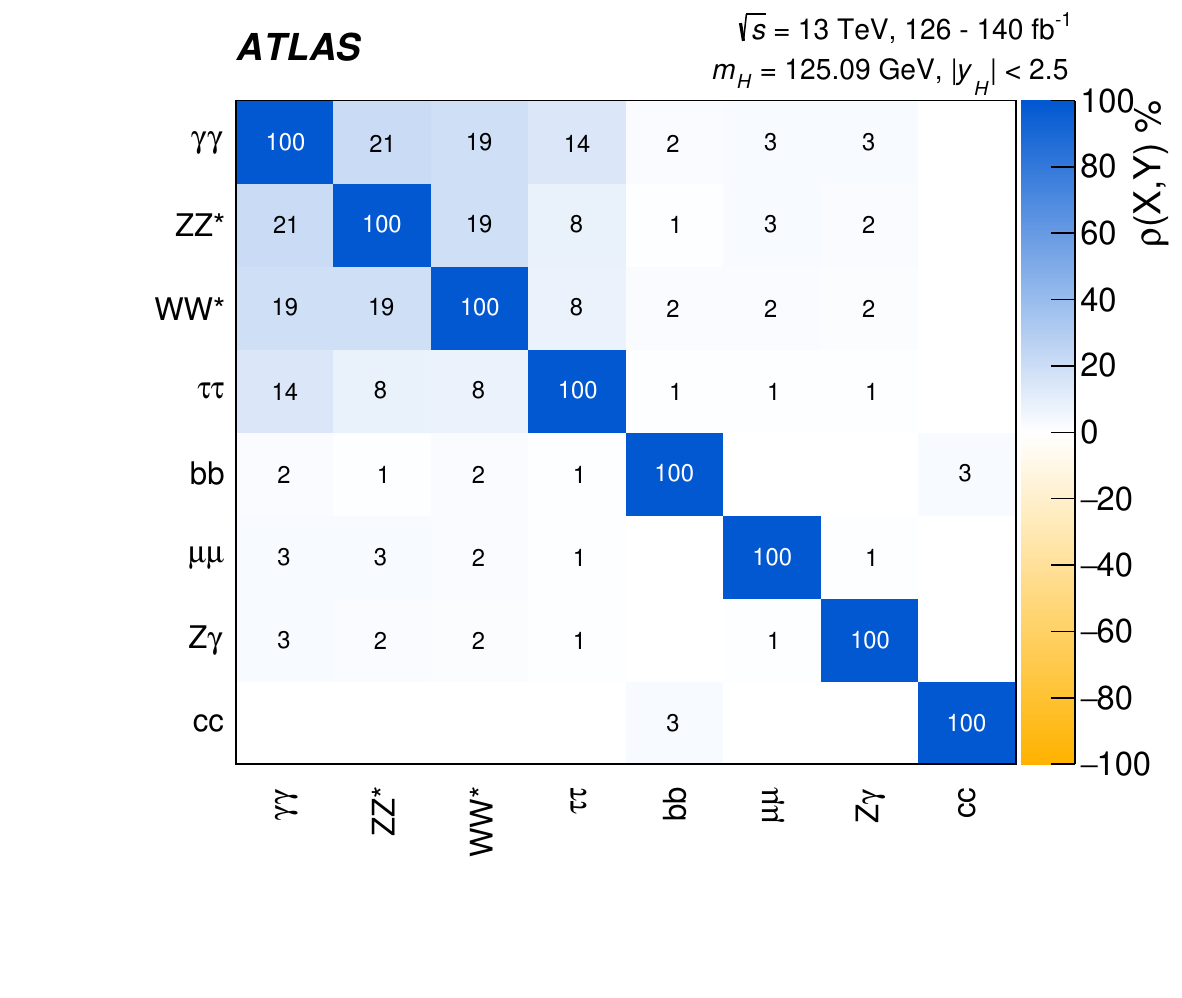}}
\subfloat[]{\includegraphics[width=0.49\columnwidth]{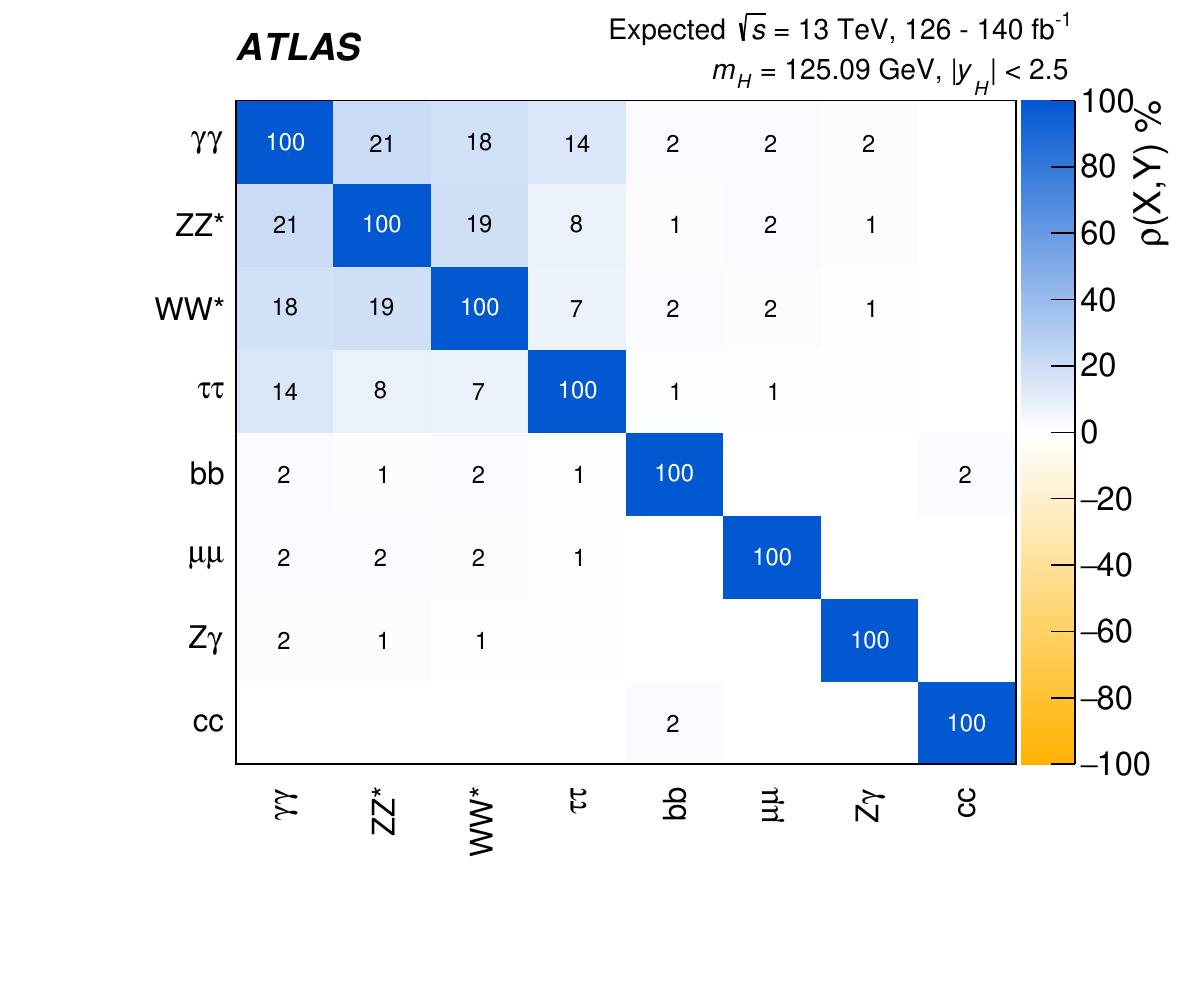}}
\end{center}
\caption{(a) Observed and (b) expected correlation matrices for the measurement of branching ratios.}
\label{fig:br:corr}
\end{figure}

\begin{figure}[tbp]
\begin{center}
\subfloat[]{\includegraphics[width=0.49\columnwidth]{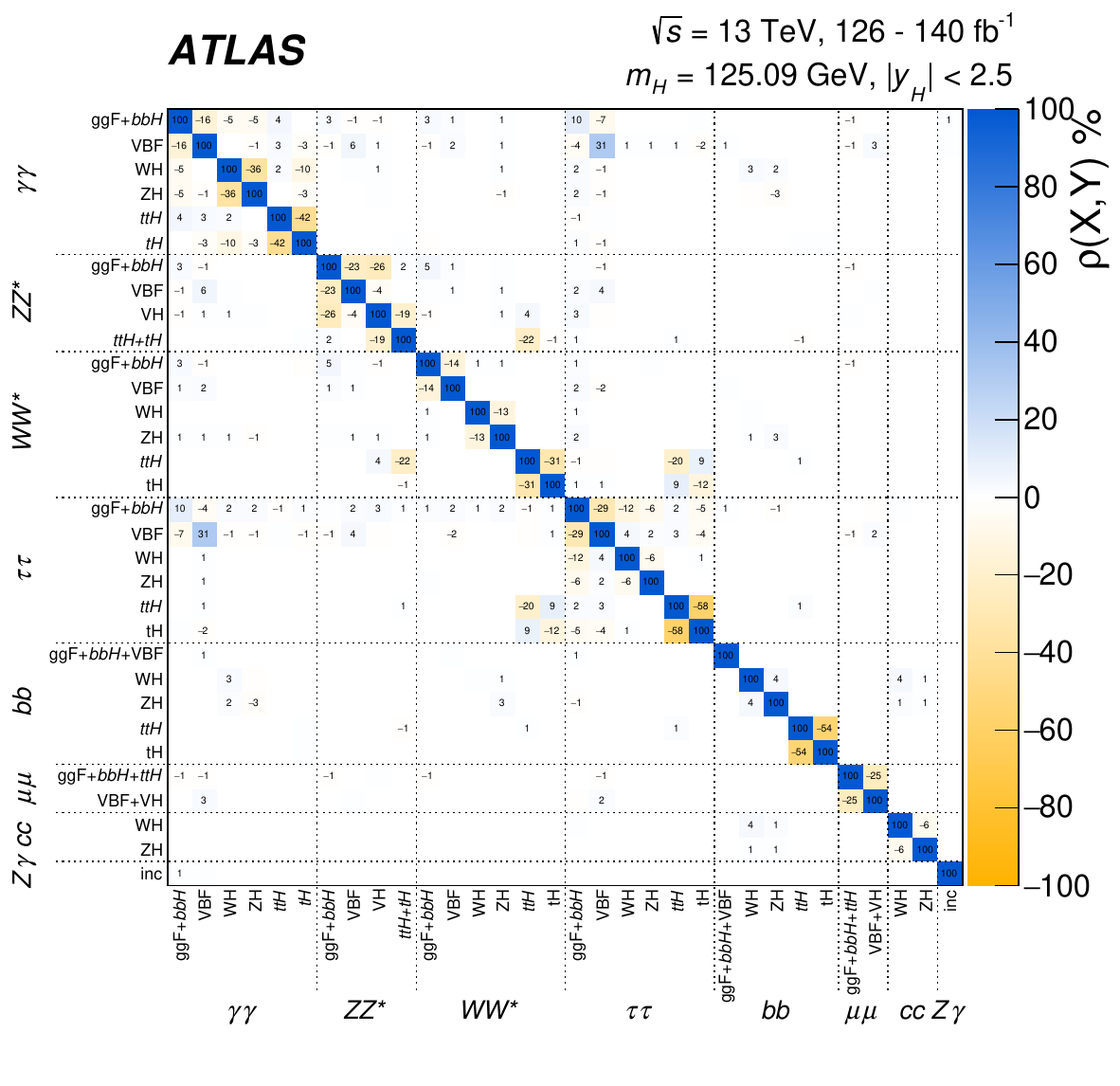}}
\subfloat[]{\includegraphics[width=0.49\columnwidth]{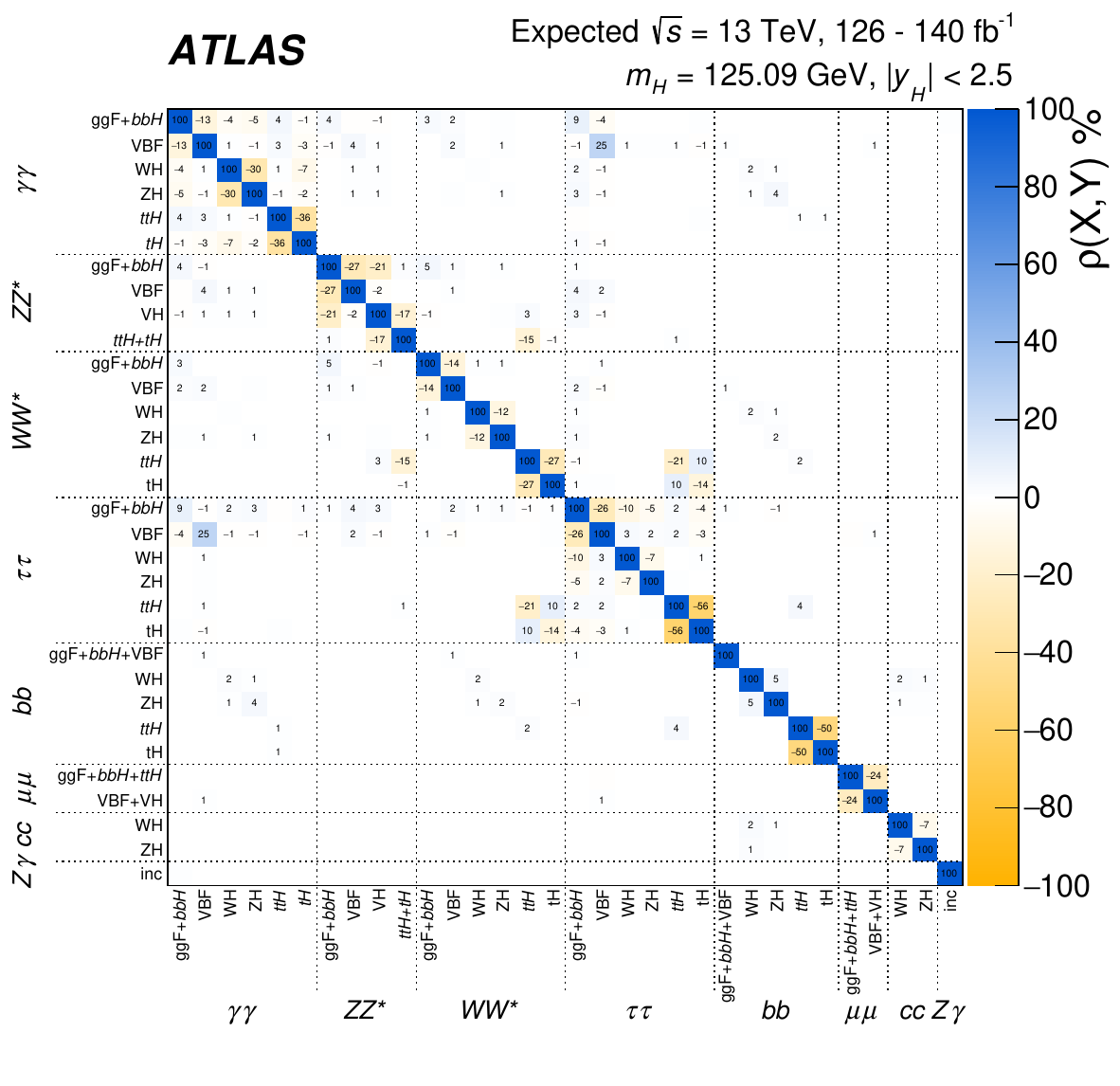}}
\end{center}
\caption{(a) Observed and (b) expected correlation matrices for the measurement of products of production cross-sections and branching ratios.}
\label{fig:PxD:corr}
\end{figure}

Expected results under the SM hypothesis for the Higgs boson production cross-sections described in Section~\ref{sec:prodXS}, assuming branching ratios to correspond to SM expectations, are shown in Figure~\ref{fig:xs:exp}. Expected values of the branching ratios described in Section~\ref{sec:decaymodes} are shown in Figure~\ref{fig:br:exp}, assuming SM values for production cross-sections. Finally, observed and expected results for the 32 combinations $(\sigma_i \times \BR_f)$ of production cross-sections $\sigma_i$ and branching ratios $\BR_f$ described in Section~\ref{sec:proddecay} are shown in Figure~\ref{fig:PxD:details}.
Observed and expected correlation matrices for the production cross-section results are shown in Figure~\ref{fig:xs:corr}, for branching ratio results in Figure~\ref{fig:br:corr} and for products of cross-sections and branching ratios in Figure~\ref{fig:PxD:corr}.


\FloatBarrier


\FloatBarrier

\clearpage
\section{Coupling modifier parameterisation}
\label{sec:kappa_defs}

The parameterisation used to express Higgs boson production cross-sections and branching ratios in terms of the coupling modifiers introduced in Section~\ref{sec:kappa} is shown in Table~\ref{tab:kappa_param}.

\begin{table}[tbp]
\caption{Parameterisations of Higgs boson production cross-sections $\sigma_i$ at 13 \TeV, partial decay widths $\Gamma^f$, and the total width $\Gamma_H$, normalised to their SM values, as functions of the coupling-strength modifiers~$\kappa$. The effect of invisible and undetected decays is not considered in the expression for $\Gamma_H$.
For effective $\kappa$~parameters associated with loop processes, the resolved scaling in terms of the modifications of the Higgs boson couplings to the fundamental SM particles is given. The coefficients are derived following the methodology in Ref.~\cite{YR3}.}
\begin{center}
\renewcommand{\arraystretch}{1.3}
\begin{tabular}{lcccl}
\toprule
\multirow{2}{*}{Production}  & \multirow{2}{*}{Loops}  & Main & Effective & \multirow{2}{*}{Resolved modifier}                                 \\
&       &             interference                  & modifier                      &                                                                    \\
\midrule
$\sigma({\ggF})$          & \chk & $t$--$b$ & $\kappa_g^2$
& $1.040\,\kappa_t^2\,{+\,0.002}\,\kappa_b^2\,{-\,0.038} \, \kappa_t \kappa_b\,{-\,0.005} \, \kappa_t \kappa_c$ \\
$\sigma({\VBF})$          & -    & -        & -            & $0.733\,\kappa_W^2 + 0.267\,\kappa_Z^2$                  \\
$\sigma({\qqZH})$         & -    & -        & -            & $\kappa_Z^2$ \\
\multirow{2}{*}{$\sigma({\ggZH})$} & \multirow{2}{*}{\chk} & \multirow{2}{*}{$t$--$Z$} & \multirow{2}{*}{-}
& $2.456\,\kappa_Z^2 \,{+    \,0.456}\,\kappa_t^2 \,{-\,1.903}\,\kappa_Z \kappa_t$    \\
& & & & $\; -\,0.011\,\kappa_Z \kappa_b \,{+\,0.003}\,\kappa_t \kappa_b$ \\
$\sigma({\WH})$           & -    & -        & -            & $\kappa_W^2$ \\
$\sigma({\ttH})$          & -    & -        & -            & $\kappa_t^2$ \\
$\sigma({\WtH})$          & -    & $t$--$W$ & -            & $2.909\,\kappa_t^2 + 2.310\,\kappa_W^2 - 4.220\,\kappa_t \kappa_W$  \\
$\sigma({\tHq})$          & -    & $t$--$W$ & -            & $2.633\,\kappa_t^2 + 3.578\,\kappa_W^2 - 5.211\,\kappa_t \kappa_W$  \\
$\sigma({\bbH})$          & -    & -        & -            & $\kappa_b^2$ \\
\midrule
\multicolumn{5}{l}{Partial decay width}                                                                                           \\
\midrule
$\Gamma^{bb}$             & -    & -        & -            & $\kappa_b^2$                                                        \\
$\Gamma^{\ww}$            & -    & -        & -            & $\kappa_W^2$                                                        \\
$\Gamma^{gg}$             & \chk & $t$--$b$ & $\kappa_g^2$ & $1.111\,\kappa_t^2 + 0.012\,\kappa_b^2 - 0.1229\, \kappa_t \kappa_b$   \\
$\Gamma^{\tau\tau}$       & -    & -        & -            & $\kappa_\tau^2$                                                     \\
$\Gamma^{\zz}$            & -    & -        & -            & $\kappa_Z^2$                                                        \\
$\Gamma^{cc}$             & -    & -        & -            & $\kappa_c^2$                                        \\
\multirow{3}{*}{$\Gamma^{\yy}$} & \multirow{3}{*}{\chk} & \multirow{3}{*}{$t$--$W$} & \multirow{3}{*}{$\kappa_\gamma^2$}
& $1.589\,\kappa_W^2 + 0.    072\,\kappa_t^2 - 0.674\,\kappa_W \kappa_t$    \\
& & & & $\; + 0.009\,\kappa_W \kappa_\tau + 0.008\,\kappa_W \kappa_b$ \\
& & & & $\; - 0.002\,\kappa_t \kappa_b - 0.002\,\kappa_t \kappa_\tau$ \\
$\Gamma^{Z\gamma}$        & \chk & $t$--$W$ & $\kappa_{Z\gamma}^2$ & $1.118\,\kappa_W^2 - 0.125\,\kappa_W \kappa_t + 0.004\,\kappa_t^2 + 0.003\,\kappa_W \kappa_b$ \\
$\Gamma^{ss}$             & -    & -        & -            & $\kappa_s^2\ (= \kappa_b^2)$                                        \\
$\Gamma^{\mu\mu}$         & -    & -        & -            & $\kappa_\mu^2$                                                      \\
\midrule
\multicolumn{4}{l}{Total width} \\
\midrule
\multirow{3}{*}{$\Gamma_H$} & \multirow{3}{*}{\chk} & \multirow{3}{*}{-} & \multirow{3}{*}{$\kappa_H^2$}
& $0.581\,\kappa_b^2 + 0.215\,\kappa_W^2 + 0.082\,\kappa_g^2 + 0.063\,\kappa_\tau^2 $   \\
& & & & $\; + 0.026\,\kappa_Z^2 + 0.029\,\kappa_c^2 + 0.0023\,\kappa_\gamma^2$                \\
& & & & $\; + 0.0015\,\kappa_{(Z\gamma)}^2 + 0.0004\,\kappa_s^2 + 0.00022\,\kappa_\mu^2$      \\
\bottomrule
\end{tabular}
\end{center}
\label{tab:kappa_param}
\end{table}


\FloatBarrier

\clearpage

\section{Additional Higgs boson coupling modifier results}
\label{app:kappa}

Table~\ref{tab:kappa_ratio_obs_noRcb} shows the observed values of the model-independent coupling-ratio parameters for the model with $\kappa_c = \kappa_t$ ($p_{\text{SM}} = 70\%$), complementing the results for the model with free $\lambda_{cb}$ presented in Table~\ref{tab:kappa_ratio_obs}.

\begin{figure}[tbp]
\begin{center}
\includegraphics[width=0.75\columnwidth]{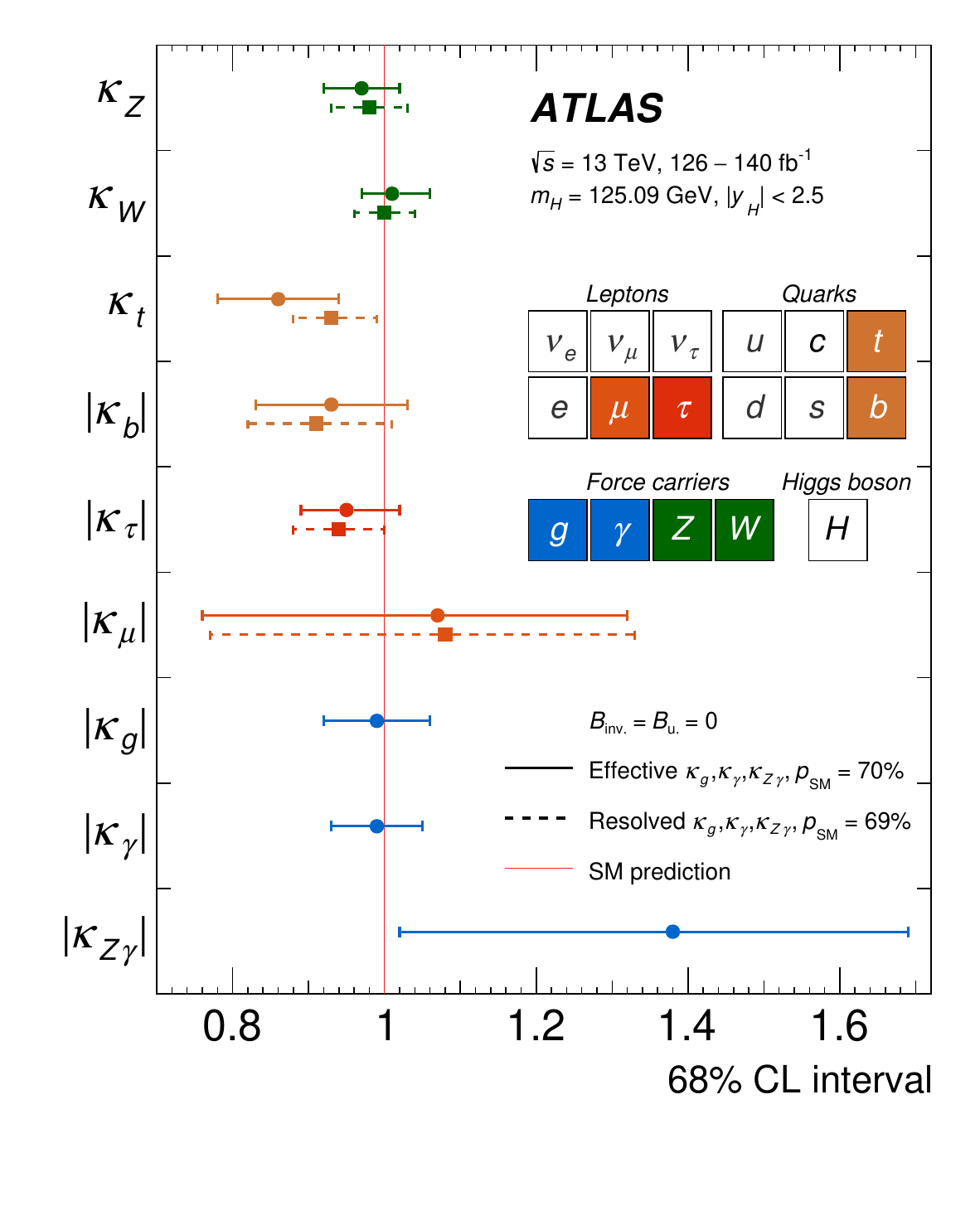}
\end{center}
\caption{Observed values of the Higgs boson coupling modifiers in the generic $\kappa$ framework.
Results are shown for the effective parameterisation of the loop-induced couplings $\kappa_g$,
$\kappa_\gamma$ and $\kappa_{Z\gamma}$ (round markers and solid error bars), and for the resolved parameterisation in which these
effective couplings are expressed in terms of the coupling modifiers of the particles contributing
to the corresponding loops (square markers and dashed error bars). The intervals correspond to 68\% CL uncertainties.
The SM prediction is indicated by the vertical red line. Results are shown under
the assumption that $\kappa_c = \kappa_t$ and $B_\mathrm{inv} = B_\mathrm{u} = 0$.}
\label{fig:kappa:generic_effective_resolved}
\end{figure}

\begin{table}[tbp]
\caption{Definitions and observed values of the model-independent coupling-ratio parameters for the model with $\kappa_c = \kappa_t$, together with the expected total uncertainty (the SM central value is 1 for all parameters by construction). Observed uncertainties are decomposed into contributions from data statistics (stat.), experimental systematic uncertainties (exp.), signal theory uncertainties (sig.\ theo.) and background theory uncertainties (bkg.\ theo.). All results are consistent with the SM ($p_{\text{SM}} = 70\%$).}
\label{tab:kappa_ratio_obs_noRcb}
\centering
\renewcommand{\arraystretch}{1.5}
\resizebox{\textwidth}{!}{%
\begin{tabular}{llcccccc}
\toprule
Parameter & Definition & Observed & Stat. & Exp. & Sig.\ theo. & Bkg.\ theo. & Expected unc. \\
\midrule
$\kappa_{gZ}$         & $\kappa_g\kappa_Z/\kappa_H$ & $1.003$\numpmerr{+0.052}{-0.051} & ${\scriptstyle \pm 0.038}$ & \numpmerr{+0.015}{-0.013} & \numpmerr{+0.031}{-0.029} & \numpmerr{+0.008}{-0.009} & ${\scriptstyle \pm 0.05}$ \\
$\lambda_{tg}$        & $\kappa_t/\kappa_g$         & $0.866$\numpmerr{+0.091}{-0.086} & ${\scriptstyle \pm 0.060}$ & \numpmerr{+0.028}{-0.030} & \numpmerr{+0.053}{-0.044} & \numpmerr{+0.033}{-0.031} & ${\scriptstyle \pm 0.09}$ \\
$\lambda_{Zg}$        & $\kappa_Z/\kappa_g$         & $0.975$\numpmerr{+0.083}{-0.077} & \numpmerr{+0.054}{-0.052} & \numpmerr{+0.031}{-0.029} & \numpmerr{+0.047}{-0.042} & \numpmerr{+0.028}{-0.026} & \numpmerr{+0.09}{-0.08} \\
$\lambda_{WZ}$        & $\kappa_W/\kappa_Z$         & $1.041$\numpmerr{+0.056}{-0.053} & \numpmerr{+0.044}{-0.042} & \numpmerr{+0.020}{-0.022} & \numpmerr{+0.019}{-0.016} & \numpmerr{+0.019}{-0.018} & ${\scriptstyle \pm 0.05}$ \\
$\lambda_{\gamma Z}$  & $\kappa_\gamma/\kappa_Z$    & $1.025$\numpmerr{+0.058}{-0.055} & \numpmerr{+0.050}{-0.048} & \numpmerr{+0.023}{-0.022} & \numpmerr{+0.014}{-0.013} & ${\scriptstyle \pm 0.010}$ & \numpmerr{+0.06}{-0.05} \\
$\lambda_{\tau Z}$    & $\kappa_\tau/\kappa_Z$      & $0.979$\numpmerr{+0.072}{-0.067} & \numpmerr{+0.054}{-0.051} & \numpmerr{+0.029}{-0.032} & \numpmerr{+0.033}{-0.025} & \numpmerr{+0.018}{-0.016} & ${\scriptstyle \pm 0.07}$ \\
$\lambda_{bZ}$        & $\kappa_b/\kappa_Z$         & $0.966$\numpmerr{+0.094}{-0.087} & \numpmerr{+0.069}{-0.065} & ${\scriptstyle \pm 0.032}$ & \numpmerr{+0.040}{-0.033} & \numpmerr{+0.038}{-0.035} & ${\scriptstyle \pm 0.09}$ \\
$\lambda_{\mu\tau}$   & $\kappa_\mu/\kappa_\tau$    & $1.131$\numpmerr{+0.270}{-0.332} & \numpmerr{+0.257}{-0.323} & \numpmerr{+0.059}{-0.062} & \numpmerr{+0.050}{-0.039} & \numpmerr{+0.027}{-0.026} & \numpmerr{+0.28}{-0.38} \\
$\lambda_{Z\gamma Z}$ & $\kappa_{Z\gamma}/\kappa_Z$ & $1.426$\numpmerr{+0.314}{-0.374} & \numpmerr{+0.289}{-0.347} & \numpmerr{+0.098}{-0.127} & \numpmerr{+0.071}{-0.053} & \numpmerr{+0.011}{-0.009} & \numpmerr{+0.4}{-0.6} \\
\bottomrule
\end{tabular}}
\end{table}
For the general coupling modifier models described in Section~\ref{sec:kappa}, results in the effective and resolved parameterisations assuming $\kappa_c = \kappa_t$ are shown in Figure~\ref{fig:kappa:generic_effective_resolved}. The compatibility of the effective and resolved measurements with the SM prediction corresponds to $p$-values of $p_{\mathrm{SM}} = 70\%$ and $p_{\mathrm{SM}} = 69\%$, respectively.

\begin{figure}[tbp]
\begin{center}
\subfloat[]{\includegraphics[width=0.49\columnwidth]{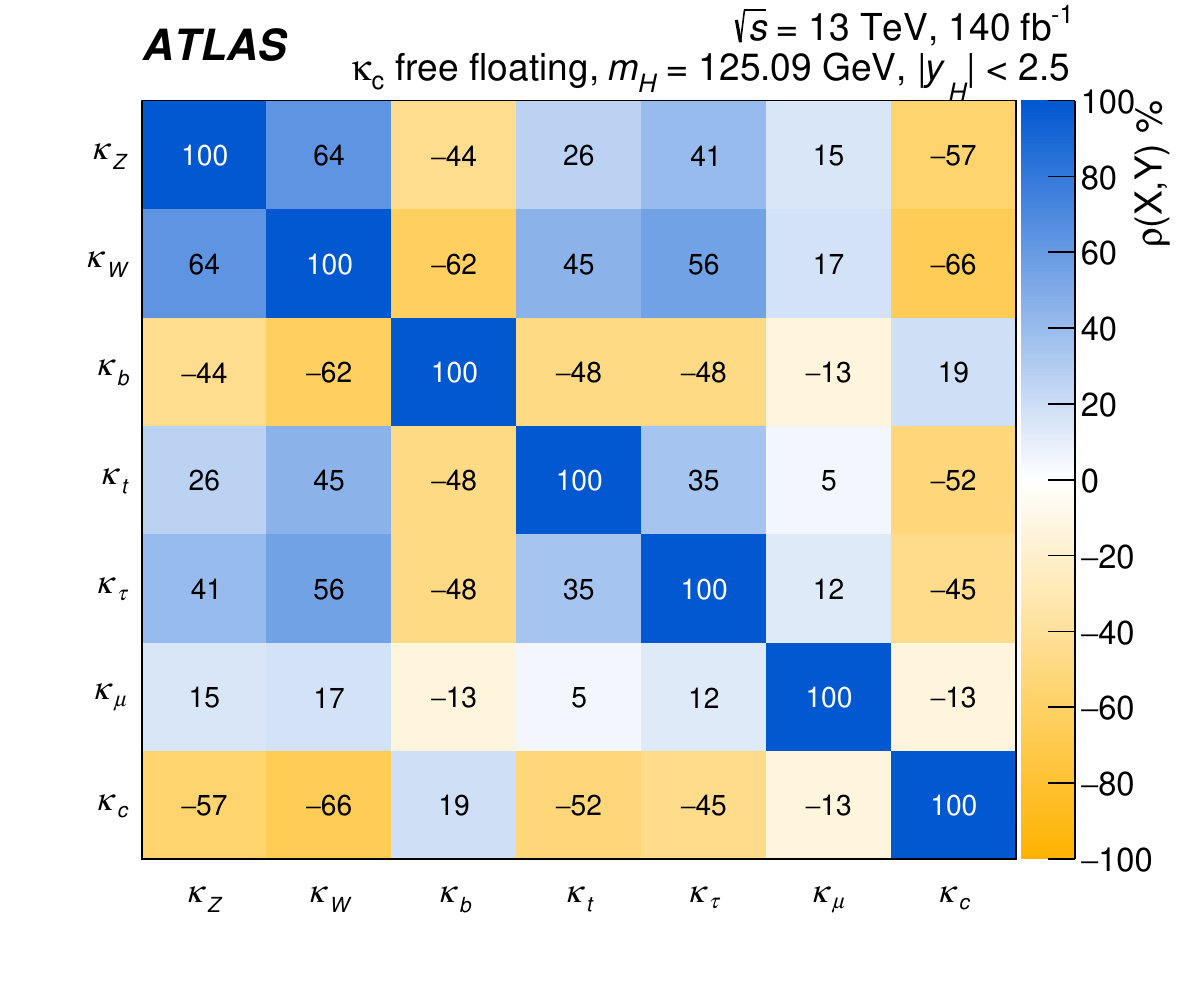}}
\subfloat[]{\includegraphics[width=0.49\columnwidth]{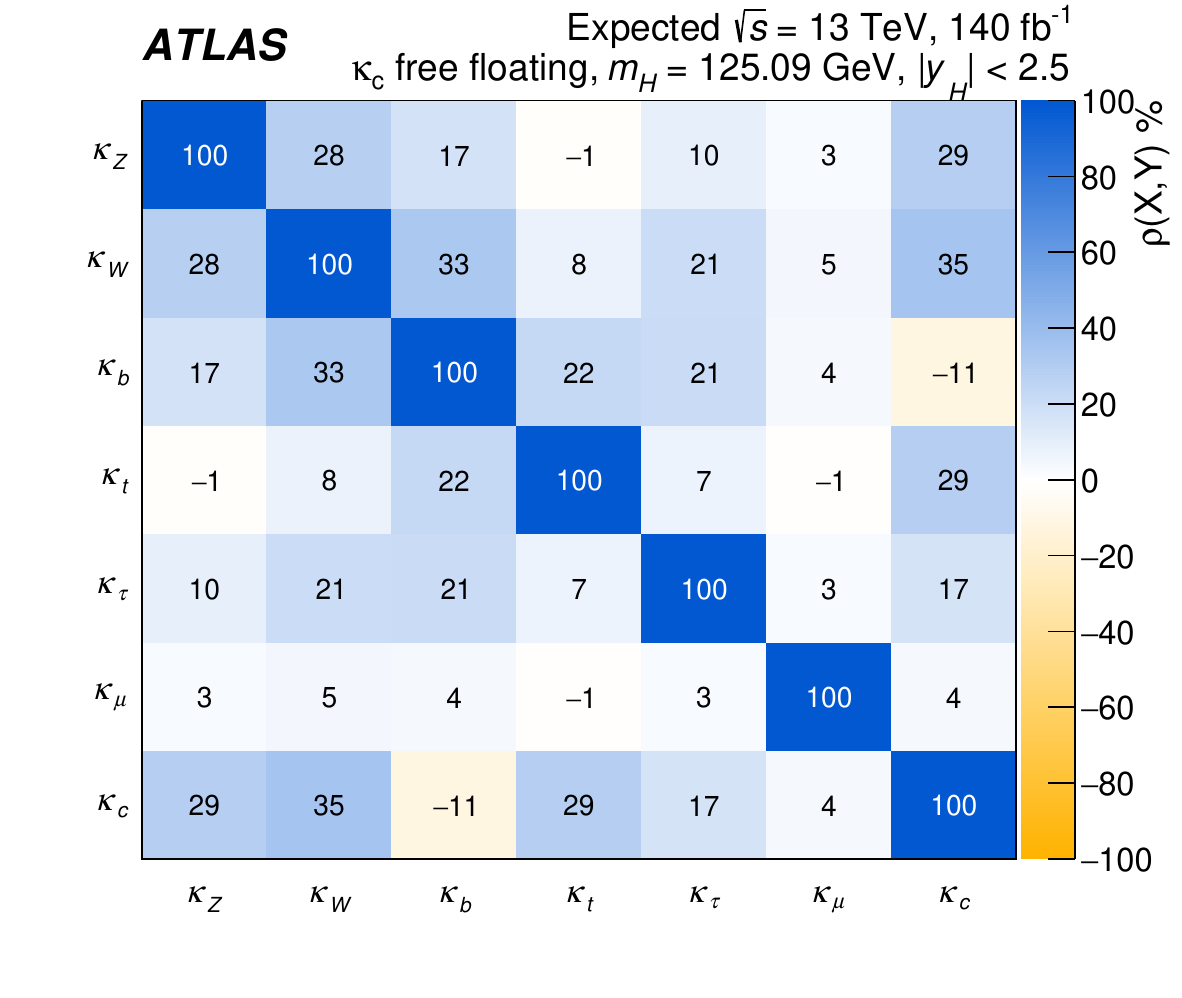}}
\end{center}
\caption{(a) Observed and (b) expected correlation matrices for the measurement of Higgs boson coupling modifiers in the resolved parameterisation with $\kappa_c$ included as a free parameter. Expected results are obtained from an Asimov dataset generated with $\kappa_b = 1$ but the observed value of $\kappa_b$ is negative, related to the fact that the combination has almost no sensitivity to its sign. The observed and expected correlation coefficients of $\kappa_b$ with other parameters therefore have opposite signs.}
\label{fig:corr_resolved_with_kc}
\end{figure}

\begin{figure}[tbp]
\begin{center}
\subfloat[]{\includegraphics[width=0.49\columnwidth]{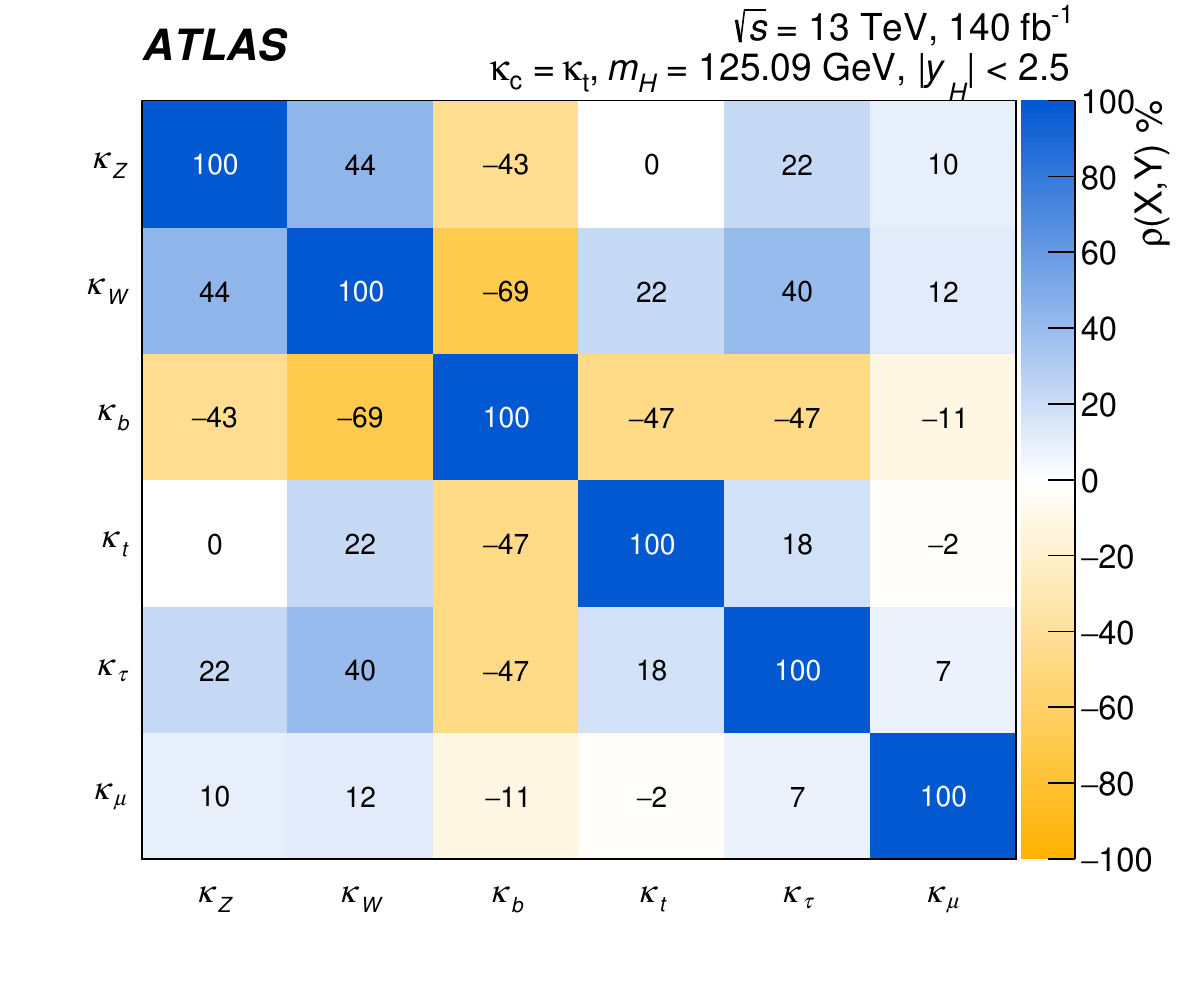}}
\subfloat[]{\includegraphics[width=0.49\columnwidth]{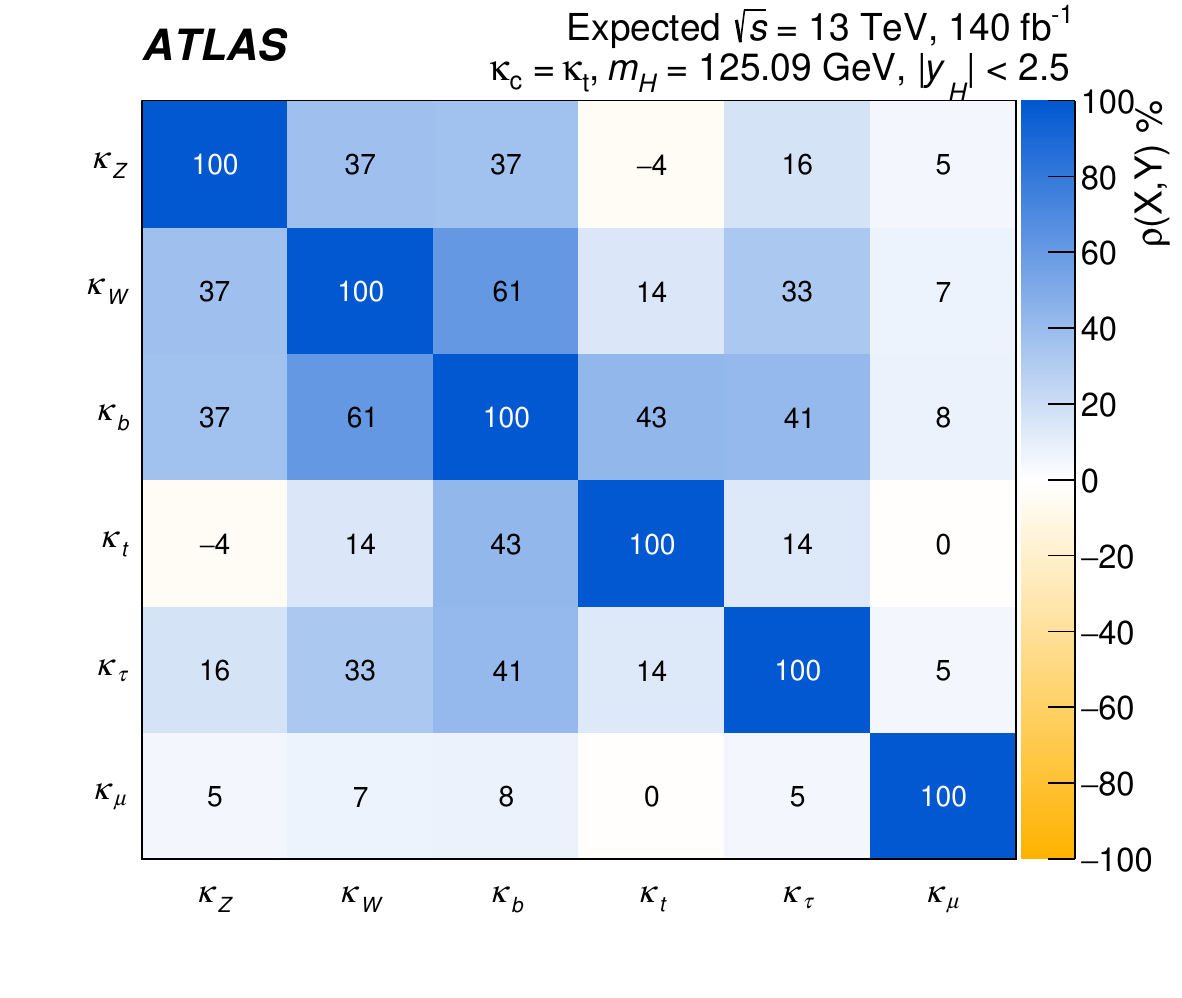}}
\end{center}
\caption{(a) Observed and (b) expected correlation matrices for the measurement of Higgs boson coupling modifiers in the resolved parameterisation, assuming $\kappa_c = \kappa_t$. Expected results are obtained from an Asimov dataset generated with $\kappa_b = 1$ but the observed value of $\kappa_b$ is negative, related to the fact that the combination has almost no sensitivity to its sign. The observed and expected correlation coefficients of $\kappa_b$ with other parameters therefore have opposite signs.}
\label{fig:corr_resolved_no_kc}
\end{figure}

\begin{figure}[tbp]
\begin{center}
\subfloat[]{\includegraphics[width=0.49\columnwidth]{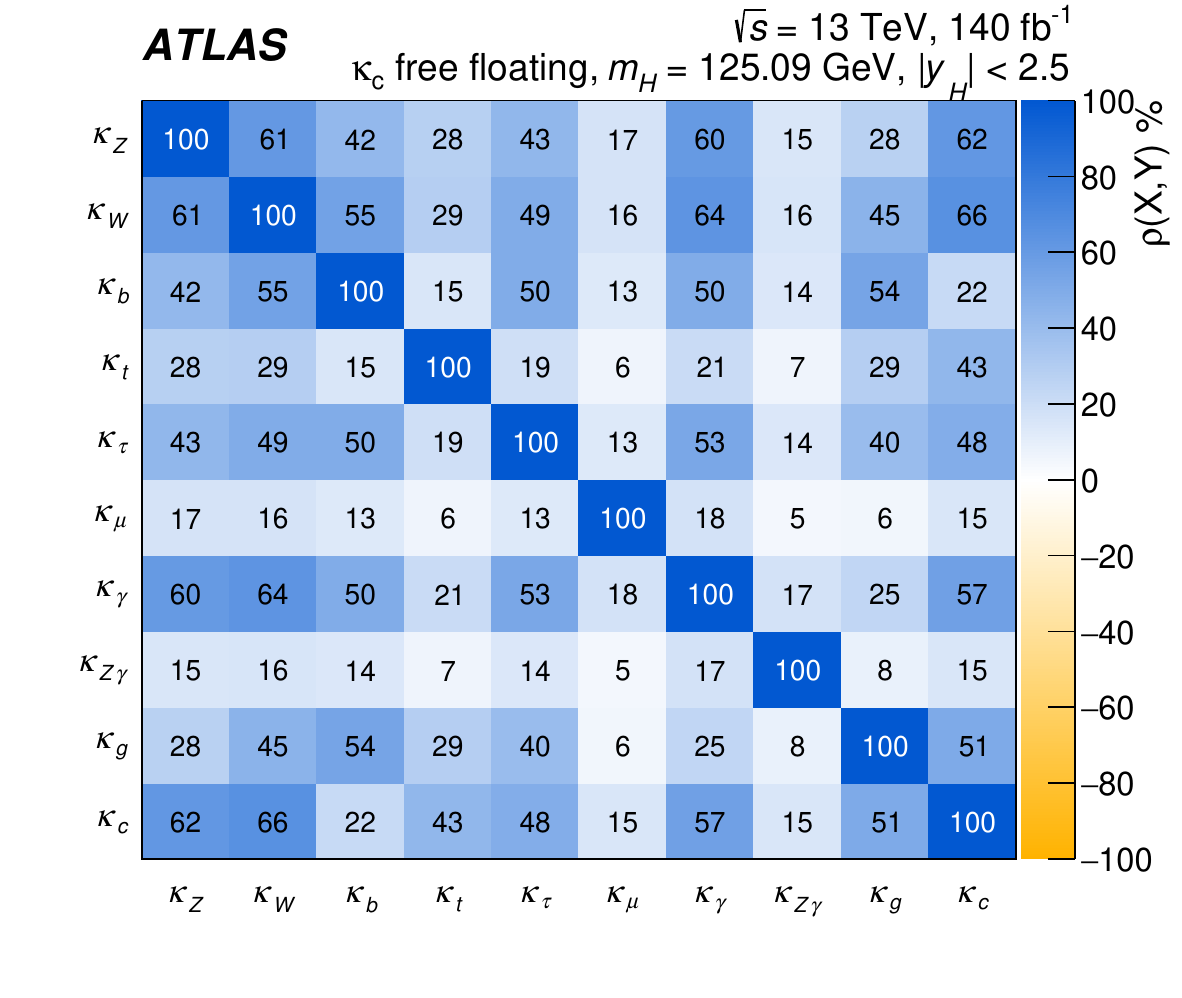}}
\subfloat[]{\includegraphics[width=0.49\columnwidth]{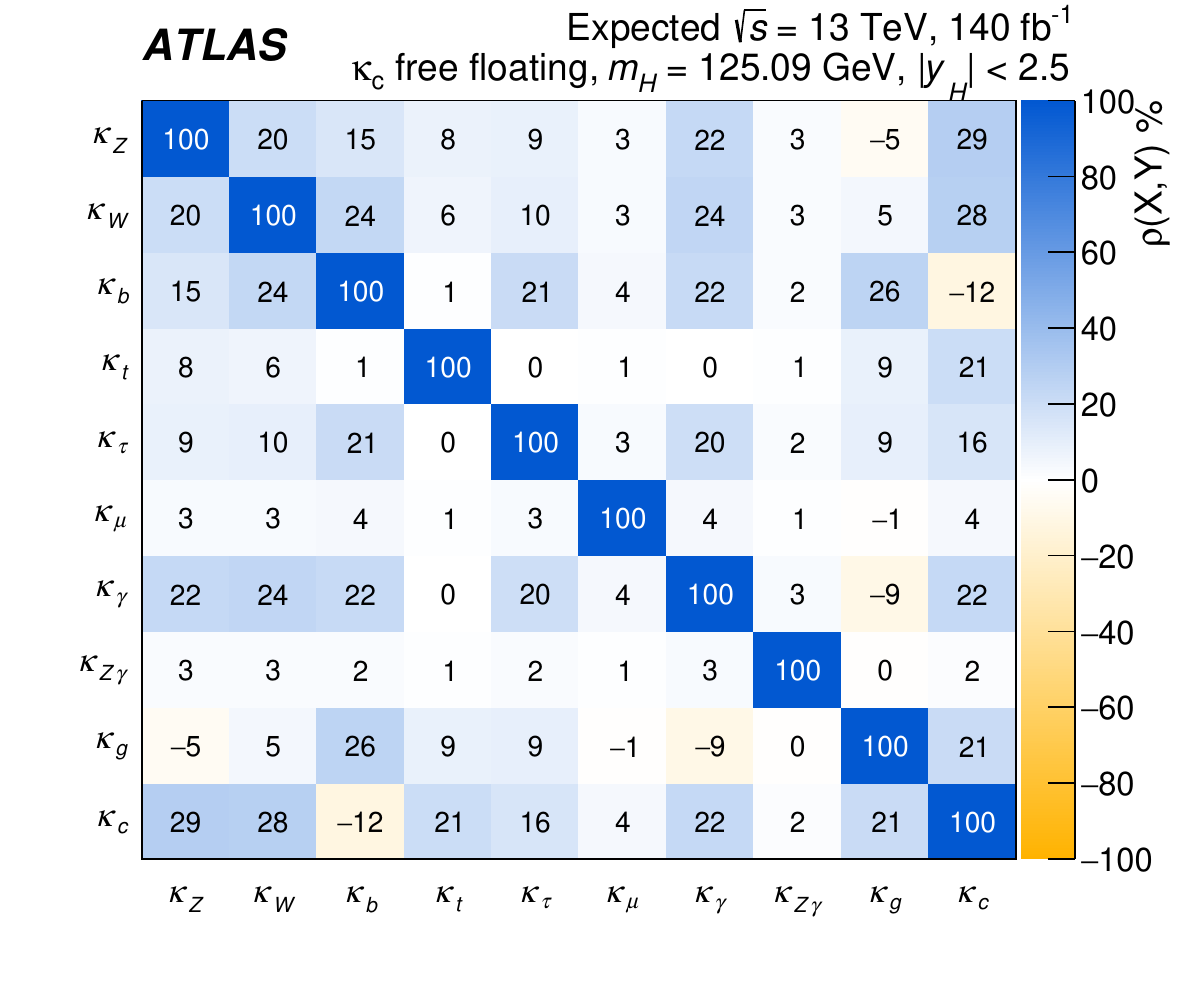}}
\end{center}
\caption{(a) Observed and (b) expected correlation matrices for the measurement of Higgs boson coupling modifiers in the effective parameterisation with $\kappa_c$ included as a free parameter.}
\label{fig:corr_effective_with_kc}
\end{figure}

\begin{figure}[tbp]
\begin{center}
\subfloat[]{\includegraphics[width=0.49\columnwidth]{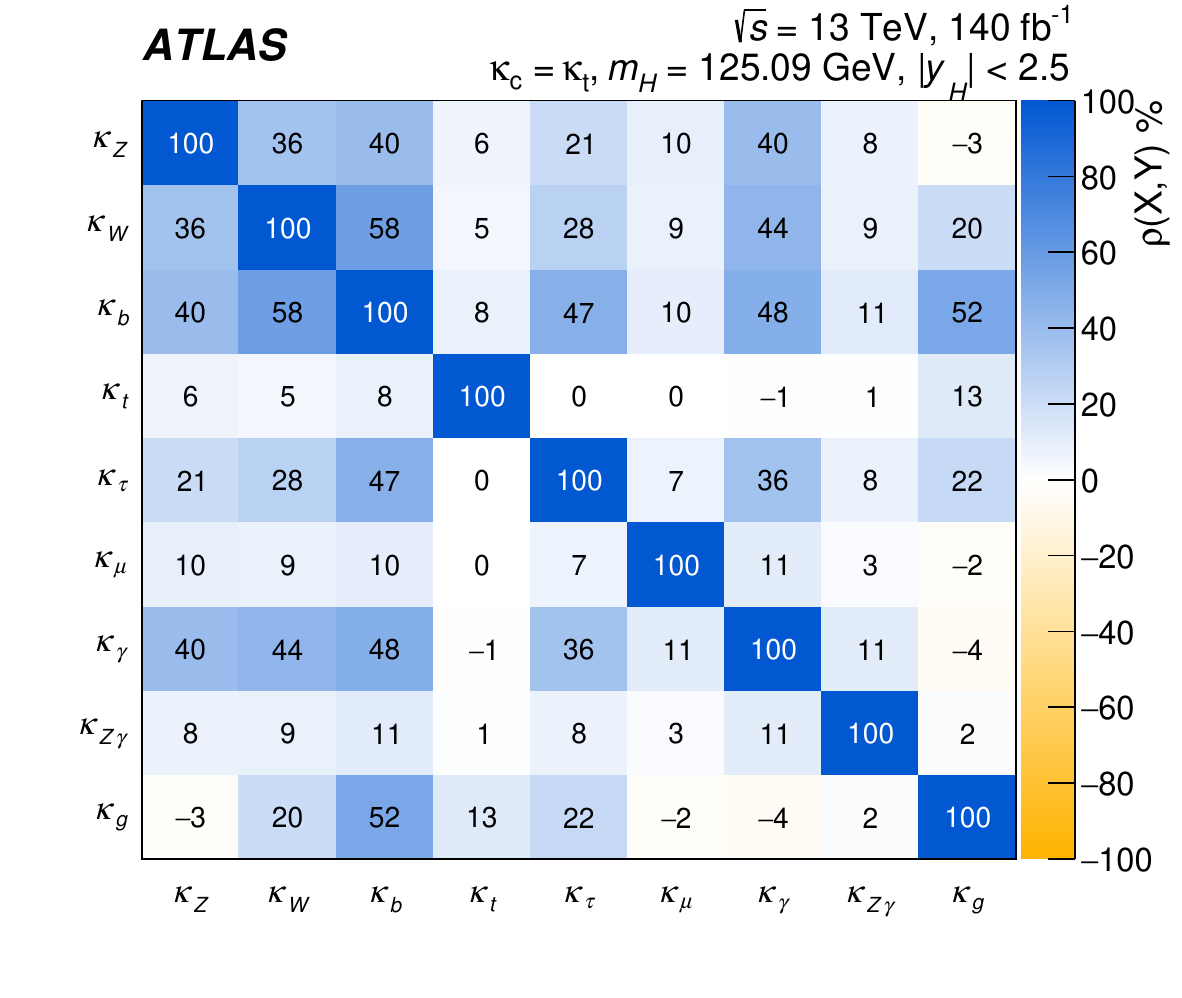}}
\subfloat[]{\includegraphics[width=0.49\columnwidth]{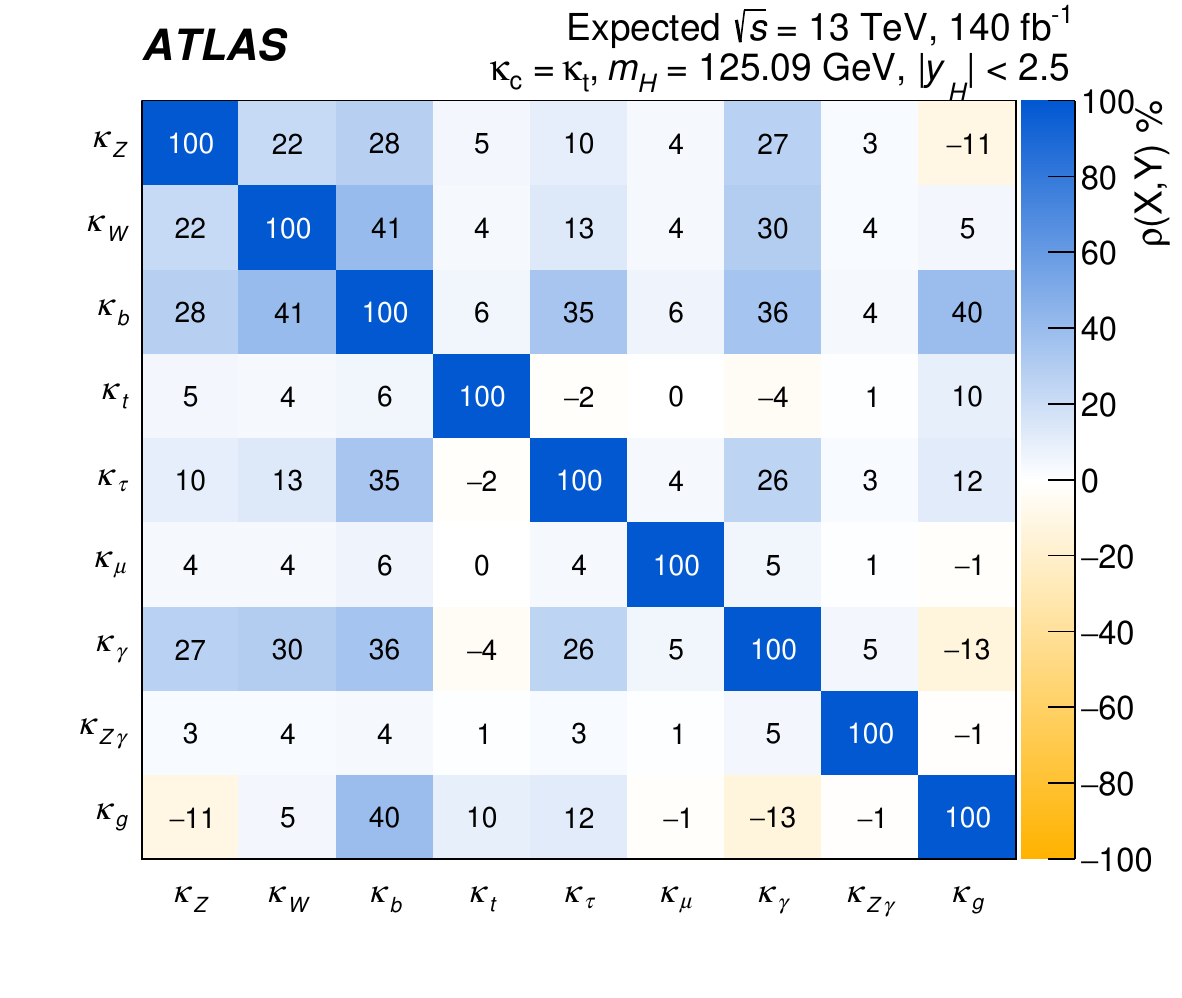}}
\end{center}
\caption{(a) Observed and (b) expected correlation matrices for the measurement of Higgs boson coupling modifiers in the effective parameterisation, assuming $\kappa_c = \kappa_t$.}
\label{fig:corr_effective_no_kc}
\end{figure}

Expected results under the SM hypothesis are shown in Table~\ref{tab:kappas_exp}. Tables~\ref{tab:kappas_exp_breakdown} and~\ref{tab:kappas_exp_5B5F_breakdown} show the decomposition of the total uncertainties in the $\kappa$ parameters into contributions from data statistics (stat.), experimental systematic uncertainties (exp.), signal theory uncertainties (sig.\ theo.) and background theory uncertainties (bkg.\ theo.) for the two models with free $\kappa_c$ or with $\kappa_c = \kappa_t$, for the resolved and effective parameterisations respectively. Observed and expected correlation matrices for the resolved model with $\kappa_c$ included as a measurement parameter, the resolved model with $\kappa_c = \kappa_t$, and the effective model (both with $\kappa_c$ included as a free parameter and with $\kappa_c = \kappa_t$) are shown respectively in Figures~\ref{fig:corr_resolved_with_kc},~\ref{fig:corr_resolved_no_kc}, ~\ref{fig:corr_effective_with_kc} and ~\ref{fig:corr_effective_no_kc}.

\begin{figure}[tbp]
\begin{center}
\subfloat[]{\includegraphics[width=0.49\columnwidth]{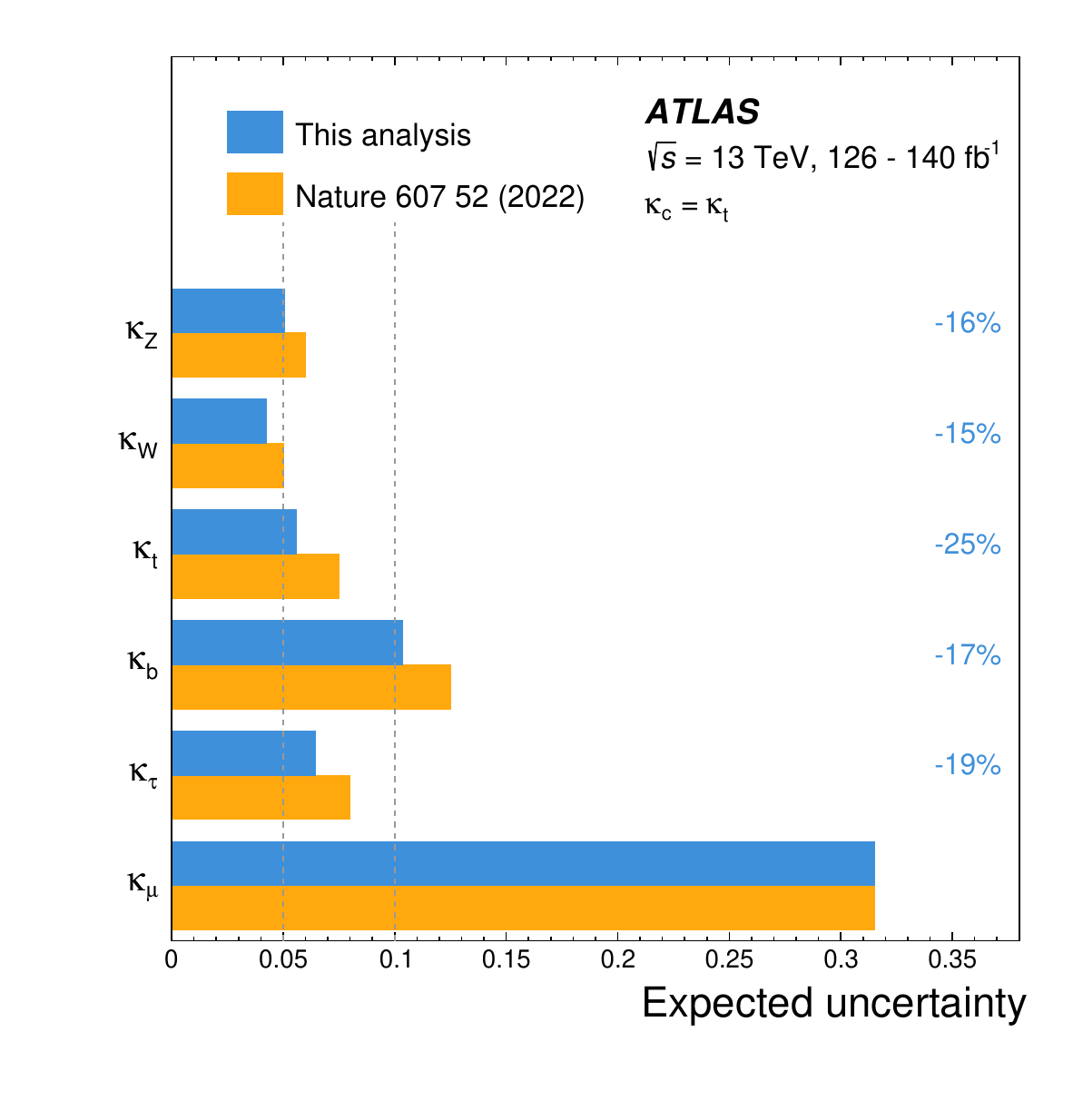}}
\subfloat[]{\includegraphics[width=0.49\columnwidth]{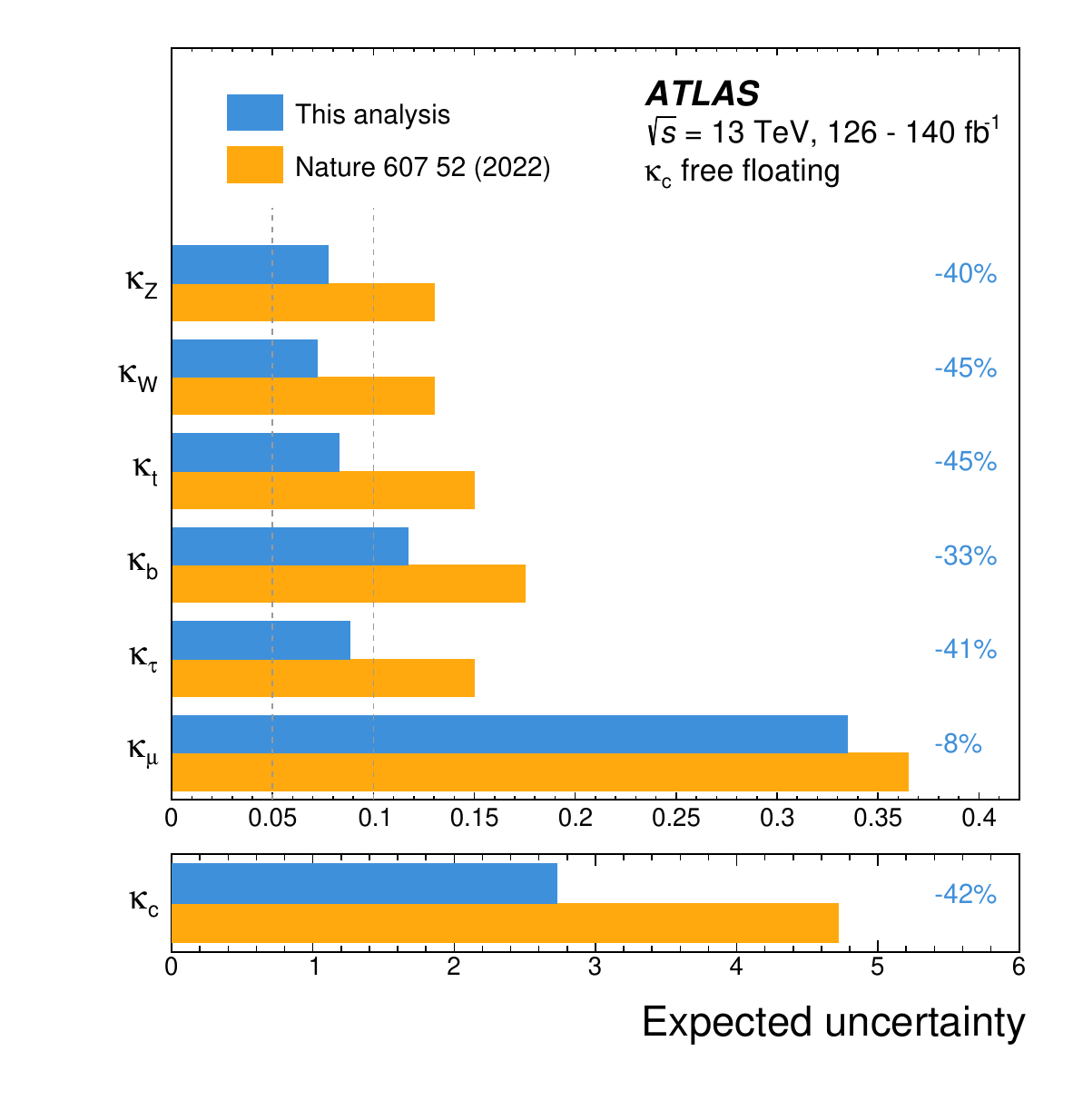}}
\end{center}
\caption{Expected uncertainties in the coupling modifiers in the resolved parameterisation (a) assuming $\kappa_c = \kappa_t$ and (b) with $\kappa_c$ included as a free parameter, for the results presented in this work (blue bars) and those of Ref.~\cite{HIGG-2021-23} (orange bars).
The numbers on the right side show the percentage reduction in the uncertainties achieved relative to Ref.~\cite{HIGG-2021-23}.}
\label{fig:exp_kappa_unc_resolved}
\end{figure}

\begin{figure}[tbp]
\begin{center}
\includegraphics[width=0.75\columnwidth]{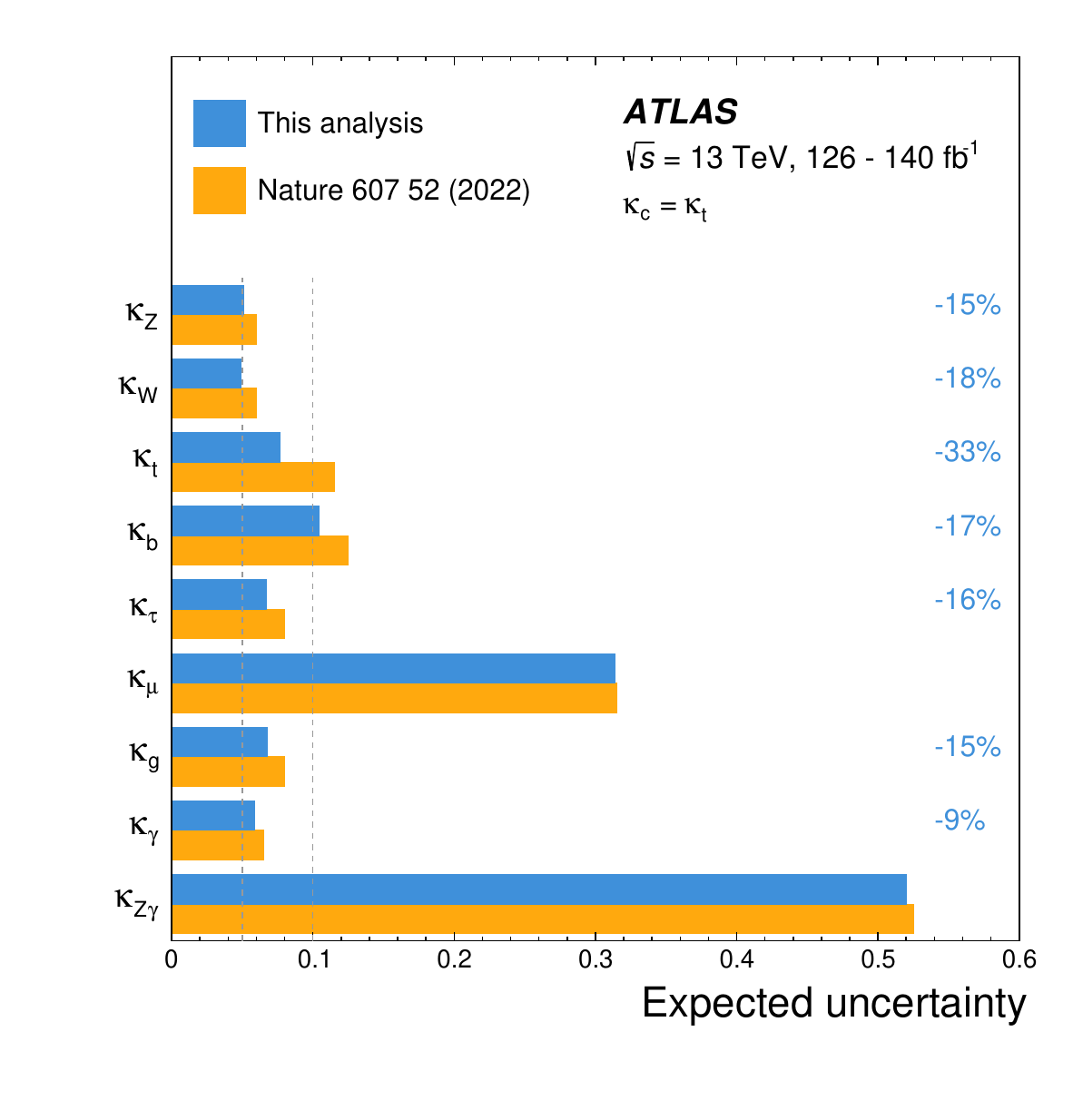}
\end{center}
\caption{Expected uncertainties in the coupling modifiers in the effective parameterisation, for the results presented in this work (blue bars) and those of Ref.~\cite{HIGG-2021-23} (orange bars).
The numbers on the right side show the percentage reduction in the uncertainties achieved relative to Ref.~\cite{HIGG-2021-23}. }
\label{fig:exp_kappa_unc_effective}
\end{figure}

Expected uncertainties in the coupling modifiers in the resolved parameterisation are shown in Figure~\ref{fig:exp_kappa_unc_resolved} and in the effective parameterisation in Figure~\ref{fig:exp_kappa_unc_effective}, in both cases together with the corresponding values in Ref.~\cite{HIGG-2021-23}.
\begin{DIFnomarkup}
\begin{table}[tbp]
\caption{Expected values of Higgs boson coupling modifiers. The second and third columns correspond to the resolved parameterisation, in which the modifiers $\kappa_g, \kappa_\gamma$ and $\kappa_{Z\gamma}$ are expressed as functions of the other modifiers. The fourth and fifth columns correspond to the effective parameterisation in which they are treated as independent parameters. In the second and fourth columns, $\kappa_c$ is included as a free parameter; in the third and fifth columns it is set equal to $\kappa_t$.}
\label{tab:kappas_exp}
\renewcommand{\arraystretch}{1.5}
\begin{center}
\begin{tabular}{lcccc}
\toprule
\multirow{2}{*}{Parameter} & \multicolumn{2}{c}{Resolved $\kappa_g, \kappa_\gamma, \kappa_{Z\gamma}$} &
\multicolumn{2}{c}{Effective $\kappa_g, \kappa_\gamma, \kappa_{Z\gamma}$} \\ \cmidrule{2-5}
& Free $\kappa_c$ & $\kappa_c = \kappa_t$ & Free $\kappa_c$ & $\kappa_c = \kappa_t$  \\ \midrule
$\kappa_Z$            & $1.00$\numpmerr{+0.09}{-0.06} & $1.00\,{\scriptstyle \pm 0.05}$        & $1.00$\numpmerr{+0.09}{-0.06}   & $ 1.00\,{\scriptstyle \pm 0.05}$        \\
$\kappa_W$            & $1.00$\numpmerr{+0.09}{-0.06} & $1.00\,{\scriptstyle \pm 0.04}$        & $1.00$\numpmerr{+0.09}{-0.06}   & $ 1.00\,{\scriptstyle \pm 0.05}$        \\ \midrule
$\kappa_t$            & $1.00$\numpmerr{+0.10}{-0.07} & $1.00\,{\scriptstyle \pm 0.06}$        & $1.00$\numpmerr{+0.11}{-0.09}   & $ 1.00\,{\scriptstyle \pm 0.08}$        \\
$|\kappa_b|$          & $1.00$\numpmerr{+0.13}{-0.11} & $1.00$\numpmerr{+0.10}{-0.09} & $1.00$\numpmerr{+0.14}{-0.11}   & $1.00$\numpmerr{+0.11}{-0.10} \\
$|\kappa_\tau|$       & $1.00$\numpmerr{+0.10}{-0.08} & $1.00$\numpmerr{+0.07}{-0.06} & $1.00$\numpmerr{+0.10}{-0.08}   & $ 1.00\,{\scriptstyle \pm 0.07}$        \\
$|\kappa_\mu|$        & $1.00$\numpmerr{+0.28}{-0.38} & $1.00$\numpmerr{+0.28}{-0.38} & $1.00$\numpmerr{+0.30}{-0.38}   & $1.00$\numpmerr{+0.28}{-0.37} \\ \midrule
$|\kappa_g|$            &                    --- &                    --- & $1.00$\numpmerr{+0.11}{-0.08}   & $ 1.00\,{\scriptstyle \pm 0.07}$ \\
$|\kappa_\gamma|$       &                    --- &                    --- & $1.00$\numpmerr{+0.10}{-0.07}         & $ 1.00\,{\scriptstyle \pm 0.06}$ \\
$|\kappa_{Z\gamma}|$ &                    --- &                    --- & $1.00$\numpmerr{+0.40}{-0.65}   & $1.00$\numpmerr{+0.39}{-0.65} \\ \midrule
$|\kappa_c|$ (95\% CL)& $< 4.2$                 &                   ---& $<4.2$         &                 ---     \\
\bottomrule
\end{tabular}
\end{center}
\end{table}
\end{DIFnomarkup}

\begin{DIFnomarkup}
\begin{table}[tbp]
\caption{Decomposition of the expected uncertainties in the coupling modifiers into contributions from data statistics (stat.), experimental systematic uncertainties (exp.), signal theory uncertainties (sig.\ theo.) and background theory uncertainties (bkg.\ theo.), for the resolved parameterisation with $\kappa_c$ as a free parameter or $\kappa_c = \kappa_t$. The decomposition for $|\kappa_c|$ shows one-sided uncertainties at 95\% CL; for all other rows, the uncertainties are shown at 68\% CL. Central values are given in Table~\ref{tab:kappas_exp}.}
\label{tab:kappas_exp_breakdown}
\centering
\renewcommand{\arraystretch}{1.5}
\resizebox{\textwidth}{!}{%
\begin{tabular}{lcccc|cccc}
\toprule
& \multicolumn{4}{c}{Resolved $\kappa_g, \kappa_\gamma, \kappa_{Z\gamma}$ — Free $\kappa_c$} & \multicolumn{4}{c}{Resolved $\kappa_g, \kappa_\gamma, \kappa_{Z\gamma}$ — $\kappa_c = \kappa_t$} \\
\midrule
Parameter & Stat. & Exp. & Sig.\ theo. & Bkg.\ theo. & Stat. & Exp. & Sig.\ theo. & Bkg.\ theo. \\
\midrule
$\kappa_Z$      & \numpmerr{+0.066}{-0.051} & \numpmerr{+0.039}{-0.022} & \numpmerr{+0.026}{-0.023} & \numpmerr{+0.040}{-0.023} & \numpmerr{+0.037}{-0.038} & \numpmerr{+0.019}{-0.018} & \numpmerr{+0.023}{-0.020} & \numpmerr{+0.019}{-0.019}\\
$\kappa_W$      & \numpmerr{+0.063}{-0.042} & \numpmerr{+0.038}{-0.022} & ${\scriptstyle \pm 0.023}$ & \numpmerr{+0.039}{-0.021} & \numpmerr{+0.029}{-0.028} & \numpmerr{+0.018}{-0.017} & \numpmerr{+0.021}{-0.020} & \numpmerr{+0.018}{-0.017}\\
\midrule
$\kappa_t$      & \numpmerr{+0.065}{-0.044} & \numpmerr{+0.042}{-0.028} & \numpmerr{+0.039}{-0.036} & \numpmerr{+0.044}{-0.026} & ${\scriptstyle \pm 0.031}$ & \numpmerr{+0.022}{-0.024} & \numpmerr{+0.038}{-0.032} & \numpmerr{+0.022}{-0.022}\\
$|\kappa_b|$    & \numpmerr{+0.079}{-0.068} & \numpmerr{+0.054}{-0.043} & \numpmerr{+0.059}{-0.049} & \numpmerr{+0.060}{-0.047} & \numpmerr{+0.058}{-0.057} & ${\scriptstyle \pm 0.039}$ & \numpmerr{+0.056}{-0.046} & \numpmerr{+0.046}{-0.044}\\
$|\kappa_\tau|$ & \numpmerr{+0.070}{-0.055} & \numpmerr{+0.046}{-0.035} & \numpmerr{+0.036}{-0.033} & \numpmerr{+0.041}{-0.023} & \numpmerr{+0.043}{-0.042} & ${\scriptstyle \pm 0.031}$ & \numpmerr{+0.033}{-0.030} & \numpmerr{+0.021}{-0.019}\\
$|\kappa_\mu|$  & \numpmerr{+0.278}{-0.370} & \numpmerr{+0.077}{-0.043} & \numpmerr{+0.035}{-0.019} & \numpmerr{+0.050}{-0.037} & \numpmerr{+0.252}{-0.367} & \numpmerr{+0.113}{-0.084} & \numpmerr{+0.030}{-0.012} & \numpmerr{+0.023}{-0.009}\\
\midrule
$|\kappa_c|$ (95\% CL)   & $+2.3$             & $ +0.8$              & $+1.1$              & $ +1.6$ & \multicolumn{4}{c}{---} \\
\bottomrule
\end{tabular}}
\end{table}
\end{DIFnomarkup}

\begin{DIFnomarkup}
\begin{table}[tbp]
\caption{Decomposition of the expected uncertainties in the coupling modifiers for the effective parameterisation with $\kappa_c$ as a free parameter or $\kappa_c = \kappa_t$, using the same breakdown as Table~\ref{tab:kappas_exp_breakdown}. The decomposition for $|\kappa_c|$ shows one-sided uncertainties at 95\% CL; for all other rows, the uncertainties are shown at 68\% CL. Central values are given in Table~\ref{tab:kappas_exp}.}
\label{tab:kappas_exp_5B5F_breakdown}
\centering
\renewcommand{\arraystretch}{1.5}
\resizebox{\textwidth}{!}{%
\begin{tabular}{lcccc|cccc}
\toprule
& \multicolumn{4}{c}{Effective $\kappa_g, \kappa_\gamma, \kappa_{Z\gamma}$ — Free $\kappa_c$} & \multicolumn{4}{c}{Effective $\kappa_g, \kappa_\gamma, \kappa_{Z\gamma}$ — $\kappa_c = \kappa_t$} \\
\midrule
Parameter & Stat. & Exp. & Sig.\ theo. & Bkg.\ theo. & Stat. & Exp. & Sig.\ theo. & Bkg.\ theo. \\
\midrule
$\kappa_Z$             & \numpmerr{+0.065}{-0.048} & \numpmerr{+0.038}{-0.027} & \numpmerr{+0.026}{-0.023} & \numpmerr{+0.039}{-0.023} & \numpmerr{+0.037}{-0.038} & \numpmerr{+0.019}{-0.018} & \numpmerr{+0.023}{-0.020} & \numpmerr{+0.019}{-0.019}\\
$\kappa_W$             & \numpmerr{+0.064}{-0.042} & \numpmerr{+0.041}{-0.032} & \numpmerr{+0.026}{-0.024} & \numpmerr{+0.041}{-0.025} & ${\scriptstyle \pm 0.032}$ & \numpmerr{+0.021}{-0.020} & \numpmerr{+0.023}{-0.022} & \numpmerr{+0.021}{-0.021}\\
\midrule
$\kappa_t$             & \numpmerr{+0.071}{-0.047} & ${\scriptstyle \pm 0.042}$ & \numpmerr{+0.053}{-0.047} & \numpmerr{+0.046}{-0.033} & \numpmerr{+0.043}{-0.044} & \numpmerr{+0.024}{-0.027} & \numpmerr{+0.053}{-0.045} & \numpmerr{+0.030}{-0.030}\\
$|\kappa_b|$           & \numpmerr{+0.090}{-0.075} & \numpmerr{+0.056}{-0.051} & \numpmerr{+0.055}{-0.046} & \numpmerr{+0.064}{-0.049} & \numpmerr{+0.070}{-0.067} & \numpmerr{+0.043}{-0.040} & \numpmerr{+0.051}{-0.043} & \numpmerr{+0.048}{-0.045}\\
$|\kappa_\tau|$        & \numpmerr{+0.073}{-0.058} & \numpmerr{+0.048}{-0.036} & \numpmerr{+0.036}{-0.032} & \numpmerr{+0.043}{-0.024} & \numpmerr{+0.046}{-0.045} & \numpmerr{+0.033}{-0.032} & \numpmerr{+0.033}{-0.029} & \numpmerr{+0.022}{-0.020}\\
$|\kappa_\mu|$         & \numpmerr{+0.283}{-0.369} & \numpmerr{+0.063}{-0.057} & \numpmerr{+0.028}{-0.000} & \numpmerr{+0.050}{-0.037} & \numpmerr{+0.270}{-0.368} & \numpmerr{+0.054}{-0.069} & \numpmerr{+0.019}{-0.000} & \numpmerr{+0.024}{-0.010}\\
\midrule
$|\kappa_g|$             & \numpmerr{+0.071}{-0.048} & \numpmerr{+0.045}{-0.040} & \numpmerr{+0.044}{-0.041} & \numpmerr{+0.044}{-0.026} & \numpmerr{+0.042}{-0.040} & \numpmerr{+0.029}{-0.028} & \numpmerr{+0.042}{-0.037} & \numpmerr{+0.024}{-0.022}\\
$|\kappa_\gamma|$        & \numpmerr{+0.070}{-0.050} & \numpmerr{+0.046}{-0.039} & \numpmerr{+0.028}{-0.027} & \numpmerr{+0.041}{-0.022} & \numpmerr{+0.043}{-0.042} & \numpmerr{+0.029}{-0.026} & \numpmerr{+0.026}{-0.024} & \numpmerr{+0.019}{-0.018}\\
$|\kappa_{Z\gamma}|$  & \numpmerr{+0.383}{-0.624} & \numpmerr{+0.097}{-0.185} & \numpmerr{+0.056}{-0.000} & \numpmerr{+0.051}{-0.000} & \numpmerr{+0.364}{-0.590} & \numpmerr{+0.114}{-0.271} & \numpmerr{+0.050}{-0.000} & \numpmerr{+0.039}{-0.000}\\
\midrule
$|\kappa_c|$ (95\% CL)   & $+2.3$             & $ +0.9$              & $+1.1$              & $ +1.6$ & \multicolumn{4}{c}{---} \\
\bottomrule
\end{tabular}}
\end{table}
\end{DIFnomarkup}


Expected results of the scenarios in which the couplings to vector bosons are unified ($\kappa_V = \kappa_W = \kappa_Z$) are shown in Table~\ref{tab:kappas_bsm_exp}. In the two scenarios with BSM contributions to Higgs boson decays, the branching ratios to invisible ($B_{\text{inv}}$) and undetected ($B_{\text{u}}$) final states are left free.
\begin{DIFnomarkup}
\begin{table}[tbp]
\caption{Expected values of Higgs boson coupling modifiers in the effective parameterisation with unified vector boson couplings ($\kappa_V = \kappa_W = \kappa_Z$) and $\kappa_c$ left floating, for the three scenarios shown in Figure~\ref{fig:kappaV_effective}. \textit{Scenario~1}: no BSM contributions to Higgs boson decays, except for those encoded in the coupling modifiers. \textit{Scenario~2}: invisible ($B_{\text{inv}}$) and undetected ($B_{\text{u}}$) branching ratios free, $\kappa_V \leq 1$ assumed. \textit{Scenario~3}: $B_{\text{inv}}$ and $B_{\text{u}}$ branching ratios free, measurements of \Hinv\ and off-shell Higgs boson production included, no $\kappa_V \leq 1$ constraint. Uncertainties are reported at 68\% CL and upper limits at 95\% CL.}
\label{tab:kappas_bsm_exp}
\centering
\renewcommand{\arraystretch}{1.5}
\begin{tabular}{lccc}
\toprule
Parameter & Scenario 1 ($B_{\text{inv}}=B_{\text{u}}=0$) & Scenario 2 ($\kappa_V\leq 1$, Hinv) & Scenario 3 (Hinv+off-shell) \\
\midrule
$\kappa_V$        & $1.00$\numpmerr{+0.09}{-0.06} & $1.00$\numpmerr{+0.00}{-0.06} & $1.00$\numpmerr{+0.19}{-0.06} \\
$\kappa_t$        & $1.00$\numpmerr{+0.11}{-0.09} & $1.00$\numpmerr{+0.08}{-0.09} & $1.00$\numpmerr{+0.21}{-0.09} \\
$|\kappa_b|$      & $1.00$\numpmerr{+0.14}{-0.11} & $1.00$\numpmerr{+0.08}{-0.11} & $1.00$\numpmerr{+0.21}{-0.11} \\
$|\kappa_\tau|$   & $1.00$\numpmerr{+0.10}{-0.08} & $1.00$\numpmerr{+0.06}{-0.03} & $1.00$\numpmerr{+0.20}{-0.08} \\
$|\kappa_\mu|$    & $1.00$\numpmerr{+0.30}{-0.38} & $1.00$\numpmerr{+0.27}{-0.38} & $1.00$\numpmerr{+0.38}{-0.37} \\
$|\kappa_g|$        & $1.00$\numpmerr{+0.10}{-0.08} & $1.00\,{\scriptstyle \pm 0.07}$ & $1.00$\numpmerr{+0.20}{-0.08} \\
$|\kappa_\gamma|$   & $1.00$\numpmerr{+0.10}{-0.07} & $1.00$\numpmerr{+0.05}{-0.06} & $1.00$\numpmerr{+0.20}{-0.07} \\
$|\kappa_{Z\gamma}|$& $1.00$\numpmerr{+0.40}{-0.65} & $1.00$\numpmerr{+0.38}{-0.65} & $1.00$\numpmerr{+0.48}{-0.59} \\
$|\kappa_c|$      & $< 3.99$               & $< 2.59$                & $< 3.85$               \\
\midrule
$B_{\text{inv}}$  & $0$ (fixed)  & $< 8\%$   & $< 7\%$ \\
$B_{\text{u}}$    & $0$ (fixed)  & $< 20\%$   & $< 45\%$ \\
\bottomrule
\end{tabular}
\end{table}
\end{DIFnomarkup}

The total branching ratio $B_{\text{BSM}}$ to BSM final states can also be considered as a free parameter. In this scenario, the vector boson couplings are unified ($\kappa_V = \kappa_W = \kappa_Z$) and $\kappa_c$ is left free in the fit, and measurements of off-shell Higgs boson production are combined with on-shell measurements to provide a constraint on the total width of the Higgs boson, so that the $\kappa_V \le 1$ assumption is not required. The observed and expected results for this scenario are shown in Table~\ref{tab:kappas_bsm_BRBSM}.

\begin{table}[tbp]
\caption{Observed and expected values of Higgs boson coupling modifiers in the effective parameterisation with unified vector boson couplings ($\kappa_V = \kappa_W = \kappa_Z$), $\kappa_c$ and the total BSM branching ratio $B_{\text{BSM}}$ free to vary, measurements of \Hinv\ and off-shell Higgs boson production included, and no $\kappa_V \leq 1$ constraint. Uncertainties are reported at 68\% CL and upper limits at 95\% CL.}
\label{tab:kappas_bsm_BRBSM}
\centering
\renewcommand{\arraystretch}{1.5}
\resizebox{\textwidth}{!}{%
\begin{tabular}{lcccccc}
\toprule
& \multicolumn{3}{c}{Observed} & \multicolumn{3}{c}{Expected} \\
\cmidrule{2-4}\cmidrule{5-7}
Parameter & Total & Stat. & Syst. & Total & Stat. & Syst. \\
\midrule
$\kappa_V$           & $0.99$\numpmerr{+0.11}{-0.06} & \numpmerr{+0.10}{-0.04} & \numpmerr{+0.05}{-0.04} & $1.00$\numpmerr{+0.15}{-0.06} & \numpmerr{+0.13}{-0.04} & \numpmerr{+0.08}{-0.04} \\
$\kappa_t$           & $0.86$\numpmerr{+0.13}{-0.09} & \numpmerr{+0.10}{-0.06} & \numpmerr{+0.08}{-0.06} & $1.00$\numpmerr{+0.18}{-0.09} & \numpmerr{+0.14}{-0.06} & \numpmerr{+0.11}{-0.07} \\
$|\kappa_b|$         & $0.92$\numpmerr{+0.12}{-0.10} & \numpmerr{+0.08}{-0.07} & \numpmerr{+0.09}{-0.07} & $1.00$\numpmerr{+0.17}{-0.11} & \numpmerr{+0.14}{-0.08} & \numpmerr{+0.09}{-0.08} \\
$|\kappa_\tau|$      & $0.95$\numpmerr{+0.12}{-0.08} & \numpmerr{+0.10}{-0.06} & \numpmerr{+0.07}{-0.05} & $1.00$\numpmerr{+0.16}{-0.08} & \numpmerr{+0.14}{-0.06} & \numpmerr{+0.09}{-0.05} \\
$|\kappa_\mu|$       & $1.07$\numpmerr{+0.28}{-0.30} & \numpmerr{+0.27}{-0.29} & \numpmerr{+0.07}{-0.08} & $1.00$\numpmerr{+0.35}{-0.37} & \numpmerr{+0.33}{-0.36} & \numpmerr{+0.11}{-0.07} \\
$|\kappa_g|$         & $0.98$\numpmerr{+0.12}{-0.07} & \numpmerr{+0.10}{-0.05} & \numpmerr{+0.06}{-0.05} & $1.00$\numpmerr{+0.15}{-0.08} & \numpmerr{+0.13}{-0.05} & \numpmerr{+0.08}{-0.05} \\
$|\kappa_\gamma|$    & $0.99$\numpmerr{+0.12}{-0.07} & \numpmerr{+0.11}{-0.05} & \numpmerr{+0.06}{-0.05} & $1.00$\numpmerr{+0.16}{-0.07} & \numpmerr{+0.14}{-0.05} & \numpmerr{+0.09}{-0.04} \\
$|\kappa_{Z\gamma}|$ & $1.34$\numpmerr{+0.33}{-0.35} & \numpmerr{+0.30}{-0.33} & ${\scriptstyle \pm 0.12}$ & $1.00$\numpmerr{+0.44}{-0.59} & \numpmerr{+0.41}{-0.55} & \numpmerr{+0.15}{-0.21} \\
$|\kappa_c|$ (95\% CL) & $< 3.49$ & \multicolumn{2}{c}{---} & $< 3.75$ & \multicolumn{2}{c}{---} \\
\midrule
$B_{\text{BSM}}$ (95\% CL) & $< 35\%$ & \multicolumn{2}{c}{---} & $< 38\%$ & \multicolumn{2}{c}{---} \\
\bottomrule
\end{tabular}}
\end{table}

\FloatBarrier
\section{Measurement of inclusive production cross-sections and ratios of branching ratios}

\begin{figure}[tbp]
\begin{center}
\includegraphics[width=0.75\columnwidth]{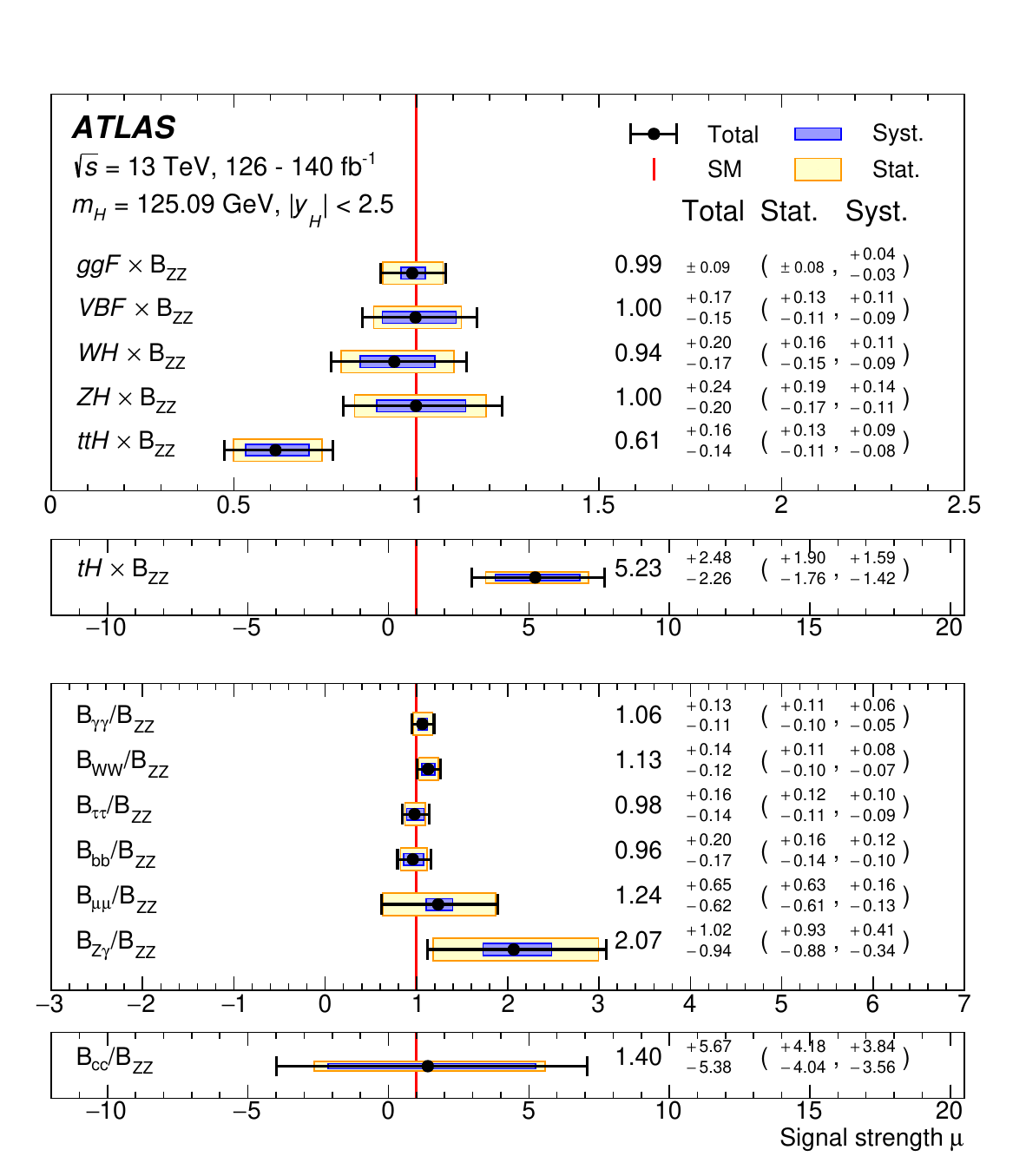}
\end{center}
\caption{Observed values and uncertainties in the cross-sections and ratios of branching ratios in the model described in the text, relative to their SM predictions. Total uncertainties, statistical uncertainties (Stat.) and systematic uncertainties (Syst.) are shown. The vertical line indicates the SM expectation.}
\label{fig:mb:Gamma:obs}
\end{figure}

\begin{figure}[tbp]
\begin{center}
\includegraphics[width=0.75\columnwidth]{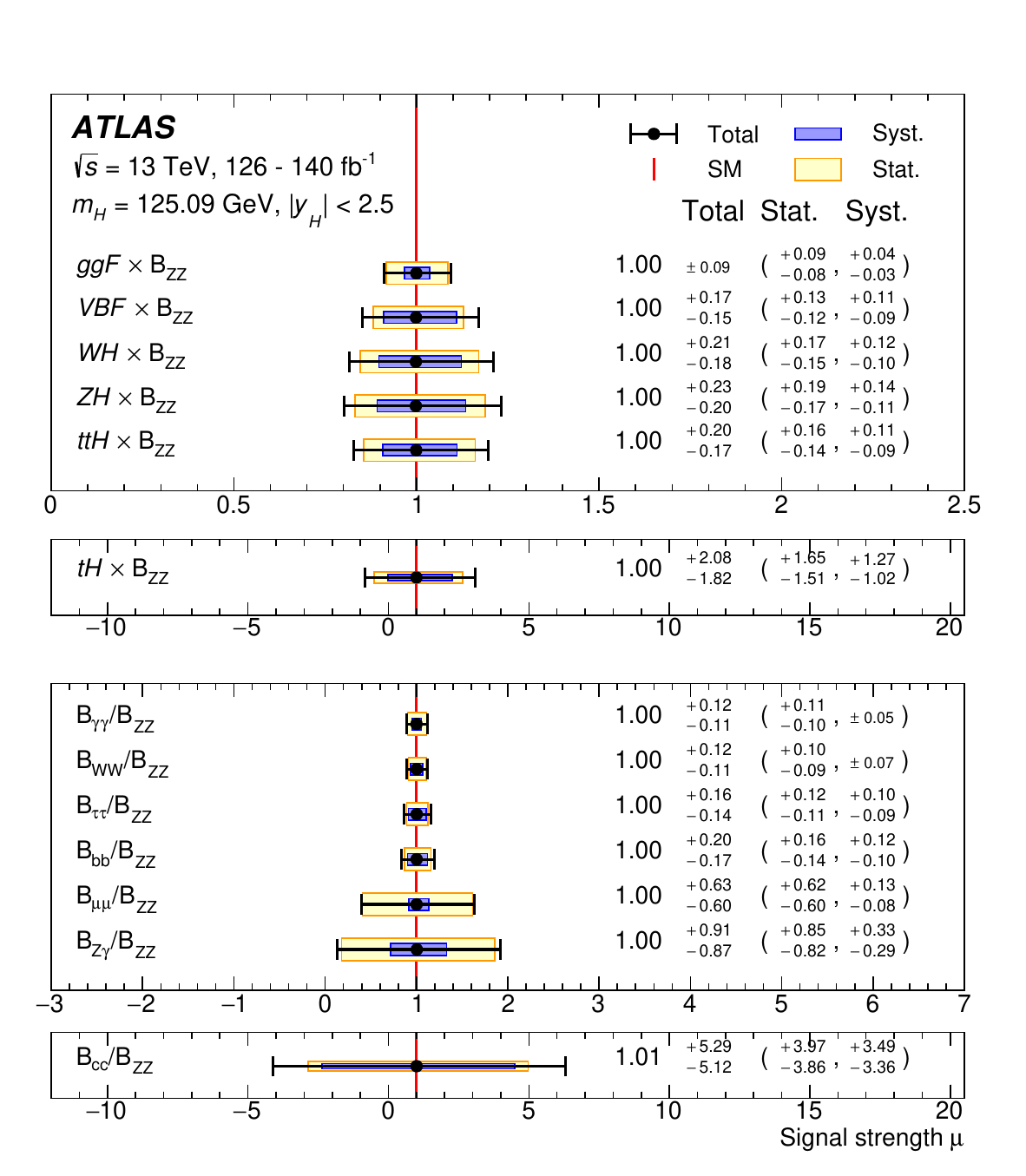}
\end{center}
\caption{Expected values and uncertainties in the cross-sections and ratios of branching ratios in the model described in the text, relative to their SM predictions. The results are obtained from an Asimov dataset generated under the SM hypothesis. Total uncertainties, statistical uncertainties (Stat.) and systematic uncertainties (Syst.) are shown. The vertical line indicates the SM expectation.}
\label{fig:mb:Gamma:exp}
\end{figure}

\begin{figure}[tbp]
\begin{center}
\includegraphics[width=0.75\columnwidth]{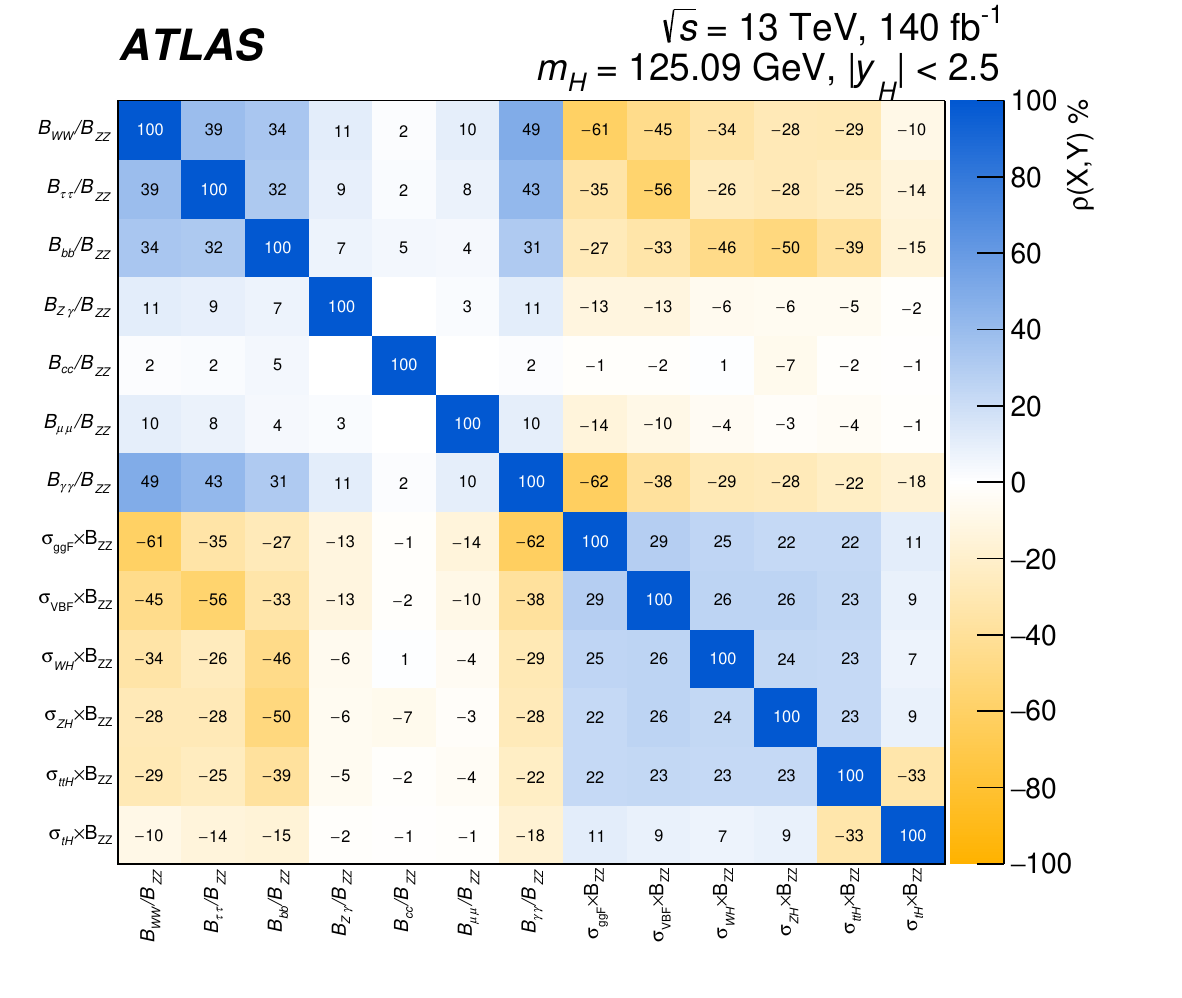}
\end{center}
\caption{Observed correlation coefficients for the measured cross-sections and ratios of branching ratios in the model described in the text.}
\label{fig:mb:Gamma:corr}
\end{figure}

\begin{figure}[tbp]
\begin{center}
\includegraphics[width=0.75\columnwidth]{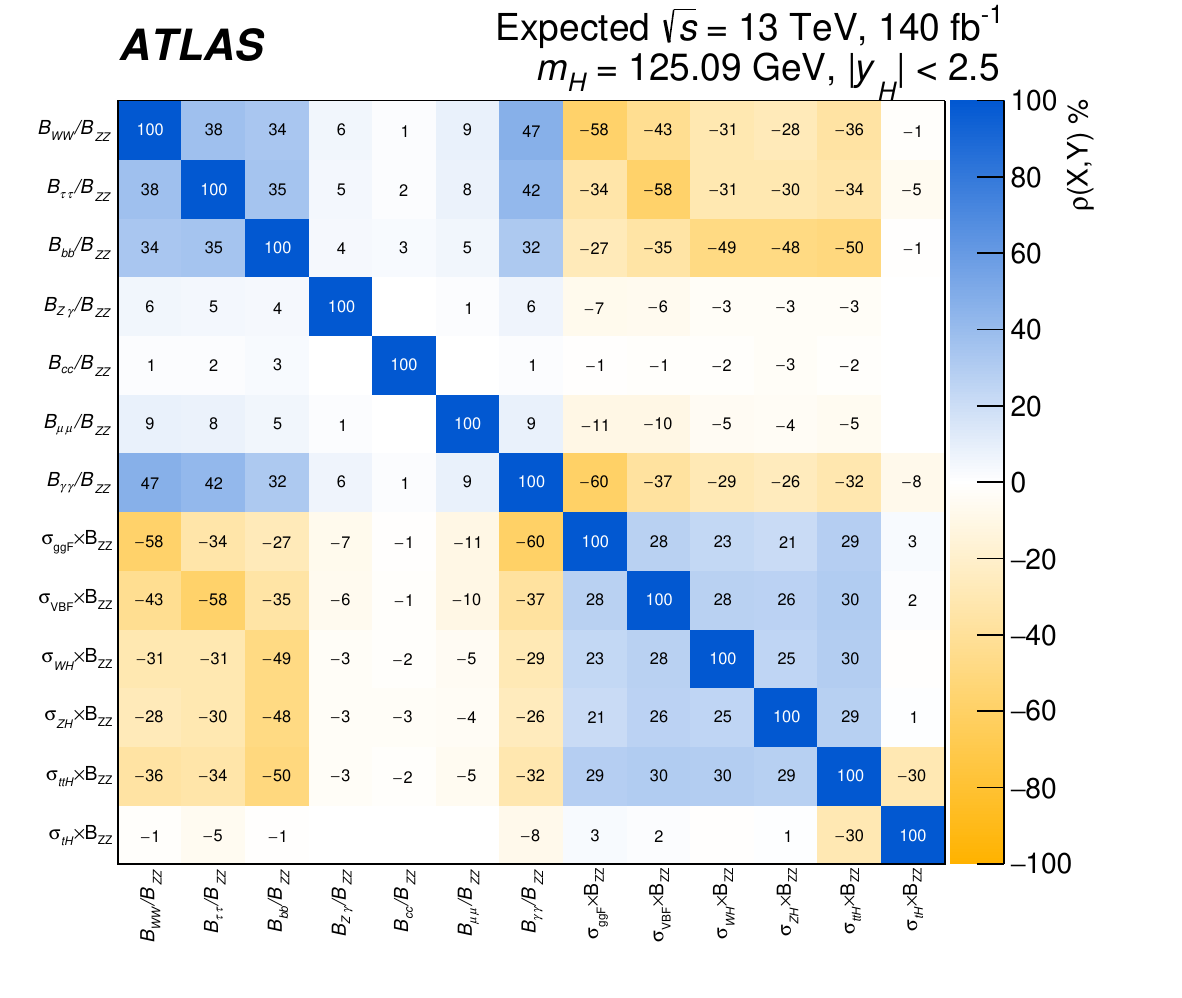}
\end{center}
\caption{Expected correlation coefficients for the measured cross-sections and ratios of branching ratios in the model described in the text. The correlations are obtained from an Asimov dataset generated under the SM hypothesis.}
\label{fig:mb:Gamma:corr:exp}
\end{figure}

This section describes a measurement of inclusive Higgs boson production and decay rates parameterised using the cross-sections $\sigma_i^{ZZ}$ in the \Hzz\ final state and ratios of branching ratios ${\Gamma_f}/{\Gamma_{\zz}}$. The model corresponds to the one described in Section~\ref{sec:mb}, except that ${\Gamma_{bb}}/{\Gamma_{\zz}}$ is considered as a free parameter and not expressed as a function of the $b$-quark mass. The observed values and uncertainties in the parameters are shown in Figure~\ref{fig:mb:Gamma:obs}, and the corresponding expected results in Figure~\ref{fig:mb:Gamma:exp}. The observed and expected correlations between the measured parameters are shown in Figures~\ref{fig:mb:Gamma:corr} and~\ref{fig:mb:Gamma:corr:exp}, respectively.

\FloatBarrier
\section{Polynomial coefficients for the $\Gamma_{bb}/\Gamma_{ZZ}$ parameterisation}
\label{app:mb}

The numerical values of the polynomial coefficients used in Eq.~\eqref{eq:GammabbGammaZZ} in Section~\ref{sec:mb}, parametrising the expression of $\Gamma_{bb}/\Gamma_{\zz}$ as a function of $m_b(m_H)$, are provided in Table~\ref{tab:mb:param}.
\begin{table}[tbp]
\caption{Polynomial coefficient values for the parameterisation of $\Gamma_{bb}/\Gamma_{\zz}$ as a function of $m_b(m_H)$.}
\label{tab:mb:param}
\centering
\begin{tabular}{cccccc}
\toprule
$\left(\frac{\Gamma_{bb}}{\Gamma_{\zz}}\right)_{\text{SM}}$ & $a_1$ & $a_2$ & $a_3$ & $a_4$ & $a_5$ \\
\midrule
22.00 & \num{-0.07086} & 1.144 & \num{-0.1361} & 0.06038 & \num{-0.01167} \\
\bottomrule
\end{tabular}
\end{table}

\FloatBarrier
\section{Systematic uncertainty decomposition in the measurement of the Higgs boson width}
\label{app:width}
The effect of the main sources of systematic uncertainties in the measurement of the Higgs boson width under the first scenario described in Section~\ref{sec:width} is found in Table~\ref{tab:unc_breakdown_hww_hzz}.

\begin{table}[tbp]
\caption{Absolute systematic uncertainties in the measurement of the Higgs boson width in the combination of analyses targeting off-shell and on-shell Higgs boson production in the \HsZZ\ and \HsWW\ final states. Uncertainties are estimated using the shifted auxiliary observable method, allowing variations to be summed in quadrature. Uncertainties are quoted at 68\% CL.}
\label{tab:unc_breakdown_hww_hzz}
\begin{center}
\renewcommand{\arraystretch}{1.4}
\begin{tabular}{lc}
\toprule
\textbf{Uncertainty source} & \textbf{Absolute impact on $\Gamma_H/\Gamma_H^{\text{SM}}$} \\
\midrule
Electron uncertainties & \numpmerr{+0.02}{-0.02} \\
Muon uncertainties & \numpmerr{+0.01}{-0.01} \\
Jet uncertainties & \numpmerr{+0.03}{-0.04} \\
Fakes uncertainties & \numpmerr{+0.01}{-0.01} \\
Luminosity & \numpmerr{+0.01}{-0.01} \\
\midrule
Total experimental & \numpmerr{+0.04}{-0.05} \\
\midrule
Signal modelling & \numpmerr{+0.10}{-0.12} \\
Background modelling & \numpmerr{+0.02}{-0.03} \\
\midrule
Total modelling & \numpmerr{+0.10}{-0.12} \\
\midrule
\textbf{Systematic uncertainty} & \numpmerr{+0.11}{-0.13} \\
\textbf{Statistical uncertainty} & \numpmerr{+0.41}{-0.40} \\
\midrule
\textbf{Total uncertainty} & \numpmerr{+0.50}{-0.42} \\
\bottomrule
\end{tabular}
\end{center}
\end{table}

\FloatBarrier
\section{Additional Higgs boson self-coupling results}
\label{app:kappa_lambda}

The expected 68\% CL and 95\% CL intervals on \kl\ obtained in the various configurations described in Section~\ref{sec:kappa:kl} are listed in Table~\ref{tab:kl_results:exp}.
\begin{table}[tbp]
\caption{Expected best-fit values of \kl and 68\% and 95\% CL intervals in various measurement configurations. In the first three rows (Single-Higgs, Di-Higgs, H+HH others fixed), the remaining coupling modifiers ($\kappa_V$, $\kappa_t$, $\kappa_b$, $\kappa_\tau$) are fixed to their SM values. In the fourth row (H+HH generic) these modifiers are free in the fit alongside \kl; the shift in the best-fit value between these two scenarios is due to the correlations between the \kl\ and the other coupling modifiers through their impact on single-$H$ event rates.}
\label{tab:kl_results:exp}
\centering
\renewcommand{\arraystretch}{1.2}
\begin{tabular}{lcc}
\toprule
Fit & 68\% CL interval & 95\% CL interval \\
\midrule
Single-Higgs          & $[-2.17,\;6.41]$ & $[-4.35,\;10.56]$ \\
Di-Higgs              & $[-0.55,\;5.69]$  & $[-1.71,\;7.18]$ \\
H+HH (others fixed)  & $[-0.44,\;4.64]$ & $[-1.55,\;6.80]$ \\
H+HH (generic)       & $[-0.49,\;4.96]$ & $[-1.62,\;6.88]$ \\
\bottomrule
\end{tabular}
\end{table}

\FloatBarrier


\FloatBarrier

\clearpage

\section{Expected results and correlation matrices of STXS measurements}
\label{app:stxs}

\begin{figure}[tbp]
\begin{center}
\subfloat[]{\includegraphics[width=0.49\columnwidth]{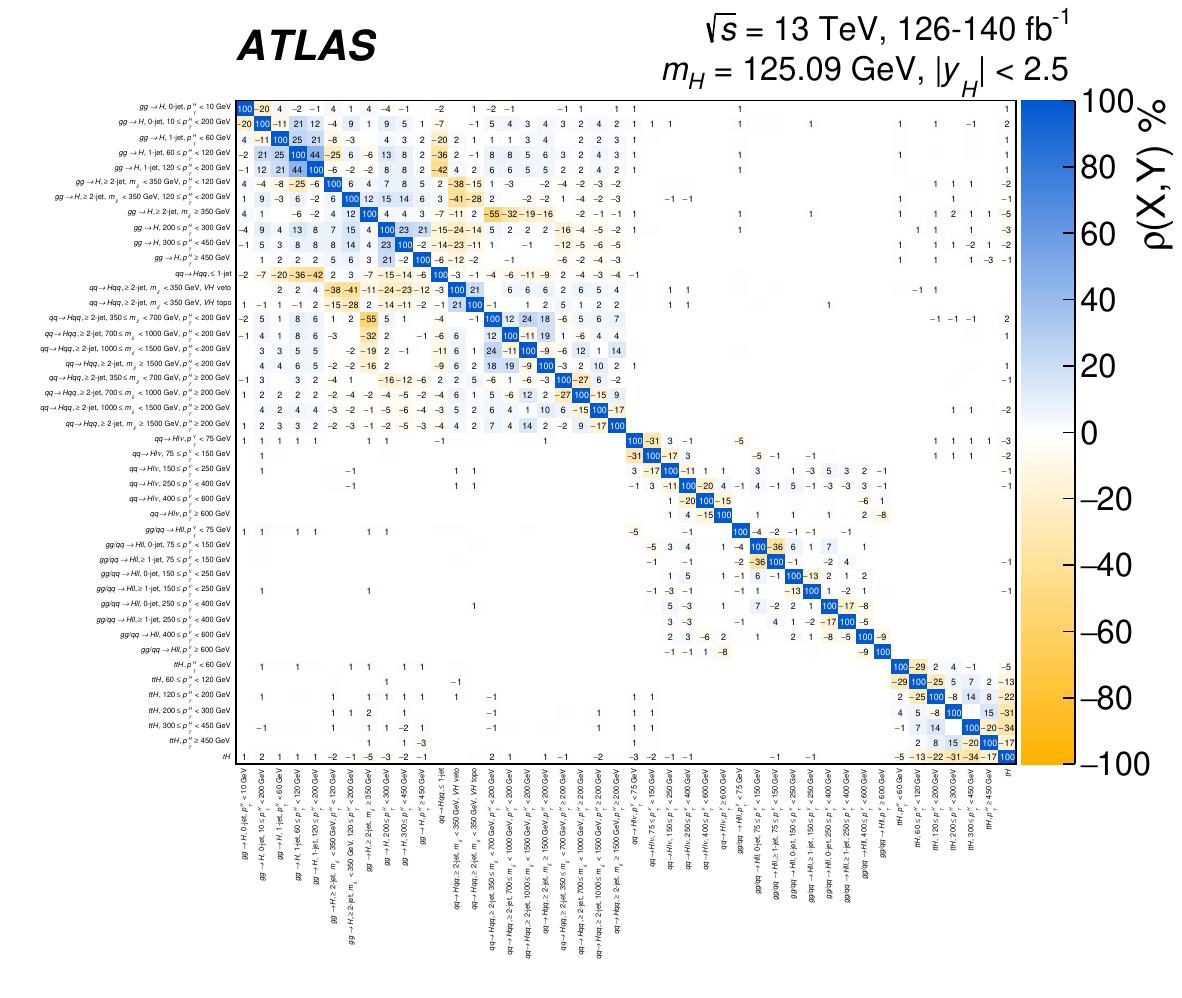}}
\subfloat[]{\includegraphics[width=0.49\columnwidth]{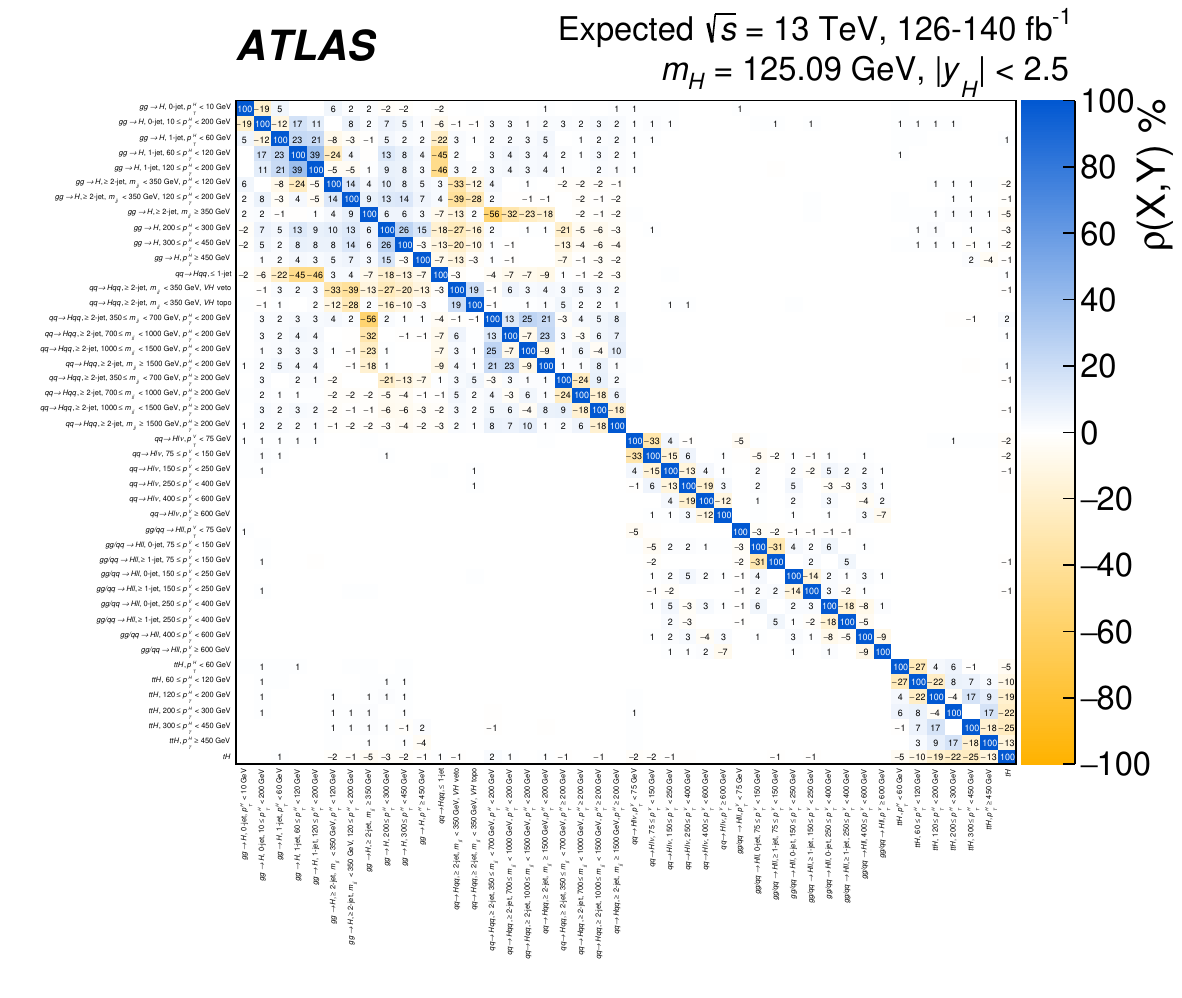}}
\end{center}
\caption{(a) Observed and (b) expected correlation matrices for the measurement of Higgs boson production cross-sections in the STXS framework.}
\label{fig:stxs:fixedBR:corr}
\end{figure}

\begin{figure}[tbp]
\begin{center}
\subfloat[]{\includegraphics[width=0.49\columnwidth]{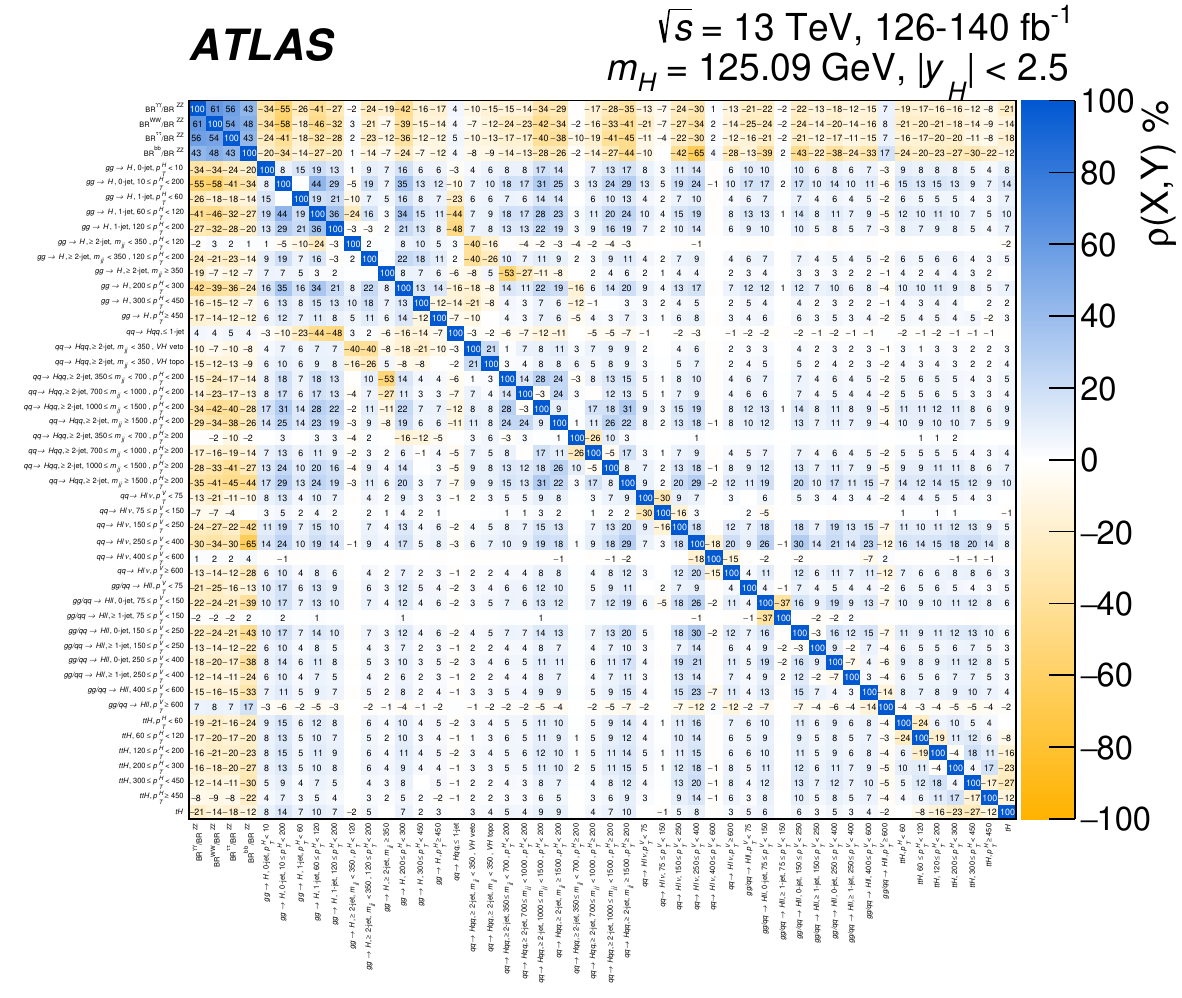}}
\subfloat[]{\includegraphics[width=0.49\columnwidth]{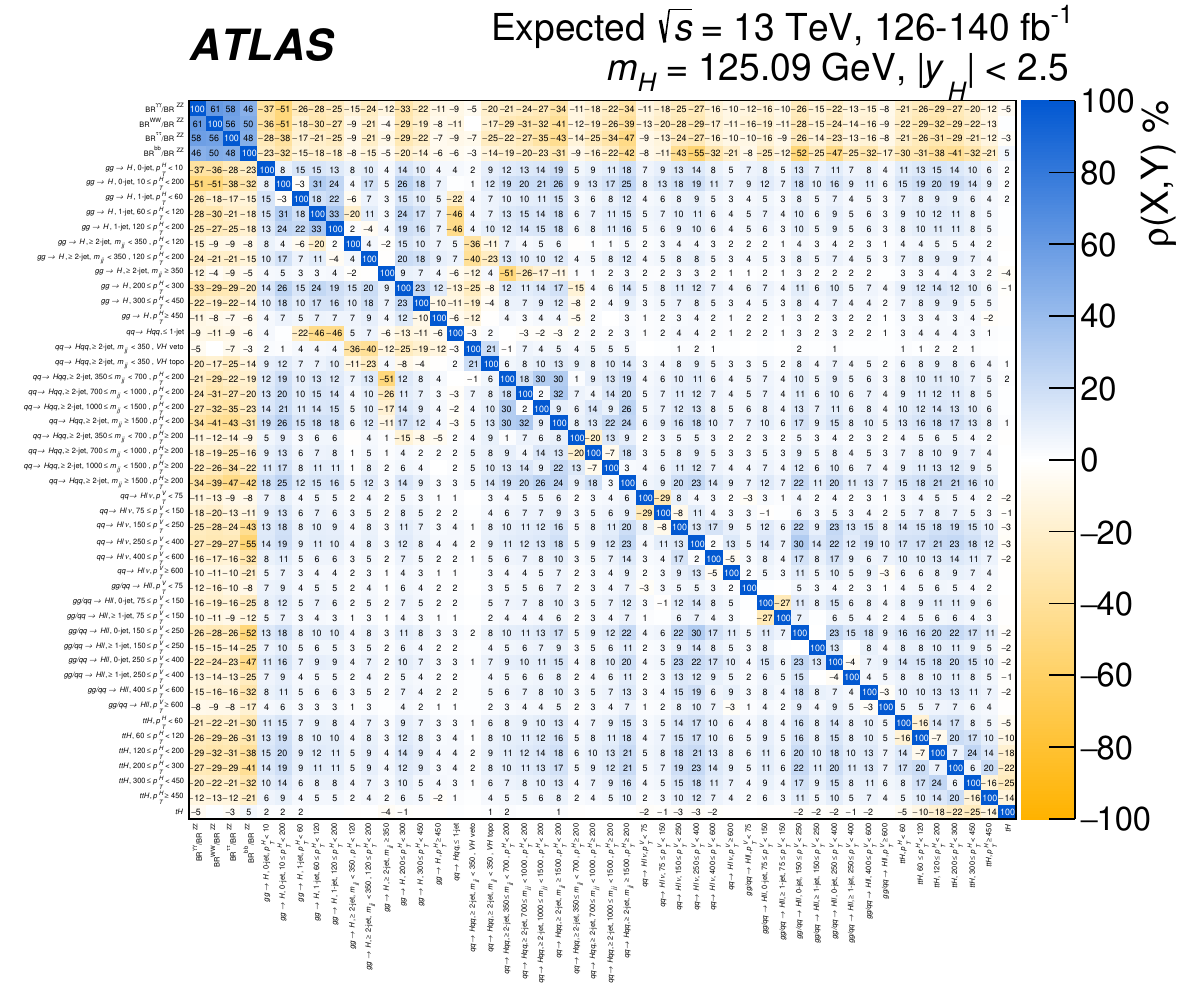}}
\end{center}
\caption{(a) Observed and (b) expected correlation matrices for the measurement of STXS production cross-sections in the \Hzz\ channel and ratios of branching ratios.}
\label{fig:stxs:ratioBR:corr}
\end{figure}

The expected results for the STXS measurements presented in Section~\ref{sec:stxs} are shown in Table~\ref{tab:stxs:fixedBR:exp} for the model with branching ratios fixed to their SM expectations are shown, and in Figure~\ref{tab:stxs:ratioBR:exp} for the model with production cross-sections in the \Hzz\ channel and ratios of branching ratios.
The corresponding observed and expected correlation matrices are shown respectively in Figure~\ref{fig:stxs:fixedBR:corr} and Figure~\ref{fig:stxs:ratioBR:corr}.

\begin{table}[ht]
\caption{Expected results for the production cross-section in each measured STXS region, assuming SM values for the Higgs boson branching ratios. The values for the \ggtoH\ process also include the contributions from \bbH\ production.
The total uncertainties are decomposed into components for data statistics (Stat.) and systematic uncertainties (Syst.).
SM predictions are also shown for each quantity with their total uncertainties.
}
\centering
\renewcommand{\arraystretch}{1.2}
\resizebox{0.8\textwidth}{!}{
\begin{tabular}{l S[table-format=-5.1,round-mode=none] lll S[table-format=5.2,round-mode=none] l}
\toprule
\multirow{2}{*}{STXS region} & \multicolumn{1}{c}{Value} & \multicolumn{3}{c}{Uncertainty [fb]} & \multicolumn{2}{c}{SM prediction} \\
&  \multicolumn{1}{c}{[fb]}                & Total   & Stat.                & Syst.   & \multicolumn{2}{c}{[fb]}   \\
\midrule
\ggHjPt{0}{}{10}{} &  6600 & ${\scriptstyle \pm \numRP[-2]{1200.0}}$ & ${\scriptstyle \pm \numRP[-2]{1100.0}}$ & ${\scriptstyle \pm \numRP[-2]{500.0}}$ &  6600 &  ${\scriptstyle \pm \numRP[-2]{900.0}}$ \\
\ggHjPt{0}{10}{}{} &  20600 & \errRP{-2}{2200.0}{-2100.0} & ${\scriptstyle \pm \numRP[-2]{1800.0}}$ & \errRP{-2}{1300.0}{-1200.0} &  20600 &  ${\scriptstyle \pm \numRP[-2]{1500.0}}$ \\
\ggHjPt{1}{}{60}{} &  6500 & ${\scriptstyle \pm \numRP[-2]{1600.0}}$ & ${\scriptstyle \pm \numRP[-2]{1300.0}}$ & ${\scriptstyle \pm \numRP[-2]{900.0}}$ &  6500 &  ${\scriptstyle \pm \numRP[-2]{900.0}}$ \\
\ggHjPt{1}{60}{120}{} &  4500 & \errRP{-2}{1100.0}{-1000.0} & ${\scriptstyle \pm \numRP[-2]{900.0}}$ & \errRP{-2}{600.0}{-500.0} &  4500 &  ${\scriptstyle \pm \numRP[-2]{600.0}}$ \\
\ggHjPt{1}{120}{200}{} &  700 & \errRP{-2}{300.0}{-200.0} & ${\scriptstyle \pm \numRP[-2]{200.0}}$ & ${\scriptstyle \pm \numRP[-2]{100.0}}$ &  750 &  ${\scriptstyle \pm \numRP[-1]{130.0}}$ \\
\ggHmPt{}{350}{}{120} &  3000 & ${\scriptstyle \pm \numRP[-2]{1300.0}}$ & ${\scriptstyle \pm \numRP[-2]{1100.0}}$ & ${\scriptstyle \pm \numRP[-2]{700.0}}$ &  3000 &  ${\scriptstyle \pm \numRP[-2]{600.0}}$ \\
\ggHmPt{}{350}{120}{200} &  900 & ${\scriptstyle \pm \numRP[-2]{400.0}}$ & ${\scriptstyle \pm \numRP[-2]{300.0}}$ & ${\scriptstyle \pm \numRP[-2]{200.0}}$ &  900 &  ${\scriptstyle \pm \numRP[-2]{200.0}}$ \\
\ggHmPt{350}{}{}{200} &  900 & \errRP{-2}{700.0}{-600.0} & ${\scriptstyle \pm \numRP[-2]{600.0}}$ & ${\scriptstyle \pm \numRP[-2]{300.0}}$ &  900 &  ${\scriptstyle \pm \numRP[-2]{200.0}}$ \\
\ggHPt{200}{300}{} &  460 & ${\scriptstyle \pm \numRP[-1]{130.0}}$ & ${\scriptstyle \pm \numRP[-1]{110.0}}$ & \errRP{-1}{70.0}{-60.0} &  500 &  ${\scriptstyle \pm \numRP[-2]{100.0}}$ \\
\ggHPt{300}{450}{} &  110 & \errRP{-1}{50.0}{-40.0} & ${\scriptstyle \pm \numRP[-1]{40.0}}$ & \errRP{-1}{20.0}{-10.0} &  110 &  ${\scriptstyle \pm \numRP[-1]{30.0}}$ \\
\ggHPt{450}{}{} &  20 & ${\scriptstyle \pm \numRP[-1]{20.0}}$ & ${\scriptstyle \pm \numRP[-1]{20.0}}$ & \errRP{-1}{10.0}{-0.0} &  18 &  ${\scriptstyle \pm \numRP[0]{5.0}}$ \\
\midrule
\ewqqH, $\le 1$-jet &  2200 & \errRP{-2}{1500.0}{-1400.0} & \errRP{-2}{1400.0}{-1300.0} & \errRP{-2}{600.0}{-500.0} &  2160 &  ${\scriptstyle \pm \numRP[-1]{60.0}}$ \\
\ewqqH, \VBF-enriched &  700 & ${\scriptstyle \pm \numRP[-2]{900.0}}$ & ${\scriptstyle \pm \numRP[-2]{800.0}}$ & \errRP{-2}{400.0}{-300.0} &  740 &  ${\scriptstyle \pm \numRP[-1]{20.0}}$ \\
\ewqqH, \VH-enriched &  500 & ${\scriptstyle \pm \numRP[-2]{200.0}}$ & ${\scriptstyle \pm \numRP[-2]{200.0}}$ & ${\scriptstyle \pm \numRP[-2]{100.0}}$ &  510 &  ${\scriptstyle \pm \numRP[-1]{20.0}}$ \\
\HqqmPt{350}{700}{}{200} &  500 & ${\scriptstyle \pm \numRP[-2]{200.0}}$ & ${\scriptstyle \pm \numRP[-2]{200.0}}$ & ${\scriptstyle \pm \numRP[-2]{100.0}}$ &  540 &  ${\scriptstyle \pm \numRP[-1]{10.0}}$ \\
\HqqmPt{700}{1000}{}{200} &  260 & ${\scriptstyle \pm \numRP[-1]{90.0}}$ & ${\scriptstyle \pm \numRP[-1]{80.0}}$ & ${\scriptstyle \pm \numRP[-1]{40.0}}$ &  260 &  ${\scriptstyle \pm \numRP[-1]{10.0}}$ \\
\HqqmPt{1000}{1500}{}{200} &  220 & ${\scriptstyle \pm \numRP[-1]{70.0}}$ & \errRP{-1}{70.0}{-60.0} & ${\scriptstyle \pm \numRP[-1]{30.0}}$ &  220 &  ${\scriptstyle \pm \numRP[-1]{10.0}}$ \\
\HqqmPt{1500}{}{}{200} &  220 & ${\scriptstyle \pm \numRP[-1]{50.0}}$ & ${\scriptstyle \pm \numRP[-1]{40.0}}$ & \errRP{-1}{30.0}{-20.0} &  220 &  ${\scriptstyle \pm \numRP[-1]{10.0}}$ \\
\HqqmPt{350}{700}{200}{} &  44 & \errRP{0}{38.0}{-34.0} & \errRP{0}{36.0}{-33.0} & \errRP{0}{11.0}{-10.0} &  44 &  ${\scriptstyle \pm \numRP[0]{1.0}}$ \\
\HqqmPt{700}{1000}{200}{} &  29 & \errRP{0}{16.0}{-14.0} & \errRP{0}{15.0}{-14.0} & \errRP{0}{5.0}{-4.0} &  29 &  ${\scriptstyle \pm \numRP[0]{1.0}}$ \\
\HqqmPt{1000}{1500}{200}{} &  33 & \errRP{0}{13.0}{-12.0} & \errRP{0}{12.0}{-11.0} & \errRP{0}{4.0}{-3.0} &  33 &  ${\scriptstyle \pm \numRP[0]{1.0}}$ \\
\HqqmPt{1500}{}{200}{}{} &  41 & \errRP{0}{10.0}{-9.0} & ${\scriptstyle \pm \numRP[0]{9.0}}$ & \errRP{0}{4.0}{-3.0} &  41 &  ${\scriptstyle \pm \numRP[0]{1.0}}$ \\
\midrule
\HlnPt{}{75}{} &  215 & \errRP{0}{153.0}{-142.0} & \errRP{0}{145.0}{-135.0} & \errRP{0}{51.0}{-46.0} &  215 &  ${\scriptstyle \pm \numRP[0]{8.0}}$ \\
\HlnPt{75}{150}{} &  134 & \errRP{0}{71.0}{-66.0} & \errRP{0}{52.0}{-50.0} & \errRP{0}{49.0}{-43.0} &  134 &  ${\scriptstyle \pm \numRP[0]{5.0}}$ \\
\HlnPt{150}{250}{} &  41 & \errRP{0}{16.0}{-15.0} & \errRP{0}{11.0}{-10.0} & ${\scriptstyle \pm \numRP[0]{11.0}}$ &  41 &  ${\scriptstyle \pm \numRP[0]{2.0}}$ \\
\HlnPt{250}{400}{} &  10 & ${\scriptstyle \pm \numRP[0]{3.0}}$ & ${\scriptstyle \pm \numRP[0]{3.0}}$ & \errRP{0}{2.0}{-1.0} &  10.0 &  ${\scriptstyle \pm \numRP[1]{0.4}}$ \\
\HlnPt{400}{600}{} &  1.8 & \errRP{1}{1.2}{-1.1} & \errRP{1}{1.1}{-1.0} & \errRP{1}{0.6}{-0.5} &  1.8 &  ${\scriptstyle \pm \numRP[1]{0.1}}$ \\
\HlnPt{600}{}{} &  0.3 & \errRP{1}{0.4}{-0.3} & \errRP{1}{0.4}{-0.3} & ${\scriptstyle \pm \numRP[1]{0.1}}$ &  0.34 &  ${\scriptstyle \pm \numRP[2]{0.02}}$ \\
\midrule
\HllnnPt{}{75}{} &  112 & \errRP{0}{86.0}{-67.0} & \errRP{0}{82.0}{-65.0} & \errRP{0}{26.0}{-16.0} &  112 &  ${\scriptstyle \pm \numRP[0]{8.0}}$ \\
\HllnnPtj{75}{150}{0} &  51 & \errRP{0}{30.0}{-29.0} & \errRP{0}{25.0}{-24.0} & \errRP{0}{18.0}{-16.0} &  51 &  ${\scriptstyle \pm \numRP[0]{4.0}}$ \\
\HllnnPtj{75}{150}{1+} &  36 & \errRP{0}{38.0}{-36.0} & ${\scriptstyle \pm \numRP[0]{33.0}}$ & \errRP{0}{18.0}{-16.0} &  36 &  ${\scriptstyle \pm \numRP[0]{5.0}}$ \\
\HllnnPtj{150}{250}{0} &  15 & \errRP{0}{6.0}{-5.0} & ${\scriptstyle \pm \numRP[0]{5.0}}$ & ${\scriptstyle \pm \numRP[0]{3.0}}$ &  15 &  ${\scriptstyle \pm \numRP[0]{2.0}}$ \\
\HllnnPtj{150}{250}{1+} &  17 & \errRP{0}{12.0}{-11.0} & ${\scriptstyle \pm \numRP[0]{10.0}}$ & \errRP{0}{6.0}{-5.0} &  17 &  ${\scriptstyle \pm \numRP[0]{3.0}}$ \\
\HllnnPtj{250}{400}{0} &  2.9 & \errRP{1}{1.3}{-1.2} & \errRP{1}{1.2}{-1.1} & \errRP{1}{0.5}{-0.4} &  2.9 &  ${\scriptstyle \pm \numRP[1]{0.4}}$ \\
\HllnnPtj{250}{400}{1+} &  4.2 & \errRP{1}{3.5}{-3.2} & \errRP{1}{3.2}{-3.0} & \errRP{1}{1.3}{-1.1} &  4.2 &  ${\scriptstyle \pm \numRP[1]{0.8}}$ \\
\HllnnPt{400}{600}{} &  1.1 & \errRP{1}{0.8}{-0.7} & \errRP{1}{0.7}{-0.6} & \errRP{1}{0.3}{-0.2} &  1.1 &  ${\scriptstyle \pm \numRP[1]{0.1}}$ \\
\HllnnPt{600}{}{} &  0.2 & \errRP{1}{0.3}{-0.2} & \errRP{1}{0.3}{-0.2} & ${\scriptstyle \pm \numRP[1]{0.1}}$ &  0.18 &  ${\scriptstyle \pm \numRP[2]{0.01}}$ \\
\midrule
\ttHPt{}{60}{} &  120 & ${\scriptstyle \pm \numRP[-1]{50.0}}$ & ${\scriptstyle \pm \numRP[-1]{50.0}}$ & \errRP{-1}{30.0}{-20.0} &  120 &  ${\scriptstyle \pm \numRP[-1]{20.0}}$ \\
\ttHPt{60}{120}{} &  180 & ${\scriptstyle \pm \numRP[-1]{60.0}}$ & ${\scriptstyle \pm \numRP[-1]{50.0}}$ & ${\scriptstyle \pm \numRP[-1]{30.0}}$ &  180 &  ${\scriptstyle \pm \numRP[-1]{20.0}}$ \\
\ttHPt{120}{200}{} &  130 & ${\scriptstyle \pm \numRP[-1]{40.0}}$ & ${\scriptstyle \pm \numRP[-1]{30.0}}$ & ${\scriptstyle \pm \numRP[-1]{20.0}}$ &  130 &  ${\scriptstyle \pm \numRP[-1]{20.0}}$ \\
\ttHPt{200}{300}{} &  53 & \errRP{0}{18.0}{-17.0} & \errRP{0}{16.0}{-15.0} & \errRP{0}{9.0}{-8.0} &  53 &  ${\scriptstyle \pm \numRP[0]{7.0}}$ \\
\ttHPt{300}{450}{} &  19 & ${\scriptstyle \pm \numRP[0]{9.0}}$ & ${\scriptstyle \pm \numRP[0]{8.0}}$ & ${\scriptstyle \pm \numRP[0]{5.0}}$ &  19 &  ${\scriptstyle \pm \numRP[0]{3.0}}$ \\
\ttHPt{450}{}{} &  5 & \errRP{0}{5.0}{-4.0} & ${\scriptstyle \pm \numRP[0]{4.0}}$ & ${\scriptstyle \pm \numRP[0]{2.0}}$ &  5 &  ${\scriptstyle \pm \numRP[0]{1.0}}$ \\
\tH &  80 & \errRP{-1}{180.0}{-160.0} & \errRP{-1}{140.0}{-130.0} & \errRP{-1}{110.0}{-90.0} &  80 &  ${\scriptstyle \pm \numRP[-1]{10.0}}$ \\
\bottomrule
\end{tabular}}
\label{tab:stxs:fixedBR:exp}
\end{table}


\begin{table}[ht]
\caption{Expected results for the production cross-section in each measured STXS region in the \Hzz\ decay process and ratios of branching ratios to $B_{\zz}$. The values for the \ggtoH\ process also include the contributions from \bbH\ production.
The total uncertainties are decomposed into components for data statistics (Stat.) and systematic uncertainties (Syst.).
SM predictions are also shown for each quantity with their total uncertainties. The relative uncertainty in $B_{\ww} / B_{\zz}$ is smaller than $10^{-3}$ and is not shown in the table.
}
\centering
\renewcommand{\arraystretch}{1.2}
\resizebox{0.7\textwidth}{!}{
\begin{tabular}{l S[table-format=-3.4,round-mode=none] lll S[table-format=3.5,round-mode=none] l}
\toprule
\multirow{2}{*}{Parameter} & \multicolumn{1}{c}{Value} & \multicolumn{3}{c}{Uncertainty} & \multicolumn{2}{c}{SM prediction} \\
&                 & Total   & Stat.  & Syst.   &  &  \\
\midrule
$B_{\yy} / B_{\zz}$ &  0.086 & \errRP{3}{0.011}{-0.009} & \errRP{3}{0.01}{-0.009} & \errRP{3}{0.005}{-0.004} &  0.086 &  ${\scriptstyle \pm \numRP[3]{0.001}}$ \\
$B_{\ww} / B_{\zz}$ &  8.1 & \errRP{1}{1.0}{-0.9} & \errRP{1}{0.9}{-0.8} & \errRP{1}{0.6}{-0.5} &  8.1 & - \\
$B_{\bb} / B_{\zz}$ &  22.0 & \errRP{1}{5.0}{-4.1} & \errRP{1}{3.9}{-3.3} & \errRP{1}{3.2}{-2.5} &  22.0 &  ${\scriptstyle \pm \numRP[1]{0.7}}$ \\
$B_{\tautau} / B_{\zz}$ &  2.37 & \errRP{2}{0.4}{-0.34} & \errRP{2}{0.32}{-0.27} & \errRP{2}{0.23}{-0.2} &  2.37 &  ${\scriptstyle \pm \numRP[2]{0.02}}$ \\
\midrule
\ggHjPt{0}{}{10}{} &  175 & \errRP{0}{37.0}{-34.0} & \errRP{0}{34.0}{-32.0} & \errRP{0}{13.0}{-12.0} &  175 &  ${\scriptstyle \pm \numRP[0]{23.0}}$ \\
\ggHjPt{0}{10}{}{} &  545 & \errRP{0}{71.0}{-65.0} & \errRP{0}{63.0}{-59.0} & \errRP{0}{32.0}{-28.0} &  545 &  ${\scriptstyle \pm \numRP[0]{41.0}}$ \\
\ggHjPt{1}{}{60}{} &  172 & \errRP{0}{44.0}{-42.0} & \errRP{0}{38.0}{-36.0} & \errRP{0}{23.0}{-22.0} &  172 &  \errRP{0}{23.0}{-24.0} \\
\ggHjPt{1}{60}{120}{} &  119 & \errRP{0}{30.0}{-28.0} & \errRP{0}{26.0}{-24.0} & \errRP{0}{15.0}{-13.0} &  119 &  ${\scriptstyle \pm \numRP[0]{16.0}}$ \\
\ggHjPt{1}{120}{200}{} &  19.7 & \errRP{1}{7.1}{-6.6} & \errRP{1}{6.2}{-5.8} & \errRP{1}{3.5}{-3.1} &  19.7 &  ${\scriptstyle \pm \numRP[1]{3.4}}$ \\
\ggHmPt{}{350}{}{120} &  78 & \errRP{0}{36.0}{-35.0} & \errRP{0}{31.0}{-30.0} & \errRP{0}{19.0}{-17.0} &  78 &  ${\scriptstyle \pm \numRP[0]{16.0}}$ \\
\ggHmPt{}{350}{120}{200} &  25 & \errRP{0}{11.0}{-10.0} & \errRP{0}{10.0}{-9.0} & \errRP{0}{5.0}{-4.0} &  24.9 &  ${\scriptstyle \pm \numRP[1]{5.7}}$ \\
\ggHmPt{350}{}{}{200} &  23 & \errRP{0}{18.0}{-17.0} & \errRP{0}{16.0}{-15.0} & \errRP{0}{9.0}{-7.0} &  23.2 &  ${\scriptstyle \pm \numRP[1]{5.5}}$ \\
\ggHPt{200}{300}{} &  12.1 & \errRP{1}{3.8}{-3.4} & \errRP{1}{3.3}{-3.0} & \errRP{1}{1.9}{-1.5} &  12.1 &  ${\scriptstyle \pm \numRP[1]{2.7}}$ \\
\ggHPt{300}{450}{} &  2.8 & \errRP{1}{1.3}{-1.2} & \errRP{1}{1.2}{-1.1} & \errRP{1}{0.5}{-0.4} &  2.81 &  ${\scriptstyle \pm \numRP[2]{0.71}}$ \\
\ggHPt{450}{}{} &  0.47 & \errRP{2}{0.57}{-0.45} & \errRP{2}{0.55}{-0.44} & \errRP{2}{0.15}{-0.08} &  0.47 &  ${\scriptstyle \pm \numRP[2]{0.14}}$ \\
\midrule
\ewqqH, $\le 1$-jet &  57 & \errRP{0}{40.0}{-37.0} & \errRP{0}{37.0}{-34.0} & \errRP{0}{16.0}{-15.0} &  57.1 &  ${\scriptstyle \pm \numRP[1]{1.7}}$ \\
\ewqqH, \VBF-enriched &  19 & \errRP{0}{25.0}{-23.0} & \errRP{0}{23.0}{-21.0} & \errRP{0}{10.0}{-9.0} &  19.4 &  ${\scriptstyle \pm \numRP[1]{0.6}}$ \\
\ewqqH, \VH-enriched &  13.5 & \errRP{1}{6.7}{-6.1} & \errRP{1}{5.9}{-5.5} & \errRP{1}{3.0}{-2.6} &  13.5 &  ${\scriptstyle \pm \numRP[1]{0.5}}$ \\
\HqqmPt{350}{700}{}{200} &  14.1 & \errRP{1}{6.1}{-5.6} & \errRP{1}{5.3}{-4.9} & \errRP{1}{3.0}{-2.8} &  14.1 &  ${\scriptstyle \pm \numRP[1]{0.4}}$ \\
\HqqmPt{700}{1000}{}{200} &  6.8 & \errRP{1}{2.7}{-2.4} & \errRP{1}{2.4}{-2.1} & \errRP{1}{1.2}{-1.0} &  6.77 &  ${\scriptstyle \pm \numRP[2]{0.2}}$ \\
\HqqmPt{1000}{1500}{}{200} &  5.9 & \errRP{1}{2.1}{-1.9} & \errRP{1}{1.9}{-1.7} & \errRP{1}{1.0}{-0.8} &  5.92 &  ${\scriptstyle \pm \numRP[2]{0.17}}$ \\
\HqqmPt{1500}{}{}{200} &  5.7 & \errRP{1}{1.6}{-1.4} & \errRP{1}{1.4}{-1.2} & \errRP{1}{0.8}{-0.6} &  5.70 &  ${\scriptstyle \pm \numRP[2]{0.18}}$ \\
\HqqmPt{350}{700}{200}{} &  1.17 & \errRP{2}{1.04}{-0.91} & \errRP{2}{0.99}{-0.87} & \errRP{2}{0.32}{-0.27} &  1.169 &  ${\scriptstyle \pm \numRP[3]{0.032}}$ \\
\HqqmPt{700}{1000}{200}{} &  0.78 & \errRP{2}{0.44}{-0.38} & \errRP{2}{0.42}{-0.37} & \errRP{2}{0.14}{-0.1} &  0.778 &  ${\scriptstyle \pm \numRP[3]{0.022}}$ \\
\HqqmPt{1000}{1500}{200}{} &  0.86 & \errRP{2}{0.37}{-0.32} & \errRP{2}{0.35}{-0.31} & \errRP{2}{0.13}{-0.1} &  0.860 &  ${\scriptstyle \pm \numRP[3]{0.025}}$ \\
\HqqmPt{1500}{}{200}{}{} &  1.1 & ${\scriptstyle \pm \numRP[1]{0.3}}$ & \errRP{1}{0.3}{-0.2} & ${\scriptstyle \pm \numRP[1]{0.1}}$ &  1.07 &  ${\scriptstyle \pm \numRP[2]{0.03}}$ \\
\midrule
\HlnPt{}{75}{} &  5.7 & \errRP{1}{4.2}{-3.8} & \errRP{1}{3.9}{-3.6} & \errRP{1}{1.4}{-1.2} &  5.68 &  \errRP{2}{0.21}{-0.22} \\
\HlnPt{75}{150}{} &  3.6 & \errRP{1}{2.0}{-1.8} & \errRP{1}{1.5}{-1.4} & \errRP{1}{1.3}{-1.1} &  3.55 &  ${\scriptstyle \pm \numRP[2]{0.15}}$ \\
\HlnPt{150}{250}{} &  1.1 & \errRP{1}{0.5}{-0.4} & ${\scriptstyle \pm \numRP[1]{0.3}}$ & ${\scriptstyle \pm \numRP[1]{0.3}}$ &  1.087 &  \errRP{3}{0.046}{-0.047} \\
\HlnPt{250}{400}{} &  0.27 & \errRP{2}{0.11}{-0.09} & \errRP{2}{0.09}{-0.08} & \errRP{2}{0.06}{-0.04} &  0.265 &  ${\scriptstyle \pm \numRP[3]{0.012}}$ \\
\HlnPt{400}{600}{} &  0.048 & \errRP{3}{0.035}{-0.03} & \errRP{3}{0.03}{-0.026} & \errRP{3}{0.018}{-0.014} &  0.047 &  ${\scriptstyle \pm \numRP[3]{0.002}}$ \\
\HlnPt{600}{}{} &  0.009 & \errRP{3}{0.011}{-0.009} & \errRP{3}{0.011}{-0.008} & \errRP{3}{0.004}{-0.002} &  0.0090 &  ${\scriptstyle \pm \numRP[4]{0.0006}}$ \\
\midrule
\HllnnPt{}{75}{} &  3.0 & \errRP{1}{2.3}{-1.8} & \errRP{1}{2.2}{-1.7} & \errRP{1}{0.7}{-0.4} &  2.95 &  ${\scriptstyle \pm \numRP[2]{0.21}}$ \\
\HllnnPtj{75}{150}{0} &  1.35 & \errRP{2}{0.86}{-0.77} & \errRP{2}{0.71}{-0.64} & \errRP{2}{0.49}{-0.43} &  1.35 &  ${\scriptstyle \pm \numRP[2]{0.12}}$ \\
\HllnnPtj{75}{150}{1+} &  0.9 & ${\scriptstyle \pm \numRP[1]{1.0}}$ & ${\scriptstyle \pm \numRP[1]{0.9}}$ & \errRP{1}{0.5}{-0.4} &  0.94 &  ${\scriptstyle \pm \numRP[2]{0.12}}$ \\
\HllnnPtj{150}{250}{0} &  0.4 & \errRP{1}{0.2}{-0.1} & \errRP{1}{0.2}{-0.1} & ${\scriptstyle \pm \numRP[1]{0.1}}$ &  0.408 &  ${\scriptstyle \pm \numRP[3]{0.061}}$ \\
\HllnnPtj{150}{250}{1+} &  0.44 & \errRP{2}{0.33}{-0.3} & \errRP{2}{0.29}{-0.26} & \errRP{2}{0.17}{-0.14} &  0.443 &  ${\scriptstyle \pm \numRP[3]{0.084}}$ \\
\HllnnPtj{250}{400}{0} &  0.077 & \errRP{3}{0.04}{-0.032} & \errRP{3}{0.035}{-0.03} & \errRP{3}{0.018}{-0.012} &  0.077 &  ${\scriptstyle \pm \numRP[3]{0.01}}$ \\
\HllnnPtj{250}{400}{1+} &  0.11 & \errRP{2}{0.1}{-0.09} & \errRP{2}{0.09}{-0.08} & \errRP{2}{0.04}{-0.03} &  0.112 &  ${\scriptstyle \pm \numRP[3]{0.021}}$ \\
\HllnnPt{400}{600}{} &  0.028 & \errRP{3}{0.022}{-0.018} & \errRP{3}{0.02}{-0.017} & \errRP{3}{0.01}{-0.007} &  0.0284 &  ${\scriptstyle \pm \numRP[4]{0.0023}}$ \\
\HllnnPt{600}{}{} &  0.0048 & \errRP{4}{0.0075}{-0.0058} & \errRP{4}{0.0071}{-0.0056} & \errRP{4}{0.0024}{-0.0015} &  0.00481 &  ${\scriptstyle \pm \numRP[5]{0.00032}}$ \\
\midrule
\ttHPt{}{60}{} &  3.12 & \errRP{2}{1.55}{-1.38} & \errRP{2}{1.35}{-1.22} & \errRP{2}{0.76}{-0.64} &  3.12 &  ${\scriptstyle \pm \numRP[2]{0.42}}$ \\
\ttHPt{60}{120}{} &  4.71 & \errRP{2}{1.8}{-1.61} & \errRP{2}{1.59}{-1.44} & \errRP{2}{0.84}{-0.71} &  4.70 &  ${\scriptstyle \pm \numRP[2]{0.52}}$ \\
\ttHPt{120}{200}{} &  3.34 & \errRP{2}{1.18}{-1.04} & \errRP{2}{1.03}{-0.92} & \errRP{2}{0.59}{-0.5} &  3.34 &  ${\scriptstyle \pm \numRP[2]{0.41}}$ \\
\ttHPt{200}{300}{} &  1.399 & \errRP{3}{0.55}{-0.479} & \errRP{3}{0.485}{-0.428} & \errRP{3}{0.26}{-0.215} &  1.39 &  ${\scriptstyle \pm \numRP[2]{0.2}}$ \\
\ttHPt{300}{450}{} &  0.503 & \errRP{3}{0.27}{-0.241} & \errRP{3}{0.234}{-0.21} & \errRP{3}{0.135}{-0.118} &  0.503 &  \errRP{3}{0.082}{-0.081} \\
\ttHPt{450}{}{} &  0.142 & \errRP{3}{0.133}{-0.118} & \errRP{3}{0.118}{-0.105} & \errRP{3}{0.062}{-0.053} &  0.142 &  ${\scriptstyle \pm \numRP[3]{0.026}}$ \\
\tH &  2.24 & \errRP{2}{4.82}{-4.2} & \errRP{2}{3.82}{-3.5} & \errRP{2}{2.93}{-2.31} &  2.24 &  \errRP{2}{0.15}{-0.29} \\
\bottomrule
\end{tabular}}
\label{tab:stxs:ratioBR:exp}
\end{table}


%

\FloatBarrier
\section{Measurements of STXS cross-sections per decay channel}
\label{app:stxs_br}

\begin{figure}[tbp]
\begin{center}
\subfloat[]{\includegraphics[width=0.49\columnwidth]{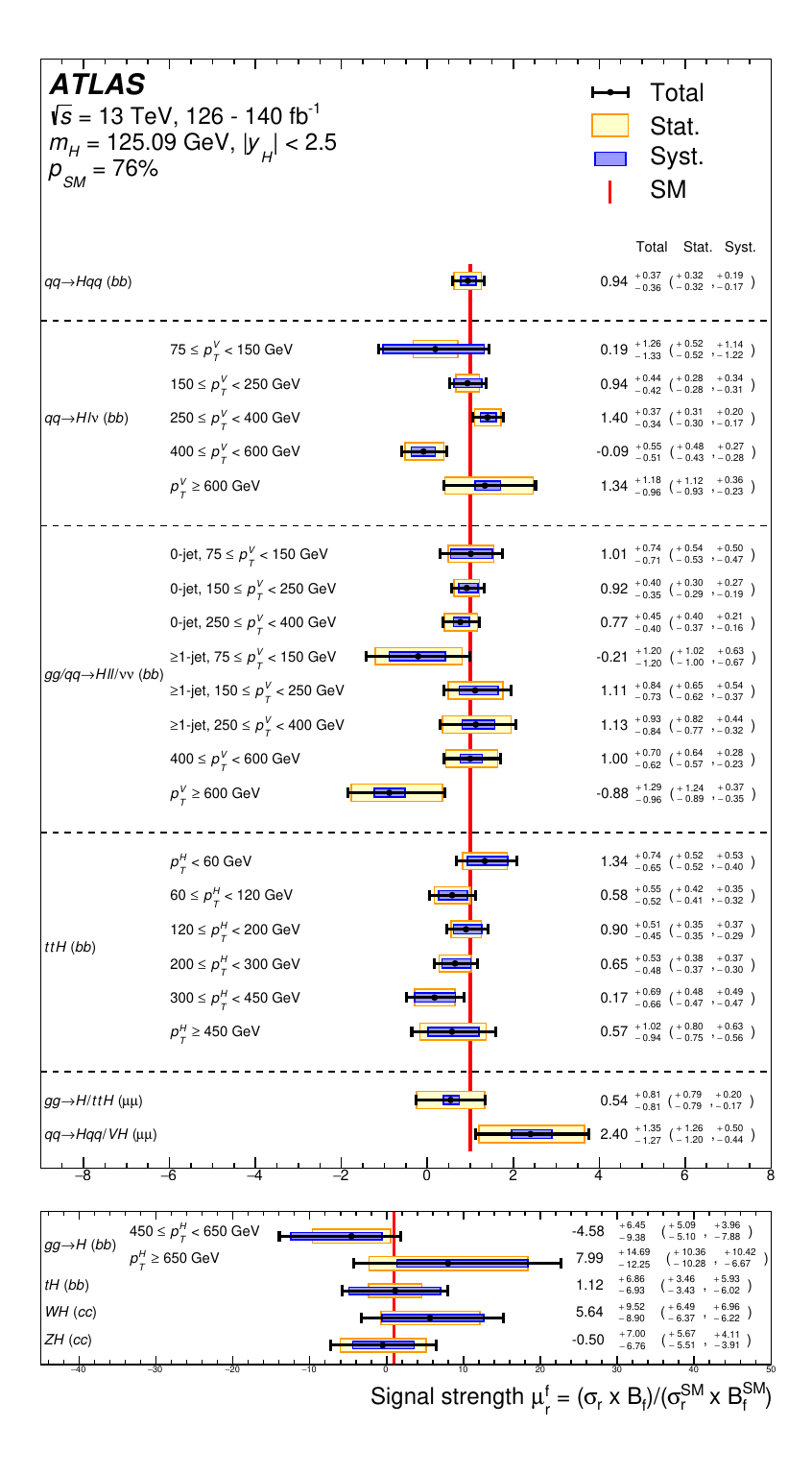}}
\subfloat[]{\includegraphics[width=0.49\columnwidth]{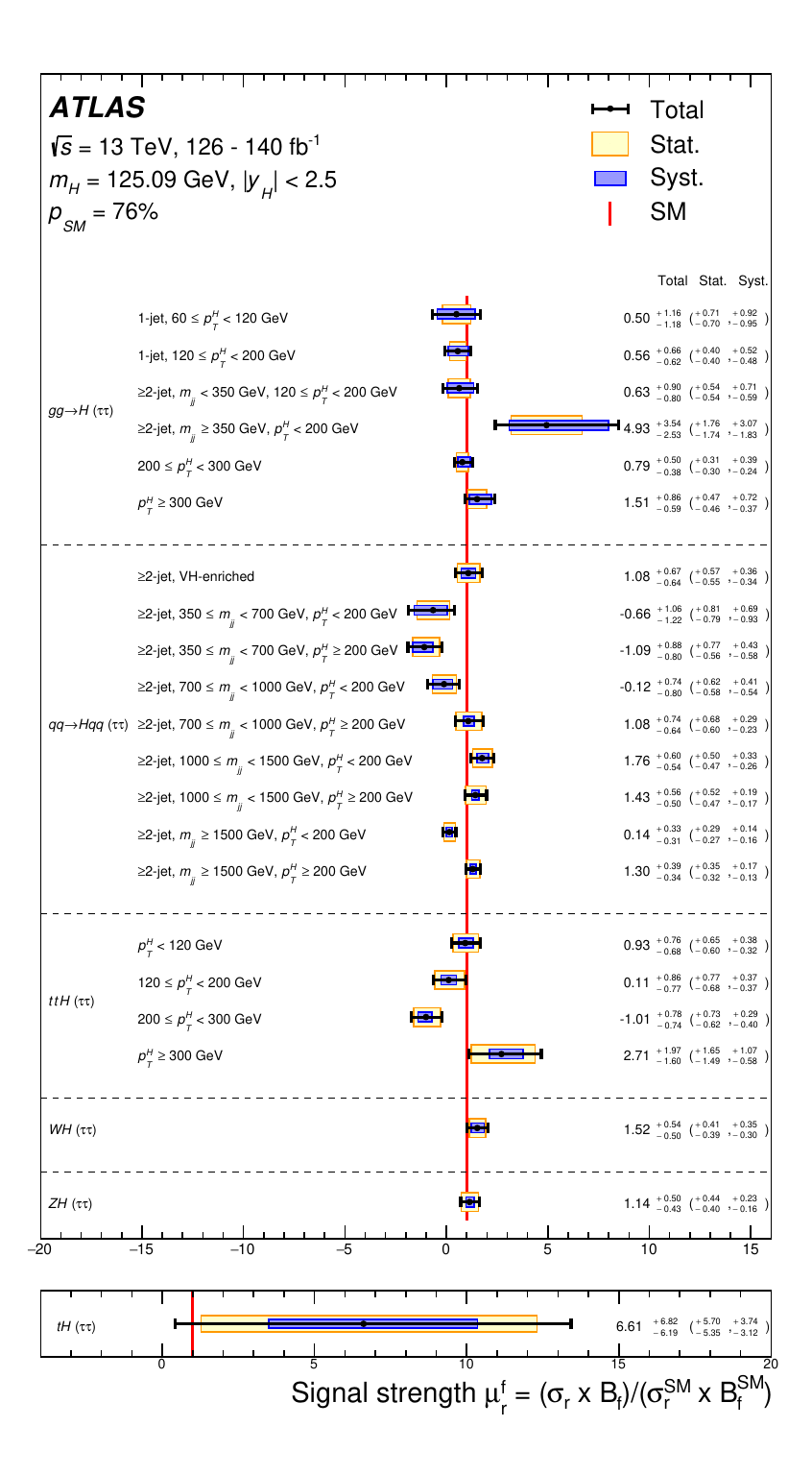}}
\end{center}
\caption{Observed best-fit values and uncertainties of the signal strengths for STXS production in the (a) \Hbb, \Hcc\ and \Hmm\ and (b) \Htt\ decay processes, relative to their SM predictions.}
\label{fig:stxs:br:1}
\end{figure}

\begin{figure}[tbp]
\begin{center}
\subfloat[]{\includegraphics[width=0.49\columnwidth]{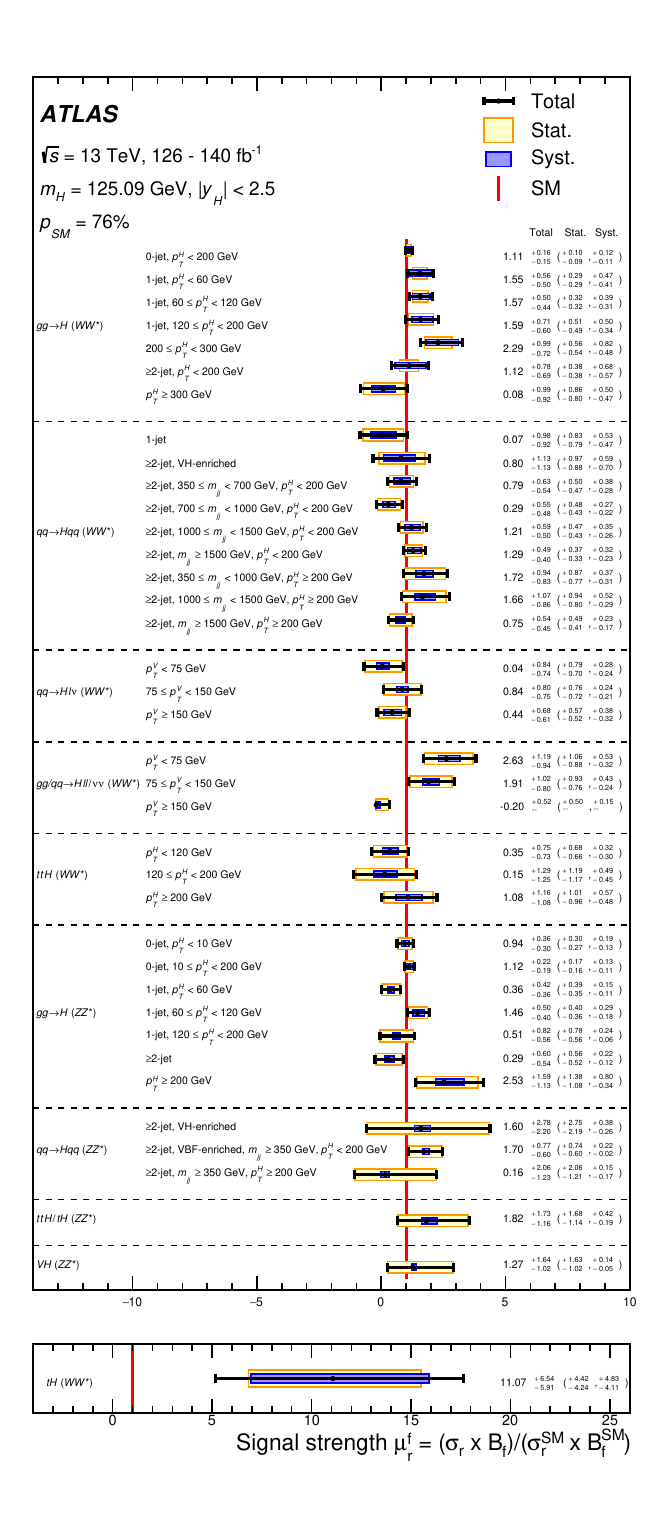}}
\subfloat[]{\includegraphics[width=0.49\columnwidth]{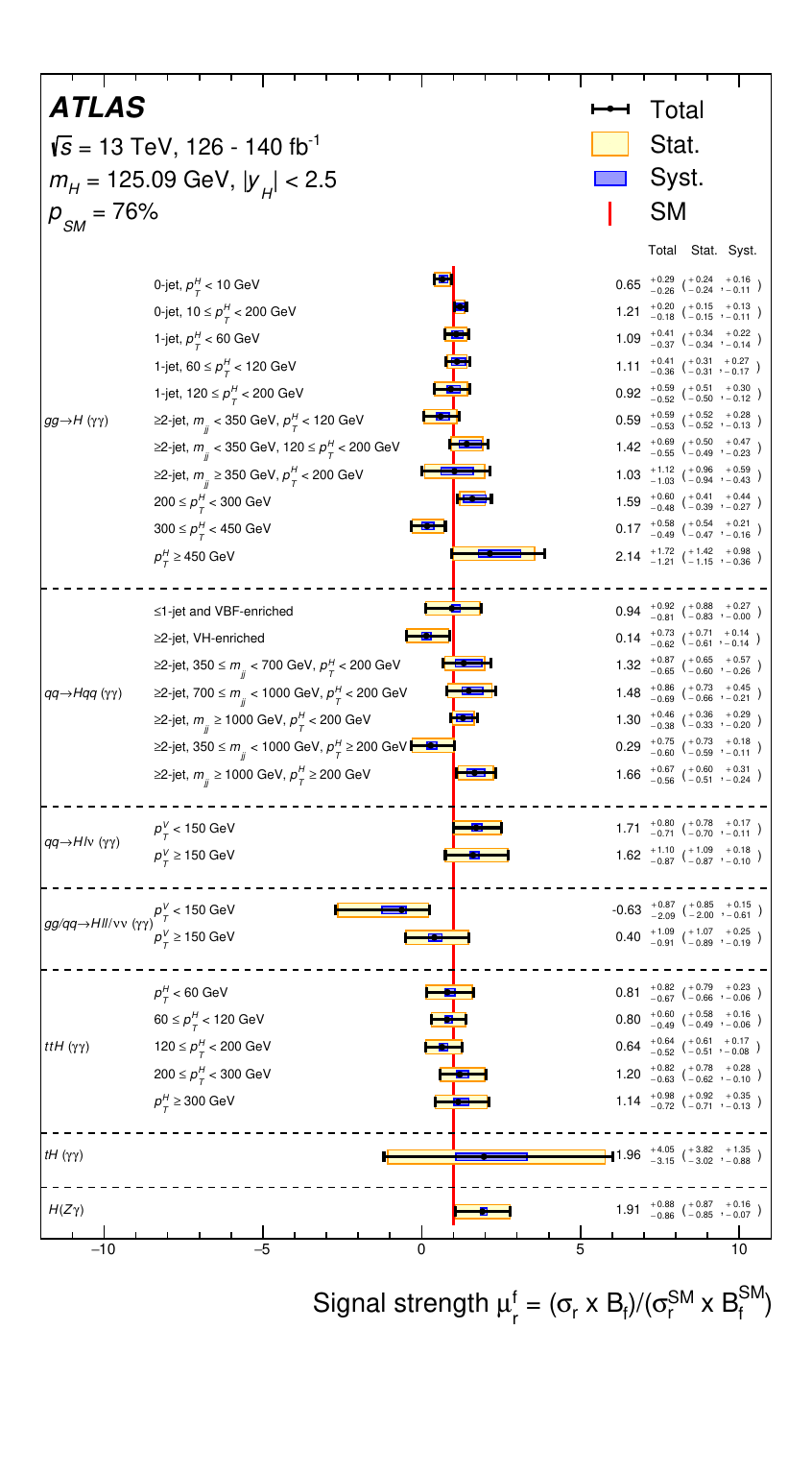}}
\end{center}
\caption{Observed best-fit values and uncertainties of the signal strengths for STXS production in the (a) \Hww\ and \Hzz\ and (b) \Hyy\ and \HZy\ decay processes, relative to their SM predictions.}
\label{fig:stxs:br:2}
\end{figure}

\begin{figure}[tbp]
\begin{center}
\includegraphics[width=0.9\columnwidth]{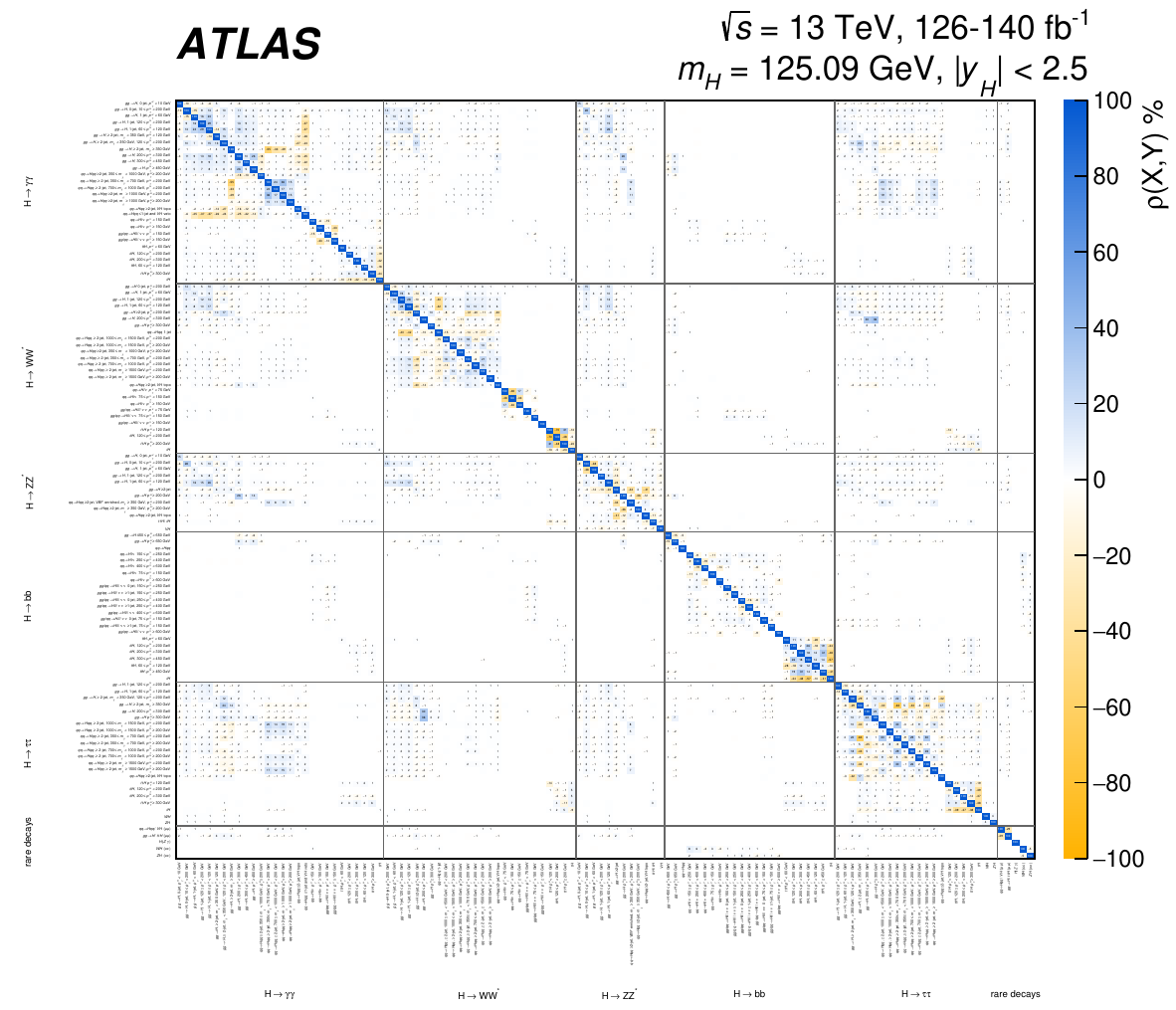}
\end{center}
\caption{Observed correlation matrix for the measurement of STXS production cross-sections in the \Hyy, \Hww, \Hzz, \Hbb, \Htt, \Hmm, \HZy\ and \Hcc\ decay modes.}
\label{fig:stxs:br:corr}
\end{figure}

This section presents a measurement of separate sets of STXS cross-sections in each of the \Hbb, \Hww, \Htt, \Hcc, \Hyy, \HZy\ and \Hmm\ decay modes. The granularity of the STXS binning used in each decay process is based on the one presented in Section~\ref{sec:stxs}, adapted to the sensitivity of the measurements in each decay process. In cases where the analysis of the decay process does not provide sensitivity to an STXS region, this region is either merged together with a neighboring region or its event rate is fixed to its SM prediction. Conversely, finer bins are introduced in some processes with particular sensitivity in some regions. The scheme used in each decay mode is presented in Table~\ref{tab:stxs:br_scheme}. In the \Hbb\ decay process, the $\ggtoH$, $\ptV \ge \qty{450}{\GeV}$ region is further split into separate regions for $\ptV < \qty{650}{\GeV}$ and $\ptV \ge \qty{650}{\GeV}$. In the \HZy\ process, all STXS bins are merged in a single region.
\begin{table}[tbp]
\caption{Definitions of the STXS regions for each decay process when measuring event rates in individual production and decay channels. When STXS regions are merged in the measurement, the entry indicates the name of the merged region. The cross-sections of regions marked \emph{Fixed to SM} are fixed to their SM predictions, within SM uncertainties. In the \Hbb\ decay process, the $\ggtoH$, $\ptV \ge \qty{450}{\GeV}$ region is further split into separate regions for $\ptV < \qty{650}{\GeV}$ and $\ptV \ge \qty{650}{\GeV}$. Unlabelled regions use the same binning as listed in the first column. In the \HZy\ process, all STXS bins are merged in a single region.}
\centering
\renewcommand{\arraystretch}{1.4}
\resizebox{0.9\textwidth}{!}{
\begin{tabular}{lccccccc}
\toprule
STXS region                & \Hbb & \Hww & \Htt & \Hcc & \Hzz & \Hyy & \Hmm \\
\midrule
\ggHjPt{0}{}{10}{}         & \multirow{10}{*}{Fixed to SM} & \multirow{2}{*}{\ggHj{0}} & \multirow{3}{*}{Fixed to SM} & \multirow{12}{*}{Fixed to SM} & & & \multirow{12}{*}{$\ggF + \ttH$} \\
\ggHjPt{0}{10}{}{}         & & & & & & & \\
\cmidrule{3-3}
\ggHjPt{1}{}{60}{}         & & & & & & & \\
\cmidrule{4-4}
\ggHjPt{1}{60}{120}{}      & & & & & & & \\
\ggHjPt{1}{120}{200}{}     & & & & & & & \\
\cmidrule{3-4}
\cmidrule{6-6}
\ggHmPt{}{350}{}{120}      & & \multirow{3}{*}{\ggHmPt{}{350}{}{200}} & Fixed to SM & & \multirow{3}{*}{\ggHj{2+}} & & \\
\cmidrule{4-4}
\ggHmPt{}{350}{120}{200}   & & & & & & & \\
\ggHmPt{350}{}{}{200}      & & & & & & & \\
\cmidrule{3-3}
\cmidrule{6-6}
\ggHPt{200}{300}{}         & & & & & \multirow{4}{*}{\ggHPt{200}{}{}} & & \\
\cmidrule{3-4}
\ggHPt{300}{450}{}         & & \multirow{3}{*}{\ggHPt{300}{}{}} & \multirow{3}{*}{\ggHPt{300}{}{}} & & & & \\
\cmidrule{2-2}
\multirow{2}{*}{\ggHPt{450}{}{}}  & \ggHPt{450}{650}{} & & & & & & \\
& \ggHPt{650}{}{}    & & & & & & \\
\midrule
\ewqqH, $\le 1$-jet        &  \multirow{11}{*}{\ewqqH} & & \multirow{2}{*}{Fixed to SM} & \multirow{11}{*}{Fixed to SM} & \multirow{2}{*}{Fixed to SM}  & Fixed to SM & \multirow{11}{*}{$\VBF+\VH$} \\
\cmidrule{3-3}
\cmidrule{7-7}
\ewqqH, \VBF-enriched      &  & Fixed to SM & & & & & \\
\cmidrule{3-4}
\cmidrule{6-6}
\ewqqH, \VH-enriched       &  & & & & & & \\
\cmidrule{6-7}
\HqqmPt{350}{700}{}{200}   &  & & & & \multirow{4}{*}{\HqqmPt{350}{}{}{200}} & & \\
\HqqmPt{700}{1000}{}{200}  &  & & & & & & \\
\cmidrule{7-7}
\HqqmPt{1000}{1500}{}{200} &  & & & & & \multirow{2}{*}{\HqqmPt{1000}{}{}{200}} & \\
\HqqmPt{1500}{}{}{200}     &  & & & & & & \\
\cmidrule{3-3}
\cmidrule{6-6}
\cmidrule{7-7}
\HqqmPt{350}{700}{200}{}   &  & \multirow{2}{*}{\HqqmPt{350}{1000}{200}{}} & & & \multirow{4}{*}{\HqqmPt{350}{}{200}{}} & \multirow{2}{*}{\HqqmPt{350}{1000}{200}{}} & \\
\HqqmPt{700}{1000}{200}{}  &  & & & & & & \\
\cmidrule{3-3}
\cmidrule{7-7}
\HqqmPt{1000}{1500}{200}{} &  & & & & & \multirow{2}{*}{\HqqmPt{1000}{}{200}{}} & \\
\HqqmPt{1500}{}{200}{}{}   &  & & & & & & \\
\midrule
\HlnPt{}{75}{}             & Fixed to SM & & \multirow{6}{*}{\qqtoHln} & \multirow{6}{*}{\qqtoHln} & \multirow{6}{*}{\VH} & \multirow{2}{*}{\HlnPt{}{150}{}} & \multirow{6}{*}{$\VBF+\VH$}\\
\cmidrule{2-2}
\HlnPt{75}{150}{}          &  & & & & & & \\
\cmidrule{3-3}
\cmidrule{7-7}
\HlnPt{150}{250}{}         &  & \multirow{4}{*}{\HlnPt{150}{}{}} & & & & \multirow{4}{*}{\HlnPt{150}{}{}} & \\
\HlnPt{250}{400}{}         &  & & & & & & \\
\HlnPt{400}{600}{}         &  & & & & & & \\
\HlnPt{600}{}{}            &  & & & & & & \\
\midrule
\HllnnPt{}{75}{}           & Fixed to SM & & \multirow{9}{*}{\pptoHllnn} & \multirow{9}{*}{\pptoHllnn} & \multirow{9}{*}{\VH} & \multirow{3}{*}{\HllnnPtj{}{150}{}} & \\
\cmidrule{2-3}
\HllnnPtj{75}{150}{0}      &  & \multirow{2}{*}{\HllnnPtj{75}{150}{}} & & & & & \\
\HllnnPtj{75}{150}{1+}     &  & & & & & & \\
\cmidrule{3-3}
\cmidrule{7-7}
\HllnnPtj{150}{250}{0}     &  & \multirow{6}{*}{\HllnnPtj{150}{}{}} & & & & \multirow{6}{*}{\HllnnPtj{150}{}{}} & \\
\HllnnPtj{150}{250}{1+}    &  & & & & & & \\
\HllnnPtj{250}{400}{0}     &  & & & & & & \\
\HllnnPtj{250}{400}{1+}    &  & & & & & & \\
\HllnnPt{400}{600}{}       &  & & & & & & \\
\HllnnPt{600}{}{}          &  & & & & & & \\
\midrule
\ttHPt{}{60}{}             &  & \multirow{2}{*}{\ttHPt{0}{120}{}} & \multirow{2}{*}{\ttHPt{0}{120}{}} & \multirow{6}{*}{Fixed to SM} & \multirow{7}{*}{$\ttH+\tH$} & & \multirow{7}{*}{$\ggF + \ttH$} \\
\ttHPt{60}{120}{}          &  & & & & & & \\
\cmidrule{3-4}
\ttHPt{120}{200}{}         &  & & & & & & \\
\cmidrule{3-3}
\ttHPt{200}{300}{}         &  & \multirow{3}{*}{\ttHPt{200}{}{}} & & & & & \\
\cmidrule{4-4}
\cmidrule{7-7}
\ttHPt{300}{450}{}         &  & & \multirow{2}{*}{\ttHPt{300}{}{}} & & & \multirow{2}{*}{\ttHPt{300}{}{}} & \\
\ttHPt{450}{}{}            &  & & & & & & \\
\cmidrule{3-4}
\cmidrule{7-7}
\tH                        &  & & & & & & \\
\bottomrule
\end{tabular}}
\label{tab:stxs:br_scheme}
\end{table}
The observed best-fit values and uncertainties for the signal strength relative to the SM prediction in each measurement region are presented in Figures~\ref{fig:stxs:br:1} and~\ref{fig:stxs:br:2}. The correlation matrix is shown in Figure~\ref{fig:stxs:br:corr}.

\FloatBarrier


\FloatBarrier

\clearpage

\section{Additional material on EFT measurements}
\label{app:eft}

\begin{figure}[tbp]
\centering
\includegraphics[width=.95\textwidth]{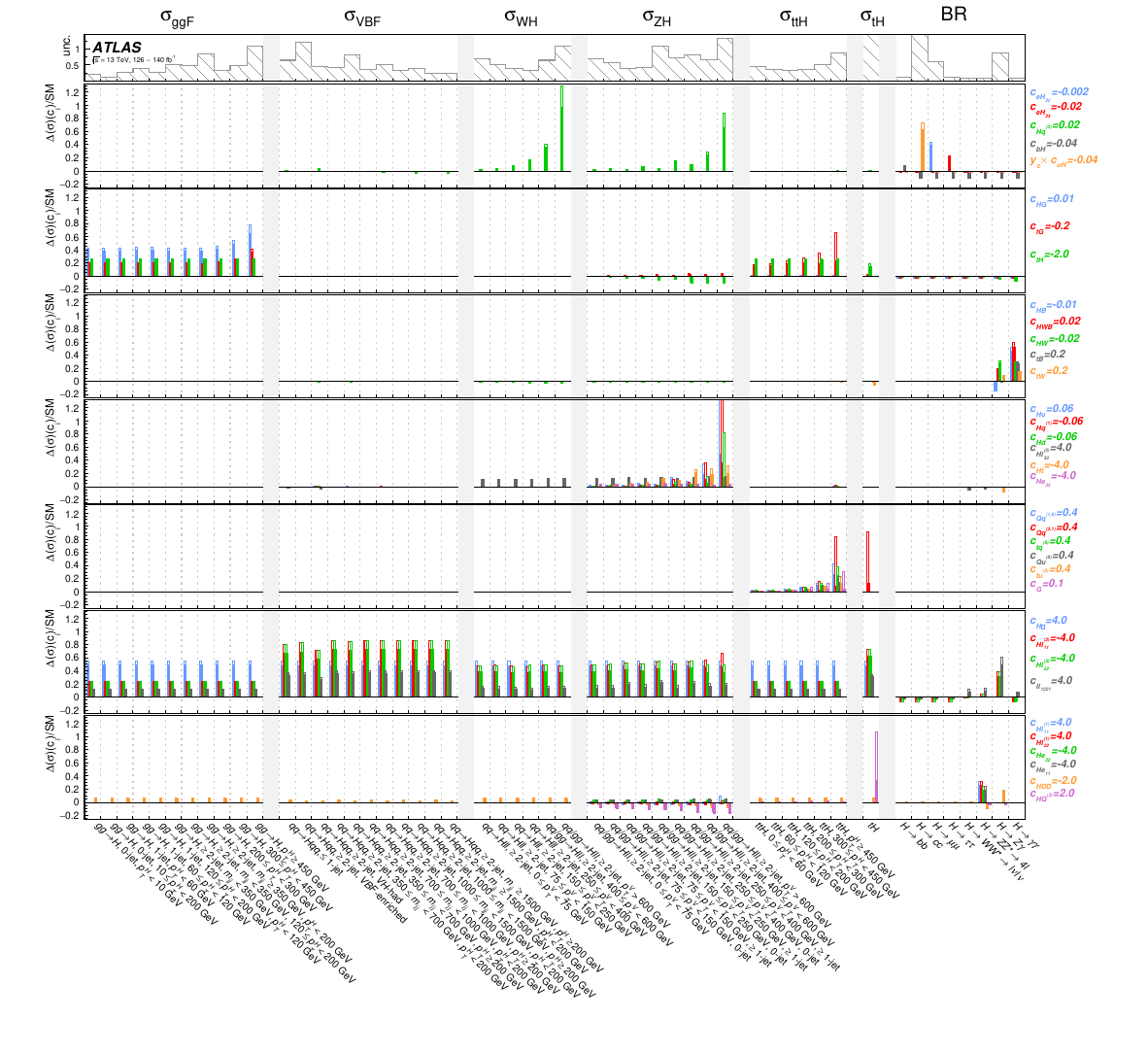}
\caption{Relative change in the rate of each production and decay process entering the combination compared to the SM reference, for representative values
of the EFT parameters in the Warsaw basis. Filled bars indicate the changes due to linear terms, while the empty bars also include the contributions of quadratic terms. The symmetrised statistical uncertainty for each observable are shown in the top panel.}
\label{fig:eft:impacts:warsaw}
\end{figure}

\begin{figure}[tbp]
\centering
\subfloat[]{\includegraphics[width=0.49\columnwidth]{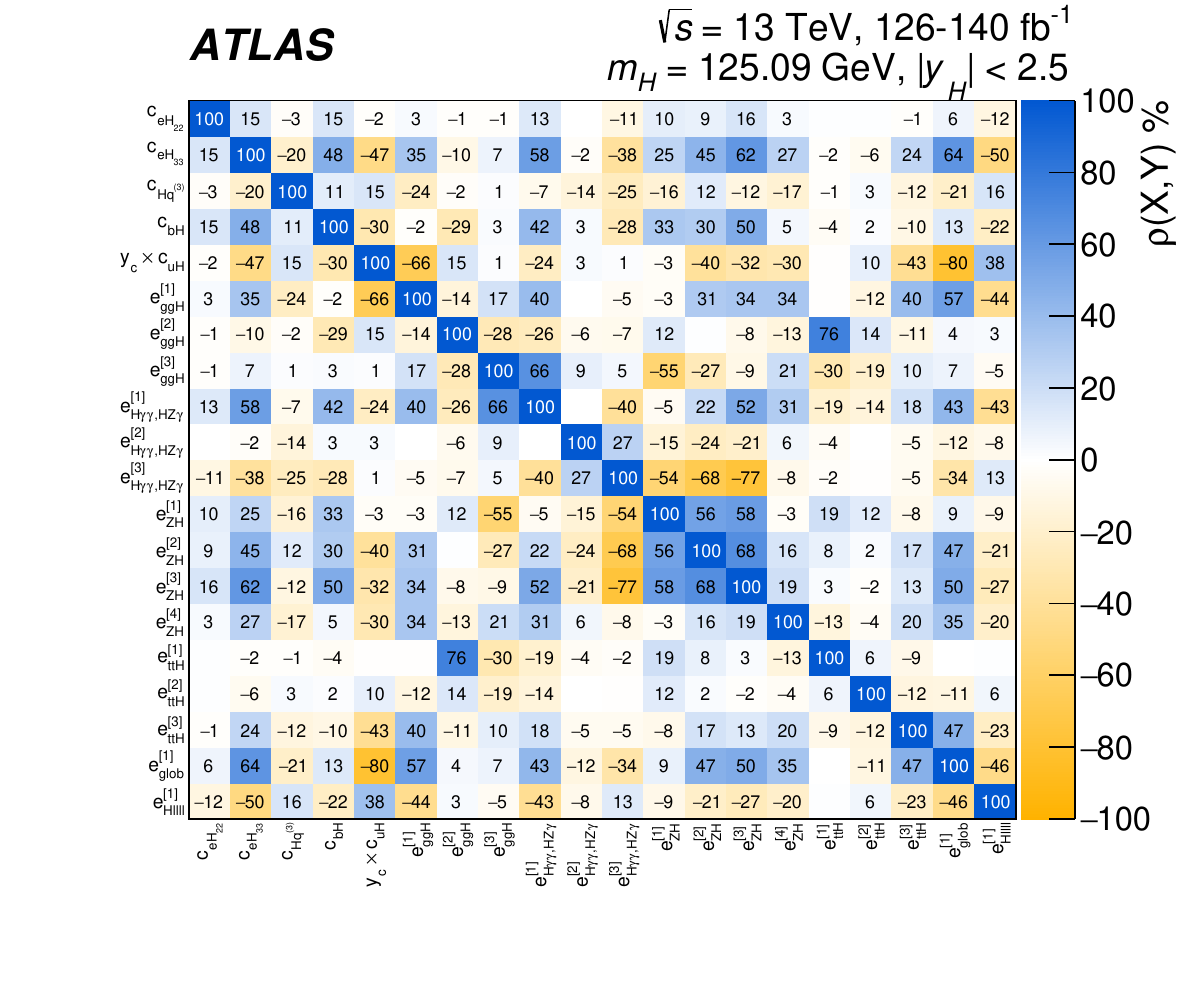}}
\subfloat[]{\includegraphics[width=0.49\columnwidth]{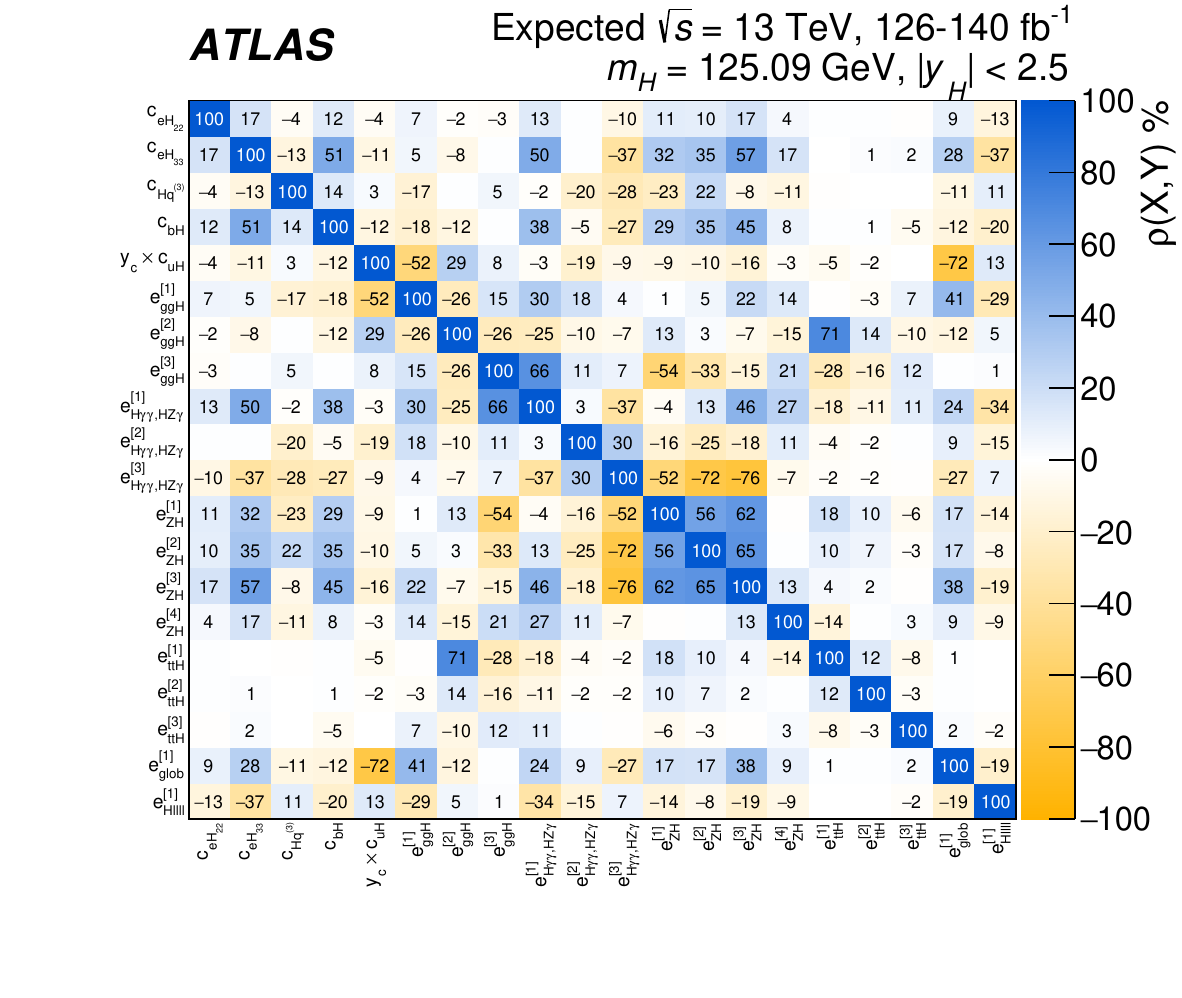}}
\caption{(a) Observed and (b) expected correlation matrices in the measurement of EFT parameters in the linear parameterisation.}
\label{fig:eft:corr}
\end{figure}

\begin{figure}[tbp]
\centering
\includegraphics[width=.95\textwidth]{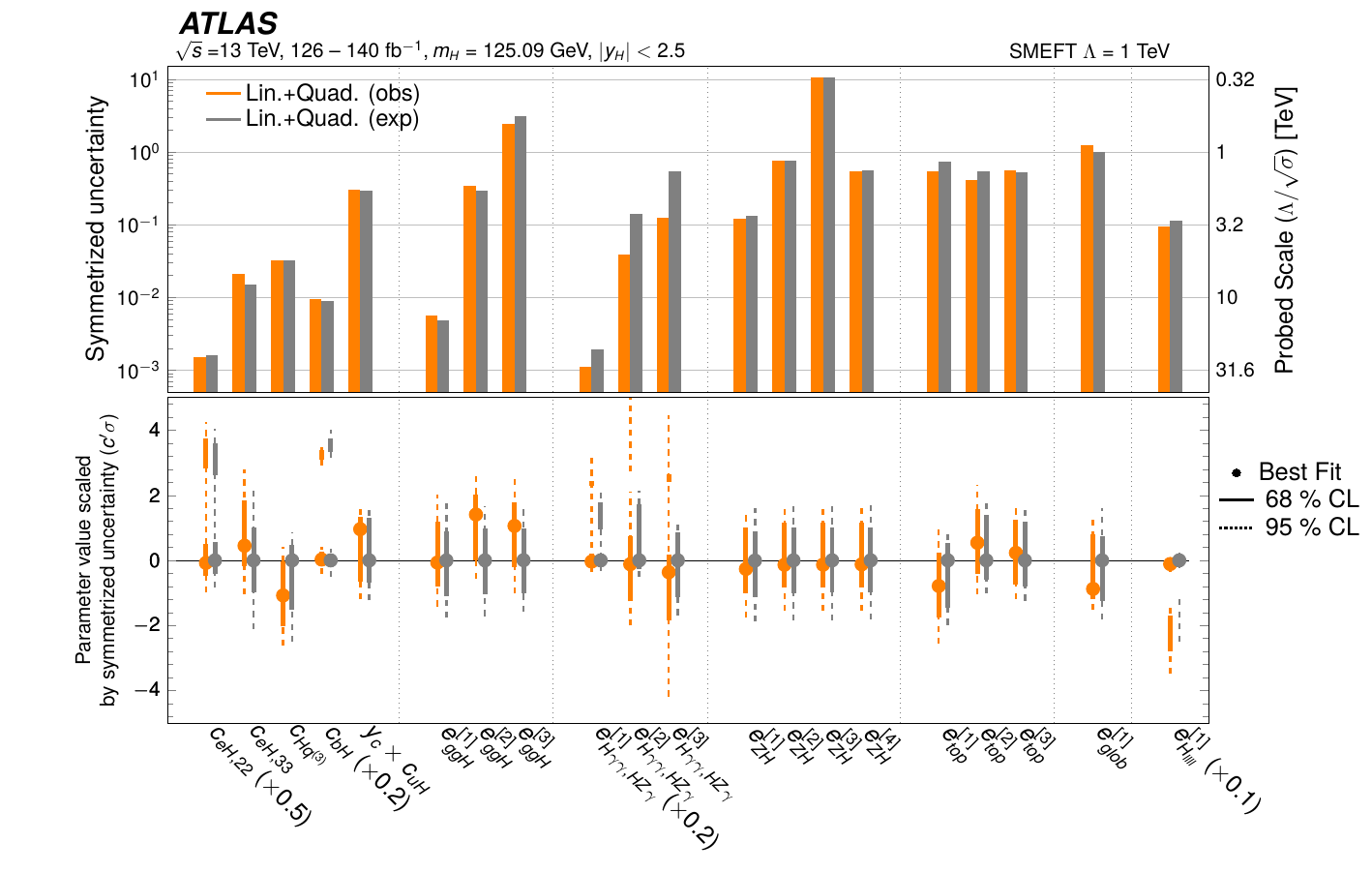}
\caption{Fitted values of the EFT measurement parameters for the linear+quadratic parameterisation. The top panel shows the symmetrised uncertainty in each parameter, and the bottom panel the best-fit value and uncertainty of each parameter divided by its symmetrised uncertainty. Observed and expected results are shown. Confidence intervals are shown for 68\% CL (solid lines) and 95\% CL (dashed lines). In the top panel, the left axis shows the uncertainty $\sigma$ on the parameter assuming $\Lambda = \qty{1}{\TeV}$ for the EFT scale, while the right axis shows the probed EFT scale estimated as $\Lambda/\sqrt{\sigma}$.}
\label{fig:eft:quad:exp}
\end{figure}

Figure~\ref{fig:eft:impacts:warsaw} shows the relative change in STXS cross-sections and branching ratios induced by representative values of EFT parameters in the Warsaw basis. In each case, the other parameters are set to zero. The content is similar to that of Figure~\ref{fig:eft:impacts:ev}, except that for the latter the parameters are defined in the measurement basis. Figure~\ref{fig:eft:corr} shows the observed and expected correlation matrices in the fit to the linear parameterisation. Figure~\ref{fig:eft:quad:exp} shows the observed and expected results obtained in the linear+quadratic parameterisation.

\section{Simplified likelihood for the EFT measurement}
\label{app:mvg}

A simplified description of the measurement of EFT parameters of Section~\ref{sec:eft} is obtained using a Gaussian approximation for the underlying measurement of the signal strengths $\mu_r^f$ in STXS bins. The $\mu_r^f$ measurement is described by a multivariate Gaussian distribution with central values given by the best-fit values shown in Figures~\ref{fig:stxs:br:1} and~\ref{fig:stxs:br:2} and a covariance matrix built from the symmetrised uncertainties shown in these figures and the correlation matrix shown in Figure~\ref{fig:stxs:br:corr} in Appendix~\ref{app:stxs_br}. The simplified likelihood is then obtained by expressing $\mu_r^f$ in terms of the EFT parameters $c_n$ using Eq.~\eqref{eq:eft:param_mu} and rotating to the measurement basis shown in Figure~\ref{fig:eft:evs}, in the same way as for the full likelihood used in Section~\ref{sec:eft}.

\begin{figure}[tbp]
\begin{center}
\includegraphics[width=0.9\columnwidth]{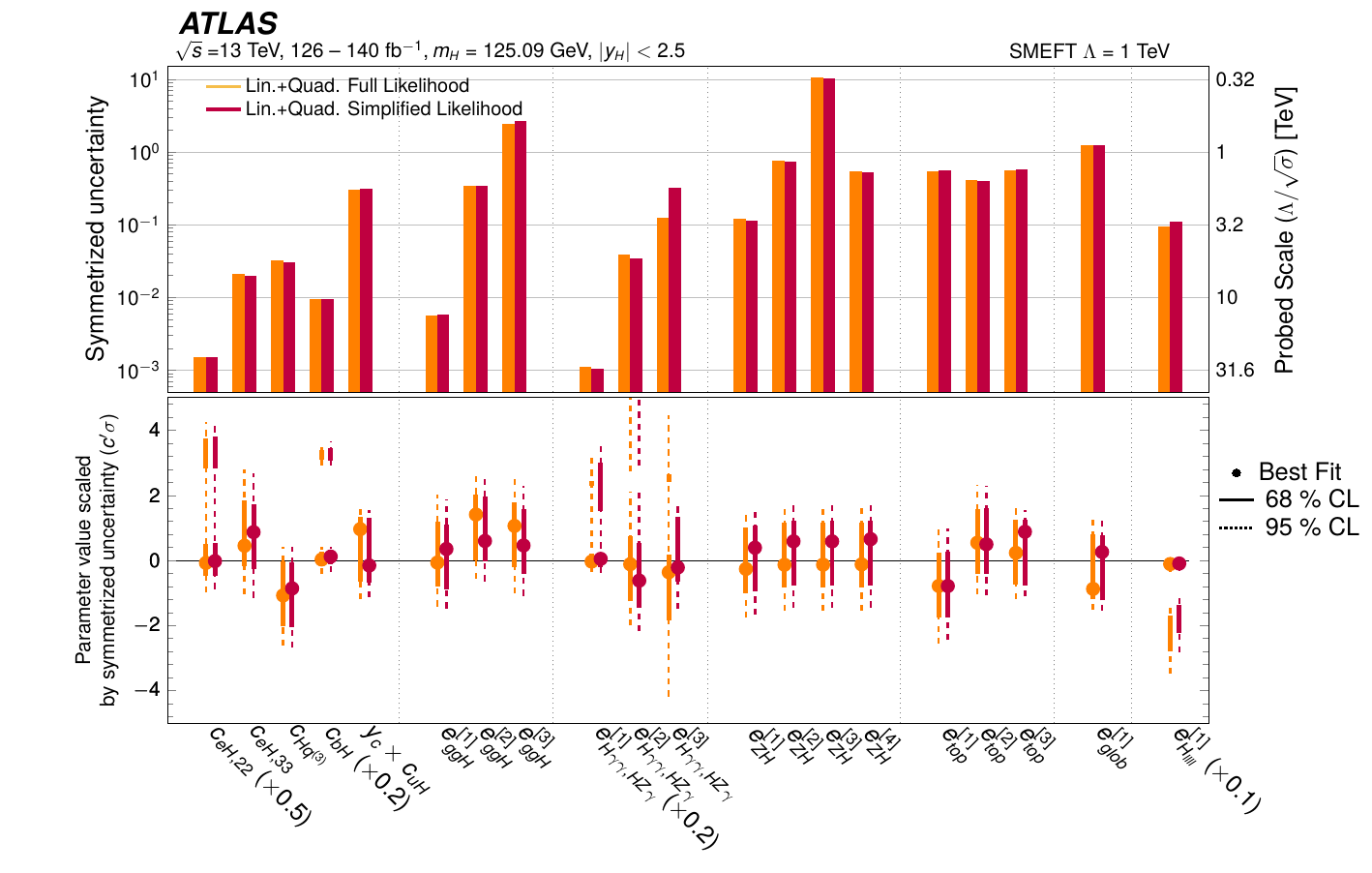}
\end{center}
\caption{Fitted values of the EFT measurement parameters obtained using a simplified description of the likelihood function and using the full likelihood models. In both cases the linear+quadratic parameterisation is used. The top panel shows the symmetrised uncertainty in each parameter, and the bottom panel the best-fit value and uncertainty of each parameter divided by its symmetrised uncertainty. Confidence intervals are shown for 68\% CL (solid lines) and 95\% CL (dashed lines). In the top panel, the left axis shows the uncertainty $\sigma$ on the parameter assuming $\Lambda = \qty{1}{\TeV}$ for the EFT scale, while the right axis shows the probed EFT scale estimated as $\Lambda/\sqrt{\sigma}$.}
\label{fig:eft:mvg}
\end{figure}

The Gaussian approximation is applied at the level of the measurement of the $\mu_r^f$ and not the EFT parameters themselves, so that non-Gaussian effects due to the linear nature of the parameterisation, in particular for the linear+quadratic model, are accounted for. Deviations from Gaussianity include the effect of small expected event yields in some of the measurement bins and that of systematic uncertainties. Observed results obtained in the linear+quadratic model using the simplified likelihood are shown in Figure~\ref{fig:eft:mvg} alongside the results from the full likelihood. Good agreement is observed in the best-fit values and uncertainties reported by the two models.
The simplified likelihood provides much quicker evaluation times, on the order of a few seconds compared to several hours for the full likelihood, which should facilitate its use for the reinterpretation of these results in the context of other models of BSM phenomena.

\FloatBarrier


\FloatBarrier

\clearpage
\printbibliography
\FloatBarrier

\clearpage
 
\begin{flushleft}
\hypersetup{urlcolor=black}
{\Large The ATLAS Collaboration}

\bigskip

\AtlasOrcid[0000-0002-6665-4934]{G.~Aad}$^\textrm{\scriptsize 102}$,
\AtlasOrcid[0000-0001-7616-1554]{E.~Aakvaag}$^\textrm{\scriptsize 17}$,
\AtlasOrcid[0000-0002-5888-2734]{B.~Abbott}$^\textrm{\scriptsize 121}$,
\AtlasOrcid[0000-0002-0287-5869]{S.~Abdelhameed}$^\textrm{\scriptsize 83b}$,
\AtlasOrcid[0000-0002-1002-1652]{K.~Abeling}$^\textrm{\scriptsize 54}$,
\AtlasOrcid[0000-0001-5763-2760]{N.J.~Abicht}$^\textrm{\scriptsize 48}$,
\AtlasOrcid[0000-0002-8496-9294]{S.H.~Abidi}$^\textrm{\scriptsize 30}$,
\AtlasOrcid[0009-0003-6578-220X]{M.~Aboelela}$^\textrm{\scriptsize 44}$,
\AtlasOrcid[0000-0002-9987-2292]{A.~Aboulhorma}$^\textrm{\scriptsize 36e}$,
\AtlasOrcid[0000-0001-5329-6640]{H.~Abramowicz}$^\textrm{\scriptsize 154}$
% ...
\bigskip
\\

% ...
$^{17}$Department for Physics and Technology, University of Bergen, Bergen; Norway.\\
% ...
$^{30}$Physics Department, Brookhaven National Laboratory, Upton NY; United States of America.\\
% ...
$^{36}$$^{(a)}$Facult\'e des Sciences Ain Chock, Universit\'e Hassan II de Casablanca;$^{(b)}$Facult\'{e} des Sciences, Universit\'{e} Ibn-Tofail, K\'{e}nitra;$^{(c)}$Facult\'e des Sciences Semlalia, Universit\'e Cadi Ayyad, LPHEA-Marrakech;$^{(d)}$LPMR, Facult\'e des Sciences, Universit\'e Mohamed Premier, Oujda;$^{(e)}$Facult\'e des sciences, Universit\'e Mohammed V, Rabat;$^{(f)}$Institute of Applied Physics, Mohammed VI Polytechnic University, Ben Guerir; Morocco.\\
% ...
$^{44}$Physics Department, Southern Methodist University, Dallas TX; United States of America.\\
% ...
$^{48}$Fakult\"{a}t Physik, Technische Universit{\"a}t Dortmund, Dortmund; Germany.\\
% ...
$^{54}$II. Physikalisches Institut, Georg-August-Universit\"{a}t G\"ottingen, G\"ottingen; Germany.\\
% ...
$^{83}$$^{(a)}$Khalifa University of Science and Technology, Abu Dhabi;$^{(b)}$New York University Abu Dhabi, Abu Dhabi;$^{(c)}$United Arab Emirates University, Al Ain;$^{(d)}$University of Sharjah, Sharjah; United Arab Emirates.\\
% ...
$^{102}$CPPM, Aix-Marseille Universit\'e, CNRS/IN2P3, Marseille; France.\\
% ...
$^{121}$Homer L. Dodge Department of Physics and Astronomy, University of Oklahoma, Norman OK; United States of America.\\
% ...
$^{154}$Raymond and Beverly Sackler School of Physics and Astronomy, Tel Aviv University, Tel Aviv; Israel.\\
% ...
$^{a}$ Also at Affiliated with an institute formerly covered by a cooperation agreement with CERN.\\
$^{b}$ Also at An-Najah National University, Nablus; Palestine.\\
$^{c}$ Also at Borough of Manhattan Community College, City University of New York, New York NY; United States of America.\\
$^{d}$ Also at Center for Interdisciplinary Research and Innovation (CIRI-AUTH), Thessaloniki; Greece.\\
$^{e}$ Also at Centre of Physics of the Universities of Minho and Porto (CF-UM-UP); Portugal.\\
$^{f}$ Also at CERN, Geneva; Switzerland.\\
% ...

\end{flushleft}


%

\FloatBarrier
\clearpage

\end{document}